\documentclass[12pt]{article}

\ifx\pdfoutput\undefined
\usepackage[dvips,bookmarks]{hyperref}
\else
\usepackage{hyperref}
\fi
\hypersetup{colorlinks=false,bookmarksopen,bookmarksnumbered,citecolor=blue,
   pdfstartview=FitH}

\usepackage{latexsym}
\usepackage{amssymb,amsfonts,amsmath}
\usepackage{graphicx} 
\usepackage{indentfirst}
\usepackage{bbm}
\usepackage{amssymb}
\usepackage{verbatim}
\usepackage{amsmath, amsthm,amssymb}
\usepackage{mathrsfs}
\usepackage{hyperref}
\usepackage{amsfonts}
\usepackage{dsfont}
\usepackage{slashed, tensor}
\usepackage{booktabs}
\usepackage{graphicx}
\usepackage{mathrsfs}

\parskip=\medskipamount

\newcommand {\cA}{{\cal A}}
\newcommand {\cB}{{\cal B}}
\newcommand {\cC}{{\cal C}}
\newcommand {\cD}{{\cal D}}
\newcommand {\cE}{{\cal E}}
\newcommand {\cF}{{\cal F}}
\newcommand {\cG}{{\cal G}}
\newcommand {\cH}{{\cal H}}
\newcommand {\cI}{{\cal I}}

\newcommand {\cK}{{\cal K}}
\newcommand {\cL}{{\cal L}}
\newcommand {\cM}{{\cal M}}
\newcommand {\cN}{{\cal N}}
\newcommand {\cO}{{\cal O}}

\newcommand {\cR}{{\cal R}}

\newcommand {\cT}{{\cal T}}
\newcommand {\cU}{{\cal U}}
\newcommand {\cV}{{\cal V}}
\newcommand {\cW}{{\cal W}}

\newcommand {\cY}{{\cal Y}}

\def\a{\alpha}
\def\b{\beta}
\def\c{\chi}
\def\d{\delta}

\def\g{\gamma}
\def\G{\Gamma}

\def\k{\kappa}
\def\l{\lambda}

\def\o{\omega}

\def\q{\theta}
\def\r{\rho}
\def\s{\sigma}
\def\t{\tau}

\def\x{\xi}
\def\z{\zeta}
\def\D{\Delta}
\def\F{\Phi}
\def\J{\Psi}
\def\L{\Lambda}
\def\O{\Omega}
\def\P{\Pi}

\def\S{\Sigma}
\def\U{\Upsilon}
\def\X{\Xi}

\def\tr{{\rm tr}}

\def\ri{{\rm i}}
\def\re{{\rm e}}

\newcommand{\ad}{{\dot{\alpha}}}                           
\newcommand{\bd}{{\dot{\beta}}}                            
\newcommand{\ve}{\varepsilon}                            

\newcommand{\pa}{\partial}                           
\newcommand{\hf}{\frac12}

\newcommand{\vf}{\varphi}

\newcommand{\be}{\begin{equation}}
\newcommand{\ee}{\end{equation}}
\newcommand{\bea}{\begin{eqnarray}}
\newcommand{\eea}{\end{eqnarray}}
\newcommand{\non}{\nonumber}
\newcommand{\ba}{\begin{array}}
\newcommand{\ea}{\end{array}}

\newcommand{\1}{{\underline{1}}}
\newcommand{\2}{{\underline{2}}}

\newcommand{\bm}[1]{\mbox{\boldmath$#1$}}

\def\double #1{#1{\hbox{\kern-2pt $#1$}}}

\newcommand{\hm}{{\hat{m}}}
\newcommand{\hn}{{\hat{n}}}
\newcommand{\hp}{{\hat{p}}}
\newcommand{\hq}{{\hat{q}}}
\newcommand{\ha}{{\hat{a}}}
\newcommand{\hb}{{\hat{b}}}
\newcommand{\hc}{{\hat{c}}}
\newcommand{\hd}{{\hat{d}}}
\newcommand{\he}{{\hat{e}}}
\newcommand{\hM}{{\hat{M}}}

\newcommand{\hA}{{\hat{A}}}
\newcommand{\hB}{{\hat{B}}}
\newcommand{\hC}{{\hat{C}}}
\newcommand{\hD}{{\hat{D}}}
\newcommand{\hE}{{\hat{E}}}

\newcommand{\hal}{{\hat{\a}}}
\newcommand{\hbe}{{\hat{\b}}}
\newcommand{\hga}{{\hat{\g}}}
\newcommand{\hde}{{\hat{\d}}}
\newcommand{\hrh}{{\hat{\rho}}}
\newcommand{\hta}{{\hat{\tau}}}

\newcommand{\de}{{\nabla}}
\newcommand{\deb}{{\bar{\nabla}}}

\newcommand{\bbD}{{\mathbb {D}}}

\newcommand{\bsubeq}{\begin{subequations}}
\newcommand{\esubeq}{\end{subequations}}

\newcommand{\ul}{\underline}
\newcommand{\eps}{{\ve}}

\newcommand{\eol}{\notag \\}
\newcommand{\rd}{\mathrm d}
\newcommand{\gD}{{\mathbb D}}

\newcommand{\veps}{\varepsilon}

\numberwithin{equation}{section}

\newcommand{\RM}{R(M)}
\newcommand{\RD}{R(\mathbb D)}

\newcommand{\RJ}{R(J)}
\newcommand{\RS}{R(S)}
\newcommand{\RK}{R(K)}
\newcommand{\sRM}{\mathscr{R}(M)}
\newcommand{\sRD}{\mathscr{R}(\mathbb D)}

\newcommand{\sRJ}{\mathscr{R}(J)}
\newcommand{\sRS}{\mathscr{R}(S)}
\newcommand{\sRK}{\mathscr{R}(K)}

\newcommand{\pz}{{(0)}}
\newcommand{\pu}{{(1)}}
\newcommand{\pmu}{{(-1)}}
\newcommand{\pd}{{(2)}}
\newcommand{\pmd}{{(-2)}}

\newcommand{\pn}{{(n)}}

\newcommand{\pq}{{(4)}}
\newcommand{\pmq}{{(-4)}}

\newcommand{\sfH}{{\frak{V}}}
\newcommand{\tfH}{\mathscr H}
\newcommand{\tfB}{\mathscr B}
\newcommand{\scT}{{\mathscr{T}}}

\begin{document}

\begin{titlepage}
\begin{flushright}
Nikhef-2014-046\\
October, 2014 \\
\end{flushright}
\vspace{5mm}

\begin{center}
{\Large \bf 
Conformal supergravity in five dimensions: \\
New approach and applications}
\\ 
\end{center}

\begin{center}

{\bf
Daniel Butter${}^{a}$, Sergei M. Kuzenko${}^{b}$, Joseph Novak${}^{b}$, \\
and
Gabriele Tartaglino-Mazzucchelli${}^{b}$
} \\
\vspace{5mm}

\footnotesize{
${}^{a}${\it Nikhef Theory Group \\
Science Park 105, 1098 XG Amsterdam, The Netherlands}}
~\\
\texttt{dbutter@nikhef.nl}\\
\vspace{2mm}

\footnotesize{
${}^{b}${\it School of Physics M013, The University of Western Australia\\
35 Stirling Highway, Crawley W.A. 6009, Australia}}  
~\\
\texttt{joseph.novak,\,gabriele.tartaglino-mazzucchelli@uwa.edu.au}\\
\vspace{2mm}

\end{center}

\begin{abstract}
\baselineskip=14pt
We develop a new off-shell formulation for five-dimensional (5D) conformal supergravity obtained by gauging  the 
5D superconformal algebra in superspace.
An important property of the conformal superspace introduced  is that it reduces 
to the superconformal tensor calculus (formulated in the early 2000's) 
upon 
gauging away a number of superfluous fields.
On the other hand, a different gauge fixing reduces our formulation
to the SU(2) superspace of arXiv:0802.3953, which 
is suitable to describe the most general off-shell supergravity-matter couplings. 
Using the conformal superspace approach, we show how to reproduce practically 
all off-shell constructions derived so far,
 including 
the supersymmetric extensions of $R^2$ terms, thus demonstrating the power of our formulation.
Furthermore, we construct for the first time a supersymmetric completion of the Ricci 
tensor squared term using the standard Weyl multiplet coupled to an off-shell 
vector multiplet. 
In addition,  we present several procedures to generate higher-order off-shell invariants in supergravity, including  
higher-derivative ones.
The covariant projective multiplets proposed in arXiv:0802.3953 
are lifted to conformal superspace, and a manifestly superconformal action principle
is given.  We also introduce unconstrained prepotentials for the vector
multiplet, the $\cO(2)$ multiplet (i.e., the linear multiplet without central charge)
and $\cO(4+n)$ multiplets, with $n=0,1,\dots$ Superform formulations are given 
for the BF action and the non-abelian Chern-Simons action. 
Finally, we describe locally supersymmetric theories with gauged central charge in 
conformal superspace. 
\end{abstract}

\vfill

\vfill
\end{titlepage}

\newpage
\renewcommand{\thefootnote}{\arabic{footnote}}
\setcounter{footnote}{0}

\tableofcontents



\allowdisplaybreaks

\section{Introduction}

Minimal supergravity in five spacetime dimensions\footnote{Historically, 
different authors use different notations, 
$\cN=1$ or $\cN=2$, for 
5D supersymmetric theories with eight supercharges.  
We choose to use $\cN=1$
following, e.g., \cite{GG1,KL,KT-M5D1}.} 
(5D) was introduced more than three decades ago by Cremmer  \cite{Cremmer}
and independently by Chamseddine and Nicolai \cite{CN}. 
A year later, an off-shell  formulation for this 
theory was sketched by Howe \cite{Howe5Dsugra} (building on the  supercurrent multiplet constructed by him and Lindstr\"om \cite{HL}),
who used superspace techniques and provided a 5D extension of the 
so-called $\cN=2$ minimal supergravity multiplet in four dimensions
\cite{BS}. Since then, 5D minimal supergravity 
and its matter couplings have extensively been studied
at the component level, both  in on-shell \cite{GST1,GST2,GZ,CD} 
and off-shell  \cite{Zucker1,Zucker2,Zucker3,Ohashi1,Ohashi2,Ohashi3,Ohashi4,Bergshoeff1,Bergshoeff2,Bergshoeff3} settings. 
The superspace approach to general off-shell 5D $\cN=1$ supergravity-matter systems 
has been developed in \cite{KT-M_5D2,KT-M_5D3,KT-M08}.\footnote{Refs.
 \cite{KT-M_5D2,KT-M_5D3} made use of Howe's minimal supergravity multiplet
  \cite{Howe5Dsugra}. Ref. \cite{KT-M08} developed a superspace formulation for 
  conformal supergravity, which in this paper will be referred to 
  as SU(2) superspace. In five dimensions, there is only one superconformal algebra,
 $\rm F^2(4)$ \cite{Nahm}, and it corresponds to the choice $\cN=1$. 
This is why one can simply speak of 5D conformal supergravity.}

 Off-shell formulations for supergravity make the supersymmetry transformation
 laws of fields 
 model-independent and, in principle, offer a tensor calculus to generate arbitrary 
 supergravity-matter couplings. 
A non-conformal tensor calculus for 5D $\cN=1$ supergravity 
was developed by Zucker \cite{Zucker1,Zucker2} 
(see also \cite{Zucker3} for a review and applications). By making 
use 
of Howe's minimal supergravity multiplet \cite{Howe5Dsugra} 
and the supercurrent multiplet \cite{HL}
(both carefully reduced to components), 
he 
extended to five dimensions various off-shell techniques developed for 
4D $\cN=2$ matter-coupled supergravity (see, e.g., \cite{FVP} for a review).
A more complete approach is the 5D 
superconformal tensor calculus  
developed independently  by two groups: 
Fujita, Kugo, and Ohashi\footnote{Actually Refs. \cite{Ohashi1,Ohashi2} presented the 5D tensor 
calculus in which some of the superconformal symmetries ($S$ and $K$) 
are gauge fixed.} 
\cite{Ohashi1,Ohashi2,Ohashi3,Ohashi4} 
and
Bergshoeff et al. \cite{Bergshoeff1,Bergshoeff2,Bergshoeff3}. 
Among the most interesting off-shell constructions obtained by applying 
the 5D 
superconformal calculus are
(i) the non-abelian Chern-Simons action coupled to conformal supergravity \cite{Ohashi2},
(ii) the massive tensor multiplet models \cite{Ohashi4},
and (iii) the supersymmetric completions of $R^2$ terms \cite{HOT,BRS,OP1,OP2}.

Within the component approaches of 
\cite{Zucker1,Zucker2,Zucker3,Ohashi1,Ohashi2,Ohashi3,Ohashi4,Bergshoeff1,Bergshoeff2,Bergshoeff3},
hypermultiplets are either on-shell or involve a gauged central charge. 
As is well known,
such hypermultiplet realizations cannot be used to provide an off-shell formulation 
for the most general locally supersymmetric sigma model. 
It is also known that such a sigma model formulation, if it exists,
requires the
use of 
off-shell hypermultiplets possessing an infinite number of auxiliary fields.
The latter feature of the off-shell hypermultiplets makes them extremely difficult
to work with at the component level.
This problem was solved
 within the superspace approach to 5D $\cN=1$ supergravity-matter systems  \cite{KT-M_5D2,KT-M_5D3,KT-M08} 
by putting forward
the novel concept of covariant projective multiplets.
These supermultiplets are a curved-superspace extension of  
the 4D $\cN=2$ and 5D $\cN=1$ superconformal projective multiplets \cite{K06,K07}. 
The latter reduce to the off-shell projective multiplets 
pioneered by Lindstr\"om and Ro\v{c}ek 
\cite{KLR,LR1,LR2} in the 4D $\cN=2$ super-Poincar\'e case
and  generalized to the cases of 
5D $\cN=1$ Poincar\'e and anti-de Sitter supersymmetries 
in   \cite{KL} and \cite{KT-M5D1}, respectively. 
Among the most interesting covariant projective multiplets
are polar ones that have infinitely many auxiliary fields
and indeed are suitable to realize the most general locally supersymmetric sigma model.
These have never appeared within the component settings of
\cite{Zucker1,Zucker2,Zucker3,Ohashi1,Ohashi2,Ohashi3,Ohashi4,Bergshoeff1,Bergshoeff2,Bergshoeff3}. 

This paper is devoted to new applications of the superspace approach to 5D $\cN=1$ 
matter-coupled supergravity \cite{KT-M_5D2,KT-M_5D3,KT-M08}.
In order to make a better transition to the superconformal 
calculus of \cite{Ohashi1,Ohashi2,Ohashi3,Ohashi4,Bergshoeff1,Bergshoeff2,Bergshoeff3}, 
we present an extension of the superspace formulation for 5D conformal 
supergravity given in  \cite{KT-M08}. Such an extension is based on the concept 
of conformal superspace \cite{Butter4DN=1,Butter4DN=2,BKNT-M1}. 

Conformal superspace is an off-shell formulation for conformal supergravity based on
gauging the superconformal algebra in superspace. It was originally developed 
for $\cN=1$  and $\cN=2$ supergravity theories in four dimensions
\cite{Butter4DN=1,Butter4DN=2} and more recently for $\cN$-extended conformal 
supergravity in three dimensions \cite{BKNT-M1}.\footnote{In the physics literature, 
the name ``conformal space'' has been used since the 1930s.
It was Dirac \cite{Dirac} who, following Veblen \cite{Veblen}, introduced it for 
the conformal compactification of 4D Minkowski space, on which the conformal group 
acts transitively.
Since the 1980s, the name ``conformal superspace'' has also been used 
for supersymmetric extensions of this construction 
\cite{Manin} (see also \cite{K06,KPT-MvU} for more recent presentations).
We hope no confusion may occur in our usage.}
For example, one may think of the 4D $\cN=1$ or $\cN=2$ conformal superspace as 
a superspace analogue of the corresponding superconformal multiplet calculus
developed many years earlier in the component setting, see e.g. \cite{FVP} 
for a pedagogical review, 
since both approaches are gauge theories of the superconformal group.
 From a technical point of view, conformal superspace is a more general setting, 
 since the gauge superfields contain more component fields and the gauge group is much larger than 
 in the superconformal calculus. However, it turns out that 
 the former formulation
 reduces to the 
 latter 
upon gauging away a number of superfluous component fields. 
On the other hand, a different gauge fixing allows one to reduce
conformal superspace to more traditional superspace settings.
For instance, in the 4D $\cN=2$ case 
a certain gauge fixing reduces 
the conformal superspace of \cite{Butter4DN=2}  
to the so-called U(2) superspace \cite{Howe-conf}, which has been used 
to construct the most general off-shell supergravity-matter couplings
\cite{KLRT-M_4D-2}. Thus conformal superspace provides a bridge between
the component superconformal calculus and more traditional superspace formulations
for conformal supergravity. 

Recent applications of the conformal superspace approach have involved
constructing
(i) the $\cN$-extended conformal supergravity 
actions in three dimensions for $3\leq \cN \leq 6$ \cite{BKNT-M2,KNT-M}, and 
(ii) new higher-derivative invariants in 4D $\cN=2$ supergravity, 
including the Gauss-Bonnet term \cite{BdeWKL}.
This paper is the first to explore applications of conformal superspace in five dimensions.
In particular, we will demonstrate that the formalism of conformal superspace 
provides new tools to construct various composite primary multiplets
that can be used to generate higher-order off-shell invariants in supergravity, including  higher-derivative ones.

This paper is organized as follows. Section \ref{setup} describes the geometry of conformal superspace in five dimensions. In particular, 
we present the procedure in which the superconformal algebra is gauged in superspace and show how to constrain the resulting 
geometry to describe conformal supergravity, thus deriving a new off-shell formulation. 
We also
describe the Yang-Mills multiplet in conformal superspace. In section \ref{degauging} 
we show how the superspace formulation for conformal 
supergravity proposed in  \cite{KT-M08} may be viewed as a gauge-fixed version of 
conformal superspace. Section \ref{WeylMultiplet} is devoted to uncovering the component structure of conformal superspace and comparing 
it to the existing superconformal tensor calculus \cite{Ohashi3,Ohashi4, Bergshoeff1,Bergshoeff2,Bergshoeff3}. In section \ref{section5} 
we lift the covariant projective multiplets of \cite{KT-M_5D2,KT-M_5D3,KT-M08}
to conformal superspace. 
A general procedure to generate such multiplets is given. 
We also present 
a universal locally supersymmetric action principle. 
Section \ref{N1SYM_and_PS} presents prepotential formulations for the 
vector multiplet in conformal superspace. In section \ref{section7} we develop a prepotential formulation for the $\cO(2)$ multiplet and discuss its universal 
role in generating actions. We also provide a prepotential formulation for $\cO(4+n)$ multiplets. Sections \ref{section8}, \ref{ACStheory} and \ref{section9} 
are devoted to superform formulations 
of the BF, abelian and non-abelian Chern-Simons actions, respectively. In section \ref{SGCC} we describe multiplets with gauged central charge in conformal superspace 
by giving their superform formulations. In particular, the linear multiplet with central charge, two-form multiplet and large tensor multiplet are discussed. Section \ref{section11} 
is devoted to a description of the dilaton Weyl multiplet and its variants with the use of superforms. In section \ref{section12} we present 
several procedures to generate higher-order off-shell invariants in supergravity, including higher derivative ones. Concluding comments are 
given in section \ref{conclusion}.

We have included a number of technical appendices. In Appendix \ref{NC} we include a summary of 
our notation and conventions. In Appendix \ref{KVF} we derive the superconformal algebra from the algebra 
of conformal Killing supervector fields of 5D $\cN = 1$ Minkowski superspace. 
In Appendix \ref{app:ModifiedSuperspace} we give an alternative covariant derivative algebra 
based on a new vector covariant derivative with a deformed $S$-supersymmetry transformation. Appendix \ref{app:Conventions} 
describes how our component field conventions relate to those of superconformal tensor calculus. In Appendix 
\ref{HarmonicG} we give the $\cO(2)$ multiplet prepotential formulation in harmonic superspace. Appendix \ref{gauge-invariance-G} 
discusses the gauge freedom for the $\cO(2)$ multiplet. Finally, in Appendix \ref{AppendixH} we derive prepotentials for the $\cO(4+n)$ multiplets 
in harmonic superspace.


\section{Conformal superspace in five dimensions} 
\label{setup}

Conformal superspace in four  \cite{Butter4DN=1,Butter4DN=2}
and three \cite{BKNT-M1} dimensions possesses
the following key properties:
(i) it gauges the entire superconformal algebra; 
(ii) the curvature and torsion tensors may be expressed in terms of a single
primary superfield; and 
(iii) the algebra obeys the same basic constraints as those of super Yang-Mills theory.
In this section we will show how these properties may be used to 
develop conformal superspace in five dimensions. We will present the superconformal algebra 
and the geometric setup for conformal superspace based on gauging the entire algebra. We then show 
how to constrain the geometry to describe superconformal gravity
by constraining its covariant derivative algebra to be expressed in terms of
a single primary superfield, the super Weyl tensor. 
We conclude the section by discussing an application and turning on a
Yang-Mills multiplet in the conformal superspace setting.


\subsection{The superconformal algebra}

The bosonic generators of the 5D superconformal algebra 
 $\rm F^2 (4)$ \cite{Nahm} 
include the translation ($P_{\hat{a}}$), Lorentz ($M_{\hat{a}\hat{b}}$), 
special conformal ($K_{\hat{a}}$), dilatation ($\mathbb{D}$) and $\rm SU(2)$ generators ($J_{ij}$), where 
$\hat{a}, \hat{b} = 0,1,2,3,5$ and $i , j = \underline{1} , \underline{2}$. 
Their algebra is
\bsubeq \label{SCA}
\begin{align} [M_{\ha \hb} , M_{\hc \hd}] &= 2 \eta_{\hc [\ha} M_{\hb] \hd} - 2 \eta_{\hat{d} [ \ha} M_{\hb ] \hc} \ , \\
[M_{\ha \hb}, P_\hc] &= 2 \eta_{\hc [\ha} P_{\hb]} \ , \quad [\mathbb{D} , P_\ha] = P_\ha \ , \\
[M_{\ha \hb} , K_\hc] &= 2 \eta_{\hc [\ha} K_{\hb]} \ , \quad [\mathbb{D} , K_\ha] = - K_{\ha} \ , \\
[K_\ha , P_\hb] &= 2 \eta_{\ha \hb} \gD + 2 M_{\ha\hb} \ , \\
[J^{ij} , J^{kl}] &= \eps^{k(i} J^{j) l} + \eps^{l (i} J^{j) k} \ ,
\end{align}
with all other commutators vanishing. 
The superconformal algebra is obtained by extending the 
translation generator to $P_\hA = (P_\ha , Q_\hal^i)$ and the special conformal generator to 
$K_\hA = (K_\ha , S_{\hal i})$,
where $Q_\hal^i$ and $S_{\hal}^{i}$ are an imaginary and a real
pseudo-Majorana spinor, respectively (see 
Appendix \ref{NC}).\footnote{Our convention for $S_\hal^i$ is chosen to match the 4D convention
\cite{Butter4DN=2} upon dimensional reduction. This means, for example,
that contractions between $K_\hA$ and the corresponding 
gauge parameters, connections, and curvatures must be interpreted
with care: for example, $\L^\hA K_\hA$ should be understood as
$\eta^{\hal i} S_{\hal i} + \L_K^\ha K_\ha$
with $\L^\hA = (\eta^{\hal i} , \L_K^\ha )$, while $\xi^\hA P_\hA = \xi^\hal_i Q_\hal^i + \xi^\ha P_\ha$
with $\xi^\hA=(\xi^\hal_i,\xi^\ha)$.}
The fermionic generator $Q_\hal^i$ obeys the algebra
\begin{align} \{ Q_\hal^i , Q_\hbe^j \} &= - 2 \ri \,\eps^{ij} (\G^{\hc})_{\hal\hbe} P_{\hc} \ , \quad [Q^i_\hal , P_\ha ] = 0 \ , \quad [\gD , Q_\hal^i ] = \hf Q_\hal^i \ , \\
[M_{\hal \hbe} , Q_\hga^i ] &= \eps_{\hga (\hal} Q_{\hbe)}^i \ , \quad [J^{ij} , Q_\hal^k ] = \eps^{k (i} Q_\hal^{j)} \ ,
\end{align}
while the generator $S_\hal^i$ obeys the algebra
\begin{align} \{ S_\hal^i , S_\hbe^j \} &= - 2 \ri \,\eps^{ij} (\G^{\hc})_{\hal\hbe} K_{\hc} \ , \quad [S_{\hal i} , K_\ha ] = 0 \ , \quad [\gD , S_{\hal i} ] = - \hf S_{\hal i} \ ,\\
[M_{\hal \hbe} , S_\hga^i ] &= \eps_{\hga (\hal} S_{\hbe)}^i \ , \quad [J^{ij} , S_\hal^k ] = \eps^{k (i} S_\hal^{j)} \ .
\end{align}
Finally, the (anti-)commutators of $K_{\hA}$ with $P_\hA$ are
\begin{align} {[} K_\ha , Q_\hal^i {]} &=  \ri (\G_\ha)_\hal{}^\hbe S_{\hbe}^i \ , \quad [S_{\hal i} , P_\ha ] = \ri (\G_\ha)_\hal{}^\hbe Q_{\hbe i} \ , \\
\{ S_{\hal i} , Q_\hbe^j \} &= 2 \eps_{\hal \hbe} \d_i^j \gD - 4 \d_i^j M_{\hal\hbe} + 6 \eps_{\hal\hbe} J_i{}^j \ .
\end{align}
\esubeq
One may explicitly check that the (anti-)commutation relations \eqref{SCA}
are consistent with the  Jacobi identities and thus define a superalgebra.  
A shorter way to convince oneself of the algebraic structure required
is to notice that  the (anti-)commutation relations \eqref{SCA}
follow from the algebra of conformal Killing supervector fields of 5D $\cN=1$ 
Minkowski superspace \cite{K06}, 
see Appendix \ref{KVF} for the technical details.


\subsection{Gauging the superconformal algebra} \label{Gauging}

To perform our gauging procedure, we begin with a curved 5D $\cN = 1$ superspace
 $\cM^{5|8}$ parametrized by
local bosonic $(x)$ and fermionic coordinates $(\theta_i)$:
\be z^{\hat{M}} = (x^{\hat{m}}, \ \q^{\hat{\mu}}_i) \ ,
\ee
where ${\hat{m}} = 0, 1,2,3, 5$, $\hat{\mu} = 1, \cdots,4$ and $i = \1, \2$. In order to describe supergravity it is necessary to 
introduce a vielbein and appropriate connections. However the gauging of the superconformal algebra is made  
non-trivial due to the fact that the graded commutator of
$K_{\hat{A}}$ with $P_{\hat{A}}$ contains generators other
than $P_{\hat{A}}$.
This requires some of the connections to transform under $K_{\hat{A}}$ into the vielbein.
To perform the gauging we will  follow closely the approach
given in \cite{Butter4DN=1, Butter4DN=2, BKNT-M1}.

We denote by $X_{\ul a}$ the closed subset of generators that do not
contain the $P_{\hat{A}}$ generators. The superconformal algebra takes the form of
a semidirect product algebra
\bsubeq
\begin{align}
[X_{\underline{a}} , X_{\underline{b}} \} &= -f_{\underline{a} \underline{b}}{}^{\underline{c}} X_{\underline{c}} \ , \\
[X_{\underline{a}} , P_{\hat{B}} \} &= -f_{\underline{a} {\hat{B}}}{}^{\underline{c}} X_{\underline{c}}
	- f_{\underline{a} {\hat{B}}}{}^{\hat{C}} P_{\hat{C}}
	\ , \\
[P_{\hat{A}} , P_{\hat{B}} \} &= -f_{{\hat{A}} {\hat{B}}}{}^{{\hat{C}}} P_{\hat{C}}
	\ ,
\end{align}
\esubeq
where $f_{{\hat{A}}{\hat{B}}}{}^{\hat{C}}$ contains only the constant torsion tensor
$f_{\hal}^i{}_\hbe^j{}^\hc = \scT_\hal^i{}_\hbe^j{}^\hc = 2 \ri \,\ve^{ij} (\G^\hc)_{\hal\hbe}$.
The gauge group associated with the superalgebra generated by $X_{\ul a}$ will be denoted $\cH$. 
Now we associate with each generator $X_{\underline{a}} = (M_{\ha\hb},J_{ij},\bbD, S_{\hal i}, K_\ha)$ a connection one-form 
$\omega^{\underline{a}} = (\O^{\ha\hb},\F^{ij},B,\frak{F}^{\hal i},\frak{F}^{\ha})= \rd z^\hM \omega_\hM{}^{\underline{a}}$ 
and with $P_{\hat{A}}$ the vielbein 
$E^{\hat{A}} = (E^\hal_i , E^\ha) = \rd z^{\hat{M}} E_\hM{}^\hA$. 
Their $\cH$-gauge transformations are postulated to be
\begin{subequations} \label{VGCTR}
\begin{align}
\d_{\cH} E^\hA &= E^\hB \L^{\underline{c}} f_{\underline{c} \hB}{}^\hA \ , \\
\d_{\cH} \omega^{\underline{a}} &= \rd \L^{\underline{a}} 
+ E^\hB \L^{\underline{c}} f_{\underline{c} \hB}{}^{\underline{a}} 
+ \omega^{\underline{b}} \L^{\underline{c}} f_{\underline{c} \underline{b}}{}^{\underline{a}} \ ,
\end{align}
\end{subequations}
with
$\L^{\underline{a}}$ the gauge parameters. 

A superfield $\Phi$ is said to be {\it covariant} 
if it transforms under $\cH$ with no derivatives on the parameter $\L^{\underline{a}}$
\be \d_{\cH} \Phi = \L \Phi := \L^{\underline{a}} X_{\underline{a}} \Phi \ .
\ee
A superfield $\Phi$ is said to be \emph{primary} if it is annihilated by the special
conformal generators, $K_\hA \Phi = 0$.
From the algebra \eqref{SCA}, we see that if a superfield is annihilated by $S$-supersymmetry,
then it is necessarily primary.

Given a covariant superfield $\Phi$, it is obvious that $\pa_\hM \Phi$ is not itself covariant. We are led to introduce
the {\it covariant derivative}
\be\label{eq:covD}
\nabla = \rd - \omega^{\underline{a}} X_{\underline{a}} \ , \quad \nabla = E^\hA \nabla_\hA \ .
\ee
Its transformation is found to be
\be
\d_{\cH} (\nabla_\hA \Phi) = (-1)^{\eps_\hA \eps_{\underline{b}}} \L^{\underline{b}} \nabla_\hA X_{\underline{b}} \Phi - \L^{\underline{b}} f_{\underline{b} \hA}{}^\hC \nabla_\hC \Phi 
- \L^{\underline{b}} f_{\underline{b} \hA}{}^{\underline{c}} X_{\underline{c}} \Phi \ ,
\ee
with no derivatives on the gauge parameter $\L^{\underline{a}}$. 
Rewriting this as 
$\d_{\cH} (\nabla_\hA \Phi) = \L^{\underline{b}} X_{\underline{b}} \nabla_\hA \Phi$, 
we immediately derive the operator relation
\be
[ X_{\underline{b}} , \nabla_\hA \} = -f_{\underline{b} \hA}{}^\hC \nabla_\hC
	- f_{\underline{b} \hA}{}^{\underline{c}} X_{\underline{c}} \ .
\ee
The torsion and curvature tensors appear in the commutator of two covariant derivatives,
\be
[ \nabla_\hA , \nabla_\hB \} = - \scT_{\hA\hB}{}^\hC \nabla_\hC - \mathscr{R}_{\hA\hB}{}^{\underline{c}} X_{\underline{c}} \ ,
\ee
where the torsion and curvature tensors
are defined, respectively, by
\begin{subequations} \label{TRexpComp}
\begin{align}
\scT^\hA &:= \hf E^\hC \wedge E^\hB \scT_{\hB\hC}{}^\hA = \rd E^\hA - E^\hC \wedge \omega^{\underline{b}} \,f_{\underline{b} \hC}{}^\hA \ , \\
\mathscr{R}^{\underline{a}} &:= \hf E^\hC \wedge E^\hB \mathscr{R}_{\hB\hC}{}^{\underline{a}} = \rd \omega^{\underline{a}}
	- E^\hC \wedge \omega^{\underline{b}} \, f_{\underline{b} \hC}{}^{\underline{a}}
	- \hf \omega^{\underline{c}} \wedge \omega^{\underline{b}} \,
		f_{\underline{b} \underline{c}}{}^{\underline{a}} \ .
\end{align}
\end{subequations}
Using the definition of curvature and torsion \eqref{TRexpComp} 
together with the vielbein and connection transformation rules \eqref{VGCTR}, we find
\begin{subequations}
\begin{align}
\d_\cH \scT^\hA &= \scT^\hC \L^{\underline{b}} f_{\underline{b} \hC}{}^\hA
	- E^\hC \wedge E^\hB \L^{\underline{a}} f_{\underline{a} \hB}{}^{\underline{f}} f_{\underline{f} \hC}{}^\hA \ , \\
\d_{\cH} \mathscr{R}^{\underline{a}} &=
	\mathscr{R}^{\underline{c}} \L^{\underline{b}} f_{\underline{b} \underline{c}}{}^{\underline{a}}
	+ \scT^\hC \L^{\underline{b}} f_{\underline{b} \hC}{}^{\underline{a}}
	- E^\hD \wedge E^\hC \L^{\underline{b}}
		f_{\underline{b} \hC}{}^{\underline{f}} f_{\underline{f} \hD}{}^{\underline{a}} \ ,
\end{align}
\end{subequations}
indicating that the torsion and curvature superfields are covariant.
Writing the transformation rules as 
$\d_\cH \scT^\hA = \L^{\underline{a}} X_{\underline{a}} \scT^\hA$, 
$\d_\cH \mathscr{R}^\hA  = \L^{\underline{a}} X_{\underline{a}} \mathscr{R}^\hA$ and 
$\d_\cH E^\hA = \L^{\underline{b}} X_{\underline{b}} E^\hA$ 
leads to the action of $X_{\underline{a}}$ on the torsion and curvature:
\begin{subequations}
\begin{align}
X_{\underline{a}} \scT_{\hB\hC}{}^\hD =&
- (-1)^{\eps_{\underline{a}} (\eps_{\hB}+\eps_{\hC})} 
\scT_{\hB\hC}{}^{\hat{E}} f_{\hat{E} \underline{a}}{}^\hD
- 2 f_{\underline{a} [\hB}{}^{\hat{E}} \scT_{|\hat{E}| \hC\}}{}^\hD
- 2 f_{\underline{a} [\hB}{}^{\underline{e}} f_{|\underline{e}| \hC\}}{}^\hD \ , \\
X_{\underline{a}} \mathscr{R}_{\hB\hC}{}^{\underline{d}} =&
- (-1)^{\eps_{\underline{a}} (\eps_{\hB}+\eps_{\hC})} 
\Big(\scT_{\hB\hC}{}^{\hat{E}} f_{\hat{E} \underline{a}}{}^{\underline{d}}
+ \mathscr{R}_{\hB\hC}{}^{\underline{e}} f_{\underline{e}\underline{a} }{}^{\underline{d}}\Big)
- 2 f_{\underline{a} [\hB}{}^{\hat{E}} \mathscr{R}_{|\hat{E}| \hC \}}{}^{\underline{d}} \non\\
&- 2 f_{\underline{a} [\hB}{}^{\underline{e}} f_{|\underline{e}| \hC \}}{}^{\underline{d}} \ .
\end{align}
\end{subequations}

One can show that the above results are the necessary conditions for the Jacobi identity involving two $\nabla$'s
\be 0 = [ X_{\underline{a}} , [ \nabla_\hB , \nabla_\hC \} \} +
	\text{(graded cyclic permutations)}
\ee
to be identically satisfied. The Bianchi identities
\be 0 = [ \nabla_\hA , [ \nabla_\hB , \nabla_\hC \} \} +
	\text{(graded cyclic permutations)}
\ee
can also be shown to be satisfied identically.
Therefore, we have a consistent algebraic structure
\begin{subequations}\label{2.20}
\begin{align}
[X_{\underline{a}} , X_{\underline{b}} \} &
= -f_{\underline{a} \underline{b}}{}^{\underline{c}} X_{\underline{c}} \ , \\
[X_{\underline{a}} , \nabla_\hB \} &= - f_{\underline{a} \hB}{}^\hC \nabla_\hC -f_{\underline{a} \hB}{}^{\underline{c}} X_{\underline{c}} \ , \label{eq:XwithNabla} \\
[\nabla_\hA , \nabla_\hB \} &= -\scT_{\hA\hB}{}^\hC \nabla_\hC - \mathscr{R}_{\hA\hB}{}^{\underline{c}} X_{\underline{c}} \ ,
\end{align}
\end{subequations}
which satisfies all the Jacobi identities.
In the flat space limit the curvature vanishes and the torsion becomes the usual constant torsion, 
so that the algebra \eqref{2.20} exactly matches the 
superconformal algebra that we started with, in which $P_\hA$ is replaced with $\nabla_\hA$. 
The curved case involves a so-called \emph{soft algebra}, where some of the
structure constants have been replaced by structure functions, corresponding
to the introduction of torsion and curvature.
The superconformal algebra is then said to be ``gauged'' in this sense.

The full set of operators $(\nabla_\hA, X_{\ul a})$ generates the conformal supergravity gauge group $\cG$.
The form of the covariant derivative suggests that we should extend the usual
diffeomorphisms $\delta_{\textrm{gct}}$ into \emph{covariant diffeomorphisms}
\begin{align}
\delta_{\rm cgct}(\xi^\hA) := \delta_{\textrm{gct}} (\xi^\hA E_\hA{}^\hM) - \delta_{\cH}(\xi^\hA \omega_\hA{}^{\ul a})~,
\end{align}
where $\delta_{\textrm{gct}}(\xi^\hM)$
acts on scalars under diffeomorphisms as
\begin{align} \delta_{\textrm{gct}} \Phi = \xi^\hM \partial_\hM \Phi \ .
\end{align}
The full conformal supergravity gauge group $\cG$ is then generated by
\begin{align}
\cK = \xi^\hC \nabla_\hC + \L^{\ul a} X_{\ul a}~.
\end{align}
If a superfield $\Phi$ is a scalar under diffeomorphisms and covariant under the group $\cH$,
then its transformation under the full supergravity gauge group $\cG$ is
\begin{align}
\d_\cG \Phi = \cK \Phi = \xi^\hC \nabla_\hC \Phi + \L^{\ul a} X_{\ul a} \Phi~.
\end{align}
It is a straightforward exercise to show that the vielbein and connection one-forms
transform as
\begin{subequations}\label{eq:deltaConn}
\begin{align}
\delta_\cG E^\hA &= \rd \xi^\hA + E^\hB \L^{\ul c} f_{\ul c \hB}{}^\hA
	+ \omega^{\ul b} \xi^{\hC} f_{\hC \ul b}{}^{\hA}
	+ E^\hB \xi^{\hC} \scT_{\hC \hB}{}^\hA~, \\
\delta_\cG \omega^{\ul a} &= \rd \L^{\ul a}
	+ \omega^{\ul b} \L^{\ul c} f_{\ul c \ul b}{}^{\ul a}
	+ \omega^{\ul b} \xi^{\hC} f_{\hC \ul b}{}^{\ul a}
	+ E^\hB \L^{\ul c} f_{\ul c \hB}{}^{\ul a}
	+ E^\hB \xi^{\hC} R_{\hC \hB}{}^{\ul a}~.
\end{align}
\end{subequations}
${}$From this definition, one can check that the covariant derivative transforms as
\begin{align}\label{TransCD}
\delta_\cG \nabla_\hA = [\cK,\nabla_\hA]
\end{align}
provided we interpret
\begin{subequations}\label{eq:nablaParams}
\begin{align}
\nabla_\hA \xi^\hB &:= E_\hA \xi^\hB + \omega_\hA{}^{\ul c} \xi^\hD f_{\hD \ul c}{}^\hB~, \\
\nabla_\hA \L^{\ul b} &:= E_\hA \L^{\ul b}
	+ \omega_\hA{}^{\ul c} \xi^\hD f_{\hD \ul c}{}^{\ul b}
	+ \omega_\hA{}^{\ul c} \L^{\ul d} f_{\ul d \ul c}{}^{\ul b}
	~.
\end{align}
\end{subequations}

We can summarize the superspace geometry of conformal supergravity
as follows. The covariant derivatives have the form
\be
\nabla_\hA = E_\hA - \o_\hA{}^{\underline b} X_{\underline b} 
= E_\hA - \hf \Omega_\hA{}^{\ha\hb} M_{\ha\hb} - \Phi_\hA{}^{kl} J_{kl} - B_\hA \mathbb D - \mathfrak{F}_\hA{}^\hB K_\hB \ .
\ee
The action of the generators on the covariant derivatives, eq.~\eqref{eq:XwithNabla},
resembles that for the $P_\hA$ generators given in \eqref{SCA}.
The supergravity gauge group is generated by local transformations of the form
\eqref{TransCD}
where
\bea
\cK &=& \xi^\hC \nabla_\hC + \hf \L^{\hc\hd} M_{\hc\hd} + \L^{kl} J_{kl} + \s \mathbb D 
+ \L^\hA K_\hA  
\eea
and the gauge parameters satisfy natural reality conditions. 
The covariant derivatives satisfy the (anti-)commutation relations
\begin{align}
[ \nabla_\hA , \nabla_\hB \}
	&= -\scT_{\hA\hB}{}^\hC \nabla_\hC
	- \frac{1}{2} \sRM_{\hA\hB}{}^{\hc\hd} M_{\hc\hd}
	- \sRJ_{\hA\hB}{}^{kl} J_{kl}
	\non \\ & \quad
	- \sRD_{\hA\hB} \mathbb D
	- \sRS_{\hA\hB}{}^{\hga k} S_{\hga k}
	- \sRK_{\hA\hB}{}^\hc K_\hc~,
\end{align}
where the torsion and curvature tensors are given by
\begin{subequations} \label{torCurExp}
\bea
\scT^\ha &=& \rd E^\ha + E^\hb \wedge \Omega_\hb{}^\ha + E^\ha \wedge B \ , \\
\scT{}^\hal_i &=& \rd E^\hal_i + 2 E^\hbe_i \wedge \Omega_\hbe{}^\hal + \hf E^\hal_i \wedge B - E^{\hal j} \wedge \Phi_{ji} - \ri \, E^\hc \wedge \mathfrak{F}^\hbe_i (\G_\hc)_\hbe{}^\hal \ ,~~~~~~~~~~~ \\
\sRD &=& \rd B + 2 E^\ha \wedge \mathfrak{F}_\ha - 2 E^\hal_i \wedge \mathfrak{F}_\hal^i \ , \\
\sRM^{\ha\hb} &=& \rd \Omega^{\ha\hb} + \Omega^{\ha\hc} \wedge \Omega_\hc{}^\hb - 4 E^{[\ha} \wedge \mathfrak{F}^{\hb]} - 4 E^\hal_j \wedge \mathfrak{F}^{\hbe j} (\S^{\ha\hb})_{\hal\hbe} \ , \\
\sRJ^{ij} &=& \rd \Phi^{ij} - \Phi^{k (i} \wedge \Phi^{j)}{}_k + 6 E^{\hal (i} \wedge \mathfrak{F}_{\hal}^{j)} \ , \\
\sRK^\ha &=& \rd \mathfrak{F}^\ha + \mathfrak{F}^\hb \wedge \Omega_\hb{}^\ha - \mathfrak{F}^\ha \wedge B - \ri \mathfrak{F}^{\hal k} \wedge \mathfrak{F}_{\hbe k} (\G^\ha)_\hal{}^\hbe \ , \\
\sRS^{\hal i}  &=& \rd \mathfrak{F}^{\hal i} + 2 \mathfrak{F}^{\hbe i} \wedge \Omega_\hbe{}^\hal - \hf \mathfrak{F}^{\hal i} \wedge B - \mathfrak{F}^{\hal j} \wedge \Phi_j{}^i - \ri E^{\hbe i} \wedge \mathfrak{F}^\hc (\G_\hc)_\hbe{}^\hal     \ .
\eea
\end{subequations}


\subsection{Conformal supergravity}

In the conformal superspace approach to supergravity 
in four \cite{Butter4DN=1, Butter4DN=2} and three \cite{BKNT-M1} dimensions, 
the entire covariant derivative algebra may be expressed in terms of a single primary superfield: the super Weyl tensor for
$D = 4$ 
and the super Cotton tensor for $D=3$.
We will seek a similar solution in $D=5$ in terms of a single primary superfield,
the  super Weyl tensor $W_{\hal\hbe} = W_{\hbe\hal}$ \cite{KT-M08}.

In the three- and four-dimensional cases the second ingredient to describe conformal supergravity 
was to realize that 
the right constraints for the covariant derivative  were such that
their algebra obeyed the same constraints as super Yang-Mills theory.
Guided by the structure of 5D $\cN=1$ super Yang-Mills theory \cite{KL,HL,Zupnik99}, we impose the constraint 
$\{ \nabla_\hal^{(i} , \nabla_\hbe^{j)} \} = 0$, which is equivalent to
the spinor derivative anti-commutation relation
\bsubeq\label{algebra-W}
\begin{align} \{ \nabla_\hal^i , \nabla_\hbe^j \} &= - 2 \ri \eps^{ij} (\G^\hc)_{\hal\hbe} \nabla_\hc - 2 \ri \eps^{ij} \eps_{\hal \hbe} 
\mathscr{W} \ ,
\label{algebra-W-1}
\end{align}
where $\mathscr{W}$ is some operator taking values in the superconformal algebra.
The Bianchi identities give the other commutators
\begin{align}
[\nabla_\ha , \nabla_\hbe^j ] &= (\G_\ha)_\hbe{}^\hga [\nabla_\hga^j , \mathscr{W}] \ , 
\label{algebra-W-2}\\
[\nabla_\ha , \nabla_\hb] &= - \mathscr{F}_{\ha \hb} 
= 
\frac{\ri}{4} (\S_{\ha\hb})^{\hal\hbe} \{ \nabla_\hal^k , [\nabla_{\hbe k} ,\mathscr{W}] \}
\label{algebra-W-3}
\end{align}
\esubeq
and the additional constraint
\be \{ \nabla_\hal^{(i} , [ \nabla_\hbe^{j)} ,\mathscr{W}] \} = \frac{1}{4} \eps_{\hal \hbe} \{ \nabla^{\g (i} , [\nabla_\g^{j)} ,\mathscr{W} ] \}  \ . \label{WalbeBI}
\ee

In analogy to conformal superspace in four dimensions \cite{Butter4DN=1, Butter4DN=2}, we constrain the form of the operator $\mathscr{W}$ to be
\be 
\mathscr{W} = W^{\hal \hbe} M_{\hal \hbe} + W(S)^{\hal i} S_{\hal i} + W(K)^\hb K_\hb \ ,
\ee
where $W_{\hal\hbe}$ is a symmetric dimension-1 primary superfield.
One can show that the Bianchi identity \eqref{WalbeBI} is identically satisfied for
\be 
\mathscr{W}= W^{\hal\hbe} M_{\hal\hbe} 
- \frac{1}{10} (\nabla_\hbe^i W^{\hal \hbe}) S_{\hal i} 
- \frac{1}{4} (\nabla^\ha W_{\ha\hb}) K^\hb \ ,
\label{def-W}
\ee
provided $W_{\hal\hbe}$ satisfies
\be \nabla_\hga^k W_{\hal\hbe} = \nabla_{(\hal}^k W_{\hbe \hga )} + \frac{2}{5} \eps_{\hga (\hal} \nabla^{\hde k} W_{\hbe ) \hde} \ . \label{WBI}
\ee

It is convenient to introduce higher dimension descendant superfields constructed 
from spinor derivatives of $W_{\hal \hbe}$. At dimension-3/2, we introduce
\bsubeq \label{2.37}
\begin{gather}
W_{\hal \hbe \hga}{}^k := \nabla_{(\hal}^k W_{\hbe \hga )} \ , \quad X_\hal^i := \frac{2}{5} \nabla^{\hbe i} W_{\hbe\hal} \ ,
\end{gather}
and at dimension-2, we choose
\begin{gather}
W_{\hal \hbe \hga \hde} := \nabla_{(\hal}^k W_{\hbe \hga \hde) k} \ , \quad 
X_{\hal \hbe}{}^{ij} := \nabla_{(\hal}^{(i} X_{\hbe)}^{j)} = - \frac{1}{4} \nabla^{\hga (i} \nabla_\hga^{j)} W_{\hal\hbe} \ , \\
Y := \ri \nabla^{\hga k} X_{\hga k} \ .
\end{gather}
\esubeq
One can check that only these superfields and their vector derivatives
appear upon taking successive spinor derivatives of $W_{\hal \hbe}$.
Specific relations we will need later are given below:
\bsubeq \label{eq:Wdervs}
\bea
\nabla_{\hga}^k W_{\hal\hbe} 
&=& W_{\hal\hbe\hga}{}^k + \eps_{\hga (\hal} X_{\hbe)}^k \ , 
\\
\nabla_\hal^i X_\hbe^j
&=&
 X_{\hal \hbe}{}^{ij} 
 + \frac{\ri}{8} \eps^{ij}\Big(
  \eps_{\hal\hbe} Y
+ 4 \eps^{\ha\hb\hc\hd\he} (\S_{\ha\hb})_{\hal\hbe} \nabla_\hc W_{\hd\he} 
- 4 (\G^\hb)_{\hal\hbe} \nabla^\ha W_{\ha \hb}
\Big) \ , ~~~~~~~~~
\\
\nabla_\hal^i W_{\hbe \hga \hde}{}^j
&=&
 - \hf \eps^{ij} \Big(
 W_{\hal \hbe \hga \hde} +3 \ri \nabla_{\hal (\hbe} W_{\hga\hde)} 
- \frac{3 \ri}{4} \eps^{\ha\hb\hc\hd\he} \eps_{\hal (\hbe} (\S_{\ha\hb})_{\hga\hde )} \nabla_\hc W_{\hd\he}
\Big) 
\non\\
&&
- \frac{3}{2} \eps_{\hal (\hbe} X_{\hga \hde)}{}^{ij} 
\ , \\
\nabla_\hal^i W_{\hbe\hga\hde\hrh} 
&=&
- 4 \ri \nabla_{\hal (\hbe} W_{\hga \hde \hrh )}{}^i 
-12 \ri \eps_{\hal (\hbe} \Big(
\nabla_\hga{}^\hta W_{\hde \hrh ) \hta}{}^i 
+ W_{\hga \hde } X_{\hrh)}^i - 2 W_{\hga}{}^\hta W_{\hde \hrh) \hta}{}^i
\Big) \ ,~~~~~~~~~
 \\
\nabla_\hal^i X_{\hbe \hga}{}^{jk} 
&=&
 \eps^{i(j} \Big( 2 \ri \nabla_\hal{}^\hde W_{\hbe\hga\hde}{}^{k)} + 2 \ri \nabla_{(\hbe}{}^\hde W_{\hga ) \hal \hde}{}^{k)} 
- \ri \nabla_{\hal (\hbe} X_{\hga )}^{k)} - \ri \eps_{\hal (\hbe} \nabla_{\hga)}{}^\hde X_{\hde}^{k)} 
\non\\
&&
~~~~~~
+ 6 \ri W_{(\hal \hbe} X_{\hga)}^{k)} - 12 \ri  W_{(\hal}{}^{\hde} W_{\hbe \hga ) \hde}{}^{k)} \Big) \ ,
~~~~~~~ \\
\nabla_{\hal}^i Y 
&=& 8 \nabla_\hal{}^\hga X_{\hga}^i 
\ .
\eea
\esubeq

These descendant superfields transform under $S$-supersymmetry as
\begin{align}
S_{\hal i} W_{\hbe\hga\hde}{}^j &= 6 \d^j_i \eps_{\hal (\hbe} W_{\hga \hde)} \ , \qquad
S_{\hal i} X_\hbe^j = 4 \d_i^j W_{\hal\hbe}~, \eol
S_{\hal i} W_{\hbe\hga\hde\hrh} &= 24 \eps_{\hal (\hbe} W_{\hga\hde \hrh)}{}_i \ , \qquad
S_{\hal i} Y = 8 \ri X_{\hal i} ~, \eol
S_{\hal i} X_{\hbe\hga}{}^{jk} &= - 4 \d_i^{(j} W_{\hal\hbe\hga}{}^{k)} + 4 \d_i^{(j} \eps_{\hal (\hbe} X_{\hga)}^{k)} \ .
\end{align}

In terms of these superfields, we can now construct the algebra of
covariant derivatives for 5D conformal supergravity:
\bsubeq \label{ACDer}
\bea
\{ \nabla_\hal^i , \nabla_\hbe^j \} &= &
- 2 \ri \eps^{ij} (\G^\hc)_{\hal\hbe} \nabla_\hc 
- 2 \ri \eps^{ij} \eps_{\hal \hbe} W^{\hga\hde} M_{\hga\hde} 
- \frac{\ri}{2} \eps^{ij} \eps_{\hal \hbe} X^{\hga k} S_{\hga k} 
\non\\
&&+ \frac{\ri}{2} \eps^{ij} \eps_{\hal \hbe} (\nabla^\ha W_{\ha\hb}) K^\hb \ , \\
{[}\nabla_\ha , \nabla_\hbe^j {]} &=& (\G_\ha)_\hbe{}^\hga \Big( W_{\hga\hde} \nabla^{\hde j} + \hf X_\hga^j \gD 
+ W_{\hga\hde\hrh}{}^j M^{\hde\hrh} + \frac{3}{2} X_\hga^k J_k{}^j \non\\
&&
- \frac{1}{4} (\nabla_\hga^j X^\hde_k) S_\hde^k 
+ \frac{\ri}{4} (\G^\hc)_\hga{}^\hde (\nabla^{\hb} W_{\hb\hc}) S_\hde^j 
- \frac{1}{4}( \nabla_\hga^j \nabla^\hc W_{\hc \hb} )K^\hb \Big) \ , \\
{[}\nabla_\ha , \nabla_\hb{]} &=& - \scT_{\ha\hb}{}^\hc \de_\hc - \scT_{\ha\hb}{}^\hga_k \de_\hga^k
- \hf \sRM_{\ha\hb}{}^{\hc\hd} M_{\hc\hd} - \sRJ_{\ha\hb}{}^{ij} J_{ij} - \sRD_{\ha\hb} \gD 
\non\\
&& - \sRS_{\ha\hb}{}^{\hga k} S_{\hga k} - \sRK_{\ha\hb}{}^\hc K_\hc \ ,
\label{eq:Curv3}
\eea
\esubeq
where
{\allowdisplaybreaks
\bsubeq
\begin{align}
\scT_{\ha\hb}{}^\hc &= - \hf \eps_{\ha\hb}{}^{\hc\hd\he} W_{\hd\he} \ , \\
\scT_{\ha\hb}{}^\hga_k &= - \frac{\ri}{2} \nabla^\hga_k W_{\ha\hb}
	= - \frac{\ri}{2} (\S_{\ha\hb})_{\hal\hbe} W^{\hal\hbe\hga}{}_k
	- \frac{\ri}{2} (\S_{\ha\hb})^{\hbe \hga} X_{\hbe k} \ ,\label{eq:Tors2}  \\
\sRM_{\ha\hb}{}^{\hc\hd} &= - \frac{\ri}{4} (\S_{\ha\hb})^{\hal\hbe} (\S^{\hc\hd})^{\hga\hde}
		\nabla_\hal^k \nabla_{\hbe k} W_{\hga\hde} 
	+ \frac{\ri}{10} \nabla^{\hga k} \nabla^\hde_k W_{\hga\hde} \d_{[\ha}^\hc \d^\hd_{\hb]}
	\non\\ &\quad
	+ 2 \nabla_{\ha'} W_{\hb'\hc'} \eps^{\ha'\hb'\hc' [\hc}{}_{[\ha} \d_{\hb]}^{\hd]}
	- W_{\ha\hb} W^{\hc\hd} \non\\
	&= - \frac{\ri}{4} (\S_{\ha\hb})^{\hal\hbe} (\S^{\hc\hd})^{\hga\hde} W_{\hal\hbe\hga\hde}
	+ \frac{1}{8} Y \d_{[\ha}^\hc \d_{\hb]}^\hd
	- W_{\ha\hb} W^{\hc\hd}
	+ \hf \nabla_{\ha'} W_{\hb'\hc'} \eps^{\ha'\hb'\hc' [\hc}{}_{[\ha} \d_{\hb]}^{\hd]}
	\non\\&\quad
	+ \frac{1}{4} \eps_{\ha\hb}{}^{\hc\hd {\hat{e}}} \nabla^{\hat{f}} W_{\hat{f} \hat{e}}
	+ \frac{1}{4} \eps^{\hc\hd \hat{e} \hat{f}}{}_{[\ha} \nabla_{\hb ]} W_{\hat{e} \hat{f}}
	- \hf \nabla^{\hat{e}} W^{\hat{f} [\hat{c}} \eps^{\hat{d}]}{}_{\ha\hb\hat{e}\hat{f}}\ , \\
\sRJ_{\ha\hb}{}^{kl}
	&= - \frac{3 \ri}{4} (\S_{\ha\hb})^{\hal\hbe} X_{\hal\hbe}{}^{kl}
	= \frac{3 \ri}{16} (\S_{\ha\hb})^{\hal\hbe} \nabla^{\hga (k} \nabla_{\hga}^{l)} W_{\hal\hbe} \ , \\
\sRD_{\ha\hb} &= - \frac{\ri}{4} (\S_{\ha\hb})^{\hal\hbe} \nabla_\hal^k X_{\hbe k} = - \hf \eps_{\ha\hb\hc\hd\he} \nabla^\hc W^{\hd\he} \ ,\\
\sRS_{\ha\hb}{}^{\hga k} &= - \frac{\ri}{16} (\S_{\ha\hb})^{\hal\hbe} \Big( \nabla_\hal^j \nabla_{\hbe j} X^{\hga k} 
+ 2 \ri (\G^{\hc})_\hal{}^\hga \nabla^\hd \nabla_\hbe^k W_{\hd\hc} 
- 4 \ri W_{\hal\hbe} X^{\hga k} \Big) \non\\
&= - \frac{1}{16} \eps^{\hga \hrh} (\S_{\ha\hb})^{\hal\hbe} \Big( 2 \nabla_\hal{}^\hde W_{\hbe \hde \hrh}{}^k + 6 \nabla_\hrh{}^\hde W_{\hal\hbe \hde}{}^k
+ \eps_{\hrh \hal} \nabla_{\hbe}{}^\hde X_{\hde}^k + 3 \nabla_{\hrh\hal} X_{\hbe}^k \non\\
&\quad- 12 W_{(\hrh}{}^\hde W_{\hal\hbe) \hde}{}^k + 6 W_{(\hal\hbe} X_{\hrh )}^k  \Big) - \frac{1}{4} W_{\ha\hb} X^{\hga k} \ ,\\
\sRK_{\ha\hb}{}^{\hc} &=
	\frac{\ri}{16} (\S_{\ha \hb})^{\hal\hbe}
		(\nabla^k_{(\hal} \nabla_{\hbe) k} \nabla_\hd W^{\hd \hc}
		- 4 \ri W_{\hal\hbe} \nabla_\hd W^{\hd \hc}) \non\\
	&=  \frac{\ri}{16} (\S_{\ha \hb})^{\hal\hbe}
	\Big(\nabla_\hd \nabla^k_{(\hal} \nabla_{\hbe) k} W^{\hd \hc}
	+ \frac{\ri}{8} \eps^{\hc\he\hat{f}\hat{g}\hat{h}} (\S_{\hat{e} \hat{f}})_{\hal\hbe} Y W_{\hat{g} \hat{h}}
	\non\\&\quad
	- (\G^\hc)_\hde{}^\hrh W^{\hga\hde} \nabla_\hga^k \nabla_{\hal k} W_{\hbe \hrh} 
	+ (\G^\hc)_\hal{}^\hrh W^{\hga\hde} \nabla_\hga^k \nabla_{\hbe k} W_{\hde \hrh} 
	\non\\&\quad
	+ 2 \ri W_{\hal\hbe}\nabla_\he W^{\he\hc} 
	 - \ri (\S^{\he\hat{f}})_{\hal\hbe} W_{\hat{f}}{}^\hc\nabla^{\hat{g}} W_{\hat{g} \hat{e}} 
	+ 3 \ri (\S^{\hc\he})_{\hal\hbe}  W_{\hat{e}}{}^{\hat{g}}\nabla^{\hat{f}} W_{\hat{f} \hat{g}}
	\non\\&\quad
	- 6 \ri (\S^{\he \hat{f}})_{\hal\hbe}W^{\hat{g} \hc} \nabla_{[\he} W_{\hat{f} \hat{g}]} 
	- 3 (\G^\hc)^{\hga \hde} X_{\hga}^k W_{\hal\hbe\hde k} - 3 (\G^\hc)_\hal{}^\hde X^{\hga k} W_{\hbe\hga\hde k}
	\non\\&\quad
	- 2 (\G^\hc)^{\hde \hrh} W_{\hde (\hal}{}^{\hga k} W_{\hbe) \hrh \hga k}
	- 2 (\G^\hc)_\hal{}^\hrh W_\hbe{}^{\hga\hde k} W_{\hga \hde \hrh k}
	- 4 \ri W_{\hal\hbe} \nabla_\hd W^{\hd \hc}
	\Big) \ .
\end{align}
\esubeq}Despite
possessing a larger structure group, the covariant derivative algebra is more compact than that of SU(2) superspace 
\cite{KT-M08}. 
This provides a significant advantage in performing superspace calculations.


\subsection{Full superspace actions}
Given the geometry we have described, it is immediately apparent that one
may construct an action principle involving a full superspace integral
\begin{align}
S[\cL] = \int\rd^{5|8}z\,  E\, \cL~,\qquad\rd^{5|8}z:=\rd^5x\,\rd^8\q~, \qquad
E := {\rm Ber}(E_\hM{}^\hA)~,
\end{align}
where $\cL$ is a primary superspace Lagrangian of dimension $+1$.

For later applications, it will be important to know the rule
for integrating by parts in full superspace. It is given by
\begin{align}
\int \rd^{5|8}z\, E \ (-1)^{\eps_\hA} \nabla_\hA V^\hA 
= 
\int \rd^{5|8}z\, E \ 
\Big\{
&
-(-1)^{\eps_\hA}\Big( \frak{F}_\hA{}^\hb K_\hb V^\hA
+ \frak{F}_\hA{}^{\hbe k} S_{\hbe k} V^\hA\Big)
 \non\\
&
+\ri \frak{F}_{\hga k}{}^{\hbe k} V^\ha (\G_\ha)_\hbe{}^\hga
\Big\} \ ,
\end{align}
where $V^\hA$ transforms as a Lorentz and SU(2) tensor with
$\bbD V^\ha =  0$ and $\bbD V^\hal_i = \hf V^\hal_i$.

In the special case where $V^\hA$ corresponds to an $S$-invariant vector field
$V = V^\hA E_\hA = V^\hA E_\hA{}^M \partial_\hM$, which requires
\be S_\hbe^i V^\hal_j = - \ri \d^i_j V^\ha (\G_\ha)_\hbe{}^\hal \ , \quad S_\hbe^j V^\ha = 0 \ ,
\ee
we have the simple integration rule
\be
\int \rd^{5|8}z\, E \ (-1)^{\eps_\hA} \nabla_\hA V^\hA  = 0 \ .
\ee

\subsection{Gravitational composite $\cO(2)$ multiplet} 

As an application of the formalism introduced, we will construct a
composite superfield that may be used to generate a supersymmetric completion of an
$R^2$ term. This composite superfield is constructed in terms of the super Weyl
tensor as follows:
\begin{align}
H^{ij}_{\rm Weyl} : = - \frac{\ri}{2} W^{\hal\hbe\hga\,i} W_{\hal\hbe\hga}{}^{j} 
	+ \frac{3\ri}{2} \,W^{\hal\hbe} X_{\hal \hbe}{}^{ij}
	- \frac{3\ri}{4} \, X^{\hal i} X_\hal^{j}
	= H^{ji}_{\rm Weyl}~, 
	\label{2.43}
\end{align}
where we have used the definitions \eqref{2.37}. 
This superfield is real in the sense that 
$\overline{H^{ij}_{\rm Weyl}}=\ve_{ik} \ve_{jl} H^{kl}_{\rm Weyl}$.
One can check that $H^{ij}_{\rm Weyl}$  is primary and obeys the constraint
\bea
\nabla_\hal^{(i} H^{jk)}_{\rm Weyl} = 0~.
\eea
It corresponds exactly to the composite
multiplet $L^{ij}[\mathbf W^2]$ constructed by
Hanaki, Ohashi, and Tachikawa \cite{HOT}.

This is an example of a covariant real $\cO(2)$ multiplet, which will be introduced
in section \ref{section5}. The structure of \eqref{2.43} is completely analogous 
to that of the composite $\cO(2)$ multiplet associated with the Yang-Mills multiplet
given in \cite{KL}, see the next subsection. 
The supersymmetric $R^2$-invariant of \cite{HOT} may be constructed straightforwardly in
superspace using \eqref{2.43} and the BF action.


\subsection{Turning on the Yang-Mills multiplet}
\label{SYM}
 
Let us conclude this subsection by presenting a Yang-Mills  multiplet
in conformal superspace. To describe such a non-abelian vector multiplet, 
the covariant derivative $\nabla = E^\hA \nabla_\hA $ has to be replaced with 
a gauge covariant one, 
\bea
\bm \nabla = E^\hA \bm \nabla_\hA \ , \quad {\bm\nabla}_\hA := \nabla_\hA 
- \ri \bm V_\hA
~.
\label{SYM-derivatives}
\eea
Here the  gauge connection one-form $\bm V = E^\hA \bm V_\hA$ 
takes its values in the Lie algebra 
of the Yang-Mills gauge group, $G_{\rm YM}$, with its (Hermitian) generators 
commuting with all the generators of the superconformal algebra. 
The gauge covariant derivative algebra is
\bea
[{\bm \nabla}_\hA, {\bm \nabla}_\hB\} 
&=&
 -\scT_{\hA\hB}{}^\hC{\bm \nabla}_\hC
-\hf \sRM_{\hA\hB}{}^{\hc\hd} M_{\hc\hd}
-\sRJ_{\hA\hB}{}^{kl} J_{kl}
- \sRD_{\hA\hB} \mathbb D 
\non\\
&&
 - \sRS_{\hA\hB}{}^{\hga k}S_{\hga k}
	- \sRK_{\hA\hB}{}^\hc K_\hc
	- \ri \bm F_{\hA\hB} \ ,
\eea
where the torsion and curvatures are those of conformal superspace 
but with $\bm F_{\hA\hB}$ corresponding 
to the gauge covariant field strength two-form 
$\bm F = \hf E^\hB \wedge E^\hA \bm F_{\hA\hB}$.
The field strength $\bm F_{\hA\hB}$ 
satisfies the Bianchi identity
\be \bm \nabla \bm F = 0  \quad \Longleftrightarrow \quad
\bm \nabla_{[\hA} \bm F_{\hB\hC\}} 
+ \scT_{[\hA\hB}{}^\hD \bm F_{|\hD| \hC\}} = 0
~.
\ee
The Yang-Mills gauge transformation acts on the gauge covariant 
derivatives $\bm \nabla_\hA$ and a matter  superfield $U$ (transforming 
in some representation of the gauge group) 
as
\be 
\bm \nabla_\hA ~\rightarrow~ \re^{\ri  \bm  \t} \bm \nabla_\hA \re^{- \ri \bm \t } , 
\qquad  U~\rightarrow~ U' = \re^{\ri  \bm  \t} U~, 
\qquad \bm \t^\dag = \bm \t \ ,
\label{2.2}
\ee
where the Hermitian gauge parameter ${\bm \t} (z)$ takes its values in the Lie algebra 
of $G_{\rm YM}$. 
This implies that the gauge one-form and the field strength transform as follows:
\bea 
\bm V ~\rightarrow ~\re^{\ri \bm \t} \bm V \re^{-\ri \bm \t} 
+ \ri \,\re^{\ri \bm \t} \rd \, \re^{- \ri \bm \t} \ , \qquad 
\bm F ~\rightarrow ~ \re^{\ri \bm \t} \bm F \re^{- \ri \bm \t} \ .
\eea

As in the flat case \cite{HL} (see also \cite{Zupnik99,KL}), 
some components of the field strength have to be constrained 
in order to describe an irreducible multiplet. 
In conformal superspace 
the right
constraint is
\bsubeq \label{YMsuperformF}
\bea 
\bm F_{\hat{\a}}^i{}_{\hat{\b}}^j 
&=& 
2 \ri \eps^{ij} \eps_{\hat{\a} \hat{\b}} \bm W \ ,
\eea
which fixes the remaining components of the field strengths to be
\bea
\bm F_{\hat{a}}{}_{\hat{\b}}^j 
&=& 
- (\G_{\hat{a}})_{\hat{\b}}{}^{\hat{\g}} \bm\nabla_{\hat{\g}}^j \bm W \ , 
\\
\bm F_{\hat{a}\hat{b}} 
&=& 
- \frac{\ri}{4} (\S_{\hat{a}\hat{b}})^{\hat{\a} \hat{\b}} 
\big(\bm\nabla^k_{(\hat{\a}} \bm\nabla_{\hat{\b}) k} - 4 \ri  W_{\hal\hbe}\big) \bm W 
\ ,
\eea
\esubeq
where the superfield $\bm W$ is Hermitian, 
$\bm W^\dag = \bm W$, and obeys the Bianchi identity
\bea 
\bm\nabla_{\hat{\a}}^{(i} \bm\nabla_{\hat{\b}}^{j)} \bm W 
= \frac{1}{4} \eps_{\hat{\a} \hat{\b}} \bm\nabla^{\hat{\g} (i} \bm\nabla_{\hat{\g}}^{j)} 
\bm W \ .
\label{vector-Bianchi}
\eea
Moreover, $\bm W$ is a conformal primary of dimension 1,
 $S_{\hal}^i \bm W=0$ and
 $\bbD \bm W=\bm W$.

Now let $T_I$ be the Hermitian generators of the gauge group $G_{\rm YM}$.
The gauge connection $\bm V_{\hA}$ and the field strengths 
$\bm F_{\hA\hB}$ and $\bm W$ can be decomposed as 
$\bm V_{\hA}= V_{\hA}{}^I T_I  $, 
$\bm F_{\hA\hB} = F_{\hA\hB}{}^I T_I$  and $\bm W = W^I T_I$. 
For a single abelian vector multiplet, we will use  
$V_{\hA} $, $ F_{\hA\hB}$  and $W$. 

It is helpful to introduce the following descendant superfields constructed from spinor derivatives of $\bm W$:
\begin{align}
\bm \l_\hal^i := - \ri \bm\nabla_\hal^i \bm W \ , \qquad
\bm X^{ij} := \frac{\ri}{4} \bm\nabla^{\hal (i} \bm\nabla_\hal^{j)} \bm W 
= - \frac{1}{4} \bm\nabla^{\hal (i} \bm \l_\hal^{j)} \ .
\end{align}
The above superfields together with
\be 
\bm F_{\hal\hbe} = - \frac{\ri}{4} \bm\nabla^k_{(\hal} \bm\nabla_{\hbe) k} \bm W 
- W_{\hal \hbe} \bm W 
= \frac{1}{4} \bm\nabla_{(\hal}^k \bm \l_{\hbe) k} - W_{\hal \hbe} \bm W
\ee
satisfy the following useful identities:
\bsubeq \label{VMIdentities}
\begin{align}
\bm\nabla_\hal^i \bm \l_\hbe^j 
&= - 2 \eps^{ij} \big(\bm F_{\hal \hbe} + W_{\hal\hbe} \bm W\big) 
- \eps_{\hal\hbe} \bm X^{ij} - \eps^{ij} \bm\nabla_{\hal\hbe} \bm W \ , \\
\bm\nabla_\hal^i \bm F_{\hbe\hga} 
&=
 - \ri \bm\nabla_{\hal (\hbe} \bm \l_{\hga)}^i 
 - \ri \eps_{\hal (\hbe} \bm\nabla_{\hga )}{}^\hde \bm \l_\hde^i 
- \ri W_{\hbe\hga} \bm \l_\hal^i - W_{\hal\hbe \hga}{}^i \bm W 
- \eps_{\hal (\hbe} X_{\hga)}^i \bm W \ , \\
\bm\nabla_\hal^i \bm X^{jk} &= 2 \ri \eps^{i (j} 
\Big(
\bm\nabla_\hal{}^\hbe \bm \l_\hbe^{k)} 
+ W_{\hal\hbe} \bm \l^{\hbe k)} 
- \frac{\ri}{2} X_\hal^{k)} \bm W
- \ri [\bm W , \bm \l_\hal^{k)}] 
\Big) 
\ .
\end{align}
\esubeq
The $S$-supersymmetry generator acts on these descendants as
\begin{align}
S_\hal^i \bm \l_\hbe^j &= - 2 \ri \eps_{\hal\hbe} \eps^{ij} \bm W \ , \qquad
S_\hal^i \bm F_{\hbe\hga} = 4 \eps_{\hal (\hbe} \bm \l_{\hga)}^i \ , \qquad
S_\hal^i \bm X^{jk} = - 2 \eps^{i (j} \bm \l_\hal^{k)} \ .
\end{align}

Now consider a primary composite superfield $\bm H^{ij}_{\rm YM}$ 
that is quadratic in the generators of the gauge group and is defined by
\bea  \label{YML}
\bm H^{ij}_{\rm YM} &=& 
\ri (\bm \nabla^{\hat \a (i} \bm W ) \, \bm\nabla^{j)}_{\hat \a} \bm W 
+ \frac{\ri}{4} \Big\{ \bm W \,, 
 \bm\nabla^{\hal (i } \bm\nabla^{j)}_\hal \bm W \Big\} \non\\
 &=& \{ \bm W , \bm X^{ij}\} - \ri \bm \l^{\hal (i} \bm \l_\hal^{j)} \ .
\eea
Its important property is
\bea
\bm\nabla^{(i}_{\hat \a} \bm H^{jk)}_{\rm YM} =0~.
\eea
In the rigid superspace limit, $\bm H^{ij}_{\rm YM}$ reduces to the composite superfield introduced 
in \cite{KL}. Associated with $\bm H^{ij}_{\rm YM}$ is the  gauge singlet
$H^{ij}_{\rm YM}:= \tr\, \bm H^{ij}_{\rm YM}$, which is a primary superfield constrained by 
$\nabla^{(i}_{\hat \a} H^{jk)}_{\rm YM} =0$.
This  is an example of a covariant  $\cO(2)$ multiplet defined in section
\ref{section5}.


\section{From conformal to SU(2) superspace}
\label{degauging}

The superspace structure we have presented in the previous section involves, as 
in four and three dimensions \cite{Butter4DN=1, Butter4DN=2, BKNT-M1},
the gauging of the entire superconformal algebra in order to describe conformal supergravity.
Traditionally, however, conformal supergravity has been described in superspace in a different
manner: local component scale and special 
conformal transformations were encoded in
super Weyl transformations.
This was exactly the approach taken previously
in \cite{KT-M08} where 5D conformal supergravity was described by
gauging $\rm SO(4,1) \times \rm SU(2)$, corresponding to the Lorentz and
$R$-symmetry groups, with additional super Weyl transformations realized non-linearly.
As in the introduction, we refer to the latter formulation of conformal supergravity 
as SU(2) superspace.

The relation between these two approaches mirrors the simpler non-supersymmetric
situation. Conformal gravity may be described as the gauge theory of the conformal
algebra, with a vielbein, Lorentz, dilatation, and special conformal connection.
Certain constraints are usually imposed so that the only independent fields are the
vielbein and dilatation connection. A special conformal transformation can be made
to eliminate the dilatation connection; upon making such a choice, one keeps the vielbein and
Lorentz connections in the covariant derivative, while discarding the special conformal
connection -- this is often called ``degauging'' the special conformal symmetry.
The dilatation symmetry survives as the usual Weyl symmetry of the vielbein, and
one recovers a formulation of conformal gravity with a vielbein alone.

As alluded to in the introduction, it is possible to ``degauge'' conformal superspace
to recover $\rm SU(2)$ superspace
in a similar way.
This is the goal of this section. The procedure
follows exactly the path laid out in the four and three dimensional
cases \cite{Butter4DN=1, Butter4DN=2, BKNT-M1}.
In particular, we will show explicitly how to recover the connections and curvatures
of SU(2) superspace and derive the form of the super Weyl transformations. 
The material in this section provides the necessary ingredients to relate results in 
conformal superspace to those of SU(2) superspace.


\subsection{Degauging to SU(2) superspace}

Let us recall that SU(2) superspace is described by a superspace vielbein,
Lorentz connection, and $\rm SU(2)_R$ connection. Conformal superspace possesses
in addition dilatation and special conformal connections;
these must be dealt with in a particular way.
The first step is to eliminate the dilatation connection. Because
the one-form $B = E^\ha B_\ha + E^{\hal}_i B_{\hal}^i$ transforms as
\begin{align}
\d_{K}(\L) B
	= -2 E{}^\ha \L_\ha - 2 E{}^\hal_i \L_\hal^i \ , \label{GFsuperWeyl}
\end{align}
under special conformal transformations, it is straightforward to impose
the gauge choice
\be B_\hA = 0 \ , \label{GCond}
\ee 
eliminating the dilatation connection entirely.
The special conformal connection $\mathfrak{F}^\hA$ remains, but its
corresponding gauge symmetry has been fixed, so we will extract it from the covariant
derivative. The resulting \emph{degauged} covariant derivatives are given by
\be
\cD_\hA := \nabla_\hA + \mathfrak{F}_\hA{}^\hB K_\hB
	= E_\hA - \frac{1}{2} \Omega_\hA{}^{\hb\hc} M_{\hb\hc} - \Phi_\hA{}^{ij} J_{ij} \ ,
\ee
and possess an $\rm SO(4, 1) \times SU(2)$ structure group. They satisfy
(anti-)commutation relations of the form\footnote{We distinguish the degauged versions of 
the torsion and curvatures with a tilde.}
\be
[\cD_\hA , \cD_\hB \} = -\widetilde{\scT}_{\hA\hB}{}^\hC \cD_\hC 
- \hf \widetilde{\mathscr{R}}_{\hA\hB}{}^{\hc\hd} M_{\hc\hd}
-\widetilde{\mathscr{R}}_{\hA\hB}{}^{kl} J_{kl} 
	~.
\ee
Because the vielbein, Lorentz, and SU(2) connections are exactly those
of conformal superspace, it is easy to give expressions for the new torsion and curvature
tensors in terms of the conformal ones using \eqref{torCurExp}.
For example, one finds for the torsion tensor,
\be\label{eq:degaugedTorsion}
\widetilde{\scT}^\ha = \scT^\ha \ , \quad
\widetilde{\scT}{}^\hal_i = \scT{}^\hal_i + \ri E^\hc \wedge \mathfrak{F}^\hbe_i (\G_\hc)_\hbe{}^\hal \ .
\ee
The special 
conformal connections $\mathfrak{F}_\hA{}^\hB$ provide new
contributions to the superfield torsion and similarly to the other curvatures.

It turns out there is actually a subtlety in this degauging procedure.
A careful examination of \eqref{eq:degaugedTorsion} shows that one recovers \emph{almost} all the same
constraints on the torsion tensor as in SU(2) superspace, except that
\be  
\widetilde{\scT}_\ha{}_{\hbe (j}{}^\hbe{}_{k)} \neq 0 
\ , \quad 
\widetilde{\scT}_{\ha \hb}{}^\hc \neq 0 
\ . 
\label{unNatTorsionComps}
\ee
In SU(2) superspace, both of these combinations are required to vanish.
The solution to this is that there is some freedom to redefine the vector components
of the Lorentz and SU(2) connections when we degauge, corresponding to a redefinition
of the vector covariant derivative of SU(2) superspace. This in turn modifies the torsion and curvature tensors.
A particular choice sets to zero the combinations \eqref{unNatTorsionComps}
and exactly reproduces the torsion and curvature tensors of SU(2) superspace.
To elaborate further, we must analyze explicitly the additional superfields introduced
by the special conformal connections $\mathfrak{F}_\hA{}^\hB$.


\subsection{The degauged special conformal connection}

In the gauge \eqref{GCond} the dilatation curvature is given by\footnote{We have lowered the index on the $K$-connection as 
$\mathfrak{F}_{\hA \hb} = \eta_{\hb\hc} \mathfrak{F}_\hA{}^\hc$ 
and 
$\mathfrak{F}_{\hA}{}_\hbe^j = \eps_{\hbe\hga} \mathfrak{F}_\hA{}^{\hga j}$.}
\be \sRD_{\hA\hB} = 2 \mathfrak{F}_{\hA\hB} (-1)^{\eps_\hB}
 - 2 \mathfrak{F}_{\hB\hA} (-1)^{\eps_\hA + \eps_\hA \eps_\hB} \ .
\ee
The vanishing of the dilatation curvature at dimension-1 constrains the special conformal connection as\footnote{The reason for 
introducing these superfields via these coefficients 
will be clear later.}
\be \mathfrak{F}_\hal^i{}_\hbe^j = - \mathfrak{F}_\hbe^j{}_\hal^i = \frac{\ri}{2} \eps_{\hal \hbe} S^{ij}  - \frac{\ri}{4} C_{\hal\hbe}{}^{ij} + \ri \eps^{ij} Y_{\hal \hbe}  \ ,
\label{dim-1-frakF}
\ee
where the superfields $S^{ij}$, $C_{\hal\hbe}{}^{ij}$, $Y_{\hal \hbe}$ 
satisfy the symmetry properties
\be S^{ij} = S^{ji} 
\ , \quad 
C_{\hal \hbe}{}^{ij} = (\G^\ha)_{\hal\hbe} C_\ha{}^{ij}
=C_{\hal \hbe}{}^{ji}
 \ , 
\quad 
Y_{\hal \hbe} = Y_{\hbe \hal} 
\ .
\ee
From here it is possible to derive the degauged covariant derivative algebra 
by computing $[ \cD_\hA , \cD_\hB \}$. An efficient way to do this is to consider 
a primary superfield $\Phi$ transforming as a tensor in some representation 
of the remainder of the superconformal algebra (compare with \cite{Butter4DN=2}). For example, 
to determine the anti-commutator of spinor derivatives we consider
\be
\{ \cD_\hal^i , \cD_\hbe^j \} \Phi = \{ \nabla_\hal^i , \nabla_\hbe^j \} \Phi
	+ \mathfrak{F}_\hal^i{}^\hC [K_\hC , \nabla_\hbe^j \} \Phi
	+ \mathfrak{F}_\hbe^j{}^\hC [K_\hC , \nabla_\hal^i \} \Phi \ .
\ee
Making use of the form of $\mathfrak{F}$ and of the superconformal algebra we find
\bea
\{ \cD_\hal^i , \cD_\hbe^j \} 
&= & 
- 2 \ri \eps^{ij} \cD'_{\hal \hbe} 
+ 3 \ri \eps_{\hal \hbe} \eps^{ij} S^{kl} J_{kl} 
- \ri \eps^{ij} C_{\hal \hbe}{}^{kl} J_{kl} 
- 12 \ri Y_{\hal \hbe}  J^{ij}
\non\\
&&
- \ri \eps_{\hal \hbe} \eps^{ij} \big(W^{\hc\hd}+Y^{\hc\hd}\big)M_{\hc\hd} 
+ \frac{\ri}{4} \eps^{ij} \eps^{\ha \hb \hc \hd \he} (\G_\ha)_{\hal\hbe} 
\big(2Y_{\hb\hc}-W_{\hb\hc}\big) M_{\hd\he} 
\non\\ 
&&
- \frac{\ri}{2} \eps^{\ha \hb \hc \hd \he} (\S_{\ha\hb})_{\hal\hbe} C_\hc{}^{ij} M_{\hd\he} + 4 \ri S^{ij} M_{\hal \hbe} 
 \ ,
\eea
where we have defined the vector covariant derivative
\be \cD'_\ha := \cD_\ha + \frac{1}{4} C_{\ha}{}^{kl} J_{kl} - \frac{1}{8} \eps_{\ha\hb\hc\hd\he} W^{\hb\hc} M^{\hd\he} 
~.
\label{cD'}
\ee
The remaining algebra of covariant derivatives 
can be similarly computed directly from degauging.
It can be seen that the 
algebra of $\cD'_\hA = (\cD'_\ha, \cD_\hal^i)$ exactly matches the one of SU(2) superspace \cite{KT-M08}
once we identify the dimension-1 torsion components 
$X_{\ha\hb}$ and $N_{\ha\hb}$ used in \cite{KT-M08} as
\be 
X_{\ha\hb} := W_{\ha\hb} + Y_{\ha\hb} 
\ , \quad 
N_{\ha\hb} := 2 Y_{\ha\hb} - W_{\ha\hb}
 \ .
 \label{X_N}
\ee
The superfields $S^{ij}$ and $C_{\ha}{}^{ij}$, 
which we introduced in \eqref{dim-1-frakF}, are equivalent to the ones used in \cite{KT-M08}.
 In particular, it turns out that the 
covariant derivative algebra for $\cD'_\hA $
 does not possess the torsion components \eqref{unNatTorsionComps}.

The curvature superfields can be shown to satisfy the dimension-3/2 identities:
\begin{subequations}
{\allowdisplaybreaks{\bea
\cD_\hga^k W_{\ha\hb}
&=&
W_{\ha\hb\hga}{}^k
+(\S_{\ha\hb})_\hga{}^\hde X_{\hde}^k
~, 
\\
\cD_\hga^k Y_{\ha\hb}
&=&
2(\G_{{[}\ha})_{\hga}{}^{\hde}\cY_{\hb{]}\hde}{}^k
+(\S_{\ha\hb})_\hga{}^\hde \cY_{\hde}^k~, 
\\
\cD_\hga^kC_{\ha}{}^{ij}&=&
-{1\over 2}(\G_\ha)_\hga{}^{\hde}\cC_\hde{}^{ijk}
-{2\over 3}\Big(
\cC_{\ha}{}_{\hga}^{(i}
-\frac{1}{2}(\G_\ha)_\hga{}^{\hde}\cC_\hde^{(i}
\Big)\ve^{j)k}~,
\\
\cD_\hga^kS^{ij}&=&
-{1\over 4}\cC_{\hga}{}^{ijk}
+\Big(
X_\hga^{(i}
+\frac{5}{2}\cY_\hga^{(i}
+\frac{5}{12}\cC_\hga^{(i}
\Big)\ve^{j)k}
~,~~~~~~~~~
\eea
}}
\end{subequations}
where
\bea
(\G^\ha)_\hal{}^\hbe W_{\ha\hb\hbe}{}^i=0
~,~~~
(\G^\ha)_\hal{}^\hbe\cY_{\ha\hbe}{}^i=0
~,~~~
(\G^\ha)_\hal{}^\hbe\cC_{\ha\hbe}{}^i=0
~,~~~
\cC_\hal{}^{ijk}
=\cC_\hal{}^{(ijk)}
~.~~~~~~
\eea
Note that the dimension-3/2 torsion is 
\bea
\widetilde{\scT}_{\ha\hb}{}_\hga^k&=&
{\ri\over 2}\cD_\hga^k W_{\ha\hb}
+{\ri\over 2}\cD_\hga^k Y_{\ha\hb}
-{\ri\over 6}(\G_{[\ha})_\hga{}^\hde\cC_{\hb]}{}_\hde^{k}
+{\ri\over 4}(\S_{\ha\hb})_\hga{}^\hde\cC_\hde^{k}~.
\label{dim-3/2-torsion}
\eea

To degauge results in conformal superspace it is useful to also have the remaining special conformal connection components 
$\frak{F}_\ha{}^{\hbe j}$ and $\frak{F}_{\ha\hb}$. They are constrained by 
the dilatation curvature as follows:\footnote{Here we raise and lower the indices on the special conformal connection using $\eps^{ij}$, 
$\eps_{\hal\hbe}$ and $\eta_{\ha\hb}$ in the usual way.}
\bsubeq
\begin{align}
- \frac{1}{5} (\G_\ha)_\hbe{}^\hga \cD^{\hde j} W_{\hga \hde} &= - 2 \frak{F}_\ha{}_\hbe^j - 2 \frak{F}_\hbe^j{}_\ha \ , \label{frakFconst1} \\
- \hf \eps_{\ha\hb\hc\hd\he} \cD^\hc W^{\hd \he} &= 4 \frak{F}_{[\ha\hb]} \label{frakFconst2} \ .
\end{align}
\esubeq

The explicit expressions for $\frak{F}_\ha{}_\hbe^j$ and $\frak{F}_{\ha\hb}$ may be found by analyzing the special conformal
curvatures
\begin{subequations} \label{Kcurvs}
\begin{align}
\sRS_{\hA\hB}{}^{\hga k} =& \ 2 \cD_{[\hA} \mathfrak{F}_{\hB \}}{}^{\hga k}
	+ \widetilde{\scT}_{\hA\hB}{}^\hD \mathfrak{F}_\hD{}^{\hga k}
	+ \ri \d_\hA{}^{\hde k} \mathfrak{F}_\hB{}^\hc (\G_\hc)_\hde{}^\hga (-1)^{\eps_\hB} \non\\
	&- \ri \hde_\hB{}^{\hde k} \mathfrak{F}_\hA{}^\hc (\G_\hc)_\hde{}^\hga (-1)^{\eps_\hB \eps_\hA+\eps_\hA} \ , \label{R(K)1} \\
\sRK_{\hA\hB}{}^\hc =& \ 2 \cD_{[\hA} \mathfrak{F}_{\hB \}}{}^\hc + \widetilde{\scT}_{\hA\hB}{}^\hD \mathfrak{F}_\hD{}^\hc
	+ \ri \mathfrak{F}_\hA{}^\hga_k \mathfrak{F}_\hB{}^{\hde k} (\G^\hc)_{\hga \hde} (-1)^{\eps_\hB} \non\\
	&
- \ri \mathfrak{F}_\hB{}^\hga_k \mathfrak{F}_\hA{}^{\hde k} (\G^\hc)_{\hga\hde} (-1)^{\eps_\hB \eps_\hA+\eps_\hA} 
\ , 
\label{R(K)2}
\end{align}
\end{subequations}
which appear in the  algebra of the conformal covariant derivatives $\nabla_\hA$. The component special conformal 
connections are given by:
\bsubeq
\begin{align} \frak{F}_\ha{}_\hbe^j &= - \frac{1}{10} (\G_\ha)_\hbe{}^\hga \cD_{\hga k} S^{jk} - \frac{1}{18} (\S_{\ha\hb})_\hbe{}^\hga \cD_{\hga k} C^{\hb jk}
- \frac{1}{24} \eps_{\ha\hb\hc\hd\he} (\S^{\hd\he})_\hbe{}^\hga \cD_\hga^j Y^{\hb\hc} \non\\
&\quad- \frac{1}{12} (\G^\hb)_\hbe{}^\hga \cD_\hga^j Y_{\ha\hb} 
+ \frac{1}{18} \cD_{\hbe k} C_\ha{}^{jk}
+ \frac{1}{30} (\G_\ha)_\hbe{}^\hga \cD^{\hde j} W_{\hga\hde}
    \non\\
&=    \frac{1}{2} \cY_\ha{}_\hbe^j
-\frac{1}{12} \cC_\ha{}_\hbe^j 
- \frac{1}{16} (\G_\ha)_\hbe{}^\hga \cC_\hga^j 
- \frac{1}{8} (\G_\ha)_\hbe{}^\hga \cY_\hga^j 
~, \\
\frak{F}_\hbe^j{}_\ha &= - \frak{F}_\ha{}_\hbe^j + \frac{1}{10} (\G_\ha)_\hbe{}^\hga \cD^{\hde j} W_{\hga \hde} =
- \frak{F}_\ha{}_\hbe^j 
 + \frac{1}{4} (\G_\ha)_\hbe{}^\hga X_\hga^j \ , \\
  \frak{F}_{\ha\hb} 
 &=
\frac{\ri}{288} \eta_{\ha\hb} [\cD^{\hal i} , \cD_\hal^j] S_{ij} 
+ \frac{\ri}{576} \eta_{\ha\hb} [\cD^\hal_i , \cD^\hbe_j] C_{\hal\hbe}{}^{ij} 
- \frac{\ri}{128} (\G_{(\ha})^{\hal\hbe} [\cD_\hal^i , \cD_\hbe^j] C_{\hb)}{}_{ij}
\non\\
&
 - \frac{\ri}{96} \eta_{\ha\hb} [\cD_\hal^k , \cD_{\hbe k}] Y^{\hal\hbe} 
 - \frac{\ri}{48} (\S^{\hc}{}_{(\ha})^{\hal\hbe} [\cD_\hal^k , \cD_{\hbe k}] Y_{\hb) \hc} 
 + \frac{\ri}{240} \eta_{\ha\hb} [\cD_\hal^k , \cD_{\hbe k}] W^{\hal\hbe} 
 \non\\
&
- \frac{1}{8} \eta_{\ha\hb} S^{kl} S_{kl} + \frac{1}{16} C_\ha{}^{kl} C_\hb{}_{kl} 
- \frac{1}{32} \eta_{\ha\hb} C^{\hc kl} C_{\hc kl} \non\\
&
 + \frac{1}{2} Y_{(\ha}{}^\hc Y_{\hb ) \hc} 
- \frac{1}{8} \eta_{\ha\hb} Y^{\hc\hd} Y_{\hc\hd} 
- \frac{1}{8} \eps_{\ha\hb\hc\hd\he} \cD^\hc W^{\hd\he}  \ .
\end{align}
\esubeq

The above results provide us with the ingredients needed to degauge conformal superspace to SU(2) superspace. For example, one 
finds the commutator
\bea
{[}\cD'_\ha,\cD_{\hbe}^j{]}
&=&
- {1\over 2} \Big{[}
(Y_{\ha\hb} + W_{\ha\hb}) (\Gamma^{\hat{b}})_{\hbe}{}^{\hga} \d^j_k
+{1\over 4}\,\ve_{\ha\hb\hc\hd\he} (2 Y^{\hd\he} - W^{\hd\he}) (\Sigma^{\hb\hc})_{\hbe}{}^{\hga} \d^j_k 
\non\\
&&~~~~
-(\Gamma_{\hat{a}})_{\hbe}{}^{\hga}S^j{}_k
- (\S_{\ha \hb})_{\hbe}{}^{\hga}C^\hb{}^j{}_k
\Big{]}
\cD_{\hga}^k
\non\\
&&
+ \frac{1}{2} \Big{[}
 (\G_\ha)_\hbe{}^\hga W^{\hc\hd}{}_\hga^k 
+\d^{[\hc}_\ha\Big(\frac{1}{3}  \cC^{\hd]}{}_\hbe{}^j 
- 2 \cY^{\hd]}{}_\hbe{}^j 
+ \frac{1}{2}  (\G^{\hd]})_\hbe{}^\hga\big(\cC_\hga^j +2\cY_\hga^j +2 X_\hga^j\big)
\Big)
 \non\\
 &&~~~~~~
+(\S^{\hc\hd})_{\hbe}{}^{\hga}\Big(
 2  \cY_\ha{}_\hga{}^j
-\frac{1}{3}  \cC_\ha{}_\hga{}^j \Big)
+\frac{1}{8}\ve_{\ha}{}^{\hc\hd\he\hat{f}}(\S_{\he\hat{f}})_\hbe{}^\hga (\cC_\hga^j +2\cY_\hga^j )
 \Big{]} M_{\hc\hd}
\non\\
&&
+\Big{[}
3\cY_{\ha}{}_\hbe^{(k}\ve^{l)j}
-{1\over 3}\cC_{\ha}{}_{\hbe}^{(k}\ve^{l)j} 
+ \frac{1}{8} (\G_\ha)_\hbe{}^\hga \cC_\hga{}^{jkl} 
-{11\over 24}(\G_\ha)_{\hbe}{}^\hga\cC_{\hga}^{(k}\ve^{l)j} 
\non\\
&&~~~~
 -{3\over 4}(\G_\ha)_{\hbe}{}^\hga (\cY_\hga^{(k} + 2 X_\hga^{(k}) \ve^{l)j}
\Big{]}
J_{kl}
~,
\eea
which agrees with \cite{KT-M08} up to field redefinitions. One can also derive the $[\cD'_\ha, \cD'_\hb]$ commutator, 
which we will not need for this paper.


\subsection{The conformal origin of the super Weyl transformations}

We have just shown that SU(2) superspace is a degauged version of conformal superspace, in which the 
dilatation connection is gauged away. Although the dilatations and special conformal transformations are not 
manifestly realized, the dilatation symmetry has not been fixed. The symmetry remains as additional nonlinear 
transformations, known as super Weyl transformations. Their presence in SU(2) superspace ensures that it describes 
conformal supergravity. Below we show how to recover the super Weyl transformations from the degauging 
of conformal superspace.

Suppose we have gauge fixed the dilatation connection to vanish by using
the special conformal symmetry.
If we now perform a dilatation with parameter $\s$,
we must accompany it with an additional $K_\hA$ transformation with $\s$-dependent
parameters $\L^\hA(\s)$ to maintain the gauge $B_\hA=0$, which requires
\be \big( \d_K(\L(\s)) + \d_{\mathbb D} (\s ) \big) B_\hA = 0 \ .
\ee
Using the transformation rule \eqref{TransCD}, we find
\be
\L^\ha(\s) = \hf \cD^\ha \s \ , \quad \L^{\hal i}(\s) = - \hf \cD^{\hal i} \s \ .
\ee

Note that all primary superfields $\Phi$ transform homogeneously
\be
\d_K(\L(\s)) \Phi + \d_{\mathbb D} (\s ) \Phi = \d_{\mathbb D} (\s ) \Phi = w \s \Phi \ ,
\ee
where $w$ is the dimension of $\Phi$, $\mathbb D \Phi =  w \Phi$. 
For example, the super Weyl tensor transforms as
\be \d_\s W_{\hal\hbe} = \s W_{\hal\hbe} \ .
\label{sW_W}
\ee

The super Weyl transformations of the degauged covariant derivatives $\cD_\hA$ and 
the special conformal connection can be read from
\be
\d_\s \nabla_\hA = \d_\s \cD_\hA - \d_\s \mathfrak{F}_\hA{}^\hB K_\hB = \d_K(\L(\s)) \nabla_\hA + \d_{\mathbb D} (\s ) \nabla_\hA~,
\ee
implying that the super Weyl transformations of $\cD_\hA$ are
\begin{subequations}
\bea
\d_\s \cD_\hal^i&=&\hf \s\cD_\hal^i+2(\cD^{\hga i}\s)M_{\hga\hal}-3(\cD_{\hal k}\s)J^{ki}~,
\label{sW1} \\
\d_\s \cD_\ha&=&
\s\cD_\ha
+\frac{\ri}{2} (\G_\ha)^{\hga\hde}(\cD_{\hga}^{k}\s)\cD_{\hde k}
- (\cD^\hb\s)M_{\ha\hb}
\label{sW2}
~,
\eea
\end{subequations}
while the super Weyl transformation of, for example, $\mathfrak{F}_\hal^i{}^{\hbe j}$ is
\be 
\d_\s \mathfrak{F}_\hal^i{}^{\hbe j} =
	\s \mathfrak{F}_\hal^i{}^{\hbe j}
	- \hf \cD_\hal^i \cD^{\hbe j} \s
	+ \frac{\ri}{2} \eps^{ij} \cD_{\hal}{}^{\hbe} \s
	= \s \mathfrak{F}_\hal^i{}^{\hbe j} - \frac{1}{4} [\cD_\hal^i , \cD^{\hbe j} ] \s \ . \label{FKsuperWeyl}
\ee
Equation~\eqref{FKsuperWeyl} 
implies
\begin{subequations}
\bea
\d_\s S^{ij}&=&\s S^{ij}
+{\ri\over 4}\,\cD^{\hal (i}\cD_{\hal}^{ j)}\s~,
\label{s-Weyl-Sij}\\
\d_\s C_{\ha}{}^{ij}&=&\s C_{\ha}{}^{ij}
+\frac{\ri}{2}\,( \G_\ha)^{\hga\hde} \cD_{\hga}^{(i}\cD_{\hde}^{j)}\s~,
\label{C-var}\\
\d_\s Y_{\ha\hb}&=&\s Y_{\ha\hb}
-{\ri\over 4}\, (\S_{\ha\hb})^{\hal\hbe}\cD_\hal^k\cD_{\hbe k}\s
~.
\label{s-Weyl-Y}
\eea
\end{subequations}


\section{The Weyl multiplet} \label{WeylMultiplet}

The 5D Weyl multiplet, constructed independently by two groups
\cite{Ohashi3, Ohashi4}
and \cite{Bergshoeff1, Bergshoeff2},
consists of the following matter content:
four fundamental one-forms -- the vielbein $e_\hm{}^\ha$,
the gravitini $\psi_\hm{}_\hal^i$,
an $\rm SU(2)$ gauge field $\cV_\hm{}^{ij}$, and a dilatation gauge
field $b_\hm$; and three covariant auxiliary fields -- a real antisymmetric
tensor $w_{\ha\hb}$, a fermion $\chi_\hal^i$, and a real auxiliary scalar $D$.
In addition, there are three composite one-forms -- 
the spin connection  $\omega_\hm{}^{\ha\hb}$, the $S$-supersymmetry connection
$\phi_\hm{}_\hal^i$, and the special conformal connection
$\frak{f}_\hm{}^\ha$ -- which are algebraically determined in terms of
the other fields by imposing constraints on some of the
curvature tensors.

In a standard component analysis, one begins by interpreting the
seven one-forms appearing above as connections for the 5D superconformal
algebra $\rm{F}^2(4)$. Associated with each connection is a two-form field
strength, constructed in the usual manner from the superalgebra
$\mathrm{F}^2(4)$.
One wishes to algebraically constrain the spin, $S$-supersymmetry,
and special conformal connections in terms of the other quantities:
this can be accomplished by constraining respectively the vielbein
curvature $R(P)_{\hm\hn}{}^\ha$, the gravitino curvature
$R(Q)_{\hm\hn}{}_\hal^i$,
and the conformal Lorentz curvature $R(M)_{\hm\hn}{}^{\ha\hb}$.
However, the remaining one-forms cannot furnish an off-shell representation
of a conformal supersymmetry algebra as the bosonic and fermionic
degrees of freedom do not match, so one is led to introduce the
additional covariant fields $w_{\ha\hb}$ (denoted $T_{\ha\hb}$ in
\cite{Bergshoeff1, Bergshoeff2}
and $v_{\ha\hb}$ in \cite{Ohashi3, Ohashi4}), $\chi_\hal{}^i$, and $D$.
At this stage, one must determine how the presence of the auxiliary
fields deforms the supersymmetry algebra, the curvatures, and the
constraints imposed on the curvatures in a self-consistent way.
In general, there is no unique
solution, and indeed, the two original groups, as well as the recent
work \cite{dWK}, each use different definitions
for supersymmetry and for the curvatures.

In contrast, the technical advantage of a superspace approach is that once the
supergeometry is completely specified and the Bianchi identities solved,
one must only specify definitions for the component fields -- their
supersymmetry transformations and the corresponding curvatures are
then completely determined. Our goal in this section is to
demonstrate precisely how this occurs for the 5D Weyl multiplet.

\newcommand{\loco}{\vert}
\newcommand{\doubar}{{{\loco}\!{\loco}}}


\subsection{Component fields and curvatures from superspace}
We begin by identifying the various component fields of the Weyl
multiplet. Let us start with the vielbein and gravitino. These
appear as the coefficients of $\rd x^\hm$ of the supervielbein
$E^\hA = (E^\ha, E^\hal_i) = \rd z^\hM\, E_\hM{}^\hA$.
It is convenient to introduce the
so-called double bar projection \cite{Baulieu:1986dp, BGG01},
denoted by $E^\hA \doubar$, that restricts to
$\q = \rd \q = 0$, corresponding to
the bosonic part $T^*\cM^5$ of the cotangent bundle
$T^*\cM^{5|8}$, where $\cM^5$ is the bosonic body 
of the curved superspace $\cM^{5|8}$.
Then we can define\footnote{We
define the gravitino with a lowered
spinor index and a raised SU(2) index. We follow similar conventions
when defining other component fields.}
\begin{align}
e{}^\ha = \rd x^\hm e_\hm{}^\ha &:= E^\ha\doubar~, \qquad
	\psi_\hal^i = \rd x^\hm \psi_\hm{}_\hal^i:= 2 E_\hal^i \doubar ~.
\end{align}
This is equivalent to defining $e_\hm{}^\ha = E_\hm{}^\ha \loco$ and
$\psi_\hm{}_\hal^i = 2\, E_\hm{}_\hal^i \loco$ where
the single vertical bar denotes the usual component projection
to $\q=0$, i.e. $V(z) \loco := V(z)\loco_{\theta = 0}$
for any superfield $V(z)$.
In like fashion the remaining fundamental and composite one-forms are
found by taking the projections of the corresponding superforms,
\begin{align}
	\cV^{ij} := \Phi{}^{ij} \doubar~, \quad
	b := B\doubar ~, \quad
\omega^{\ha\hb} := \Omega{}^{\ha\hb} \doubar ~, \quad  
\phi_\hal^i := 2 \,\frak F{}_\hal^i\doubar~, \quad
\frak{f}{}^\ha := \frak{F}{}^\ha\doubar~. 
\end{align}
The additional auxiliary fields are contained within the curvature superfield
$W_{\hal\hbe}$,
\begin{gather}\label{wcf}
w_{\hal\hbe} := W_{\hal\hbe}\loco \ , \qquad
\chi_\hal^i := \frac{3\ri}{32} X_\hal^i \vert~, \qquad
D := -\frac{3}{128} Y \loco~.
\end{gather}
The normalizations we have chosen for $\chi_\hal^i$ and $D$
coincide with the normalizations of \cite{Bergshoeff1, Bergshoeff2}
and \cite{dWK}.
The other independent components of the curvature superfield are
given by $W_{\ha \hb \hal}{}^i\loco$ and by $X_{\ha \hb}{}^{ij}\loco$,
and will turn out to be given by some of the component curvatures.

It should be mentioned that one can impose a Wess-Zumino gauge to fix
the $\q$ expansions of the super one-forms, 
so that they are completely determined by the above fields.
This ensures that the entire physical content of the superspace
geometry is accounted for. In practice, it
is usually unnecessary to do this explicitly.

Now we may determine the so-called supercovariant curvatures.
In terms of the connection one-forms, the covariant derivative
$\nabla_\ha|$ is defined by taking the double bar projection of 
equation \eqref{eq:covD}, leading to
\begin{align}
e_\hm{}^\ha \nabla_\ha| = \pa_\hm
	- \frac{1}{2} \psi_\hm{}^\hal_i \nabla_{\hal}^i \loco
	- \frac{1}{2} \omega_\hm{}^{\ha \hb} M_{\ha \hb}
	- b_\hm \mathbb D
	- \cV_\hm{}^{ij} J_{ij}
	- \frac{1}{2} \phi_\hm{}^{\hal i} S_{\hal i}
	- {\frak f}_\hm{}^{\ha} K_\ha~,
\end{align}
where we have defined the lowest component of the superspace operator 
$\nabla_\hal^i|$ such that for an arbitrary tensor superfield $U$ 
\be (\nabla_\hal^i| U)| = (\nabla_\hal^i U) | \ .
\ee
We interpret $\nabla_{\hal}^i\loco$  as the
generator of supersymmetry. In what follows we will drop the 
bar projection from $\nabla_\ha|$ when it is clear from context to 
which we are referring.

It will be convenient to also introduce the spin, dilatation, and
$\rm SU(2)$ covariant derivative
\bsubeq
\begin{align}
\cD_\hm &:= \pa_\hm
	- \frac{1}{2} \omega_\hm{}^{\hb \hc} M_{\hb \hc}
	- b_\hm \mathbb D
	- \cV_\hm{}^{ij} J_{ij}~,  \\
\cD_\ha &:= e_\ha{}^\hm \cD_\hm = e_\ha{}^\hm \pa_\hm
	- \frac{1}{2} \omega_\ha{}^{\hb \hc} M_{\hb \hc}
	- b_\ha \mathbb D
	- \cV_\ha{}^{ij} J_{ij}~,
\end{align}
\esubeq
where
\be \omega_\ha{}^{\hb \hc} := e_\ha{}^\hm \omega_\hm{}^{\hb\hc} \ , \quad
b_\ha := e_\ha{}^\hm b_\hm \ , \quad 
\cV_\ha{}^{ij} := e_\ha{}^\hm \cV_\hm{}^{ij} \ .
\ee

The supercovariant curvature tensors are given by
\begin{align}
[\nabla_\ha, \nabla_\hb]
	&= - R(P)_{\ha \hb}{}^\hc \nabla_{\hc}
	- R(Q)_{\ha \hb}{}^{\hal}_i \nabla_\hal^i\loco
	- \frac{1}{2} \RM_{\ha \hb}{}^{\hc \hd} M_{\hc \hd}
	- \RJ_{\ha\hb}{}^{ij} J_{ij}
	\eol & \quad
	- \RD_{\ha\hb} \gD 
	- \RS_{\ha\hb}{}^{\hga k} S_{\hga k}
	- \RK_{\ha\hb}{}^\hc K_\hc
\end{align}
and are found by taking the component projections of the
curvature tensors in \eqref{eq:Curv3}. We have introduced
the expressions
\begin{align}
R(P)_{\ha \hb}{}^{\hc} = \scT_{\ha \hb}{}^\hc \loco~, \qquad
R(Q)_{\ha \hb}{}_\hal^i = \scT_{\ha\hb}{}_\hal^i \loco~,
\end{align}
for the lowest components of the superspace torsion tensors
to match the usual component nomenclature.

At this stage there are two distinct expressions we can
give for each of the curvature tensors. Let us demonstrate
with $R(P)_{\ha\hb}{}^\hc$. We can write two equivalent expressions
for the double-bar projection of the torsion two-form $\scT^\hc$,
\begin{align}
\scT^\hc \doubar = \frac{1}{2} \rd x^\hn \wedge \rd x^\hm\, \scT_{\hm\hn}{}^\hc \loco
	= \rd x^\hn \wedge\rd x^\hm\, \cD_{[\hm} e_{\hn]}{}^\hc 
\end{align}
and
\begin{align}
\scT^\hc \doubar &=\hf (-1)^{\veps_\hA \veps_\hB}  E^\hA \wedge E^\hB \scT_{\hA\hB}{}^\hc \doubar \nonumber\\
	&= \frac{1}{2} \rd x^\hn \wedge \rd x^\hm \Big(e_\hm{}^\ha e_\hn{}^\hb \,\scT_{\ha\hb}{}^\hc\vert
	+ e_{[\hm}{}^\ha \,\psi_{\hn]}{}^{\hbe}_j \,\scT_{\ha}{}_{  \hbe}^j{}^\hc \vert
	- \frac{1}{4} \psi_\hm{}^\hal_i \,\psi_\hn{}^{\hbe}_j \,
		\scT_{\hal}^i{}_\hbe^j{}^\hc \vert \Big)\nonumber \\
	&= \frac{1}{2} \rd x^\hn \wedge \rd x^\hm \Big(e_\hm{}^\ha e_\hn{}^\hb \,R(P)_{\ha\hb}{}^\hc
		+ \frac{\ri}{2} \psi_{\hm j} \G^\hc \psi_\hn{}^j\Big)~.
\end{align}
Equating the two expressions provides a definition for the
supercovariant curvature $R(P)_{\ha\hb}{}^\hc$. Proceeding in this
way for the other curvature two-forms, we find the following definitions:
\begin{subequations}\label{COMPcurvatures}
\begin{align}
R(P)_{\ha\hb}{}^\hc &:=
	2\, e_\ha{}^\hm e_\hb{}^\hn \cD_{[\hm} e_{\hn]}{}^\hc
	- \frac{\ri}{2} \psi_{\ha  j} \G^\hc \psi_\hb{}^j\\
R(Q)_{\ha\hb }{}_{\hal}^i &:=
	e_\ha{}^\hm e_\hb{}^\hn \cD_{[\hm} \psi_{\hn]}{}_\hal^i
	+ \ri (\G_{[\ha } \phi_{\hb]}{}^i)_\hal
	+ \hf \,w_{\hc\hd} \, (\S^{\hc\hd} \G_{[\ha } \psi_{\hb]}{}^i){}_\hal~, \\
R(M)_{\ha\hb}{}^{\hc\hd} &:= \cR(\omega)_{\ha\hb}{}^{\hc\hd}
	+ 8 \,\delta_{[\ha }{}^{[c} \frak f_{\hb]}{}^{d]}
	- 2 \,\psi_{[\ha  j} \Sigma^{\hc\hd} \phi_{\hb]}{}^{j}
	- 2\ri (\psi_{[\ha  j} \G_{\hb]} R(Q)^{\hc\hd  j})
	\eol & \quad
	- \frac{32\ri}{3} (\psi_{[\ha  j} \G_{\hb]} \S^{\hc\hd} \chi^j)
	+ \frac{\ri}{2} \,\psi_{\ha j} \psi_\hb{}^j w^{\hc\hd}~, \\
R(J)_{\ha\hb}{}^{ij} &:= \cR(\cV)_{\ha\hb}{}^{ij}
	- 3 \,(\psi_{[\ha }{}^{(i} \phi_{\hb]}{}^{j)})
	- 16\ri  \,(\psi_{[\ha }{}^{(i} \G_{\hb]} \chi^{j)}) \ , \\
R(\bbD)_{\ha\hb} &:= 2\, e_\ha{}^\hm e_\hb{}^\hn \pa_{[\hm} b_{\hn]}
	+ 4 \,\frak f_{[\ha\hb]}
	+ (\psi_{[\ha  j} \phi_{\hb]}{}^j)
	+ \frac{16\ri}{3} (\psi_{[\ha  k} \G_{\hb]} \chi^k) ~,
\end{align}
\end{subequations}
where we have introduced
\be \psi_\ha{}^\hbe_j := e_\ha{}^\hm \psi_\hm{}^\hbe_j \ , \quad 
\phi_\ha{}^\hbe_j := e_\ha{}^\hm \phi_\hm{}^{\hbe}_j \ , \quad 
\frak{f}_\ha{}^\hb = e_\ha{}^\hm \frak{f}_\hm{}^\hb
\ee
and the curvatures
\begin{align}
\cR(\omega)_{\ha\hb}{}^{\hc\hd} &:= 2 e_\ha{}^\hm e_\hb{}^\hn (\partial_{[\hm} \omega_{\hn]}{}^{\hc\hd} - 2 \omega_{[\hm}{}^{\hc\he} \omega_{\hn]}{}_\he{}^\hd)  \ , \\
\cR(\cV)_{\ha\hb}{}^{ij} &:= 2\,e_\ha{}^\hm e_\hb{}^\hn \Big( \pa_{[\hm} \cV_{\hn]}^{ij}
	+ \cV_{[\hm}^{k (i} \cV_{\hn]}^{j)}{}_k\Big)~.
\end{align}
The supercovariant forms of $R(S)_{\ha\hb}{}_\hal^i$ and $R(K)_{\ha\hb}{}^\hc$
are a good deal more complicated, so we do not give them here.

\subsection{Analysis of the curvature constraints}
We have not yet employed the constraints imposed by superspace on the
curvatures. They are
\begin{subequations}\label{compConstraintsTR}
\begin{align}
R(P)_{\ha\hb}{}^\hc &= - \tilde{w}_{\ha\hb}{}^{\hc} \equiv
	-\hf \eps_{\ha\hb}{}^{\hc\hd\he} w_{\hd\he} ~, \label{compConstraintsTR_1}\\
(\G^\ha R(Q)_{\ha\hb}{}^{i})_\hal &= -\frac{32}{3} (\G_\hb \chi^i)_\hal ~, \label{compConstraintsTR_2}\\
R(M)_{\ha\hb}{}^{\hc\hb} &= -\frac{32}{3} \d^\hc_\ha D - w_{\ha\hd} w^{\hc\hd}
	- \nabla^\hd \tilde{w}_{\hd\ha}{}^\hc \ , \label{compConstraintsTR_3}
\end{align}
\end{subequations}
and respectively determine the spin connection, the
$S$-supersymmetry connection, and the $K$-connection. In contrast
to previous conventions employed in the literature, these are actually
$S$-invariant constraints. 
The reason for this is that the superspace operators $\nabla_{\hal}^i$
and $\nabla_\ha$ have the same algebra with $S_{\hal i}$ as one finds in
the superconformal algebra $\rm F^2(4)$.
The price one pays for this simplicity is that the composite connections
will turn out to depend rather more significantly on the auxiliary fields
$w_{\ha\hb}$, $\chi_\hal^i$ and $D$ than one might have wished.

The first constraint \eqref{compConstraintsTR_1} determines the spin connection to be
\begin{align}
\omega_{\ha\hb\hc} &= 
\omega(e)_{\ha\hb\hc} + \frac{\ri}{4} (
\psi_{\ha k} \G_\hc \psi_\hb{}^k
+ \psi_{\hc k} \G_\hb \psi_\ha{}^k
- \psi_{\hb k} \G_\ha \psi_\hc{}^k)
+ 2 b_{[\hb} \eta_{\hc ] \ha} 
- \frac{1}{4} \eps_{\ha\hb\hc}{}^{\hd\he} w_{\hd\he} \ ,
\end{align}
where $\omega(e)_{\ha\hb\hc} = -\frac{1}{2} (\cC_{\ha\hb\hc} + \cC_{\hc\ha\hb} - \cC_{\hb\hc\ha})$
is the usual spin connection of general relativity, given
in terms of the anholonomy coefficient
$\cC_{\hm\hn}{}^\ha := 2 \,\partial_{[\hm} e_{\hn]}{}^\ha$.
Note that the spin connection $\omega_{\ha\hb\hc}$
possesses torsion: in addition to the
usual contribution from the gravitino bilinears, there is additional
bosonic torsion from the auxiliary field $w_{\ha\hb}$.

From the second constraint \eqref{compConstraintsTR_2},
we find the $S$-supersymmetry connection
\begin{align}\label{eq:SConn}
\ri \,\phi_\hm{}^i 
	&=
	\frac{8}{3} \G_\hm \chi^i
	+ \frac{1}{3} (\G^{[\hat p} \d_\hm^{\hat q]}
		+ \frac{1}{4} \G_\hm \S^{\hat p\hat q})
		(\Psi_{\hp\hq}{}^i + w_{\ha\hb} \S^{\ha\hb} \G_{[\hat p} \psi_{\hat q]}{}^i) \ ,
\end{align}
where we have suppressed spinor indices for legibility and introduced
the gravitino field strength
$\Psi_{\hm\hn}{}_{ \hal}^i := 2 \,\cD_{[\hm} \psi_{\hn]}{}_\hal^i$.
Reinserting this back into the original expression for $R(Q)$, we find that
\begin{align}\label{eq:RQ}
R(Q)_{\ha\hb}{}^i
	= \frac{1}{2} \Pi_{\ha\hb}{}^{\hc\hd} \Big(
		\Psi_{\hc\hd}{}^i - w_{\hc\hd} \G^e \psi_e{}^i
		- \tilde w_{\ha\hb}{}^\hc \psi_c{}^i
	\Big)
	- \frac{16}{3} \S_{\ha\hb} \chi^i~,
\end{align}
where the spinor projection operator
\begin{align}
\Pi_{\ha\hb}{}^{\hc\hd} :=\delta_{\ha }^{[\hc} \delta_{\hb}^{\hd]}
	+ \frac{2}{3} \delta_{[\ha }^{[\hc} \G_{\hb]} \G^{\hd]}
	- \frac{1}{3} \S_{\ha\hb} \S^{\hc\hd}
~, \qquad \G^\ha \Pi_{\ha\hb}{}^{\hc\hd} = 0~, \qquad \Pi^2 = \Pi~, 
\end{align}
projects onto the $\Gamma$-traceless part of a spinor-valued two-form.
It is convenient to introduce a separate symbol $\hat R(Q)_{\ha \hb}{}_{\hal}^i$
for the first term of \eqref{eq:RQ},
\begin{align}\label{eq:hatRQv1}
\hat R(Q)_{\ha\hb}{}^i = \frac{1}{2} \Pi_{\ha\hb}{}^{\hc\hd} \Big(
		\Psi_{\hc\hd}{}^i - w_{\hc\hd} \G^\he \psi_\he{}^i
		- \tilde w_{\ha\hb}{}^\hc \psi_\hc{}^i
	\Big)~,
\end{align}
to its $\G$-traceless part. Using \eqref{eq:Tors2}, we find that one of the
remaining components of the superspace curvature is determined,
\begin{align}
W_{\hal\hbe\hga}{}^i \loco
	= \ri (\S^{\ha\hb})_{\hal\hbe} \, \hat R(Q)_{\ha \hb}{}_{ \hga}^i
	= \ri \Psi_{(\hga\hbe\hal)}{}^i
	+ \ri w_{(\hga\hbe} \psi_{\hal) \hde}{}^{\hde i}
	- \ri w_{\hde (\hal}\psi^\hde{}_{\hga \hbe)}{}^i~.
\end{align}

From the third constraint \eqref{compConstraintsTR_3}, one can show that 
\begin{align}\label{eq:KConn}
\frak f_\ha{}^\hb &= 
- \frac{2}{3} \d^\ha_\hb D
- \frac{1}{6} w_{\ha\hc} w^{\hb\hc} 
+ \frac{1}{48} \d_\ha^\hb w^{\hc\hd} w_{\hc\hd}
- \frac{1}{6} \nabla^\hc \tilde{w}_{\hc\ha}{}^\hb
- \frac{1}{6} \cR_\ha{}^\hb(\omega) + \frac{1}{48} \d_\ha^\hb \cR(\omega) \non\\
&\qquad
- \frac{\ri}{6} (\psi_\hd \G_\ha \hat R(Q)^{\hb \hd j})
+ \frac{1}{3} (\psi_{[\ha j} \S^{\hb \hc} \phi_{\hc]}{}^j) 
- \frac{1}{24} \d_\ha^\hb (\psi_{\hc j} \S^{\hc\hd} \phi_\hd{}^j) \non\\
&\qquad- \frac{\ri}{12} (\psi_{\ha j} \psi_\hc{}^j) w^{\hb\hc}
+ \frac{\ri}{96} \d_\ha^\hb (\psi_\hc{}_j \psi_\hd{}^j) w^{\hc\hd} \ ,
\end{align}
where $\cR_{\ha}{}^{\hb}(\omega) = \cR_{\ha\hc}{}^{\hb \hc}(\omega)$ and
$\cR(\omega) = \cR_\ha{}^\ha(\omega)$. In principle, one can reinsert
this expression into $R(M)_{\ha\hb}{}^{\hc\hd}$. The result is quite complicated;
we remark only that it can be written
\begin{align}
R(M)_{\ha\hb}{}^{\hc\hd} &= C(\omega)_{\ha\hb}{}^{\hc\hd}
	- \frac{4}{3} \delta_{[\ha }{}^{[\hc} w_{\hb]\he} w^{\hd] \he}
	- \frac{4}{3} \nabla^\he \tilde w_{\he[\ha}{}^{[\hc} \delta_{\hb ]}{}^{\hd]}
	\eol & \quad
	+ \delta_{[\ha }{}^{[\hc} \delta_{\hb]}{}^{\hd]}
		\Big(\frac{1}{6} w^{\hat{e}\hat{f}}w_{\hat{e}\hat{f}}
		- \frac{16}{3}  D
		\Big)
	+ (\text{explicit gravitino terms}) \ ,
\end{align}
where
$C(\omega)_{\ha\hb}{}^{\hc\hd} = \cR(\omega)_{\ha\hb}{}^{\hc\hd}
	- \frac{4}{3}\delta_{[\ha }{}^{[\hc} \cR(\omega)_{\hb]}{}^{\hd]}
	+ \frac{1}{6} \delta_{[\ha }{}^{[\hc} \delta_{\hb]}{}^{\hd]} \cR(\omega)$
is the traceless part of the tensor $\cR(\omega)_{\ha\hb}{}^{\hc\hd}$.
This is not quite the usual Weyl tensor because of the presence of
bosonic torsion in the spin connection.
The superspace expression for $\sRM_{\ha\hb}{}^{\hc\hd}$ in principle determines
$W_{\hal\hbe\hga\hde}\loco$; however, we will find a more useful form of this
expression using a different method shortly.

For the remaining dimension-2 curvatures, we find
\begin{align}
R(\mathbb D)_{\ha\hb} = -\nabla^\hc \tilde w_{\hc\ha\hb}~, \qquad
R(J)_{\ha \hb}{}^{ij} = -\frac{3\ri}{4} X_{\ha \hb}{}^{ij} \loco~.
\end{align}
The first equation is automatically satisfied upon substituting
into $R(\mathbb D)$ the expression for ${\frak f}_\hm{}^\ha$. The second equation serves
as a definition for the remaining undetermined component
$X_{\ha \hb}{}^{ij}\loco$ of the Weyl superfield.

\subsection{Supersymmetry transformations of the fundamental fields}

Here we present the complete $Q$, $S$, and $K$ transformations for
the fundamental fields of the Weyl multiplet. The transformations
of the one-forms follow from eq. \eqref{TransCD}, while those of
the covariant fields can be read off from \eqref{eq:Wdervs}:
\begin{subequations}\label{eq:WeylSUSY}
\begin{align}
\delta e_\hm{}^\ha &= \ri (\xi_j \G^\ha \psi_\hm{}^j)~, \\
\delta \psi_\hm{}_\hal^i &=
	2 \cD_\hm \xi_\hal^i
	+ w_{\hc\hd} (\Sigma^{\hc\hd} \G_\hm \xi^i)_\hal
	+ 2 \ri (\G_\hm \eta^i)_\hal~, \\
\delta \cV_\hm{}^{ij} &=
	3 \xi^{(i} \phi_\hm{}^{j)}
	+ 16 \ri \,\xi^{(i} \G_\hm \chi^{j)}
	- 3 \eta^{(i} \psi_\hm{}^{j)}~, \\
\delta b_\hm &= - \xi_k \phi_\hm{}^k
	- \frac{16\ri}{3} \,\xi_k \G_\hm \chi^k
	- \eta_k \psi_\hm{}^k
	- 2 \,e_\hm{}^\ha \L_{K\ha}~,\\
\delta w_{\ha\hb} &= 2 \ri \,\xi_i R(Q)_{\ha\hb}{}^i~, \\
\delta \chi_\hal^i &=
	\frac{1}{2} \xi_\hal^i D
	+ \frac{3}{128} (\nabla_\ha w_{\hb\hc})
		\Big(3 (\Sigma^{\hb\hc} \G^\ha \xi^i)_\hal
			+ (\G^\ha \S^{\hb\hc}\xi^i)_\hal \Big)
	\eol & \quad
	- \frac{1}{16} R(J)_{\ha\hb}{}^i{}_j (\S^{\ha\hb} \xi^j)_\hal
	- \frac{3\ri}{16} w_{\ha\hb} (\S^{\ha\hb} \eta^i)_\hal~, \\
\delta D &=
	2 \ri \,(\xi_j {\slashed{\nabla}} \chi^j)
	+ 2 \,(\eta_j \chi^j)~.
\end{align}
\end{subequations}

One can also derive the transformations of the composite one-forms
from \eqref{TransCD}. For example, the transformations for the spin
connection and the $S$-supersymmetry connection are
\begin{align}
\delta \omega_\hm{}^{\ha\hb}
	&= -\ri \,\xi_j \psi_\hm{}^j w^{\ha\hb}
	+ 2\ri \,\xi_j \G_\hm R(Q)^{\ha\hb j}
	+ \frac{32\ri}{3} \xi_j \G_\hm \S^{\ha\hb} \chi^j
	+ 2 \xi_k \S^{\ha\hb} \phi_\hm{}^k
	\eol & \quad
	- 2 \eta_j \S^{\ha\hb} \psi_\hm{}^j
	+ 4 e_\hm{}^{[\ha } \L_K^{\hb]}~, \\
\delta \phi_\hm{}_\hal^i
	&= - 2\ri \,{\frak f}_\hm{}^\ha (\G_\ha \xi^i)_\hal
	- \frac{16}{3} (\xi_j \psi_\hm{}^j) \chi_\hal^i
	- \frac{8\ri}{3} (\G_\hm \xi^i)_\hal\, D
	\eol & \quad
	- \frac{\ri}{8} (\nabla_\ha w_{\hb\hc})
		(\S^{\hb\hc} \G^\ha \G_\hm + 3 \G^\ha \S^{\hb\hc} \G_\hm)_\hal{}^\hbe \xi_\hbe^i
	+ \frac{\ri}{3} R(J)_{\ha\hb}{}^{ij}\, (\S^{\ha\hb} \G_\hm \xi_j)_\hal
	\eol & \quad
	+ 2 \cD_\hm \eta_\hal^i
	+ \ri \L_K^\hb (\G_\hb \psi_\hm{}^i)_\hal~,
\end{align}
where $\L_K^\ha$ parametrizes the special conformal transformations. 
We do not give here the transformation rule for ${\frak f}_\hm{}^\ha$ as it
is quite complicated.


\subsection{A new choice for component constraints}
As already alluded to, the component constraints \eqref{compConstraintsTR} we have found
from superspace are quite interesting from a technical standpoint:
they are $S$-invariant. This is reflected in the fact that the $S$-supersymmetry
transformations of the various one-forms are exactly those derived from
the algebra $F^2(4)$.
However, this comes with a price: we must introduce
bosonic torsion involving the field $w_{\ha\hb}$ into
the spin connection. Similarly, the $S$-supersymmetry and special
conformal connections \eqref{eq:SConn} and \eqref{eq:KConn}
include additional contributions from
the auxiliary fields. The last case is particularly inconvenient -- it
reflects the fact that $R(M)_{\ha \hb \hc \hd}$ is not just a minimally
covariantized version of the Weyl tensor, but depends additionally on
the auxiliary fields $D$, $\chi_\hal{}^i$, and $w_{\ha\hb}$. From a component
point of view, it would be more 
convenient 
to extract these dependences
so that the component fields and curvatures are as
simply defined as possible.
This will turn out to lead to a formulation that more closely resembles
those of \cite{Ohashi3,Ohashi4, Bergshoeff1,Bergshoeff2,Bergshoeff3}.

Let us begin by introducing new definitions for the composite spin,
$S$-supersymmetry, and $K$-connections:
\begin{subequations}\label{eq:NewConnections}
\begin{align}
\hat \omega_{\ha \hb \hc} &:= \omega_{\ha \hb \hc}
	+ \frac{1}{2} \tilde w_{\ha \hb \hc}~, \\
\ri \,\hat\phi_\hm{}^i 
	&:= \ri \,\phi_\hm{}^i
	- \frac{8}{3} \G_\hm \chi^i~, \\
\hat {\frak f}_{\ha}{}^{\hb}
	&:= {\frak f}_{\ha}{}^{\hb}
	+ \frac{2}{3} \delta_\ha{}^\hb \,D
	+ \frac{1}{4} w_{\ha \hd} w^{\hb\hd} 
	+ \frac{1}{4} \nabla^\hc \tilde w_{\hc\ha}{}^\hb 
	- \frac{3}{64} w^{\hc\hd}w_{\hc\hd} \delta_\ha{}^\hb~.
\end{align}
\end{subequations}
These definitions actually correspond to a redefinition of the superspace vector covariant
derivative,
\begin{align}
\hat \nabla_\ha &= \nabla_\ha
	- \frac{1}{4} \tilde W_{\ha\hb\hc} M^{\hb\hc}
	+ \frac{1}{8} X^{\hbe i} (\Gamma_\ha)_\hbe{}^\hal S_{\hal i}
	+ \frac{1}{64}\big(Y+3W^{\hb\hc}W_{\hb\hc}\big) K_\ha
	\eol & \quad
	- \frac{1}{4}( \nabla^\hc \tilde W_{\hc\ha}{}^\hb) K_\hb
	- \frac{1}{4} W_{\ha\hd} W^{\hb\hd} K_\hb
~.
\end{align}
We discuss further this superspace interpretation in
Appendix \ref{app:ModifiedSuperspace}.

The new curvatures given by the algebra $[\hat\nabla_\ha, \hat\nabla_\hb]$ are
\begin{subequations}\label{eq:hatCurvs}
\begin{align}
\hat R(P)_{\ha\hb}{}^\hc &= 2 \,e_\ha{}^\hm e_\hb{}^\hn \hat\cD_{[\hm} e_{\hn]}{}^\hc
	- \frac{\ri}{2} \psi_{\ha  j} \G^\hc \psi_\hb{}^j\ , \\
\hat R(Q)_{\ha \hb}{}_{\hal}^i &=
	e_\ha{}^\hm e_\hb{}^\hn \hat\cD_{[\hm} \psi_{\hn]}{}_\hal^i
	+ \ri (\G_{[\ha } \hat \phi_{\hb]}{}^i)_\hal
	\eol & \quad
	+ \frac{1}{8} w_{\hc\hd} \Big(
		3 (\S^{\hc\hd} \G_{[\ha })_\hal{}^\hbe - (\G_{[\ha } \S^{\hc\hd})_\hal{}^\hbe
		\Big) \psi_{\hb]}{}_\hbe^i~, \\
\hat R(M)_{\ha\hb}{}^{\hc\hd} &= \cR(\hat\omega)_{\ha\hb}{}^{\hc\hd}
	+ 8 \,\delta_{[\ha }{}^{[\hc} \hat {\frak f}_{\hb]}{}^{\hd]}
	- 2 \,\psi_{[\ha  j} \Sigma^{\hc\hd} \hat \phi_{\hb]}{}^{j}
	\eol & \quad
	+ \frac{16\ri}{3} \delta_{[\ha }{}^{[\hc}  \psi_{\hb]i} \G^{\hd]} \chi^i
	- \ri \psi_{[\ha  i} \Big(\G_{\hb]} \hat R(Q)^{\hc\hd  i} +
		2\G^{[\hc} \hat R(Q)_{\hb]}{}^{\hd] i} \Big)
	\eol & \quad
	+ \frac{\ri}{2} \psi_{\ha j} \psi_\hb{}^j w^{\hc\hd}
	- \frac{\ri}{4} (\psi_{\ha  j} \G_\he \psi_{b}{}^j) \tilde w^{\hc\hd\he} \ , \\
\hat R(J)_{\ha\hb}{}^{ij} &= \cR(\cV)_{\ha\hb}{}^{ij}
	- 3 \,\psi_{[\ha }^{(i} \hat \phi_{\hb]}{}^{j)}
	- 8\ri \,\psi_{[\ha }^{(i} \G_{\hb]} \chi^{j)} \ , \\
\hat R(\bbD)_{\ha\hb} &= 2\, e_\ha{}^\hm e_\hb{}^\hn \pa_{[\hm} b_{\hn]}
	+ 4 \,\hat {\frak f}_{[\ha\hb]}
	+ \psi_{[\ha  j} \hat \phi_{\hb]}{}^j
	+ \frac{8\ri}{3} \,\psi_{[\ha  j} \G_{\hb]} \chi^j
	 \ ,
\end{align}
\end{subequations}
where we have introduced
\be \label{hatDder}
\hat{\cD}_\ha = e_\ha{}^\hm \partial_\hm - \hf \hat\omega_\ha{}^{\hb\hc} M_{\hb\hc}
- b_\ha \bbD - \cV_\ha{}^{ij} J_{ij} \ , \quad \cD_\hm = e_\hm{}^\ha \hat{\cD}_\ha \ . 
\ee
We postpone for the moment a discussion of
$\hat R(S)_{\ha\hb}{}^{\hal i}$ and $\hat R(K)_{\ha\hb}{}^\hc$.

The curvatures turn out to obey the constraints
\begin{align} \label{componentConstTRv2}
\hat R(P)_{\ha\hb}{}^\hc &= 0~, \qquad
(\G^\ha)_\hal{}^\hbe \hat R(Q)_{\ha \hb}{}_{\hbe}^i = 0~, \qquad
\hat R(M)_{\ha\hb}{}^{\hc\hb} = 0~.
\end{align}
These coincide with the constraints usually imposed in the
component formulations and are not $S$-invariant.
This is a consequence of the redefinition of the auxiliary
connections, which deforms their $S$-supersymmetry
transformations. Equivalently, $[S_{\hal i}, \hat \nabla_\ha]$
is no longer given in superspace simply by $\ri (\G_\hal)_\hal{}^\hbe \nabla_{\hbe i}$.

The constraints are solved by
\begin{subequations}
\begin{align}
\hat\omega_{\ha\hb\hc} &= 
	\omega(e)_{\ha\hb\hc} + \frac{\ri}{4} (
	\psi_{\ha k} \G_\hc \psi_\hb{}^k
	+ \psi_{\hc k} \G_\hb \psi_\ha{}^k
	- \psi_{\hb k} \G_\ha \psi_\hc{}^k)
	+ 2 b_{[\hb} \eta_{\hc ] \ha} ~, \\
\ri \,\hat \phi_{\hm}{}^i
	&= \frac{2}{3} (\G^{[\hp} \delta_\hm{}^{\hq]} + \frac{1}{4} \G_\hm \S^{\hp\hq})
		\Big(
		\hat \cD_{[\hp} \psi_{\hq]}{}^i
		+ \frac{1}{8} w_{\hc\hd}
			\big(3 \S^{\hc\hd} \G_{[\hp} \psi_{\hq]}{}^i
			- \G_{[\hp} \S^{\hc\hd} \psi_{\hq]}{}^i
		\big)\Big)~, \\
\hat {\frak f}_\ha{}^\hb &=
	- \frac{1}{6}\cR(\hat\omega)_{\ha \hc}{}^{\hb\hc}
	+ \frac{1}{48} \delta_\ha{}^\hb \cR(\hat\omega)_{\hc\hd}{}^{\hc\hd}
	- \frac{\ri}{6} \psi_{\hc j} \G^{[\hb} \hat R(Q)_\ha{}^{\hc]j}
	- \frac{\ri}{12} \psi_{\hc j} \G_\ha \hat R(Q)^{\hb\hc j}
	\eol & \quad
	+ \frac{1}{3} \psi_{[\ha  j} \S^{\hb\hd} \hat \phi_{\hd]}{}^j
	- \frac{1}{24} \delta_\ha{}^\hb (\psi_{\hc j} \S^{\hc\hd} \hat \phi_\hd{}^j)
	- \frac{2\ri}{3} (\psi_{\ha  j} \G^\hb \chi^j)
	\eol & \quad
	- \frac{\ri}{12} \psi_{\ha j} \psi_\hc{}^j w^{\hb\hc}
	+ \frac{\ri}{24} (\psi_{\ha j} \G_\he \psi_\hd{}^j) \tilde w^{\hb\hd\he}
	\eol & \quad
	+ \frac{\ri}{192} \delta_\ha{}^\hb
		\Big(2 (\psi_{\hc j} \psi_\hd{}^j) w^{\hc\hd} -
		(\psi_{\hc j} \G_\he \psi_\hd{}^j) \tilde w^{\hc\hd \he}\Big)~.
\end{align}
\end{subequations}
One may confirm that these are equivalent to \eqref{eq:NewConnections}.

This redefinition dramatically simplifies many of the component curvatures.
As we have already seen, $\hat R(P)_{\ha\hb}{}^\hc$ vanishes.
The curvature $\hat R(Q)_{\ha\hb}{}_\hal^i$ turns out to coincide
with the identically named quantity introduced in \eqref{eq:hatRQv1}. Using the
redefined spin connection, one has
\begin{align}\label{eq:hatRQv2}
\hat R(Q)_{\ha\hb}{}^i = \frac{1}{2} \Pi_{\ha\hb}{}^{\hc\hd} \Big(
		\hat \Psi_{\hc\hd}{}^i - \frac{3}{4} w_{\hc\hd} \G^\he \psi_\he{}^i
		- \frac{3}{4} \tilde w_{\ha\hb}{}^\hc \psi_\hc{}^i
	\Big)~.
\end{align}
The curvature $\hat R(D)_{\ha\hb}$
now vanishes while $\hat R(J)_{\ha\hb}{}^{ij}$ is unchanged,
\begin{align}
\hat R(D)_{\ha \hb} = 0~, \qquad
\hat R(J)_{\ha\hb}{}^{ij} = R(J)_{\ha\hb}{}^{ij} = -\frac{3\ri}{4} X_{\ha\hb}{}^{ij} \loco~.
\end{align}
The Lorentz curvature tensor $\hat R(M)_{\ha \hb}{}^{\hc \hd}$ turns out
to be simplified the most and is given, up to terms of the form
$\psi \cD \psi$ and $\psi^2 w$, as
\begin{align}
\hat R(M)_{\ha \hb}{}^{\hc \hd} = C(\hat\omega)_{\ha \hb}{}^{\hc \hd}
	+ (\text{explicit gravitino bilinears})~,
\end{align}
where $ C(\hat\omega)_{\ha \hb}{}^{\hc \hd}$ is the Weyl tensor.
Remarkably, from eq. \eqref{eq:newRMab}, one finds the rather simple expression
\begin{align}
\hat R(M)_{\ha \hb}{}^{\hc \hd}
	&= - \frac{1}{4} (\Sigma_{\ha\hb})^{\hal\hbe}
	(\Sigma^{\hc\hd})^{\hga \hde}
	\Big( \ri W_{\hal\hbe\hga\hde} \loco
		+ 3 w_{(\hal \hbe} w_{\hga \hde)}
	\Big)
\end{align}
defining the remaining undetermined component $W_{\hal\hbe\hga\hde}\loco$
in terms of the new curvature $\hat R(M)_{\ha \hb}{}^{\hc \hd}$.
In practice, this is the most convenient definition of
$W_{\hal\hbe\hga\hde}\loco$.

In principle, one can construct expressions for
$\hat R(S)_{\ha \hb}{}_\hal^i$
and $\hat R(K)_{\ha \hb}{}^\hc$ explicitly in terms of
$\hat\phi_\hm{}_\hal^i$ and $\hat{\frak f}_\hm{}^\ha$
in analogy with \eqref{eq:hatCurvs}.
In practice, such expressions are not terribly useful since
these connections are composite quantities. Instead, we can
follow the component technique of analyzing the component Bianchi identities,
which in our case is equivalent to projecting the corresponding
superspace curvatures,
$\hat R(S)_{\ha \hb}{}_\hal^i = \hat {\mathscr{R}}(S)_{\ha\hb}{}_\hal^i\loco$
and $\hat R(K)_{\ha \hb}{}^\hc = \hat {\mathscr{R}}(K)_{\ha \hb}{}^\hc\loco$.
This results in
\begin{align}
\ri \hat R(S)_{\ha\hb}{}^i
	&= \hat{\slashed{\nabla}} \hat R(Q)_{\ha\hb}{}^i
	+ \hat\nabla^\hc (\G_{[\ha } \hat R(Q)_{\hb]\hc}{}^i)
	+ \frac{1}{8} w_{\hc\hd} \S^{\hc\hd} \hat R(Q)_{\ha\hb}{}^i
	- \frac{1}{8} w^{\hc\hd} \S_{\ha\hb} \hat R(Q)_{\hc\hd}{}^i
	\eol & \quad
	- \frac{3}{4} w^\hc{}_{[\ha } \hat R(Q)_{\hb]\hc}{}^i~, \\
\hat R(K)_{\ha\hb}{}^\hc &= \frac{1}{4} \hat \nabla_\hd \hat R(M)_{\ha\hb}{}^{\hc\hd}
	- \frac{4\ri}{3} \hat R(Q)_{\ha \hb j} \G^\hc \chi^j
	+ \frac{\ri}{2} \hat R(Q)_{\hd [\ha  j} \G_{\hb]} \hat R(Q){}^{\hc\hd j}
	\eol & \quad
	- \frac{\ri}{2} \hat R(Q)_{\ha \hd j} \G^\hc \hat R(Q)_\hb{}^{\hd j}~.
\end{align}
It is useful to note the subsidiary relations
\begin{align}
\ri \,\G^\hb \hat R(S)_{\ha\hb}{}^i
	&= -\frac{1}{2} \hat \nabla^\hb \hat R(Q)_{\ha\hb}{}^i
	- \frac{5}{8} w^{\hb\hc} \G_\hb \hat R(Q)_{\ha \hc}{}^i
	+ \frac{1}{4} w^{\hb\hc} \G_\ha \hat R(Q)_{\hb\hc}{}^i~, \eol
\ri\, \G^\ha \G^\hb \hat R(S)_{\ha\hb}{}^i &= 0~, \qquad
\hat R(K)_{\ha\hb}{}^\hb = 0~.
\end{align}

The $Q$, $S$, and $K$ transformations of the independent component
fields are unchanged from \eqref{eq:WeylSUSY}, up to the redefinitions occurring above.
These lead to
\begin{subequations} \label{eq:WeylSUSY2}
\begin{align}
\delta e_\hm{}^\ha &= \ri (\xi_j \G^\ha \psi_\hm{}^j)~, \\
\delta \psi_\hm{}_\hal^i &=
	2 \hat \cD_\hm \xi_\hal^i
	- \frac{1}{4} w_{\hc\hd} \Big(
		(\G_\hm \S^{\hc\hd})_\hal{}^\hbe 
		- 3 (\Sigma^{\hc\hd} \G_\hm)_\hal{}^\hbe \Big) \xi_\hbe^i
	+ 2 \ri \,(\G_\hm \eta^i)_\hal~, \\
\delta \cV_\hm{}^{ij} &=
	3 \xi^{(i} \hat\phi_\hm{}^{j)}
	+ 8 \ri\, \xi^{(i} \G_\hm \chi^{j)}
	- 3 \,\eta^{(i} \psi_\hm{}^{j)}~, \\
\delta b_\hm &= - \xi_k \hat\phi_\hm{}^k
	- \frac{8\ri}{3} \xi_k \G_\hm \chi^k
	- \eta_k \psi_\hm{}^k
	- 2 \,e_\hm{}^\ha \L_\ha~, \\
\delta w_{\ha\hb} &= 2\ri\,\xi_i \hat R(Q)_{\ha\hb}{}^i
	- \frac{32\ri}{3} \xi_i \S_{\ha\hb} \chi^i~, \\
\delta \chi_\hal^i &=
	\frac{1}{2} \xi_\hal^i D
	- \frac{1}{16} \hat R(J)_{\ha\hb}{}^i{}_j (\S^{\ha\hb} \xi^j)_\hal
	+ \frac{3}{128}( \hat\nabla_\ha w_{\hb\hc})
		\Big(3 (\Sigma^{\hb\hc} \G^\ha \xi^i)_\hal + (\G^\ha \S^{\hb\hc}\xi^i)_\hal \Big)
	\eol & \quad
	+ \frac{3}{256} w_{\ha\hb} w_{\hc\hd} \veps^{\ha\hb\hc\hd\he} (\G_\he \xi^i)_\hal
	- \frac{3\ri}{16} w_{\ha\hb} (\S^{\ha\hb} \eta^i)_\hal~, \\
\delta D &=
	2 \ri \,\xi_i \hat{\slashed{\nabla}} \chi^i
	+ \ri w_{\ha\hb} (\xi_i \Sigma^{\ha\hb} \chi^i)
	+ 2 \,\eta_j \chi^j~.
\end{align}
\end{subequations}

We emphasize that the supersymmetry transformations
are equivalent to \eqref{eq:WeylSUSY} and only the definition of the
composite connections have been altered.

We have already noted the resemblance between the constraints
\eqref{componentConstTRv2}
and those found in the existing literature. The supersymmetry
transformations given above turn out to coincide
very closely with those of \cite{dWK}, up to a field-dependent
$K$-transformation. The differences with the other groups are more
involved. For reference, we provide a translation table
in Appendix \ref{app:Conventions} between our conventions,
employing the redefined composite connections, and those of the other groups.


\section{The covariant projective multiplets in conformal superspace}
\label{section5}

Within the superspace approach to $\cN=1$ supergravity in five dimensions
\cite{KT-M_5D2,KT-M_5D3,KT-M08}, 
general  supergravity-matter systems are described
in terms of  covariant projective multiplets. These are curved-superspace
generalizations of the 5D superconformal projective multiplets \cite{K06}.
In this section, 
the concept of covariant projective multiplets is reformulated 
in conformal superspace, a general procedure to generate such multiplets
is given, 
and a universal  locally supersymmetric action principle is presented. 


\subsection{Covariant projective multiplets}

Let $v^i \in {\mathbb C}^2 \setminus  \{0\}$ denote 
inhomogeneous coordinates for ${\mathbb C}P^1$. 
A {\em covariant projective multiplet} of weight $n$,
$Q^{(n)}(z,v)$, is defined to be 
a conformal primary Lorentz-scalar superfield,\footnote{As a rule, we will not indicate
the $z$-dependence of  $Q^{(n)}(z,v)$.}
\bea
S_\hal^i Q^\pn=0~,
\eea
that lives on the curved superspace ${\cM}^{5|8}$, 
is holomorphic with respect to the isospinor  $v^i $ 
on an open domain of 
${\mathbb C}^2 \setminus  \{0\}$, 
and is characterized by the following properties:
\begin{itemize}
\item
it obeys the covariant analyticity constraint
\be
\de^{(1)}_{\hal} Q^{(n)}  =0~, \qquad
\de_\hal^\pu:=v_i\de_\hal^i~;
\label{ana}
\ee  
\item
it is  a homogeneous function of $v$ 
of degree $n$, that is,  
\be
Q^{(n)}(c\,v)\,=\,c^n\,Q^{(n)}(v)~, \qquad c\in 
{\mathbb C} \setminus  \{0\}~;
\label{weight}
\ee
\item
the supergravity gauge transformation \eqref{TransCD} acts on $Q^{(n)}$ 
as follows:
\bsubeq \label{4.3}
\bea
\d_\cG Q^{(n)} 
&=& \Big( \x^{{\hC}} \de_{{\hC}} + \L^{ij} J_{ij}+\s\bbD \Big) Q^{(n)} ~,  
\\ 
\L^{ij} J_{ij}  Q^{(n)}&=& -\Big(\L^{(2)} {\pa}^{(-2)} 
-n \, \L^{(0)}\Big) Q^{(n)} ~.
\label{harmult1}   
\eea
\esubeq
\end{itemize}
Here we have
introduced the differential operator
\bea
{\pa}^{(-2)} :=\frac{1}{(v,u)}u^{i}\frac{\pa}{\pa v^{i}}
~,
\label{5.28}
\eea
and also defined the parameters 
\bea
\L^{(2)} :=\L^{ij}\, v_i v_j 
~,
\qquad
\L^{(0)} :=\frac{v_i u_j }{(v,u)}\L^{ij}~,
\qquad 
(v,u):=v^iu_i
~.
\label{W2t3}
\eea
The expressions in (\ref{5.28}) and (\ref{W2t3})
involve a second isospinor  $u^{i}$ 
which is subject to
the condition $(v,u)\ne0$, but otherwise it is completely arbitrary.
The isospinors $v^i$ and $u^i$ 
 are defined to be inert under the action of the 
supergravity gauge group.
For later use, in addition to (\ref{5.28}), 
we also introduce the operators
\bea
\pa^{\pd} := (v,u) v^{i} \frac{\pa}{\pa u^{i}} 
~,~~~~~~
 \pa^{\pz}:=
 v^{i} \frac{\pa}{\pa v^{ i}} 
- u^{i} \frac{\pa}{\pa u^{i}}
~,
\label{A.26}
\eea
such that 
\bea
\label{SU(2)-de}
[ \pa^{(0)} , \pa^{(\pm 2)} ] =\pm 2 \pa^{(\pm 2)} ~, \qquad
[ \pa^{\pd} , \pa^{\pmd} ] = \pa^{\pz}~.
\eea
By construction, the superfield $Q^{(n)}$ is independent of $u$, 
i.e. $\pa  Q^{(n)} / \pa u^{i} =0$.
It is not difficult to check that the variation $\d_\cG Q^{(n)} $ defined 
by \eqref{4.3} is characterized by 
the same property, $\pa (\d_\cG Q^{(n)})/\pa u^{i} =0$,
due to (\ref{weight}). 

Since the spinor covariant derivatives satisfy 
\bea
\{\de_\hal^{(i},\de_\hbe^{j)}\}=0
~~~~~~\Longleftrightarrow~~~~~~
\{\de_\hal^\pu,\de_\hbe^\pu\}=0
~,
\label{integrability-ana}
\eea
 the analyticity constraint \eqref{ana}
 is clearly consistent with the  algebra of covariant derivatives.
However, 
we still need to check whether 
the conformal primary constraint on $Q^\pn$,  
$S_\hal^i Q^\pn=0$,
and the analyticity constraint, $\de_\hal^\pu Q^\pn=0$, are mutually consistent.
In complete analogy with the 4D $\cN=2$ supergravity analysis of 
\cite{Butter:2012xg,Butter:2014gha},  
the constraints $S_\hal^i Q^\pn=0$ and $\de_\hal^\pu Q^\pn=0$
lead to the integrability condition 
\bea
0=
\{S_{\hal}^i,\de_\hbe^\pu\}Q^\pn
=
v_j\big( 2 \eps_{\hal \hbe} \ve^{ij}\gD  + 6 \eps_{\hal\hbe} J^{ij} \big)Q^\pn
=\eps_{\hal\hbe}v^i\big( 2 \gD  -3n\big)Q^\pn~,
\eea
which 
uniquely fixes the dimension of $Q^\pn$ to be \cite{KT-M08}
\bea
\bbD Q^\pn = \frac{3 n}{2} Q^\pn ~.
\eea

The above definition of the covariant projective multiplets may be generalized by removing 
the constraint $S_\hal^i Q^\pn=0$.\footnote{Non-primary projective multiplets, which 
possess inhomogeneous super Weyl transformation laws, naturally occur within 
the SU(2) superspace approach \cite{KT-M08}.}
For instance, given a non-primary scalar  $\F$, 
the superfield $\J^{(4)} := \D^{(4)} \F $ is non-primary and analytic, 
$\nabla_\hal^{(1)} \J^{(4)} =0$, with the operator $ \D^{(4)} $
defined by \eqref{proj5.21c}.

The analyticity constraint (\ref{ana}) and the homogeneity condition (\ref{weight}) 
are consistent with the interpretation that 
the isospinor
$ v^{i} \in {\mathbb C}^2 \setminus\{0\}$ is   defined modulo the equivalence relation
$ v^{i} \sim c\,v^{i}$,  with $c\in 
 {\mathbb C} \setminus  \{0\}
$, {hence it parametrizes ${\mathbb C}P^1$}.
Therefore, the projective multiplets live in ${\cM}^{5|8} \times {\mathbb C}P^1$,
a curved five-dimensional analog of the 4D $\cN=2$ projective superspace
${\mathbb R}^{4|8} \times {\mathbb C}P^1$ \cite{KLR,LR1,LR2}.\footnote{The superspace 
${\mathbb R}^{4|8} \times {\mathbb C}P^1$ was introduced for the first time  by Rosly \cite{Rosly}.
The same superspace is at the heart of the harmonic  \cite{GIKOS,GIOS}
and projective \cite{KLR,LR1,LR2} superspace approaches.}

There exists 
a real structure on the space of projective multiplets.
Given a  weight-$n$ projective multiplet $ Q^{(n)} (v^{i})$, 
its {\it smile conjugate} $ \breve{Q}^{(n)} (v^{i})$ is defined by 
\bea
 Q^{(n)}(v^{i}) \longrightarrow  {\bar Q}^{(n)} ({\bar v}_i) 
  \longrightarrow  {\bar Q}^{(n)} \big({\bar v}_i \to -v_i  \big) =:\breve{Q}^{(n)}(v^{i})~,
\label{smile-iso}
\eea
with ${\bar Q}^{(n)} ({\bar v}_i)  :=\overline{ Q^{(n)}(v^{i} )}$
the complex conjugate of  $ Q^{(n)} (v^{i})$, and ${\bar v}_i$ the complex conjugate of 
$v^{i}$. One can show that $ \breve{Q}^{(n)} (v)$ is a weight-$n$ projective multiplet.
In particular,   $ \breve{Q}^{(n)} (v)$
obeys the analyticity constraint $\de_\hal^{(1)}\breve{Q}^{(n)} =0$,
unlike the complex conjugate of $Q^{(n)}(v) $.
One can also check that 
\bea
\breve{ \breve{Q}}^{(n)}(v) =(-1)^n {Q}^{(n)}(v)~.
\label{smile-iso2}
\eea
Therefore, if  $n$ is even, one can define real projective multiplets, 
which are constrained by $\breve{Q}^{(2n)} = {Q}^{(2n)}$.
Note that geometrically, the smile-conjugation is complex conjugation composed
with the antipodal map on the projective space ${\mathbb C}P^1$.

We now list some projective multiplets that can be  used to describe superfield 
dynamical variables.\footnote{In 4D $\cN=2$ Poincar\'e supersymmetry, 
the modern terminology for projective multiplets was introduced in \cite{G-RRWLvU}.}
A complex $\cO(m) $ multiplet, 
with $m=1,2,\dots$,   is described by a  weight-$m$ projective superfield $H^{(m)} (v)$ 
of the form:
\bea
H^{(m)} (v) &=& H^{i_1 \dots i_{m}} v_{i_1} \dots v_{i_{m}} 
~.
\eea
The analyticity constraint (\ref{ana}) is equivalent to 
\bea
\de_\hal^{(i_1} H^{i_2 \dots i_{m+1} )} =0~.
\eea
If $m$ is even, $m=2n$, we can define a real $\cO(2n) $ multiplet
obeying 
the reality condition $\breve{H}^{(2n)}  = {H}^{(2n)} $, or equivalently
\bea
\overline{ H^{i_1 \dots i_{2n}} } &=& H_{i_1 \dots i_{2n}}
=\ve_{i_1 j_1} \cdots \ve_{i_{2n} j_{2n} } H^{j_1 \dots j_{2n}} ~.
\eea
For $n>1$, 
the real $\cO(2n) $ multiplet can be used to describe an off-shell (neutral) hypermultiplet. 

The $\cO(m)$ multiplets,  $H^{(m)} (v)$, 
 are well defined on the entire projective space ${\mathbb C}P^1 $.
There also exist important projective multiplets that are defined only on an open domain 
of ${\mathbb C}P^1 $. Before introducing them, let us give a few definitions. 
We define the {\it north chart} of ${\mathbb C}P^1$ to consist of those points for which 
the first component  of $v^i = (v^{\1}, v^{\2})$ is non-zero,  $v^{\1} \neq 0$.
The north chart may be parametrized by the complex inhomogeneous coordinate 
$\z= v^{\2}/v^{\1} \in \mathbb C$. The only point of ${\mathbb C}P^1 $ outside the north 
chart is characterized by $v_\infty^i = (0, v^{\2})$ and describes an infinitely separated point.
Thus we may think of the projective space ${\mathbb C}P^1 $
as  ${\mathbb C}P^1 ={\mathbb C} \cup \{\infty \}$. 
The {\it south chart} of ${\mathbb C}P^1$ is defined to consist of those points for which 
the second component  of $v^i = (v^{\1}, v^{\2})$ is non-zero,  $v^{\2} \neq 0$.
The south chart is naturally parametrized by $1/\z$.
The intersection of the north and south charts is ${\mathbb C} \setminus \{ 0\}$.

An off-shell (charged) hypermultiplet can be described in terms of the so-called {\it arctic} 
weight-$n$ multiplet $\U^{(n)} (v)$ which is defined to be 
holomorphic  in the north chart ${\mathbb C}P^1$: 
\bea
\U^{(n)} ( v) &=&  (v^{\1})^n\, \U^{[n]} ( \z) ~, \qquad 
\U^{ [n] } ( \z) = \sum_{k=0}^{\infty} \U_k  \z^k 
~. 
\label{arctic1}
\eea
Its smile-conjugate {\it antarctic} multiplet $\breve{\U}^{(n)} (v) $, has the explicit form 
 \bea
\breve{\U}^{(n)} (v) &=& 
(v^{\2}  \big)^{n}\, \breve{\U}^{[n]}(\z) =
(v^{\1} \,\z \big)^{n}\, \breve{\U}^{[n]}(\z) ~, \qquad
\breve{\U}^{[n]}( \z) = \sum_{k=0}^{\infty}  {\bar \U}_k \,
\frac{(-1)^k}{\z^k}
\label{antarctic1}
\eea
and is holomorphic in the south chart of ${\mathbb C}P^1$.
The arctic multiplet can be  coupled to a Yang-Mills multiplet
in a complex representation of the group $G_{\rm YM}$.
The pair consisting of $\U^{[n]} ( \z)$ and $\breve{\U}^{[n]}(\z) $ 
constitutes the so-called polar weight-$n$ multiplet.

Our last example is a real {\it tropical} multiplet $\cU^{(2n)} (v) $ of weight $2n$ defined by 
\bea
\cU^{(2n)} (v) &=&\big({\rm i}\, v^{1} v^{2}\big)^n \cU^{[2n]}(\z) =
\big(v^{1}\big)^{2n} \big({\rm i}\, \z\big)^n \cU^{[2n]}(\z)~,  \non \\
\cU^{[2n]}(\z) &=& 
\sum_{k=-\infty}^{\infty} \cU_k  \z^k~,
\qquad  {\bar \cU}_k = (-1)^k \cU_{-k} ~.
\label{2n-tropica1}
\eea
This multiplet is holomorphic in the intersection of the north and south charts 
of the projective space ${\mathbb C}P^1$.


\subsection{Analytic projection operator}
\label{Projector}

In this subsection we show how to engineer covariant projective multiplets
by making use of an analytic projection operator.

Let us start with a simple observation.
Due to \eqref{integrability-ana}, the spinor covariant derivatives satisfy
\be 
\nabla^{(i}_{\hal} \nabla^j_{\hbe} \nabla^k_{\hga} \nabla^l_{\hde} \nabla^{p)}_{\hat{\rho}} 
= \nabla^{(i}_{[\hal} \nabla^j_{\hbe} \nabla^k_{\hga} \nabla^l_{\hde} \nabla^{p)}_{\hat{\rho}]} = 0 
~~~
\Longleftrightarrow
~~~
\nabla^\pu_{\hal} \nabla^\pu_{\hbe} \nabla^\pu_{\hga} \nabla^\pu_{\hde} \nabla^\pu_{\hat{\rho}} 
=0
~.
\ee
Hence, if we define the operators
\bsubeq
\bea 
\D^{ijkl} 
&:=& 
- \frac{1}{96} \eps^{\hal\hbe\hga\hde} \nabla_\hal^{(i} \nabla_\hbe^j \nabla_\hga^k \nabla_\hde^{l)}
= - \frac{1}{32} \nabla^{(ij} \nabla^{kl)}
=\D^{(ijkl)} ~,
\\
\nabla^{ij}&:=&\de^{\hal i}\de_\hal^j=\nabla^{(ij)}
~,
~~~~~~
\nabla^\pd:=\de^{\hal \pu}\de_\hal^\pu
~,
\\
\D^\pq&:=& v_iv_jv_kv_l\D^{ijkl}=-\frac{1}{32}(\nabla^\pd)^2
~,
\label{proj5.21c}
\eea
\esubeq
it clearly holds that 
\bea
\de_\hal^\pu\D^\pq
=
\D^\pq\de_\hal^\pu
=0 ~.
\label{zero-proj}
\eea 
One may prove that $\D^{ijkl}$ satisfies the relations
\bea
\de_\hal^p\D^{ijkl}
=
\frac{4}{5}\ve^{p(i}\de_{\hal q}\D^{jkl)q}
~,~~~~~~
\D^{ijkl}\de_\hal^p
=
\frac{4}{5}\ve^{p(i}\D^{jkl)q}\de_{\hal q}
~.
\label{D4D2-D2D4}
\eea
The operator $\D^\pq$ is called the \emph{analytic projection operator}.
Given any superfield $U$, the superfield $Q:=\D^\pq U$ satisfies the analyticity condition
\eqref{ana}.
On the other hand, in order for $Q$ to be a covariant projective superfield,  
$U$ has to be constrained.
In \cite{KT-M08} it was proven in 
SU(2) superspace that the right prepotential for 
a covariant weight-$n$ projective superfield is an isotwistor superfield of 
weight $(n-4)$.\footnote{The concept of isotwistor superfields was introduced
in the context of 4D $\cN=2$ supergravity \cite{KLRT-M_4D-1}.}

By definition, a weight-$n$ isotwistor superfield  $U^{(n)}$ is a primary tensor superfield (with suppressed Lorentz indices)
that  lives on  ${\cM}^{5|8}$, 
is holomorphic with respect to 
the isospinor variables $v^i $ on an open domain of 
${\mathbb C}^2 \setminus  \{0\}$, 
is a homogeneous function of $v^i$ of degree $n$,
\bsubeq
\bea
&&U^{(n)}(c\,v)\,=\,c^n\,U^{(n)}(v)~, \qquad c\in 
{\mathbb C} \setminus  \{0\}
~,
\eea
and is  characterized by the supergravity gauge transformation
\bea
\d_\cG U^{(n)} 
&=& \Big( \x^{{\hC}} \de_{{\hC}} 
+  \hf \L^{\ha\hb} M_{\ha\hb}
+\L^{ij} J_{ij} 
+\s\bbD
\Big) 
U^{(n)} ~,  ~~~
\non \\
J_{ij}  U^{(n)}&=& -\Big(v_{(i}v_{j)}{\pa}^{(-2)} 
-\frac{n}{(v,u)}v_{(i}u_{j)}\Big) U^{(n)}
~.
\label{iso2}
\eea 
\esubeq
It is clear that any weight-$n$ projective multiplet  is an isotwistor superfield, but not vice versa.
The main property in the definition of isotwistor superfields is their transformation rules under SU(2).
In principle, the definition could be extended to consider non-primary superfields.

Let  $U^{(n-4)}$ be a Lorentz-scalar isotwistor 
such that 
\bea
\bbD  U^{(n-4)}&=&
\hf (3n-4)  U^{(n-4)}~.
\label{iso3}
\eea
Then  the weight-$n$ isotwistor superfield 
\bea
Q^{(n)}:=\D^{(4)}U^{(n-4)}
\eea
satisfies all the properties of a covariant projective multiplet.
Note that $Q^{(n)}$ is clearly analytic with $\bbD Q^\pn = \frac{3 n}{2} Q^\pn$.
It is an instructive  exercise to check that $Q^{(n)}$ is primary.
We define the operators
\bea
S_\hal^\pu:=v_iS_\hal^i
~,~~~~~~
S_\hal^\pmu:=\frac{u_i}{(v,u)}S_\hal^i
~,
\eea
which satisfy
\begin{align}
\{ S_\hal^\pu , \nabla_\hbe^\pu \} = 6 \eps_{\hal\hbe} J^\pd ~,~~~~~~
\{ S_\hal^\pmu , \nabla_\hbe^\pu \} = 2 \eps_{\hal\hbe} \gD - 4 M_{\hal\hbe} +6 \eps_{\hal\hbe} J^{(0)} 
~ , 
\label{Sde}
\end{align}
where
\bsubeq
\bea
J^\pd&:=&v_iv_jJ^{ij}
~,~~~
{[}J^\pd,\de_\hal^\pu{]}=0~,~~~
J^\pd U^\pn=0
~,
\\
J^\pz&:=&\frac{v_iu_j}{(v,u)}J^{ij}
~,~~~
{[}J^\pz,\de_\hal^\pu{]}=-\hf\de_\hal^\pu
~,~~~
J^\pz U^\pn=-\frac{n}{2} U^\pn
~.
\eea
\esubeq
After some algebra, it can be proven that 
\bsubeq
\bea
{[}S_\hrh^\pu,\D^\pq{]}
&=&
 - \frac{1}{4} \eps^{\hal\hbe\hga\hde} 
\eps_{\hrh\hal} \nabla_\hbe^\pu \nabla_\hga^\pu \nabla_\hde^\pu J^{(2)}
~,
\label{S+D4}
\\
{[}S_\hrh^\pmu,\D^\pq{]}
&=&
 \frac{1}{24} \eps^{\hal\hbe\hga\hde}\nabla_\hbe^\pu\nabla_\hga^\pu \nabla_\hde^\pu \Big{[}
 \eps_{\hrh\hal}  
\big(8-2  \bbD - 6 J^{(0)}\big)
+4 M_{\hrh\hal} \Big{]}
   ~.
\label{S-D4}
\eea
\esubeq
Using these results, it immediately follows that $S_\hal^i\D^\pq U^\pn=0$.

Let us conclude this subsection by giving the expression for $\D^\pq $
in SU(2) superspace. This can be computed by simply using the degauging procedure developed 
in section \ref{degauging}.
The result is
\begin{align} 
\D^\pq U^{(n-4)} &= 
- \frac{1}{32} \cD^\pd \cD^\pd U^{(n-4)} +\frac{1}{32} \cD^\pd \big(\frak{F}^{\pd\hal  \hbe } \{S_\hbe^\pmu, \nabla_\hal^\pu\} U^{(n-4)}\big) \non\\
&\qquad+\frac{1}{32} \cD^{\hal \pu} \big(\frak{F}^\pd_\hal{}^{\hbe}
{[} S_\hbe^\pmu, \nabla^\pd{]} U^{(n-4)}\big) \non\\
&\qquad+ \frac{1}{32} \frak{F}^\pd{}^{\hal \hbe} 
\{S_\hbe^\pmu, \nabla_{\hal}^\pu \nabla^\pd\} U^{(n-4)} \ ,
\label{4.30}
\end{align}
with
\bea
\cD_\hal^\pu:=v_i\cD_\hal^i
~,~~~
\cD^\pd
:=\cD^{\pu\hal}\cD_\hal^\pu
~,~~~
\frak{F}^\pd_{\hal \hbe}
:=
v_iv_j\frak{F}_{\hal}^i{}_\hbe^j
~.
\eea
Computing  the (anti-)commutators involving $S_\hbe^\pmu $ in \eqref{4.30} 
produces new terms involving $\nabla$, which have to be degauged.  
Finally, making use of the identities
\bsubeq
\bea
\cD^\pq &:=&-\frac{1}{96} \ve^{\hal\hbe\hga\hde}\cD_\hal^\pu\cD_\hbe^\pu\cD_\hga^\pu\cD_\hde^\pu
~,
\\
\cD^\pq &=&-\frac{1}{96}\Big{[} \,
3 \cD^\pd \cD^\pd 
+ 24 (\cD^{\hal \pu} \frak{F}^\pd_{\hal\hbe})\cD^{\hbe \pu}
 + 4( \cD^{\hal \pu} \frak{F}^\pd{}^{\hbe}{}_\hbe)\cD_\hal^\pu \non\\
&&~~~~~~~
+ 24 \frak{F}^\pd_{\hal\hbe} \cD^{\hal \pu} \cD^{\hbe \pu} 
 + 4 \frak{F}^\pd{}^{\hbe }{}_\hbe \cD^\pd\Big{]}
  U^{(n-4)} \ , \\
\cD_\hal^\pu \frak{F}^\pd_{\hbe\hga}&=& \cD_{[\hal}^\pu \frak{F}^\pd_{\hbe\hga]} 
\ ,
\\
\cD_\hal^\pu S^\pd &=& \frac{1}{10} \cD^{\hbe \pu} C_{\hal\hbe}^\pd
~,\qquad
S^\pd:=v_iv_jS^{ij}~,\quad
C_\ha^\pd:=v_iv_jC_\ha^{ij}
\ ,
\eea
\esubeq
we obtain
\bea
\D^\pq U^{(n-4)} 
&=& 
\Big{[} \,\cD^\pq 
- \frac{5 \ri}{12} S^\pd \cD^\pd 
- \frac{\ri}{8} C^{\hal\hbe \pd} \cD_\hal^\pu \cD_\hbe^\pu
- \frac{\ri}{6} (\cD^{\hal \pu} C_{\hal\hbe}^\pd) \cD^{\hbe \pu} \non\\
&&~
- \frac{\ri}{20} (\cD^{\hal \pu} \cD^{\hbe \pu} C_{\hal\hbe}^\pd )
+ 3 (S^\pd)^2 
+ \frac{1}{4} C^{\ha \pd} C_{\ha}^\pd
\Big{]} U^{(n-4)} \ .
\eea
This relation determines the analytic projection operator in SU(2) superspace, 
which is a new result. 
In \cite{KT-M08},  this operator
was computed only in a super Weyl gauge in which $C_\ha^{ij} = 0$.


\subsection{The action principle}

We turn to re-formulating the supersymmetric action principle given in \cite{KT-M08}
in conformal superspace. 

Consider a Lagrangian $\cL^{(2)}$ chosen  
to be a real weight-2 projective multiplet.
Associated with $\cL^{(2)}$ is the action\footnote{In parallel with the construction
in four dimensions \cite{Butter:2014gha}, it is possible to 
integrate out half of the Grassmann coordinates
thus representing the action as an integral  of $\cL^{(2)}$
over an analytic subspace. (This is analogous to the chiral integral 
in 4D $\cN=1$ supergravity, see \cite{WB,GGRS,BK} for reviews.) 
We find it more
convenient to employ the superfield $C^{(-4)}$ to always deal with  
full superspace integrals.}
\bea
S[ \cL^{(2)} ]&=&
\frac{1}{2\pi} \oint_\g (v, \rd v)
\int \rd^{5|8}z\, E\, C^{(-4)}\cL^{(2)}~. 
\label{InvarAc}
\eea
Here the superfield $C^{(-4)}$ is required to be a Lorentz-scalar 
primary isotwistor superfield of weight $-4$ 
such that the following two conditions hold:
\bea
\D^{(4)}C^{(-4)}=1~, \qquad \bbD C^{(-4)}&=&-2C^{(-4)}~.
\label{AcComp-b}
\eea
These conditions prove to guarantee 
that the action (\ref{InvarAc}) is invariant under the full  supergravity gauge group $\cG$.
The invariance of $S[ \cL^{(2)} ]$ under the Lorentz and special conformal transformations 
is obvious, since all the superfields in the action are Lorentz-scalar primary superfields. 
Invariance under the general coordinate transformations is also trivial, while
invariance under the SU(2) transformations can be shown in complete analogy with 
the proof given in \cite{KT-M_5D3,KT-M08}. 
It remains to prove that the action is invariant under dilatations.
This simply follows from the observation
that the measure, $E$, has dimension $-1$.

All information about a dynamical system is encoded in its Lagrangian $\cL^{(2)}$.
The important point is that the action (\ref{InvarAc}) does not 
depend on  $C^{(-4)}$ if the Lagrangian $\cL^{(2)} $ is independent of $C^{(-4)}$.
To prove this statement, let us represent the Lagrangian as
$\cL^{(2)}=\D^{(4)} \cU^{(-2)}$, for some
isotwistor superfield  $\cU^{(-2)}$ of weight $-2$.
We note that
for any pair of Lorentz-scalar isotwistor superfields $\F^{(n-4)}$ and $\Psi^{(-n-2)}$
such that
\bea
\bbD \F^{(n-4)}=  \hf(3n-4) \F^{(n-4)}
~,~~~~~~
\bbD\Psi^{(-n-2)}=\hf(2-3n) \Psi^{(-n-2)}
~,
\eea
we can use integration by parts to prove the following relation
\bea
\oint_\g (v, \rd v)\int \rd^{5|8}z\, E\, 
\Big\{
\F^{(n-4)}\D^\pq \Psi^{(-n-2)}
-\Psi^{(-n-2)}\D^\pq \F^{(n-4)}
\Big\}
=0
~.
\eea
If we use this result 
and  eq.~(\ref{AcComp-b}), 
we can rewrite the action  in the form
\bea
S&=&
\frac{1}{2\pi} \oint_\g (v ,\rd v)
\int \rd^{5|8}z\, E\, \cU^{(-2)}~.
\label{InvarAc2}
\eea
This representation makes manifest the fact that  the action does not depend on $C^{(-4)}$.
Upon degauging to SU(2) superspace, 
the action \eqref{InvarAc2}  coincides with the one given in  \cite{KT-M08}.

A natural choice for  $C^{(-4)}$ is available  if the theory under consideration 
possesses an abelian  vector 
multiplet such that its field strength $W$ is nowhere vanishing.
Given $W$, we can construct a composite $\cO(2)$ projective multiplet
as
\be 
 H_{\rm VM}^\pd = \frac{\ri}{2} W \nabla^\pd W + \ri (\nabla^{\hal\pu} W) \nabla_\hal^\pu W 
=v_iv_j H_{\rm VM}^{ij}
~ ,
\label{HVM}
\ee
where $ H_{\rm VM}^{ij}$ coincides with eq.~\eqref{YML} for a single abelian vector multiplet.
By using the Bianchi identity \eqref{vector-Bianchi}, which implies
\bea
\de_\hal^\pu\de_\hbe^\pu W=\frac{1}{4}\ve_{\hal\hbe}\de^\pd W
~,~~~~~~
\de_\hal^\pu\de_\hbe^\pu\de_\hga^\pu W=0
~,
\eea
it is a simple exercise to show that $H_{\rm VM}^\pd$ is an analytic superfield, $\de_\hal^\pu H_{\rm VM}^\pd=0$.
By using \eqref{Sde}, it also simple to show that $S_\hal^\pu H_{\rm VM}^\pd=S_\hal^\pmu H_{\rm VM}^\pd=0$
 and then
$H_{\rm VM}^\pd$ is primary, $S_\hal^i H_{\rm VM}^\pd=0$.
We can then introduce
\be 
C^{(-4)} = \frac{4 W^4}{3 ( H_{\rm VM}^{(2)})^2} \ ,
\label{CWS}
\ee
which consistently defines a weight $-4$ isotwistor superfield that, due to
\bea
\D^{(4)} W^4 = \frac{3}{4} ( H_{\rm VM}^\pd)^2 
\ ,
\eea
satisfies $\D^{(4)} C^\pmq=1$.
The resulting action principle takes the form
\be 
S[\cL^\pd ] = \frac{2}{3 \pi} \oint (v , \rd v) \int \rd^{5|8}z\, E \ \frac{\cL^\pd W^4}{( H_{\rm VM}^\pd)^2} \ .
 \label{actF}
\ee
Upon degauging to SU(2) superspace, 
this action reduces to the one proposed in \cite{KT-M08}. 

We conclude this section by mentioning that 
the action (\ref{InvarAc}) 
is characterized by 
the following important property:
\begin{subequations}\label{4.42}
\bea
S\Big[G^{(2)} (\l +  \breve{\l} )\Big] =0~, 
\label{44.42a}
\eea
with 
$G^{(2)} $ a real $\cO(2)$ multiplet and $\l$ an arctic weight-zero multiplet.
Since $\l$ is arbitrary, the above relation is equivalent to 
\bea
S\big[G^{(2)} \l\big] =0~.
\label{4.42b}
\eea
\end{subequations}
A proof of \eqref{4.42} will be given in section \ref{subsection6.2}.


\section{Prepotentials for the vector multiplet}
\label{N1SYM_and_PS}

In this section we develop a prepotential formulation for 
the Yang-Mills multiplet introduced in section \ref{SYM}.
Our presentation is very similar to that given in  \cite{KT-M_2014} 
in the case of 3D $\cN=4$ conformal supergravity.
The latter was inspired by the pioneer works of Lindstr\"om and Ro\v{c}ek 
\cite{LR2} and Zupnik \cite{Zupnik87}
devoted to the  4D $\cN=2$ super Yang-Mills  theory. 

\subsection{Tropical prepotential}

The Yang-Mills multiplet in conformal superspace has been described 
in section \ref{SYM}.
The field strength $\bm W$ appears in the anti-commutator of two spinor covariant derivatives as
\bea
\{{\bm \de}_\hal^{i}, {\bm \de}_\hbe^{j}\}&=& \dots
+2\ve_{\hal\hbe} \ve^{ij} \bm W
~,
\eea
where the ellipsis stands for the purely supergravity part. 
Let us  introduce the gauge covariant operators 
\bea
{\bm \de}_\hal^{(1)}:=v_{i} {\bm \de}_\hal^{i}~. 
\label{A.6}
\eea
It may be seen that they strictly anti-commute with each other, 
\bea
\{{\bm \de}_\hal^{(1)} , {\bm \de}_\hbe^{(1)} \}=0
~.
\eea
This means that ${\bm \de}_\hal^{(1)}$ may be represented in the form:
\bea
{\bm \de}_\hal^{(1)} = \re^{\bm \O_+} \de_\hal^{(1)} \re^{-\bm \O_+}~, 
\label{A.7}
\eea
where $\bm \O_{+}$ denotes 
a Lie-algebra-valued bridge superfield of the form 
\bea
\bm \O_{+}(v)= \bm \O_{+}(\z)  = \sum_{n=0}^{\infty} \bm \O_n \z^n~, \qquad \z:=\frac{v^{\2}}{v^{\1}}~.
\label{A.8}
\eea
The bridge is a covariant weight-0 isotwistor superfield.
Another representation for ${\bm \de}_\hal^{(1)}$ follows 
by applying the smile-conjugation to \eqref{A.7}.
The result is 
\bea
{\bm \de}_\hal^{\pu } = \re^{- \bm \O_-} \de_\hal^{\pu } \re^{\bm \O_-}~, \qquad 
\bm \O_{-}(v) = \bm \O_{-} (\z) 
= \sum_{n=0}^{\infty} (-1)^n \, \bm \O_n^\dagger \,
\frac{1}{ \z^n }~.
\label{A.9}
\eea
We now introduce a Lie algebra-valued superfield $\bm V(\z)$ defined by 
\bea
{\rm e}^{\bm V} := {\rm e}^{\bm \O_-} {\rm e}^{\bm \O_+} ~, \qquad 
\bm V(v) = \bm V(\z ) 
=\sum_{n=-\infty}^{\infty} \bm V_n \z^n~, \qquad \bm V_n^\dagger = (-1)^n \bm V_{-n}~.
\eea
It may be seen from \eqref{A.7} and \eqref{A.9} that $\bm V$ is a covariant 
weight-0 projective multiplet, 
\bea
\de_\hal^{\pu } \bm V = 0~. \label{A.11}
\eea

In accordance with \eqref{A.7}, the gauge transformation law of $\bm \O_+$ is 
\bea
{\rm e}^{\bm \O_+' (\z)} =  {\rm e}^{{\rm i} \bm \t }  
{\rm e}^{\bm \O_+(\z) }  {\rm e}^{ -{\rm i} \bm \l (\z) } ~,
\eea
where the new gauge parameter $\bm \l(\z) $ is a covariant weight-zero arctic multiplet
\bea
\de_\hal^{\pu } {\bm \l} = 0~, 
\qquad 
{\bm \l}(\z) =  \sum_{n=0}^{\infty} {\bm \l}_n \z^n~.
\eea
The gauge transformation law of the tropical prepotential is 
\bea
{\rm e}^{\bm V' } =  {\rm e}^{{\rm i}\breve{\bm \l} }  {\rm e}^{\bm V }  
{\rm e}^{ -{\rm i} \bm \l } 
~.
\label{A.14} 
\eea
Hence $\bm V$ transforms under the $\l$-group only. 

\subsection{Polar hypermultiplets} 

Supersymmetric matter in arbitrary representations of the gauge group 
$G_{\rm YM}$
may be described in terms of gauge covariantly
arctic multiplets and their smile-conjugate antarctic multiplets. 

A gauge covariantly arctic multiplet of weight $n$, ${\bm \U}^{\pn} (v)$,
is defined by 
\bea
{\bm \de}^{\pu }_{ \hal} {\bm \U}^{\pn} =  0~, 
\qquad 
{\bm \U}^{\pn}(v) = (v^{\1})^n  \sum_{k=0}^{\infty} {\bm \U}_k \z^k~.
\eea
It can be represented in the form
\bea
{\bm \U}^{(n)} (v )= {\rm e}^{\bm \O_+(v)} \U^{(n)} (v)~, 
\eea
where $\U^{(n)} (v) $ is an ordinary covariant arctic multiplet of weight $n$ 
as already introduced in eq.~\eqref{arctic1}.

Computing the smile conjugate of ${\bm \U}^{\pn} (v)$ gives
a gauge covariantly antarctic multiplet of weight $n$, $\breve{\bm \U}{}^{\pn}(v) $, 
with the properties
\bea
\breve{\bm \U}{}^{\pn}  \stackrel{\longleftarrow}{{\bm \de}^{\pu }_{ \hal}} =0~, 
\qquad 
\breve{\bm \U}{}^{\pn}(v) =  (v^{\2})^n \sum_{k=0}^{\infty} (-1)^k{\bm \U}_k^\dagger \frac{1}{\z^k}~.
\eea
It can be represented in the form
\bea
\breve{\bm \U}{}^{\pn} (v )=\breve{\U}^{\pn}(v)  {\rm e}^{\bm \O_-(v)}   ~, 
\eea
where $\breve{\U}^{\pn} (v) $ is an ordinary antarctic multiplet as in eq.~\eqref{antarctic1}. 

The 
arctic multiplet of weight $n$, ${\bm \U}^{\pn}(v) $, and its smile-conjugate, 
 $\breve{\bm \U}{}^{\pn} (v) $, constitute the 
 polar multiplet of weight $n$. 
The gauge transformation laws of  ${\bm \U}^{\pn} (v) $ and  $\breve{\bm \U}{}^{\pn} (v) $
are 
\bea 
 {\bm \U}^{\pn}{}' (v)= \re^{\ri \t} {\bm \U}^{\pn}(v)~, 
 \qquad \breve{\bm \U}{}^{\pn}{}' (v)= \breve{\bm \U}{}^{\pn}  (v) \re^{-\ri \bm \t } ~.
\eea 
The gauge transformation laws of  ${ \U}^{\pn} (v) $ and  $\breve{ \U}^{\pn} (v) $
are 
\bea
{ \U}^{\pn}{}'(v) = \re^{\ri \bm \l (v)} { \U}^{\pn}(v) ~, \qquad
\breve{ \U}{}^{\pn}{}' (v)= \breve{ \U}{}^{\pn}    (v)\re^{-\ri \breve{\bm \l}(v) }~.
\eea
In the $n=1$ case, a gauge invariant hypermultiplet Lagrangian
can be constructed and is given by 
\bea
\cL^{\pd} = \ri \breve{\bm  \U}{}^{\pu} {\bm \U}^{\pu} 
= \ri \breve{ \U}{}^{\pu}  \re^{\bm V} { \U}^{\pu} ~.
\eea

\subsection{Arctic and antarctic representations}

Here we demonstrate that the Yang-Mills gauge connection $\bm V_\hA$, 
eq. \eqref{SYM-derivatives}, 
may be expressed in terms
of the tropical prepotential $\bm V (\z)$, modulo the $\t$-gauge freedom. 

Let us introduce the operator
\bea 
{\bm \de}^{\pmu}_{ \hal} := \frac{1}{(v,u)} u_{i}{ \bm \de}^{i}_{ \hal} ~.
 \eea
It can be seen that 
\bea
\{ {\bm \de}^{\pu  }_{ \hal} , {\bm \de}^{\pmu }_{ \hbe} \}
=  \dots
-2\ve_{\hal\hbe} \bm W ~.
\label{A25}
\eea
Here the ellipsis denotes  purely supergravity terms. 
Note that the  operators $\pa^\pd,\,\pa^\pmd$ and $\pa^\pz$
 are  invariant under the $\t$-group transformations and 
obey 
\bsubeq\label{A.28} 
\bea
&[\pa^{\pd} , {\bm \de}^{\pu }_{ \hal} ] =[\pa^{\pmd} , {\bm \de}^{\pmu }_{ \hal} ]=0
~,
\\
&
[\pa^{\pd} , {\bm \de}^{\pmu }_{ \hal} ] = {\bm \de}^{\pu }_{ \hal}
~,~~~
[\pa^{\pmd} , {\bm \de}^{\pu }_{ \hal} ] = {\bm \de}^{\pmu  }_{ \hal}
~,
\\
&
[\pa^{\pz} , {\bm \de}^{\pu }_{ \hal} ] = {\bm \de}^{\pu  }_{ \hal}
~,~~~
[\pa^{\pz} , {\bm \de}^{\pmu }_{ \hal} ] = -{\bm \de}^{\pmu }_{ \hal}
~.
\eea
\esubeq

When dealing with polar hypermultiplets, it is useful to introduce 
an arctic representation  defined by the transformation  
\bea
\hat \cO \to \hat{\cO}_+ := {\rm e}^{-\bm \O_+} \hat{\cO}\,{\rm e}^{\bm \O_+} ~, 
\qquad U \to U_+ := {\rm e}^{-\bm \O_+} U
\label{A.29}
\eea
applied to any gauge covariant operator $\hat \cO$ and 
matter superfield $U$.\footnote{It is assumed that the gauge transformation law of 
$\hat \cO$ is $\hat \cO \to \hat \cO '= \re^{\ri \bm \t} \hat \cO \re^{-\ri \bm \t}$, 
while $U$ transforms as in \eqref{2.2}.}
In the arctic representation, any gauge covariantly arctic multiplet 
${\bm \U}^{\pn} (v)$
becomes the ordinary arctic one, ${\U}^{\pn} (v)$, 
\bea
{\bm \U}^{\pn} (v) \to { \U}^{\pn} (v)~, \qquad
\breve{\bm \U}{}^{\pn} (v ) \to \breve{\U}{}^{\pn}(v)  {\rm e}^{\bm V(\z)}   ~,
\eea
and the gauge covariant derivatives ${\bm \de}^{\pu }_{ \hal} $ turn into 
the standard ones,
\bea
{\bm \de}^{\pu }_{ \hal} \to \de^{\pu }_{ \hal} ~. 
\eea
The important point is that the projective derivative $\pa^{\pmd} $ 
is replaced by 
the operator 
\bea
\pa^{\pmd} \to \pa_+^{\pmd} := \pa^{\pmd} 
+  {\rm e}^{-\bm \O_+} ( \pa^{\pmd}   {\rm e}^{\bm \O_+} )~,
\eea
which transforms as a covariant derivative under the $\l$-group. 
It is also important  to mention that $\pa^\pd$ remains short in the arctic representation,
$\pa_+^\pd=\pa^\pd$.
Making use of the arctic-representation version of 
\eqref{A25}
as well as the relation 
\bea
{\bm \de}^\pmu_{+\hal}
={[}\pa_+^\pmd,\de_\hal^\pu{]}
=\de_\hal^\pmu
-\de_\hal^\pu\big(\re^{-\bm \O_+}\pa^\pmd \re^{\bm \O_+}\big)
~,
\eea
we read off
\bea
\bm W_+ = \frac{1}{8} 
\de^\pd
\Big( {\rm e}^{-\bm \O_+}  \pa^{\pmd}   {\rm e}^{\bm \O_+}  \Big)
~.
\label{A.33}
\eea
Since
$\pa^{\pd} \bm W_+ =0$,
$\bm W_+ $ is independent of $u^{i}$. 
The field strength $\bm W_+$ also satisfies the property 
\bea
 \pa_+^{\pmd} \bm W_+ =0~,
 \eea
 since in the original representation $\bm W$ is independent of $v^{ i}$.
The field strength can be seen to obey the Bianchi identity 
\bea
{\bm \de}_{+\hal}^{(i}{\bm \de}_{+\hbe}^{j)}\bm W_+
= \frac{1}{4}\ve_{\hal\hbe}
{\bm \de}_{+}^{\hga(i}{\bm \de}_{+\hga}^{j)}\bm W_+ ~.
\eea

In the case of a U(1) gauge group, 
$\bm W = W T$, with $T$ the U(1) generator, 
we have   $W=W_+$ 
and eq. \eqref{A.33} turns into
\bea
W
= 
\frac{1}{8}  
\de^\pd
\pa^{\pmd}
\O_+
~.
\eea
Since $\O_+ (v) = \O_+ (\z) $, in the north chart of ${\mathbb C}P^1$
we can represent 
\bea
\pa^{\pmd}\O_+(v)
=
-\frac{1}{(v^\1)^2}
\pa_\z\O_+(\z)
~.
\eea
Taking into account 
the fact that $W$ is independent of $\z$, 
it is simple to show that
\bea
W
=
-\frac{1}{8}
\de^{\2\2}
\O_1
=
\frac{1}{8}\de^{\1\1}\O_{-1}
~.
\label{propot-YM-1}
\eea

In complete analogy with the arctic representation, eq. (\ref{A.29}), 
we can introduce the antarctic representation defined by 
\bea
\hat \cO \to \hat{\cO}_- := {\rm e}^{\bm \O_-} \hat{\cO}\,{\rm e}^{-\bm \O_-} ~, 
\qquad U \to U_- := {\rm e}^{\bm \O_-} U~.
\eea
In this representation, the super Yang-Mills field strength takes the form
\bea
\bm W_- = \frac{1}{8}
\de^{\pd}
\Big( {\rm e}^{\bm \O_-}  \pa^{\pmd}   {\rm e}^{-\bm \O_-}  \Big)~.
\eea
Comparing the above with \eqref{A.33} gives 
\bea
\bm W_- = {\rm e}^{\bm V} \bm W_+ {\rm e}^{-\bm V}~.
\eea

\subsection{Abelian field strength: Contour integral representation}

In the previous subsection, in the case of an abelian vector multiplet, we have 
derived the result \eqref{propot-YM-1}.
This expresses the 
field strength in terms of the bridge components. 
It is useful to find yet another representation given in terms of the real weight-zero tropical prepotential
\bea
V
=
\O_+
+
\O_-
~.
\eea
It turns out that the expression
\be 
W 
= 
- \frac{\ri}{16 \pi} \oint (v, \rd v) \nabla^{(-2)} V 
~,~~~~~~
\de^\pmd:= \frac{1}{(v,u)^2}u_iu_j\de^{ij}
~,
\label{defWV}
\ee
is equivalent to \eqref{propot-YM-1}. 
It is instructive to prove this statement.

First of all, the expression for $W$ in \eqref{defWV} can be shown to be independent of $u_i$.
To see this, consider a shift 
\be u_i \rightarrow u_i + \d u_i
\ee
and represent it as
\be \d u_i = (v,u)v_i \a^\pmd + u_i \b^\pz \ , 
\quad 
\a^\pmd = - \frac{u^i \d u_i }{(v, u)^2} \ , 
\quad \b^\pz = \frac{v^i \d u_i}{(v,u)} \ .
\ee
Then one can compute
\bea
 \d W 
&=&
 - \frac{\ri}{16 \pi} \oint (v, \rd v)\, \a^\pmd \{ \nabla^{\hal\pu} , \nabla_\hal^\pmu \} V = 0 
 ~ ,
\eea
which is identically zero since
\bea
 \{ \nabla_\hal^\pu , \nabla_\hbe^\pmu \} 
 =
 2 \ri   \nabla_{\hal\hbe} +2 \ri \eps_{\hal \hbe} \mathscr{W}
 ~ ,\quad \implies
 \{ \nabla^{\hal \pu} , \nabla_\hal^\pmu \} 
 =
8 \ri\mathscr{W}
~,
\eea
and  $\mathscr{W} V\equiv 0$.
In the north chart of $\mathds C P^1$ we have
\be 
V(v) = V ( \z) = \displaystyle\sum\limits_{k=-\infty}^\infty \z^k V_k \ , \quad V_k = (-1)^k \bar{V}_{-k} 
\ .
\ee
Then choosing $u_i=(0,1)$ we can represent $W$ as follows
\be 
W 
= \frac{\ri}{16 \pi} \oint \frac{\rd \z}{\z^2} \nabla^{\2\2} V(\z) 
= - \frac{1}{8} \nabla^{\2\2} V_1 
= - \frac{1}{8} \nabla^{\2\2} \O_1 
~ .
\ee
The last expression is clearly equivalent to \eqref{propot-YM-1}. 
Note that, due to the analysis of the previous subsection,
this equivalence also guarantees that $W$ defined by \eqref{defWV} is a primary superfield, 
$S_\hal^i W=0$,
satisfies the Bianchi identity $\de_\hal^\pu\de_\hbe^\pu W=\frac{1}{4}\ve_{\hal\hbe}\de^\pd W$,
and is invariant under the $\l$-group transformations
\bea 
\d V  = \l + \breve{\l} ~.
\label{5.44}
\eea
All these properties can actually be directly proven by using the integral representation \eqref{defWV}.


\subsection{Mezincescu's prepotential}

According to the analysis of section \ref{Projector}, 
we can solve the analyticity constraint on the projective prepotential $V(v)$ 
in terms of a primary real isotwistor superfield $V^{(-4)} (v)$  of weight $(-4)$  as
\bea
V=\D^{(4)}V^{(-4)}
\ , \quad
\bbD V^{(-4)} = - 2 V^{(-4)} \ , \quad
S_\hal^i V^{(-4)} = 0 \ .
\eea
The vector multiplet field strength \eqref{defWV}  then takes the form
\be 
W 
= 
- \frac{\ri}{16 \pi} \oint (v, \rd v) \de^\pmd \D^{(4)}V^{(-4)} (v)\ .
\label{defWV-2}
\ee
Making use of the identity
\bea
\de^{(-2)}\D^{(4)}
=
\frac{3}{5}v_iv_j\de_{kl}\D^{ijkl}
~,
\eea
which follows from \eqref{D4D2-D2D4}, 
we can perform the contour integral and obtain the 
following alternative expression for the 
field strength
\bea
W 
&=& 
-\frac{3\ri}{5} \oint \frac{(v ,\rd v)}{16 \pi  } v_kv_l
\de_{ij}\D^{ijkl}V^{(-4)}(v)
=
-\frac{3\ri}{40}\de_{ij}\D^{ijkl}V_{kl}
~.
\label{vector-good}
\eea
Here we have defined the superfield $V_{ij}$ as
\bea
V_{ij}:= \oint \frac{(v, \rd v)}{2 \pi  } v_iv_jV^{(-4)}(v)
~.
\label{5.49}
\eea
By construction $V_{ij}$ is a real primary superfield of dimension $-2$, $\bbD V_{ij} = - 2 V_{ij} $.
It is also possible to prove that, due to \eqref{harmult1} and the definition \eqref{5.49}, $V_{ij}$ correctly transforms as an isovector 
under SU(2) transformations.
Note that $V_{ij}$ is the analogue of Mezincescu's prepotential \cite{Mezincescu}
(see also \cite{HST} and \cite{BK11})
for the 4D $\cN=2$ abelian vector multiplet.
To conclude, we note that $V_{ij}$ is defined up to gauge transformations of the form
\bea
\d V_{kl}
=
\de_\hal^p \L^{\hal}{}_{klp}
~,~~~~~~
\L^{\hal}{}_{klp}=\L^{\hal}{}_{(klp)}
~,
\label{gauge-vector}
\eea
with the gauge parameter being $\L^{\hal}{}_{ijk}$ a primary superfield, 
\bea
S_{\hal}^{i} \L^{\hbe}{}_{jkl}=0
~,~~~~~~
\bbD\L^{\hbe}{}_{jkl}=-\frac{5}{2}\L^{\hbe}{}_{jkl}
~.
\eea
The gauge invariance follows from the fact that
$\de_{ij}\D^{ijkl}\de_\hal^p \L^{\hal}{}_{klp}=0$, as  can be proven using
\eqref{D4D2-D2D4}.

\subsection{Composite $\cO(2)$ multiplet}
\label{subsection6.6}

Consider a locally supersymmetric theory that involves an abelian vector multiplet as 
one of the dynamical multiplets. Let $S[V(v)]=S[V_{ij}]$ be the corresponding gauge invariant action. A variation of the action with respect to 
the vector multiplet 
may be represented in two different forms,
\begin{subequations}\label{5.52}
\bea
\d S
&=&
\frac{1}{2\pi} \oint (v, \rd v)
\int \rd^{5|8}z\, E\, C^{(-4)}  \mathbb  H^{(2)}  \d V 
\label{5.52a} \\
&=&
\int \rd^{5|8}z\,E\,  \mathbb H^{ij}\d V_{ij}~,
\label{5.52b}
\eea
\end{subequations}
for some real weight-2 tropical multiplet $\mathbb H^{(2)}(v)$,
\bea
\nabla^{(1)}_\hal \mathbb H^{(2)} =0~,
\eea
  and some
real isovector $\mathbb H^{ij} =\mathbb H^{ji}$, which are primary 
superfields of dimension $+3$.
The theory under consideration may also involve hypermultiplets charged
under the U(1) gauge group. We assume that these hypermultiplets
obey the corresponding equations of motion. 
Then the above variation vanishes when $\d V$ or $\d V^{ij}$ is a gauge
transformation. This property has two different, but equivalent,  manifestations.  
Firstly, the variation \eqref{5.52a} is equal to zero 
for the gauge transformation \eqref{5.44}, hence
\bea
\oint (v, \rd v)
\int \rd^{5|8}z\, E\, C^{(-4)} \mathbb  H^{(2)}  \l=0~,
\eea 
for an arbitrary weight-0 arctic multiplet $\l(v)$. This implies that 
$\mathbb H^{(2)}(v)$ is an $\cO(2)$ multiplet,
\bea
\mathbb H^{(2)}(v) = \mathbb H^{ij} v_i v_j~.
\label{5.55}
\eea
Secondly, the variation \eqref{5.52b} is equal to zero 
for the gauge transformation \eqref{gauge-vector}. This means
\bea
\int \rd^{5|8}z\, E\,  \L^\hal_{ijk} \nabla_\hal^{i}\mathbb H^{jk}=0~,
\eea
and hence 
\bea
 \nabla_\hal^{(i}\mathbb H^{jk)}=0~.
 \label{5.58}
 \eea
The superfields $\mathbb H^{(2)}(v)$ and $ \mathbb H^{ij}$ defined by eqs. \eqref{5.52a}
and \eqref{5.52b}, respectively, are related to each other 
according to \eqref{5.55}, as follows from \eqref{5.49}.
 
In summary, any gauge theory of the abelian vector multiplet possesses
a composite $\cO(2)$ multiplet, $\mathbb H^{ij}$. The equation of motion for the 
vector multiplet is $\mathbb H^{ij}=0$.


\section{The $\cO(2)$ multiplet in conformal superspace} \label{section7}

In the previous section we gave the prepotential description
of the Yang-Mills multiplet.
Here we develop a prepotential formulation for the $\cO(2)$ multiplet, 
a dual version of the hypermultiplet. 
In the 4D $\cN=2$ case, it is known that the $\cO(2)$ multiplet constraints
\bea
\nabla_\a^{(i}G^{jk)}=0~,\quad
\deb_\ad^{(i}G^{jk)}=0~, 
\eea
may be solved in conformal superspace 
in terms of a complex primary scalar $U$ of dimension  $-1$, ${\mathbb D}U =-U$,  as 
\bea
G^{ij}
=
\frac{1}{192} \Big( \nabla^{ij}\deb^{kl}\deb_{kl}\bar{U}
+
\bar{\nabla}^{ij}\de^{kl}\de_{kl}U \Big)
~,
\eea
see, e.g., \cite{BK11} for a detailed discussion.
As will be demonstrated below, an analogous six-derivative representation 
for the $\cO(2)$ multiplet exists in five dimensions, but the corresponding 
prepotential is a real dimensionless scalar.


\subsection{Prepotential formulation for the $\cO(2)$ multiplet}
\label{prep-O2}

In five dimensions, the $\cO(2) $ multiplet $G^{ij}=G^{ji}$ is characterized by the properties 
\bea
\nabla_\hal^{(i} G^{jk)}=0 ~, \qquad
S_\hal^i G^{jk}=0~, \qquad {\mathbb D}G^{jk} = 3 G^{jk}~.
\label{O(2)constraints}
\eea
We always assume $G^{ij}$ to be real, $\overline{G^{ij}} = G_{ij} = \ve_{ik} \ve_{jl}G^{kl}$.
It turns out that the constraints \eqref{O(2)constraints} may be solved in terms of  
 a primary real dimensionless scalar $\O$, 
\bea
S_\hal^i\O=0~,~~~~~~
\bbD\O=0
~,
\eea
and the solution is 
\begin{subequations}\label{def-G-0}
\bea
G^{ij}
&=&
-\frac{3\ri}{40}\D^{ijkl}\de_{kl}\O 
\label{def-G-0-b}
\eea
or, equivalently,
\bea
G^{(2)}
=v_iv_jG^{ij}
&=&
-\frac{\ri}{8}\D^{(4)}\de^{(-2)}\Omega 
\label{def-G-0-a}~.
\eea
\end{subequations}
Note that representation \eqref{def-G-0-b}
follows from \eqref{def-G-0-a}
by applying  \eqref{D4D2-D2D4}.

In Appendix \ref{HarmonicG} we 
 prove
 that the decomposition \eqref{def-G-0} is the most general solution to 
 \eqref{O(2)constraints} 
 in the flat case by making use of the harmonic superspace techniques 
 \cite{GIKOS,GIOS}.
 Here we demonstrate
 that \eqref{def-G-0} defines a primary $\cO(2)$ superfield in conformal superspace.

It follows from 
\eqref{def-G-0-a} that $G^{(2)}$ is analytic, 
$\de^\pu_\hal G^{(2)}=0$.
It is also obvious that $G^{(2)}$ has the right dimension, 
$\bbD G^{(2)}=3$, since $\O$ is dimensionless.
It is slightly more involved to check that
$S_\hal^i G^{(2)}=0$, which is equivalent to proving the two conditions
$S_\hal^\pu G^{(2)}=0$ and $S_\hal^\pmu G^{(2)} =0$.
 
Let us first consider
\bea
S_\hal^\pmu G^{(2)} 
&=&
-\frac{\ri}{8}{[}S_\hal^\pmu,\D^{(4)}{]}\de^{(-2)}\Omega
-\frac{\ri}{8}\D^{(4)}{[}S_\hal^\pmu,\de^{(-2)}{]}\Omega
~.
\eea
It is straightforward to check that the second term on the right is identically zero:
 \bea
 {[}S_\hal^\pmu,\de^{(-2)}{]}\Omega
 &=&
\big(\{S_\hal^\pmu,\de^{\hbe\pmu}\}\de_\hbe^\pmu
-\de^{\hbe\pmu}\{S_\hal^\pmu,\de_\hbe^\pmu\}\big)
\O
\non\\
&=&
6\big( J^{(-2)} \de_\hal^\pmu
+\de_\hal^\pmu J^{(-2)}\big) \O
= 0
~,
 \eea
 as a consequence of 
 \bea
{[} J^{(-2)}, \de_\hal^\pmu{]}=0~, \qquad  J^\pmd:=\frac{1}{(v,u)^2}u_iu_jJ^{ij}
~.
 \eea
It remains 
to show that ${[}S_\hal^\pmu,\D^{(4)}{]}\de^{(-2)}\Omega=0$.
Using \eqref{S-D4} we obtain
\bea
{[}S_\hal^\pmu,\D^{(4)}{]}\de^{(-2)}\Omega
&=&
 \frac{1}{24} \eps^{\hbe\hga\hde\hrh}
 \eps_{\hal\hbe}  \nabla_\hga^\pu \nabla_\hde^\pu\nabla_\hrh^\pu
\big(8-2  \bbD -6 J^{(0)}\big)
\de^{(-2)}\Omega
\equiv 0
~.~~~~~~~~~
\eea
Since $\pa^{(2)}G^{(2)}=0$, we also find $S_\hal^\pu G^{(2)}=\pa^{(2)}S_\hal^\pmu G^{(2)}=0$. 
Thus we have shown that the superfield $G^{(2)}$ defined by \eqref{def-G-0}
is primary.

A crucial property of the superfield $G^{(2)}$ defined by 
\eqref{def-G-0} is that it is invariant under  gauge transformations 
of $\O$
of the form 
\bea
\d\Omega=-\frac{\ri}{2}(\G^\ha)^{\hal\hbe}\de_\hal^i \de_\hbe^jB_\ha{}_{ij}
~,
\label{gauge-O2-0}
\eea
where the gauge parameter is assumed to have the properties
\bea
B_{\ha}{}^{ij}=B_{\ha}{}^{ji}
~,~~~~~~
S_{\hal}^{i} B_\ha{}^{jk}=0
~,~~~~~~
\bbD B_\ha{}^{ij}=-B_\ha{}^{ij}
\label{gauge-inv-O2-1}
\eea
and is otherwise arbitrary.
It is an instructive exercise to show that the variation $\d\O$ defined by 
\eqref{gauge-O2-0} and \eqref{gauge-inv-O2-1} is a  
primary dimensionless superfield.
Appendix \ref{gauge-invariance-G} is devoted to 
the proof that 
the transformation \eqref{gauge-O2-0} leaves invariant the field strength $G^{(2)}$ 
defined by \eqref{def-G-0}.


\subsection{Composite vector multiplet}

Consider a dynamical system involving an $\cO(2)$ multiplet $G^{ij}$ as one of the 
dynamical multiplets. The action may be viewed as a functional of the 
field strength, $S[G^{ij}]$, or as a gauge invariant functional, $S[\O]$, of the prepotential 
$\O$.  Giving the prepotential an infinitesimal displacement 
changes the action as follows:
\bea
\d S = \int\rd^{5|8}z\, E\,  {\mathbb W} \d \O~,
\eea
for some real scalar $\mathbb W$, which is a primary superfield of dimension $+1$.
The variation must vanish if $\d \O$ is a gauge transformation of the form
\eqref{gauge-O2-0}. This holds if $\mathbb W$ obeys the equation
\bea 
\nabla_{\hat{\a}}^{(i} \nabla_{\hat{\b}}^{j)} \mathbb W 
= \frac{1}{4} \eps_{\hat{\a} \hat{\b}} \nabla^{\hat{\g} (i} \nabla_{\hat{\g}}^{j)} \mathbb W ~,
\label{6.20}
\eea
which is the Bianchi identity for the field strength of an abelian vector multiplet, 
see  eq. \eqref{vector-Bianchi}.

In summary, any dynamical system involving an $\cO(2)$ multiplet $G^{ij}$
 possesses a composite vector multiplet, $\mathbb W$. The equation of motion for the 
$\cO(2)$ multiplet is $\mathbb W=0$.

\subsection{BF coupling}

Consider the following Lagrangian
\bea
\cL^{(2)}_{\rm BF}
= V G^{(2)}
\label{BF-proj}
\eea
that describes a BF coupling of  a vector multiplet 
and an $\cO(2)$ multiplet.
The action principle \eqref{InvarAc} with $\cL^{(2)}_{\rm BF}= V G^{(2)}$ 
will be referred to as the BF action. 

The BF action involves the tropical prepotential of the vector multiplet, $V(v^i)$, 
and the field strength of the $\cO(2)$ multiplet, $G^{(2)}$. 
It can be rewritten in a different form involving the field strength of the vector multiplet, 
$W$, and the prepotential of the $\cO(2)$ multiplet, $\O$.  
This is achieved by expressing $G^{(2)}$ in terms of $\O$ and then integrating by parts to obtain
\bea 
S[\cL^{(2)}_{\rm BF}]
&=& 
\oint \frac{(v , \rd v)}{16\pi\ri} \int \rd^{5|8}z\, E\,  C^{(-4)} V \,\D^{(+4)}\de^{(-2)}\Omega
\non \\
&=&
\int \rd^{5|8}z\, E\, \O W
~.
~~~~~~~
\eea
By using \eqref{def-G-0} and \eqref{vector-good} together with integration by parts,
 the action may be rewritten in another equivalent form that involves 
 Mezincescu's prepotential $V_{ij}$ and the field strength $G^{ij}$. 
 One obtains
\bea
S[\cL^{(2)}_{\rm BF}]
=
\int\rd^{5|8}z\, E\, 
 \O W
=
 \int \rd^{5|8}z\, E\, 
 G^{ij} V_{ij}
 ~.
 \label{SGV}
\eea
One may prove that the  functionals $\int \rd^{5|8}z\, E\, 
 \O W$ and 
 $\int \rd^{5|8}z\, E\, 
 G^{ij} V_{ij}$
are invariant under the gauge transformations \eqref{gauge-O2-0} and \eqref{gauge-vector}, respectively.


\subsection{Gauge invariance} \label{subsection6.2}

The results of the previous subsection allow us to prove the
important relation 
\eqref{44.42a}. 
For this we choose $V = \l + \breve{\l}$
 in the BF Lagrangian \eqref{BF-proj}, where $\l (v) $ is a weight-0
 arctic multiplet. Since the tropical prepotential is pure gauge, 
 the field strength vanishes, $W=0$. Then eq. \eqref{SGV} leads to 
$ S[(\l + \breve{\l})G^{(2)}] =0$, which is the required result
 \eqref{44.42a}. Since $\l$ is complex, we can replace 
 $\l (v) \to \ri \l (v)$  and obtain $ S[\ri (\l - \breve{\l})G^{(2)}] =0$. 
 These two relations lead to \eqref{4.42b}, and thus 
\bea
 \oint (v, \rd v)
\int\rd^{5|8}z\, E\, C^{(-4)} G^{(2)} \l =0~,
\label{6.14}
\eea
where $G^{(2)}(v)$ is an $\cO(2)$ multiplet and $\l(v)$ is 
a weight-0 arctic multiplet.


\subsection{Universality of the BF action}\label{subsection7.5}

The goal of  this subsection is to demonstrate that the supersymmetric action 
 \eqref{InvarAc} can be rewritten as a BF action under the assumption 
 that a special vector multiplet exists.  

Consider the action  \eqref{InvarAc} written as  \eqref{InvarAc2} with $\cU^{(-2)}$ a prepotential for 
the Lagrangian $\cL^{(2)}$.
Now let $W$ be the field strength of a compensating 
vector multiplet. 
We insert the unity $1 = W/W$
in the right hand side of \eqref{InvarAc2}
 and represent $W$ in the numerator 
according to \eqref{defWV}. 
 After that we change the order of the contour integrals and integrate $\nabla^{(-2)} $ by parts. 
Finally, we insert the unity $1= \Delta^{(4)} C^{(-4)} $ and integrate by parts. 
The final result is 
\bea
S&=&
\frac{1}{2\pi}
  \oint_\g (v, \rd v)
\int\rd^{5|8}z\, E\, 
C^{(-4)}
\,V \mathbb G^{(2)}
~,
\eea
where $V$ is the tropical  prepotential for the vector multiplet 
and the composite superfield $\mathbb G^{(2)}$  is defined by
\bsubeq
\bea
\mathbb G^{(2)}
&=&
-\frac{\ri}{8} \D^{(4)}\nabla^{(-2)} \O
~,
\\
\O
&:=&
\frac{1}{2\pi W}\oint_{\g} (v, \rd v)\,
\cU^{(-2)}
~.
\label{O0}
\eea
\esubeq
According to \eqref{def-G-0}, the superfield $\mathbb G^{(2)}$ is an $\cO(2)$ multiplet.
Note that it is possible to give some alternative expressions for $\O$ in \eqref{O0}.
Consider a weight $-4$ isotwistor superfield $\tilde{C}^{(-4)}$ such that $\D^{(4)}\tilde{C}^{(-4)}=1$.
This does not necessarily have to be equal to $C^{(-4)}$.
Given $\tilde{C}^{(-4)}$, the superfield
\bea
\cU^{(-2)}
:=\tilde{C}^{(-4)}\cL^{(2)}
~,
\eea
is a prepotential for the projective Lagrangian $\cL^{(2)}$.
Hence we have the equivalent expression
\bea
\O
&:=&
 \frac{1}{ 2\pi W}\oint_{\g} (v, \rd v)\,
\tilde{C}^{(-4)}\cL^{(2)}
~.
\label{O1}
\eea
Note that in the presence of the vector multiplet compensator a natural choice for $\tilde{C}^{(-4)}$
is given by \eqref{CWS}.
Then we find
\bea
\O
&:=&
\frac{2W^3}{3 \pi}\oint_{\g} (v, \rd v)
\frac{ \cL^{(2)}}{ (H^{(2)}_{\rm VM})^2}~.
\label{O2}
\eea


\subsection{Full superspace invariants}

Consider an invariant that can be represented as an integral 
over the full superspace $\cM^{5|8}$, 
\bea
S[\cL]=
\int \rd^{5|8}z\, E\,  \cL ~, 
\label{action7.29}
\eea
where $\cL$ is a conformal primary superfield of dimension $+1$, 
${\mathbb D } \cL = \cL$.
This invariant may be represented in the form \eqref{InvarAc}, 
in which $\cL^{(2)} $ reads
\bea
\cL^{(2)} = -\frac{2}{G^{(2)}} \D^{(4)} \big( G \cL \big) ~.
\label{6.30}
\eea
Here $\D^{(4)}$ is the covariant analytic projection operator \eqref{zero-proj}
and $G^{(2)} =v_iv_jG^{ij}$ is an $\cO (2) $ multiplet such that 
\bea 
G^2 := \hf G^{ij} G_{ij} 
\label{6.32}
\eea
is nowhere vanishing, $G \neq 0$.
The Lagrangian \eqref{6.30} is an example of a covariant rational projective 
multiplet\footnote{In the 4D $\cN=2$ 
super Poincar\'e case, rational projective multiplets were first introduced by Lindstr\"om and Ro\v{c}ek \cite{LR1}.}
in the sense that  it has the structure $H^{(4)}/G^{(2)}$, for some $\cO(4$) multiplet 
$H^{(4)} (v)$.


\subsection{Prepotentials for $\cO(4+n)$ multiplets}

Let $H^{(4)} (v)$ be an $\cO(4) $ multiplet. It may be shown that 
\bea 
H^{(4)} (v) = \D^{(4)}  \F~,  \qquad S_\hal^i \F =0~, 
\qquad \mathbb D \F =4 \F~, 
\eea
for some primary scalar prepotential $\F$, see Appendix \ref{AppendixH}.
The $\cO(4)$ multiplet in 
\eqref{6.30}, $ \D^{(4)} \big( G \cL \big)$,
 is a special case of this result. 

More generally, let $H^{(4+n)} (v)$ be an $\cO(4+n) $ multiplet, with $n=0,1,\dots $
It may be represented in the form
\bea 
H^{(4+n)} (v) = \D^{(4)}  \F^{(n)}~,  \qquad 
\F^{(n)} (v) = \F^{i_1 \dots i_{n}} v_{i_1} \dots v_{i_{n}}~,
\label{7.34}
\eea
for some primary superfield $ \F^{i_1 \dots i_{n}}$ of dimension
$\big(4 +\frac{3}{2}n \big)$,
see Appendix \ref{AppendixH}.


\section{Superform formulation for the BF action} \label{section8}

In section \ref{section7} we demonstrated the universality of the BF action
\bea \label{BFactionsection9}
S_{\rm BF}&=&
\frac{1}{2\pi} \oint_\g (v, \rd v)
\int \rd^{5|8}z\, E\, C^{(-4)}\cL^{(2)}_{\rm BF}~, \qquad 
\cL^{(2)}_{\rm BF}= V G^{(2)}~.
\eea
The component structure of $S_{\rm BF}$ is of primary importance for applications. 
For the analogous action in 4D $\cN=2$ supergravity, 
two procedures have been developed to reduce the action to components.
One of them \cite{Butter:2012xg} directly carries out the integration over the Grassmann variables in the action.  
The other approach \cite{Butter:2012ze} 
provides a superform construction for the action\footnote{This approach 
makes use of the superform formalism to construct supersymmetric invariants 
\cite{Castellani, Hasler, Ectoplasm, GGKS}.} which immediately leads 
to the component action. 
The latter  has turned out to be fruitful for various generalizations,
such as the $\cN$-extended conformal supergravity actions 
\cite{KT-M1212,BKNT-M2,KNT-M}
and the Chern-Simons actions \cite{KN13} 
in three dimensions 
and the  non-abelian Chern-Simons action in  5D $\cN=1$ Minkowski superspace
\cite{KN-CS5D}.
 Here we apply the ideas put forward in \cite{Butter:2012ze} 
 to derive a superform formulation for the action $S_{\rm BF}$. 


\subsection{Superform geometry of the $\cO(2)$ multiplet}\label{section8.1}

The $\cO(2)$ multiplet can be described by a three-form gauge potential $\cB = \frac{1}{3!} E^\hC \wedge E^\hB \wedge E^\hA \cB_{\hA\hB\hC}$ 
possessing the gauge transformation
\be \d \cB = \rd \r \ ,
\ee
where $\r$ is a 2-form gauge parameter. The corresponding field strength is
\be \Phi = \rd \cB = \frac{1}{4!} E^{\hD} \wedge E^\hC \wedge E^\hB \wedge E^\hA \Phi_{\hA \hB \hC \hD} \ ,
\ee
where
\be
\Phi_{\hA \hB \hC \hD} = 4 \nabla_{[\hA} \cB_{\hB\hC\hD\}} + 6 \scT_{[\hA \hB}{}^{\hE} \cB_{|\hE|\hC\hD\}} \ .
\ee
The field strength must satisfy the Bianchi identity
\be \nabla_{[\hA} \Phi_{\hB\hC\hD \hE \}} + 2 \scT_{[\hA \hB}{}^{\hat{F}} \Phi_{|\hat{F}|\hC\hD \hE\}} = 0 \ .
\ee

In order to describe the $\cO(2)$ multiplet we need to impose some covariant constraints on the field strength $\Phi$. 
We choose the constraints
\bsubeq \label{eq:PhiComponents}
\begin{align}
\Phi_\hal^i{}_\hbe^j{}_\hga^k{}_\hde^l
= 
\Phi_{\ha}{}_\hbe^j{}_\hga^k{}_\hde^l = 0 \ , \quad
\Phi_{\hat{a} \hat{b}}{}_\hal^i{}_\hbe^j = 8 \ri (\S_{\hat{a} \hat{b}})_{\hat{\a} \hat{\b}} G^{ij} \ ,
\end{align}
where $G^{ij} = G^{ji}$ is a dimension-3 primary superfield. The constraints allow one to 
solve for the remaining components of $\Phi$ in terms of $G^{ij}$. The solution is
\begin{align}
\Phi_{\hat{a} \hat{b} \hat{c}}{}_\hal^i &= - \frac{2}{3} \eps_{\hat{a} \hat{b} \hat{c} \hat{d} \hat{e}} (\S^{\hat{d} \hat{e}})_{\hat{\a}}{}^{\hat{\b}} \nabla_{\hat{\b} j} G^{ji} 
= - 2 \eps_{\ha\hb\hc\hd\he} (\S^{\hd \he})_\hal{}^\hbe \varphi_\hbe^i \ , \\
\Phi_{\hat{a} \hat{b} \hat{c} \hat{d}} &= \frac{\ri}{12} \eps_{\hat{a} \hat{b} \hat{c} \hat{d} \hat{e}} (\G^{\hat{e}})^{\hat{\a} \hat{\b}} \nabla_{\hat{\a}}^i \nabla_{\hat{\b}}^j G_{ij} 
\equiv \eps_{\ha\hb\hc\hd\he} \Phi^\he \ , 
\end{align}
\esubeq
where $G^{ij}$ satisfies the constraint for the $\cO(2)$ multiplet
\be \nabla_{\hat{\a}}^{(i} G^{jk)} = 0
\ee
and we have introduced the superfields
\bsubeq \label{O2superfieldComps}
\begin{align}
\varphi_\hal^i &:= \frac{1}{3} \nabla_{\hal j} G^{ij} \ , \\
F &:= \frac{\ri}{12} \nabla^{\hga i} \nabla_\hga^j G_{ij} = - \frac{\ri}{4} \nabla^{\hga k} \varphi_{\hga k} \ .
\end{align}
\esubeq
The Bianchi identities also imply the differential condition on $\Phi_\ha$
\be \nabla^\ha \Phi_\ha + 5 \ri X^{\hga k} \varphi_{\hga k} =
\hat \nabla^\ha \Phi_\ha = 0 \ .
\ee


\subsection{Superform action for the $\cO(2)$ multiplet}

The superform formulation in the previous subsection gives a geometric description for the $\cO(2)$ multiplet. 
As we will see, it is a useful ingredient in the construction of the BF action principle. Below we describe the 
general setup, the construction of the superform action and its corresponding component action.

\subsubsection{General setup}

The superform approach to constructing supersymmetric invariants 
\cite{Castellani, Hasler, Ectoplasm, GGKS}
is based on 
the use of a closed superform. 
In five-dimensional spacetime $\cM^5$, which is the body 
of the $\cN = 1$ curved superspace $\cM^{5|8}$,
the formalism requires the use of a closed five-form
\be
\frak{J} = \frac{1}{5 !} E^{\hat{E}} \wedge E^{\hat{D}} \wedge E^\hC \wedge E^\hB \wedge E^\hA \ \frak{J}_{\hA\hB\hC\hat{D}\hat{E}} 
\ ,\qquad
 \rd \frak{J} = 0 ~.  \label{JactPrin}
\ee
Given such a superform, one can construct the supersymmetric invariant
\bea 
S = \int_{\cM^5} i{}^*\frak{J} 
 \ ,
\label{ectoS}
\eea
where $i : \cM^5 \to \cM^{5|8}$ is the 
inclusion map. Invariance under arbitrary general coordinate transformations of the superspace follows from the transformation 
of $\frak{J}$, 
\be \d_{\xi} \frak{J} = \cL_{\xi} \frak{J} \equiv i_\xi \rd \frak{J} + \rd i_\xi \frak{J} = \rd i_\xi \frak{J} \ .
\ee

The closed form $\frak{J}$ is required to transform as an exact form under all 
{\it gauge symmetries}, 
\be \rd \frak{J} = \rd \Theta \ ,
\ee
which ensures eq. \eqref{ectoS} is a suitable candidate for an action. 
In conformal supergravity, suitable actions  
must be  invariant under 
the {\it standard superconformal 
transformations}.
This requires that $\frak{J}$ transforms by an exact form 
 under the standard superconformal  transformations, 
\bea
 \d_{\cH} \frak{J} =  \rd \Theta (\L^{\underline{a}} ) \ , \quad 
\L = \L^{\underline{a}} X_{\underline{a}}~.
\eea

Locally superconformal matter actions 
are usually associated 
with closed five-forms that are invariant,
\be
\d_{\cH} \frak{J} =  0~.
\ee
This is equivalent to the condition
\be X_{\underline{a}} \frak{J}_{\hA_1 \cdots \hA_p}= - f_{\underline{a} [\hA_1}{}^{\hat{D}} \frak{J}_{|\hat{D} | \hA_2 \cdots \hA_p \}} \ .
\ee
The $S$-invariance, $S_{\hal i} \frak{J} = 0$, is non-trivial and we will call a superform that is $S$-invariant a {\it primary} superform.\footnote{$S$-invariance 
automatically implies $K$-invariance.} 
In general, a 
primary $p$-form $\S$ satisfies
\be S_{\hal i} \S_{\hA_1 \cdots \hA_p} = \ri p (\G_\ha)_\hal{}^\hbe \S_{\hbe i [\hA_2 \cdots \hA_p} \d^\ha_{\hA_1\}} \ ,
\ee
which implies the condition
\be S_{\hbe j} \S_{\ha_1 \cdots\ha_n}{}_{\hal_1}^{i_1} \cdots {}^{i_{p-n}}_{\hal_{p-n}} 
= \ri n (\G_{[\ha_1})_\hbe{}^\hga \S_{\hga j \ha_2 \cdots \ha_n ]} {}_{\hal_1}^{i_1} \cdots {}^{i_{p-n}}_{\hal_{p-n}} \ . \label{primCond}
\ee


\subsubsection{Superform action for the $\cO(2)$ multiplet}

In order to construct a closed form for the action we will first consider the superform 
equation
\be \rd \S = F \wedge \Phi \ , \label{SigFPhi}
\ee
where $\S$ is some five-form and $F$ is the field strength for an abelian vector multiplet with 
gauge one-form $V$ and field strength $W$ (see section \ref{SYM}).

It turns out there exist two solutions to eq. \eqref{SigFPhi} that do not 
differ by an exact form. The first solution is
\be \S_V = V \wedge \Phi \ .
\ee
The second solution is the result of the constraints that have been imposed on the components of $F$ and $\Phi$. 
If we assume that this solution is primary then we may write the Bianchi identity in terms of the covariant derivatives as 
follows:
\be \nabla_{[\hA} \S_{\hB \hC \hD \hE \hat{F} \}} + \frac{5}{2} \scT_{[\hA \hB}{}^{\hat{G}} \S_{|\hat{G}| \hC \hD \hE \hat{F} \} } 
= \frac{5}{2} F_{[\hA \hB} \Phi_{\hC \hD \hE \hat{F} \} } \ .
\ee
Making use of the components of $F$ and $\Phi$, one finds the solution:
\bsubeq \label{eq:SigmaComponents}
\begin{align}
\S_{\hat{a} \hat{b} \hat{c}}{}_\hal^i{}_\hbe^j &=  - 4 \ri \eps_{\hat{a} \hat{b} \hat{c} \hat{d} \hat{e}} (\S^{\hat{d} \hat{e}})_{\hat{\a} \hat{\b}}  W G^{ij} \ , \\
\S_{\hat{a} \hat{b} \hat{c} \hat{d}}{}_\hal^i &= 2 \eps_{\hat{a} \hat{b} \hat{c} \hat{d} \hat{e}} (\G^{\hat{e}})_{\hat{\a}}{}^{\hat{\b}} ( W \varphi_\hbe^i + \ri \l_{\hbe j} G^{ji})  \ , \\
\S_{\hat{a} \hat{b} \hat{c} \hat{d} \hat{e}} &= - \eps_{\hat{a} \hat{b} \hat{c} \hat{d} \hat{e}} ( W F + X^{ij} G_{ij} + 2 \l^{\hga k} \varphi_{\hga k} )
\end{align}
\esubeq
with the remaining components vanishing.
Here we have made use of the following useful identities:
\bsubeq
\begin{align}
\nabla_\hal^i G^{jk} &= 2 \eps^{i(j} \varphi_\hal^{k)} \ , \\
\nabla_\hal^i \varphi_\hbe^j &= - \frac{\ri}{2} \eps^{ij} \eps_{\hal\hbe} F + \frac{\ri}{2} \eps^{ij} \Phi_{\hal\hbe} + \ri \nabla_{\hal\hbe} G^{ij} \ , \\
\nabla_\hal^i F &= - 2 \nabla_\hal{}^\hbe \varphi_\hbe^i - 6 W_{\hal\hbe} \varphi^{\hbe i} - 9 X_{\hal j} G^{ij} \ , \\
\nabla_\hal^i \Phi_{\ha} &= 4 (\S_{\ha\hb})_\hal{}^\hbe \nabla^\hb \varphi_\hbe^i - 4 (\G_\ha)_\hal{}^\hbe W_{\hbe \hga} \varphi^{\hga i} 
- 6 (\G_\ha)_\hal{}^\hbe X_{\hbe j} G^{ij} \ .
\end{align}
\esubeq
The second solution is expressed entirely in terms of $G^{ij}$ and its covariant derivatives and we will denote it 
by $\S_G$. Making use of the following identities
\bsubeq
\begin{align} S_\hal^i \varphi_\hbe^j &= - 6 \eps_{\hal \hbe} G^{ij} \ , \\
S_\hal^i F &= 6 \ri \varphi_\hal^i \ , \\
S_\hal^i \Phi_\hb &= - 8 \ri (\G_\hb)_\hal{}^\hbe \varphi_\hbe^i \ ,
\end{align}
\esubeq
one can check that $\S_G$ is primary, {\it i.e.} it satisfies eq. \eqref{primCond}. Similarly one can show $\Phi$ is 
primary and hence $\S_V$ is primary also.

It is now straightforward to construct a closed invariant five-form. One may simply take the difference between $\S_V$  and 
$\S_G$,
\be \frak{J} = \S_V - \S_G = V \wedge \Phi - \S_G \ . \label{diffSigVSigG}
\ee


\subsubsection{Component BF action}

Having derived $\frak{J}$ we can now make use of the action principle \eqref{JactPrin} to construct the corresponding 
component action. The Lagrangian is given by\footnote{The Levi-Civita tensor with world indices 
is defined as $\eps^{\hat{m}\hat{n}\hat{p}\hat{q}\hat{r}} := \eps^{\ha\hb\hc\hd\he} e_\ha{}^{\hat{m}} e_\hb{}^{\hat{n}} e_\hc{}^{\hat{p}} e_\hd{}^{\hat{q}} e_\he{}^{\hat{r}}$.}
\be
e^{-1}\, {}^* \frak{J} = \frac{1}{5!} \eps^{\hm\hn\hat{p}\hat{q}\hat{r}} \frak{J}_{\hm\hn\hat{p}\hat{q}\hat{r}}
= \frac{1}{4!} \eps^{\hm\hn\hat{p}\hat{q}\hat{r}} V_\hm \Phi_{\hn \hat{p}\hat{q}\hat{r}} - \frac{1}{5!} \eps^{\hm\hn\hat{p}\hat{q}\hat{r}} \S_{\hm\hn\hat{p}\hat{q}\hat{r}} \ ,
\ee
where
\begin{align} \label{SigComps}
\frac{1}{5!} \eps^{\hat{m}\hat{n}\hat{p}\hat{q}\hat{r}} \S_{\hat{m}\hat{n}\hat{p}\hat{q}\hat{r}}| &= 
\frac{1}{5!} \eps^{\hat{m}\hat{n}\hat{p}\hat{q}\hat{r}} E_{\hat{r}}{}^{\hat{E}} E_{\hat{q}}{}^{\hat{D}} E_{\hat{p}}{}^{\hC} E_{\hat{n}}{}^{\hB} E_{\hat{m}}{}^{\hA} \S_{\hA\hB\hC\hat{D}\hat{E}}| 
\non\\
&= \frac{1}{5!} \eps^{\ha\hb\hc\hd\he} \Big( \S_{\ha\hb\hc\hd\he}| 
+ \frac{5}{2} \psi_\ha{}^\hal_i \S_{\hb\hc\hd\he}{}_\hal^i|
- \frac{5}{2} \psi_\ha{}^\hal_i \psi_\hb{}^\hbe_j \S_{\hc\hd\he}{}_\hal^i{}_\hbe^j| \non\\
&\qquad- \frac{5}{4} \psi_\ha{}^\hal_i \psi_\hb{}^\hbe_j \psi_\hc{}^\hga_k \S_{\hd\he}{}_\hal^i{}_\hbe^j{}_\hga^k|
+ \frac{5}{16} \psi_\ha{}^\hal_i \psi_\hb{}^\hbe_j \psi_\hc{}^\hga_k \psi_\hd{}^\hde_l \S_{\he}{}_\hal^i{}_\hbe^j{}_\hga^k{}_\hde^l| \non\\
&\qquad + \frac{1}{32} \psi_\ha{}^\hal_i \psi_\hb{}^\hbe_j \psi_\hc{}^\hga_k \psi_\hd{}^\hde_l \psi_\he{}^{\hat{\r}}_p
\S_\hal^i{}_\hbe^j{}_\hga^k{}_\hde^l{}_{\hat{\r}}^p|
\Big)
\ .
\end{align}
The action is then
\begin{align} \label{LinearCompActionNCC}
S &= \int \rd^5x \,e\, \Big(
\frac{1}{4!} \veps^{\hm\hn\hp\hq\hat{r}} v_\hm \phi_{\hn\hp\hq\hat{r}}
- W F
- X^{ij} G_{ij} 
- 2 \l^{k} \varphi_{k}
\eol & \quad
+ \psi_{\hm i} \G^\hm \varphi^i W
+ \ri \psi_{\hm i} \G^\hm \l_j G^{ij}
- \ri \psi_{\hm i} \S^{\hm\hn} \psi_{\hn j} W G^{ij}
\Big)
 \non\\
&= - \int \rd^5x \,e\, \Big(
v^\ha \phi_\ha
+W F
+X^{ij} G_{ij} 
+ 2 \l^{k} \varphi_{k}
\eol & \quad
- \psi_{\ha i} \G^\ha \varphi^i W
- \ri \psi_{\ha i} \G^\ha \l_j G^{ij}
+ \ri \psi_{\ha i} \S^{\ha\hb} \psi_{\hb j} W G^{ij}
\Big) \ ,
\end{align}
where all superfields appearing in the action 
are understood as their corresponding 
spacetime projections and we have defined
\bsubeq
\begin{align}
v_\ha &:= e_\ha{}^\hm v_\hm = e_\ha{}^\hm V_\hm| \ , \quad \phi^\ha := - \frac{1}{4!} e_{\hat{r}}{}^\ha \eps^{\hm\hn\hat{p}\hat{q}\hat{r}} \phi_{\hm \hn \hat{p} \hat{q}} \ , \\
\phi_{\hm\hn\hp\hq} &:= \Phi_{\hm\hn\hp\hq}| = 4 \pa_{[\hm} b_{\hm\hn\hp]}~, \qquad
b_{\hm \hn \hat{p}} = \cB_{\hm \hn \hat{p}}| \ .
\end{align}
\esubeq
The Chern-Simons coupling between the one-form $V$ and the four-form $\Phi$
can equivalently be written
\begin{align}
S &= 
\int \rd^5x \,e\, \Big(
\frac{1}{12} \veps^{\ha\hb\hc\hd\he} f_{\ha\hb} b_{\hc\hd\he}
- W F
- X^{ij} G_{ij} 
- 2 \l^{k} \varphi_{k}
\eol & \quad
+ \psi_{\ha i} \G^\ha \varphi^i W
+ \ri \psi_{\ha i} \G^\ha \l_j G^{ij}
- \ri \psi_{\ha i} \S^{\ha\hb} \psi_{\hb j} W G^{ij}
\Big) \ ,
\end{align}
where
\begin{align} f_{\ha\hb} &:= e_\ha{}^\hm e_\hb{}^\hn f_{\hm\hn} \ , \quad f_{\hm\hn} := F_{\hm\hn}| = 2 \partial_{[\hm} v_{\hn]} \ , \quad
b_{\ha\hb\hc} := e_\ha{}^\hm e_\hb{}^\hn e_\hc{}^\hp b_{\hm\hn\hp} \ .
\end{align}

It should be mentioned that the normalization of the action \eqref{LinearCompActionNCC} has been chosen to correspond 
to the projective superspace action principle \eqref{InvarAc} with $\cL^{(2)} = V G^{(2)}$. Furthermore, the 
action \eqref{LinearCompActionNCC} 
corresponds to the BF action without central charge. We 
will give a generalization in the presence of a gauged central charge in section \ref{SGCC}.


\section{Abelian Chern-Simons theory} \label{ACStheory}

In conformal superspace, the dynamics of an abelian vector multiplet coupled to conformal 
supergravity is described by the Chern-Simons action\footnote{In the 5D $\cN =1$
super-Poincar\'e case, the off-shell abelian Chern-Simons action
was constructed for the first time by Zupnik in 
harmonic superspace \cite{Zupnik99}. The action \eqref{CSaction9.1} 
is a curved-superspace extension of the one given in \cite{KL}.}
\bea 
S_{\rm CS}&=&
\frac{1}{2\pi} \oint_\g (v, \rd v)
\int \rd^{5|8}z\, E\, C^{(-4)}\cL^{(2)}_{\rm CS}~, \qquad 
\cL^{(2)}_{\rm CS}= -\frac{1}{12}V {H}^{(2)}_{\rm VM}~,
\label{CSaction9.1}
\eea
where ${ H }^{(2)}_{\rm VM}$ denotes the
composite $\cO(2)$ multiplet defined by \eqref{HVM}.
Varying the tropical prepotential gives
\bea
\d S_{\rm CS} 
&=&
-\frac{1}{8\pi} \oint (v, \rd v)
\int \rd^{5|8}z\, E\, C^{(-4)} \d V  { H }_{\rm VM}^{(2)}~,
\label{varyingCSaction}
\eea
see section \ref{Off-shellSUGRA} for the derivation. 

A component counterpart of the action \eqref{CSaction9.1}
may be constructed using 
$ H ^{(2)}_{\rm VM}$ 
and the BF action \eqref{BFactionsection9}. 
This amounts to plugging
\be \label{compositeO(2)ACS}
G^{ij} =  H ^{ij}_{\rm VM} 
= 2 W X^{ij} - \ri \l^{\hal (i} \l_\hal^{j)}
\ee
into the component BF action \eqref{LinearCompActionNCC} and computing the 
component fields of the composite $\cO(2)$ multiplet.
This produces the component $V \wedge F \wedge F$ coupling by
treating the closed gauge-invariant four-form $F \wedge F$ as the field
strength derived from the $\cO(2)$ multiplet \eqref{compositeO(2)ACS}.
A major disadvantage 
of this approach is that the non-abelian Chern-Simons 
theory cannot be constructed in the same way. In this section we will
discuss an alternative superform construction that \emph{can} be generalized
and show how to derive it explicitly from the BF action principle.

Recall that the BF action involved constructing a closed five-form
$\mathfrak{J}$ given by
\begin{align}
\mathfrak{J}_{ H } = V \wedge \Phi - \Sigma_{ H } \ ,
\end{align}
where $\Phi$ is the four-form field strength associated with the composite
$G^{ij}$ and $\Sigma_{ H }$, constructed in section \ref{section8.1}, is a covariant
four-form which solves the equation
\begin{align}
\rd \Sigma_{ H } = F \wedge \Phi~.
\end{align}
If one now substitutes the relations
\bsubeq
\begin{align}
\varphi_\hal^i &= \ri X^{ij} \l_{\hal j} - 2 \ri F_{\hal\hbe} \l^{\hbe i} + X_\hal^i W^2
- 2 \ri W \nabla_{\hal\hbe} \l^{\hbe i} - \ri \nabla_{\hal\hbe} W \l^{\hbe i} \ , \\
F &= X^{ij} X_{ij} - F^{\ha\hb} F_{\ha\hb} 
+ 4W \nabla^\ha \nabla_\ha W + 2 (\nabla^\ha W )\nabla_\ha W 
+ 2 \ri (\nabla_\hal{}^\hbe \l_\hbe^k) \l^\hal_k \non\\
&\quad - 6 W^{\ha\hb} F_{\ha\hb} W - 5 W^{\ha\hb} W_{\ha\hb} W^2
+ Y W^2 + 6 X^{\hal i} \l_{\hal i} W\ , \\
\Phi_\ha &= - \frac{1}{2} \eps_{\ha\hb\hc\hd\he} F^{\hb\hc} F^{\hd\he}
+ 4 \nabla^\hb (W F_{\hb\ha} + \frac{3}{2} W_{\hb\ha} W^2)
+ \eps_{\ha\hb\hc\hd\he} W^{\hb\hc} (F^{\hd\he} + \frac{3}{2} W^{\hd\he} W) W \non\\
&\quad- 6 (\G_\ha)^{\hal\hbe} X_\hal^k \l_{\hbe k} W 
+ 2 \ri (\S_{\hb\ha})^{\hal\hbe} \nabla^\hb (\l_\hal^k \l_{\hbe k})
+ \frac{\ri}{2} \eps_{\ha\hb\hc\hd\he} W^{\hb\hc} (\S^{\hd\he})^{\hal\hbe} \l_\hal^k \l_{\hbe k} \ ,
\end{align}
\esubeq
into the eqs. \eqref{eq:PhiComponents} defining the superform $\Phi$
and \eqref{eq:SigmaComponents} for $\Sigma_{ H }$,
one arrives at the abelian CS action. However, as is evident from
considering the expression for $\Phi_\ha$ above, the expression involves
several derivatives which must be integrated by parts to arrive
at the conventional form of the action.

We seek instead a different closed superform $\mathfrak J$, which will
be given by
\begin{align}
\mathfrak J = \Sigma_{\rm CS} - \Sigma_R~,
\end{align}
where both $\Sigma_{\rm CS}$ and $\Sigma_R$ are solutions to the equation
\be
\rd \S = F \wedge F \wedge F \ . \label{AbelianSF^3}
\ee
The first is the Chern-Simons form,
\be
\S_{\rm CS} = V \wedge F \wedge F \ ,
\ee
while the second, $\S_R$, we will refer to as the curvature induced form. Here the curvature induced form 
is required to be a gauge-invariant {\it primary} superform constructed directly out of $W$ and its 
covariant derivatives. The Chern-Simons and curvature induced forms represent ingredients in a general procedure to 
construct Chern-Simons actions in three and five dimensions, see \cite{KT-M1212,BKNT-M2,KNT-M, KN13}. 
Gauge invariance of the corresponding action $\eqref{ectoS}$ is guaranteed by the fact that
$\S_{R}$ is gauge invariant by construction, while $\S_{\rm CS}$ transforms 
via an exact form. The advantage of this construction over the use of the BF action is that it can be 
straightforwardly generalized to the non-abelian case.

Now it turns out that $\mathfrak{J}_{ H }$ and $2 \mathfrak{J}$ describe the
same component action, with $2 \mathfrak{J}$ differing from $\mathfrak{J}$
by a total derivative (i.e. by an exact form) alluded to above. In other words,
\begin{align}
\rd \sfH = 2\mathfrak {J} - \mathfrak{J}_{ H }
	= V \wedge \Big(2 F \wedge F - \Phi\Big)
	- 2 \Sigma_R
	+ \Sigma_{ H }~,
\end{align}
for some four-form $\sfH$. It is evident we can choose
\begin{align}
\sfH = V \wedge C~,
\end{align}
for some three-form $C$ satisfying
\begin{align}
\rd C = 2 F \wedge F - \Phi~.
\end{align}
Provided there exists a gauge-invariant primary three-form $C$ that
solves this equation, then the curvature induced form is given immediately as
\begin{align}
\Sigma_R = \frac{1}{2} \Big(\S_{ H } + F \wedge C\Big)~.
\end{align}

The construction of such a three-form $C$ is straightforward. From dimensional considerations,
it is evident that $C_\hal^i{}_\hbe^j{}_\hga^k$ must vanish, while $C_\ha{}_\hbe^j{}_\hga^k$
must be proportional to $W^2$. The full solution is straightforward to derive:
\bsubeq
\begin{align}
C_\ha{}_\hbe^j{}_\hga^k &= - 4 \ri (\G_\ha)_{\hbe\hga} \eps^{jk} W^2 \ , \\
C_{\ha\hb}{}_\hga^k &= 8 \ri (\S_{\ha\hb})_\hga{}^\hde \l_\hde^k W \ , \\
C_{\ha\hb\hc} &= \eps_{\ha\hb\hc\hd\he} \Big(2 F^{\hd\he} W + \ri \l_k \S^{\hd\he} \l^k +
3 W^{\hd\he} W^2 \Big) \ .
\end{align}
\esubeq
The construction of $\S_R$ is now immediate. As required, it is given purely
in terms of $W$ and its covariant derivatives, with the nonzero components
given below:\footnote{We drop the subscript $R$ when 
referring to the components of $\S_R$ to avoid awkward notation.}
{\allowdisplaybreaks
\bsubeq
\bea
\S_{\hat{a}}{}_{\hat{\a}}^i{}_{\hat{\b}}^j{}_{\hat{\g}}^k{}_{\hat{\d}}^l &=&
 4 \Big( \eps^{ij} \eps^{kl} \big( (\G_{\hat{a}})_{{\hat{\a}}{\hat{\b}}} \eps_{{\hat{\g}}{\hat{\d}}} 
+ (\G_{\hat{a}})_{{\hat{\g}}{\hat{\d}}} \eps_{{\hat{\a}}{\hat{\b}}} \big)
+ \eps^{ik} \eps^{jl} \big( (\G_{\hat{a}})_{{\hat{\a}}{\hat{\g}}} \eps_{{\hat{\b}}{\hat{\d}}} + (\G_{\hat{a}})_{{\hat{\b}}{\hat{\d}}} \eps_{{\hat{\a}}{\hat{\g}}} \big) 
\non\\
&&~~~~
+ \eps^{il} \eps^{jk} \big( (\G_{\hat{a}})_{{\hat{\a}}{\hat{\d}}} \eps_{{\hat{\b}}{\hat{\g}}} + (\G_{\hat{a}})_{{\hat{\b}}{\hat{\g}}} \eps_{{\hat{\a}}{\hat{\d}}} \big)\Big) W^3 
 \ , \\
\S_{\hat{a} \hat{b}}{}_{\hat{\a}}^i{}_{\hat{\b}}^j{}_{\hat{\g}}^k &= &
- 12 \Big( \eps^{jk} \eps_{\hat{\b}\hat{\g}} (\S_{\hat{a}\hat{b}})_{{\hat{\a}}}{}^{\hat{\d}} \l_{\hat{\d}}^i W^2
+ \eps^{ij} \eps_{\hal\hbe} (\S_{\ha\hb})_\hga{}^{\hde} \l_\hde^k W^2
+ \eps^{ik} \eps_{\hga \hal} (\S_{\ha\hb})_\hbe{}^{\hde} \l_{\hde}^j W^2 \Big) 
\non\\
&&
+ 8 \Big( \eps^{jk} (\S_{\ha\hb})_{\hal\hbe} \l_\hga^i W^2
+ \eps^{ki} (\S_{\ha\hb})_{\hbe\hga} \l_\hal^j W^2 
+ \eps^{ij} (\S_{\ha\hb})_{\hga\hal} \l_\hbe^k W^2
\Big) \ , 
\\
\S_{\hat{a} \hat{b} \hat{c}}{}_{\hat{\a}}^i{}_{\hat{\b}}^j  
&=&
 \frac{\ri}{2} \eps^{ij} \eps_{\hat{\a} \hat{\b}} \eps_{\hat{a} \hat{b} \hat{c} \hat{d} \hat{e}} 
(\S^{\hat{d} \hat{e}})^{\hat{\g} \hat{\d}} (4 W^2 F_{\hga\hde} + 5 \ri W \l_\hga^k \l_{\hat{\d} k} + 6 W_{\hga\hde} W^3) 
\non\\
&&
- \ri \eps_{\hat{a} \hat{b} \hat{c} \hat{d} \hat{e}} (\S^{\hat{d} \hat{e}})_{\hat{\a} \hat{\b}} ( 4 W^2 X^{ij}
- 5 \ri W \l^{\hat{\g} (i} \l^{j)}_{\hat{\g}}) 
\non\\
&&
+ 6 (\S_{[\ha\hb})_{\hal\hbe} (\G_{\hc]})^{\hga\hde} \l_\hga^{(i} \l_\hde^{j)} W
+ 3 \eps^{ij} (\G_{[\ha})_{\hal\hbe} (\S_{\hb\hc]})^{\hga\hde} \l_\hga^k \l_{\hde k} W 
\non\\
&&
 - 6 \ri \eps^{ij} (\G_{[\ha})_{\hal\hbe} F_{\hb\hc]} W^2 
 \ , \\
\S_{\hat{a} \hat{b} \hat{c} \hat{d}}{}_{\hat{\a}}^i 
 &=&
  - \hf \eps_{\hat{a} \hat{b} \hat{c} \hat{d} \hat{e}}  (\G^{\hat{e}})_{\hat{\a}}{}^{\hat{\b}}
\Big( 12 \ri W F_{\hbe \hga} \l^{\hat{\g} i} - 6 \ri W X^{ij} \l_{\hat{\b} j} 
- 4 \l^{\hat{\g} (i} \l_{\hat{\g}}^{j)} \l_{\hat{\b} j} 
\non\\
&&~~~~~~~~~~~~~~~~~~~~~
- 2 X_\hbe^i W^3 
+ 6 \ri W_{\hbe \hga} W^2 \l^{\hga i} 
\Big) \non\\
&&
 + \eps_{\hat{a} \hat{b} \hat{c} \hat{d} \hat{e}} (\G^{\hat{e}})^{\hat{\b} \hat{\g}} \tr \big( \l_{\hat{\b}}^{(i} \l_{\hat{\g}}^{j)} \l_{\hat{\a} j} 
+ 3 \ri W_{\hal\hbe} W^2 \l_\hga^i \big) \non\\
&&
 - \ri \eps_{\ha\hb\hc\hd\he} W \l_\hal^i  \nabla^\he W
- 2 \ri \eps_{\ha\hb\hc\hd\he} W^2  \nabla^\he \l_{\hal}^i \non\\
&&
 - 2 \ri \eps_{\ha\hb\hc\hd\he} (\S^{\he\hat{f}})_\hal{}^\hbe W  \l_\hbe^i\nabla_{\hat{f}} W 
- 4 \ri \eps_{\ha\hb\hc\hd\he} (\S^{\he\hat{f}})_\hal{}^\hbe W^2  \nabla_{\hat{f}} \l_\hbe^i
\ , \\
\S_{\hat{a} \hat{b} \hat{c} \hat{d} \hat{e}} 
&=&
 - \frac{3}{2}  \eps_{\hat{a} \hat{b} \hat{c} \hat{d} \hat{e}} \Big( 
W X^{kl} X_{kl}
- 2 W F^{\hga\hde} F_{\hga\hde}
 - \ri X^{kl} \l^{\hde}_k \l_{\hde l}
- 2 \ri F^{\hga\hde} \l^k_\hga \l_{\hde k} \non\\
&&~~~~
 + \frac{2}{3} W( \nabla^{\hat{f}} W )\nabla_{\hat{f}} W 
+ \frac{4}{3} W^2 \nabla^{\hat{f}} \nabla_{\hat{f}} W 
+ 2 \ri W (\nabla_{\hat{\g} \hat{\d}} \l^{\hat{\g} k} ) \l^{\hat{\d}}_k \non\\
&&~~~~
+ \frac{1}{3} Y W^3 - \frac{4}{3} X^{\hga k} \l_{\hga k} W^2
- 6 W^{\hga\hde} F_{\hga\hde} W^2 - \frac{10}{3} W^{\hga\hde} W_{\hga\hde} W^3
 \Big) 
\ .
\eea
\esubeq
}

The abelian Chern-Simons action is then given by
\bea 
S_{\rm CS} = - \frac{1}{6} \int_{\cM^5} i{}^*\frak{J}
 \ ,
\label{ectoSCS}
\eea
where we have adjusted the normalization to match
\eqref{CSaction9.1}. As we will show in the next section,
it is straightforward to generalize the result for $\mathfrak{J}$
to a non-abelian vector multiplet. We will give the explicit component 
action in the next section for the non-abelian case.


\section{Non-abelian Chern-Simons theory} \label{section9}

In the non-abelian case, a closed-form expression for the Chern-Simons action 
as a superspace integral is not yet known. However, the corresponding action may be defined by 
postulating its variation\footnote{This definition is inspired by the earlier works
\cite{Zupnik99,K2006,KT-M_2014}.} 
\begin{subequations} \label{nonABELvar}
\bea
\d S_{\rm CS} &=&
-\frac{1}{8\pi} \oint_\g (v, \rd v)
\int \rd^{5|8}z\, E\, C^{(-4)} {\rm tr} \Big( \D V \cdot 
\re^{-\O_+} {\bm H}_{\rm YM}^{(2)} \re^{\O_+} \Big)~,
\eea
where we have defined
\bea
\D V: = \re^{- V } \d \re^{V}~, \qquad 
{\bm H}_{\rm YM}^{(2)} = v_i v_i {\bm H}_{\rm YM}^{ij} ~,
\eea
\end{subequations}
with the composite superfield ${\bm H}_{\rm YM}^{ij}$ given by \eqref{YML}.
Here $\D V$ is the covariantized variation of the tropical prepotential.
In the abelian case, the variation \eqref{nonABELvar} reduces to \eqref{varyingCSaction}.
In this paper, we will not elaborate on the above definition, and instead 
present a superform realization of the action. 

In the previous section we derived the closed five-form describing the abelian Chern-Simons theory and introduced two key ingredients: 
the Chern-Simons and curvature induced forms. In this section we will show how 
to generalize our approach to the non-abelian Chern-Simons theory based on a Yang-Mills multiplet 
and derive the corresponding component action. Our approach is analogous to the one 
adopted in \cite{KN-CS5D} in Minkowski superspace. We remind the reader that the Yang-Mills multiplet 
is described in section \ref{SYM}.

The appropriate closed five-form $\frak{J}$ to describe the \emph{non-abelian} Chern-Simons action 
may be found by generalizing the Chern-Simons form and the curvature induced form.
These five-forms now correspond to two solutions of the superform equation
\be
\rd \S = \langle \bm F^3 \rangle := \tr \Big( \bm F \wedge \bm F \wedge \bm F \Big) \ , \label{SFEQF3}
\ee
which is a straightforward generalization of \eqref{AbelianSF^3}. The Chern-Simons form $\S_{\rm CS}$ 
is again directly constructed out of $\bm V$, while the curvature induced form is constructed out of $\bm W$ 
and its covariant derivatives. If they transform by an exact form under the gauge group then 
their difference 
\be \frak{J} = \S_{\rm CS} - \S_{R}
\ee
will yield an appropriate closed five-form that describes the action. The Chern-Simons and curvature induced five-forms are 
discussed in more detail below.


\subsection{The Chern-Simons five-form}

The Chern-Simons form is
\be \S_{\rm CS} = \tr \Big( \bm V \wedge \bm F \wedge \bm F - \frac{\ri}{2} \bm V \wedge \bm V \wedge \bm V \wedge \bm F 
- \frac{1}{10} \bm V \wedge \bm V \wedge \bm V \wedge \bm V \wedge \bm V \Big) \ .
\ee
One can verify that it satisfies the superform equation \eqref{SFEQF3} by using
\be \bm\nabla = \rd - \ri \bm V \ , \quad \bm F = \rd \bm V + \ri \bm V \wedge \bm V \ , \quad \bm\nabla \bm F = 0 \implies \rd \bm F = \ri \bm V \wedge \bm F - \ri \bm F \wedge \bm V \ .
\ee

Since $\S_{\rm CS}$ has been constructed by extracting a total derivative from the gauge invariant superform $\langle \bm F^3 \rangle$ it must transform by 
a closed form under the gauge group. In fact, one can show it transforms by an exact form,
\be \S_{\rm CS} \rightarrow \S_{\rm CS} + \rd \ \tr \Big( \rd \t \wedge \big( \bm V \wedge \bm F - \frac{\ri}{2} \bm V \wedge \bm V \wedge \bm V \big) \Big) \ .
\ee

Note that since the gauge field $\bm V$ is primary, $\S_{\rm CS}$ is also a primary superform.


\subsection{The curvature-induced five-form}

To construct the curvature-induced five-form we look for a gauge-invariant solution to
\bea 
\rd \S &=& \tr \Big(\bm F \wedge \bm F \wedge \bm F \Big) \ . \label{CIFSE}
\eea

The condition that $\S$ is invariant allows one to express eq. \eqref{CIFSE} as
\bea
2 \bm\nabla_{[\hA} \S_{\hB\hC\hat{D}\hat{E}\hat{F}\}} + 5 \scT_{[\hA\hB}{}^{\hat{G}} \S_{|\hat{G}|\hC\hat{D}\hat{E}\hat{F}\}} 
= 30 \tr (\bm F_{[\hA\hB} \bm F_{\hC\hat{D}} \bm F_{\hat{E}\hat{F} \} }) \ .
\eea
Here we have used the fact that $\S$ is a gauge singlet\footnote{Keeping this in mind, we will use gauge covariant derivatives everywhere in this section.}
\bea 
\bm\nabla_\hA \S_{\hB\hC\hat{D}\hat{E}\hat{F}} = \nabla_\hA \S_{\hB\hC\hat{D}\hat{E}\hat{F}} \ .
\eea

The curvature induced form is defined to be a primary solution of eq. \eqref{CIFSE} that can be expressed covariantly  in terms of the vector 
multiplet field strength $\bm W$. Invariance of the curvature induced form is then guaranteed. 
On dimensional grounds, it is natural to impose the 
constraint
\be \S_\hal^i{}_\hbe^j{}_\hga^k{}_\hde^l{}_{\hat{\r}}^p = 0 \ .
\ee
Then analyzing the superform equation \eqref{CIFSE} by increasing dimension, enforcing the primary condition \eqref{primCond} and using the 
identities \eqref{VMIdentities} yields all the remaining components of the curvature induced five-form:
\bsubeq
\begin{align} \S_{\hat{a}}{}_{\hat{\a}}^i{}_{\hat{\b}}^j{}_{\hat{\g}}^k{}_{\hat{\d}}^l =& \ 4 \Big( \eps^{ij} \eps^{kl} \big( (\G_{\hat{a}})_{{\hat{\a}}{\hat{\b}}} \eps_{{\hat{\g}}{\hat{\d}}} 
+ (\G_{\hat{a}})_{{\hat{\g}}{\hat{\d}}} \eps_{{\hat{\a}}{\hat{\b}}} \big)
+ \eps^{ik} \eps^{jl} \big( (\G_{\hat{a}})_{{\hat{\a}}{\hat{\g}}} \eps_{{\hat{\b}}{\hat{\d}}} + (\G_{\hat{a}})_{{\hat{\b}}{\hat{\d}}} \eps_{{\hat{\a}}{\hat{\g}}} \big) \non\\
&+ \eps^{il} \eps^{jk} \big( (\G_{\hat{a}})_{{\hat{\a}}{\hat{\d}}} \eps_{{\hat{\b}}{\hat{\g}}} + (\G_{\hat{a}})_{{\hat{\b}}{\hat{\g}}} \eps_{{\hat{\a}}{\hat{\d}}} \big)\Big) \tr(\bm W^3) \ , \\
\S_{\hat{a} \hat{b}}{}_{\hat{\a}}^i{}_{\hat{\b}}^j{}_{\hat{\g}}^k =& 
- 12 \tr\Big( \eps^{jk} \eps_{\hat{\b}\hat{\g}} (\S_{\hat{a}\hat{b}})_{{\hat{\a}}}{}^{\hat{\d}} \bm \l_{\hat{\d}}^i \bm W^2
+ \eps^{ij} \eps_{\hal\hbe} (\S_{\ha\hb})_\hga{}^{\hde} \bm \l_\hde^k \bm W^2
+ \eps^{ik} \eps_{\hga \hal} (\S_{\ha\hb})_\hbe{}^{\hde} \bm \l_{\hde}^j \bm W^2 \Big) \non\\
&+ 8 \tr \Big( \eps^{jk} (\S_{\ha\hb})_{\hal\hbe}\bm \l_\hga^i \bm W^2
+ \eps^{ki} (\S_{\ha\hb})_{\hbe\hga} \bm \l_\hal^j \bm W^2 
+ \eps^{ij} (\S_{\ha\hb})_{\hga\hal} \bm \l_\hbe^k \bm W^2
\Big) \ , \\
\S_{\hat{a} \hat{b} \hat{c}}{}_{\hat{\a}}^i{}_{\hat{\b}}^j  =& \ \frac{\ri}{2} \eps^{ij} \eps_{\hat{\a} \hat{\b}} \eps_{\hat{a} \hat{b} \hat{c} \hat{d} \hat{e}} 
(\S^{\hat{d} \hat{e}})^{\hat{\g} \hat{\d}} \tr (4 \bm W^2 \bm F_{\hga\hde} + 5 \ri \bm W \bm \l_\hga^k \bm \l_{\hat{\d} k} + 6 W_{\hga\hde} \bm W^3) \non\\
&- \ri \eps_{\hat{a} \hat{b} \hat{c} \hat{d} \hat{e}} (\S^{\hat{d} \hat{e}})_{\hat{\a} \hat{\b}} \tr ( 4 \bm W^2 \bm X^{ij}
- 5 \ri \bm W \bm \l^{\hat{\g} (i} \bm \l^{j)}_{\hat{\g}}) \non\\
&+ 6 (\S_{[\ha\hb})_{\hal\hbe} (\G_{\hc]})^{\hga\hde} \tr (\bm \l_\hga^{(i} \bm \l_\hde^{j)} \bm W)
+ 3 \eps^{ij} (\G_{[\ha})_{\hal\hbe} (\S_{\hb\hc]})^{\hga\hde} \tr (\bm \l_\hga^k \bm \l_{\hde k} \bm W) \non\\
& - 6 \ri \eps^{ij} (\G_{[\ha})_{\hal\hbe} \tr (\bm F_{\hb\hc]} \bm W^2) \ , \\
\S_{\hat{a} \hat{b} \hat{c} \hat{d}}{}_{\hat{\a}}^i  &= - \hf \eps_{\hat{a} \hat{b} \hat{c} \hat{d} \hat{e}}  (\G^{\hat{e}})_{\hat{\a}}{}^{\hat{\b}}
\tr \Big( 6 \ri \bm W \{ \bm F_{\hbe \hga} , \bm \l^{\hat{\g} i} \} - 3 \ri \bm W \{ \bm X^{ij} , \bm \l_{\hat{\b} j} \} \non\\
&\quad- 4 \bm \l^{\hat{\g} (i} \bm \l_{\hat{\g}}^{j)} \bm \l_{\hat{\b} j} 
- 2 X_\hbe^i \bm W^3 
+ 6 \ri W_{\hbe \hga} \bm W^2 \bm \l^{\hga i} 
\Big) \non\\
& + \eps_{\hat{a} \hat{b} \hat{c} \hat{d} \hat{e}} (\G^{\hat{e}})^{\hat{\b} \hat{\g}} \tr \big( \bm \l_{\hat{\b}}^{(i} \bm \l_{\hat{\g}}^{j)} \bm \l_{\hat{\a} j} 
+ 3 \ri W_{\hal\hbe} \bm W^2 \bm \l_\hga^i \big) \non\\
& - \frac{\ri}{2} \eps_{\ha\hb\hc\hd\he} \tr(\{ \bm W , \bm \nabla^\he \bm W \} \bm \l_\hal^i)
- 2 \ri \eps_{\ha\hb\hc\hd\he} \tr (\bm W^2 \bm \nabla^\he \bm \l_{\hal}^i) \non\\
& - \ri \eps_{\ha\hb\hc\hd\he} (\S^{\he\hat{f}})_\hal{}^\hbe \tr(\{ \bm W , \bm \nabla_{\hat{f}} \bm W \} \bm \l_\hbe^i)
- 4 \ri \eps_{\ha\hb\hc\hd\he} (\S^{\he\hat{f}})_\hal{}^\hbe \tr(\bm W^2 \bm \nabla_{\hat{f}} \bm \l_\hbe^i)
\ , \\
\S_{\hat{a} \hat{b} \hat{c} \hat{d} \hat{e}} =& - \frac{3}{2}  \eps_{\hat{a} \hat{b} \hat{c} \hat{d} \hat{e}} \tr \Big( 
\bm W \bm X^{kl} \bm X_{kl}
- 2 \bm W \bm F^{\hga\hde} \bm F_{\hga\hde} 
 - \ri \bm X^{kl} \bm \l^{\hde}_k \bm \l_{\hde l}
- 2 \ri \bm F^{\hga\hde} \bm \l^k_\hga \bm \l_{\hde k} \non\\
& + \frac{2}{3} \bm W( \bm\nabla^{\hat{f}} \bm W) \bm\nabla_{\hat{f}} \bm W 
+ \frac{4}{3} \bm W^2 \bm\nabla^{\hat{f}} \bm\nabla_{\hat{f}} \bm W
+ \ri \bm W [ \bm\nabla_{\hat{\g} \hat{\d}} \bm \l^{\hat{\g} k},  \bm \l^{\hat{\d}}_k ]
- 2 \bm W^2 \bm \l^{\hat{\g} k} \bm \l_{\hat{\g} k} \non\\
&+ \frac{1}{3} Y \bm W^3 - \frac{4}{3} X^{\hga k} \bm\l_{\hga k} \bm W^2 
- 6 W^{\hga\hde} \bm F_{\hga\hde} \bm W^2 - \frac{10}{3} W^{\hga\hde} W_{\hga\hde} \bm W^3
\Big) \ .
\end{align}
\esubeq

It is worth elucidating the relation of the above curvature induced form to the one constructed in \cite{KN13} in the rigid supersymmetric case. 
Switching off the Weyl multiplet ($W_{\hal\hbe} = 0$) and replacing the covariant derivatives with their corresponding flat ones, 
\be \bm \nabla_\hA \quad \rightarrow \quad \bm \cD_{\hA} = \cD_{\hA} - \ri \bm V_{\hA} \ ,
\ee
gives
\bsubeq
\bea
\S_{\hat{a}}{}_{\hat{\a}}^i{}_{\hat{\b}}^j{}_{\hat{\g}}^k{}_{\hat{\d}}^l &=& \ 4 \Big( \eps^{ij} \eps^{kl} \big( (\G_{\hat{a}})_{{\hat{\a}}{\hat{\b}}} \eps_{{\hat{\g}}{\hat{\d}}} 
+ (\G_{\hat{a}})_{{\hat{\g}}{\hat{\d}}} \eps_{{\hat{\a}}{\hat{\b}}} \big)
+ \eps^{ik} \eps^{jl} \big( (\G_{\hat{a}})_{{\hat{\a}}{\hat{\g}}} \eps_{{\hat{\b}}{\hat{\d}}} + (\G_{\hat{a}})_{{\hat{\b}}{\hat{\d}}} \eps_{{\hat{\a}}{\hat{\g}}} \big) \non\\
&&+ \eps^{il} \eps^{jk} \big( (\G_{\hat{a}})_{{\hat{\a}}{\hat{\d}}} \eps_{{\hat{\b}}{\hat{\g}}} + (\G_{\hat{a}})_{{\hat{\b}}{\hat{\g}}} \eps_{{\hat{\a}}{\hat{\d}}} \big)\Big) \tr(\bm W^3) \ , \\
\S_{\hat{a} \hat{b}}{}_{\hat{\a}}^i{}_{\hat{\b}}^j{}_{\hat{\g}}^k &=& 
- 12 \tr\Big( \eps^{jk} \eps_{\hat{\b}\hat{\g}} (\S_{\hat{a}\hat{b}})_{{\hat{\a}}}{}^{\hat{\d}} \bm \l_{\hat{\d}}^i \bm W^2
+ \eps^{ij} \eps_{\hal\hbe} (\S_{\ha\hb})_\hga{}^{\hde} \bm \l_\hde^k \bm W^2
+ \eps^{ik} \eps_{\hga \hal} (\S_{\ha\hb})_\hbe{}^{\hde} \bm \l_{\hde}^j \bm W^2 \Big) \non\\
&&+ 8 \tr \Big( \eps^{jk} (\S_{\ha\hb})_{\hal\hbe}\bm \l_\hga^i \bm W^2
+ \eps^{ki} (\S_{\ha\hb})_{\hbe\hga} \bm \l_\hal^j \bm W^2 
+ \eps^{ij} (\S_{\ha\hb})_{\hga\hal} \bm \l_\hbe^k \bm W^2
\Big) \ , \\
\S_{\hat{a} \hat{b} \hat{c}}{}_{\hat{\a}}^i{}_{\hat{\b}}^j  
&=& \ \frac{\ri}{2} \eps^{ij} \eps_{\hat{\a} \hat{\b}} \eps_{\hat{a} \hat{b} \hat{c} \hat{d} \hat{e}} 
(\S^{\hat{d} \hat{e}})^{\hat{\g} \hat{\d}} \tr (4 \bm W^2 \bm F_{\hga\hde} + 5 \ri \bm W \bm \l_\hga^k \bm \l_{\hat{\d} k}) \non\\
&&- \ri \eps_{\hat{a} \hat{b} \hat{c} \hat{d} \hat{e}} (\S^{\hat{d} \hat{e}})_{\hat{\a} \hat{\b}} \tr ( 4 \bm W^2 \bm X^{ij}
- 5 \ri \bm W \bm \l^{\hat{\g} (i} \bm \l^{j)}_{\hat{\g}}) \non\\
&&+ 6 (\S_{[\ha\hb})_{\hal\hbe} (\G_{\hc]})^{\hga\hde} \tr (\bm \l_\hga^{(i} \bm \l_\hde^{j)} \bm W)
+ 3 \eps^{ij} (\G_{[\ha})_{\hal\hbe} (\S_{\hb\hc]})^{\hga\hde} \tr (\bm \l_\hga^k \bm \l_{\hde k} \bm W) \non\\
&& - 6 \ri \eps^{ij} (\G_{[\ha})_{\hal\hbe} \tr (\bm F_{\hb\hc]} \bm W^2) \ , \\
\S_{\hat{a} \hat{b} \hat{c} \hat{d}}{}_{\hat{\a}}^i  
&=& - \hf \eps_{\hat{a} \hat{b} \hat{c} \hat{d} \hat{e}}  (\G^{\hat{e}})_{\hat{\a}}{}^{\hat{\b}}
\tr \Big( 6 \ri W \{ \bm F_{\hbe \hga} , \bm \l^{\hat{\g} i} \} - 3 \ri \bm W \{ \bm X^{ij} , \bm \l_{\hat{\b} j} \}
- 4 \bm \l^{\hat{\g} (i} \bm \l_{\hat{\g}}^{j)} \bm \l_{\hat{\b} j} 
\Big) \non\\
&& + \eps_{\hat{a} \hat{b} \hat{c} \hat{d} \hat{e}} (\G^{\hat{e}})^{\hat{\b} \hat{\g}} \tr \big( \bm \l_{\hat{\b}}^{(i} \bm \l_{\hat{\g}}^{j)} \bm \l_{\hat{\a} j} 
 \big) \non\\
&& - \frac{\ri}{2} \eps_{\ha\hb\hc\hd\he} \tr(\{ \bm W , \bm \cD^\he \bm W \} \bm \l_\hal^i)
- 2 \ri \eps_{\ha\hb\hc\hd\he} \tr (\bm W^2 \bm \cD^\he \bm \l_{\hal}^i) \non\\
&& - \ri \eps_{\ha\hb\hc\hd\he} (\S^{\he\hat{f}})_\hal{}^\hbe \tr(\{ \bm W , \bm \cD_{\hat{f}} \bm W \} \bm \l_\hbe^i)
- 4 \ri \eps_{\ha\hb\hc\hd\he} (\S^{\he\hat{f}})_\hal{}^\hbe \tr(\bm W^2 \bm \cD_{\hat{f}} \bm \l_\hbe^i)
\ ,~~~~~~~~~~~ \\
\S_{\hat{a} \hat{b} \hat{c} \hat{d} \hat{e}} 
&=& - \frac{3}{2}  \eps_{\hat{a} \hat{b} \hat{c} \hat{d} \hat{e}} \tr \Big( 
\bm W \bm X^{kl} \bm X_{kl}
- 2 \bm W \bm F^{\hga\hde} \bm F_{\hga\hde} 
 - \ri \bm X^{kl} \bm \l^{\hde}_k \bm \l_{\hde l}
- 2 \ri \bm F^{\hga\hde} \bm \l^k_\hga \bm \l_{\hde k} \non\\
&&\quad 
+ \frac{2}{3} \bm W (\bm\cD^{\hat{f}} \bm W) \bm\cD_{\hat{f}} \bm W 
+ \frac{4}{3} \bm W^2 \bm\cD^{\hat{f}} \bm\cD_{\hat{f}} \bm W
+ \ri \bm W [ \bm\cD_{\hat{\g} \hat{\d}} \bm \l^{\hat{\g} k},  \bm \l^{\hat{\d}}_k ]
\non\\
&&\quad
- 2 \bm W^2 \bm \l^{\hat{\g} k} \bm \l_{\hat{\g} k} \Big) \ ,
~~~~~~~~~
\eea
\esubeq
where
\be
\bm \l_\hal^i := - \ri \bm \cD_\hal^i \bm W \ , \quad
\bm X^{ij} := - \frac{1}{4} \bm \cD^{\hal (i} \bm \l_\hal^{j)} \ , \quad
\bm F_{\hal\hbe} = \frac{1}{4} \bm \cD_{(\hal}^k \bm \l_{\hbe) k} \ .
\ee
The above curvature induced form agrees with the one found in \cite{KN13} up to the addition of an 
exact five-form
\be
\S_{\rm exact}= \rd N \ , \label{TDerW}
\ee
with
\bsubeq
\begin{align}
N_\hal^i{}_\hbe^j{}_\hga^k{}_\hde^l &= 0 \ , \quad
N_{\ha}{}_\hbe^j{}_\hga^k = 0 \ , \quad
N_{\ha\hb}{}_\hga^k{}_\hde^l = 0 \ , \\
N_{\ha\hb\hc}{}_\hde^l &= - \ri \eps_{\ha\hb\hc\hd\he} (\S^{\hd\he})_\hde{}^\hga \tr \big( \bm W^2 \bm \l_\hga^l \big) \ , \\
N_{\ha\hb\hc\hd} &= 0 \ .
\end{align}
\esubeq
Ignoring boundary terms, the exact form does not change the corresponding action.

It is worth noting that although we can add a total derivative constructed out of $\bm W$ and its 
covariant derivatives, the curvature induced form is uniquely fixed in conformal supergravity. In particular, 
a primary generalization of \eqref{TDerW} in supergravity does not exist.


\subsection{The non-abelian Chern-Simons action}


Making use of the superform $\S$ one can construct a closed five-form in 5D from which one can derive a supersymmetric action. 
We now make use of the closed form,
\be
\frak{J} := \S_{\rm CS} - \S_{R}
\ee
and the action principle \eqref{ectoSCS} together with the formula \eqref{SigComps}.
We find the action\footnote{Here we understand the superfield $\bm W$ and the superfields constructed from its covariant derivatives 
as their corresponding component fields. It should be clear from context which we are referring to.}
{
\allowdisplaybreaks
\begin{align}
S &= \int \rd^5x \,e\, \tr \Big\{
	- \frac{1}{24} \eps^{\hat{a}\hat{b}\hat{c}\hat{d}\hat{e}} \bm v_{\hat{a}} \bm f_{\hat{b}\hat{c}} \bm f_{\hat{d}\hat{e}} 
	- \frac{\ri}{24} \eps^{\hat{a}\hat{b}\hat{c}\hat{d}\hat{e}} \bm v_{\ha} \bm v_{\hb} \bm v_{\hc} \bm f_{\hd\he}
	-  \frac{1}{60} \eps^{\hat{a}\hat{b}\hat{c}\hat{d}\hat{e}} \bm v_{\ha} \bm v_{\hb} \bm v_{\hc} \bm v_{\hd} \bm v_{\he} \non\\
	&\qquad
	- \frac{1}{4} \bm W \bm F_{\hat{a} \hat{b}} \bm F^{\hat{a} \hat{b}} + \frac{1}{4} \bm W \bm X^{ij} \bm X_{ij}
	+ \frac{\ri}{4} \bm F_{\hat{a} \hat{b}} (\bm \l^k \S^{\hat{a} \hat{b}} \bm \l_k) - \frac{\ri}{4} \bm X_{ij} (\bm \l^i \bm \l^j) \non\\
	&\qquad
	+ \frac{\ri}{4} \bm W (\bm \l^k \overleftrightarrow{\! \not \! \! {\bm \nabla}} \bm \l_k)
	+ \frac{1}{6} \bm W({\bm \nabla}^\ha \bm W) {\bm \nabla}_\ha \bm W 
	+ \frac{1}{3} \bm W^2 {\bm \nabla}^\ha {\bm \nabla}_\ha \bm W 
	- \frac{32}{9} D \bm W^3 \non\\
	&\qquad
	+ \frac{32 \ri}{9} \c^{k} \bm \l_{k} \bm W^2
	- \frac{3}{4} W^{\ha\hb} \bm F_{\ha\hb} \bm W^2 
	- \frac{5}{12} W^{\ha\hb} W_{\ha\hb} \bm W^3
	- \hf \bm W^2 \bm \l^k \bm \l_k
	\eol & \qquad
	- \frac{\ri}{8} (\psi_{\ha i} \G^\ha \S^{\hb\hc} \bm \l^{i}) 
		\Big(\{ \bm F_{\hb\hc} , \bm W \} + \bm W^2  W_{\hb\hc} \Big) 
		- \frac{\ri}{24} \psi_{\ha i} \G^\ha \G^\hb \bm \l^{i} \{ \bm W , \bm \nabla_\hb \bm W\} \non\\
	&\qquad - (\psi_{\ha i} \G^\ha \bm \l_{j})
		\Big( \frac{\ri}{8} \{ \bm X^{ij}, \bm W \} 
		+ \frac{1}{6} \bm \l^{(i} \bm \l^{j)} \Big)
		 + \frac{8 \ri}{9} (\psi_{\ha i} \G^\ha \c^i) \bm W^3 
	\eol & \qquad
	- \frac{\ri}{6} \psi_{\ha i} \G^\ha {{\slashed \nabla}} \bm \l^{i} \bm W^2
	- \frac{1}{12} (\psi_{\ha i} \bm \l_j) (\bm \l^{(i} \G^\ha \bm \l^{j)})
	+ \frac{\ri}{8} (\psi_{\hc i} \S^{\ha\hb} \G^\hc \bm \l^i) \bm W^2 W_{\ha\hb}
	\eol & \qquad
	+ \frac{\ri}{24} (\psi_\ha{}_k \psi_\hb{}^k) \big(
	2 \bm W^2 \bm F^{\ha\hb}
	- 3 \ri \bm \l^k \S^{\ha\hb} \bm \l_{k} \bm W
	+ 3 W^{\ha\hb} \bm W^3
	\big)
	\non\\&\qquad
	+ \frac{\ri}{48} \eps^{\ha\hb\hc\hd\he} (\psi_\ha{}_k \G_\hb \psi_\hc{}^{k}) \bm F_{\hd\he} \bm W^2
	+ \frac{\ri}{12} (\psi_\ha{}_i \S^{\ha\hb} \psi_\hb{}_j) \big(
	2 \bm W^2 \bm X^{ij} 
	- 3 \ri \bm W \bm \l^{(i} \bm \l^{j)} \big)
	\eol & \qquad
	+ \frac{1}{12} (\psi_{\ha i} \S^{\ha\hb} \bm \l^{i}) \{\bm W, \psi_{\hb j} \bm \l^{j}\}
	+ \frac{1}{16} \eps^{\ha\hb\hc\hd\he} (\psi_\ha{}_k \psi_\hb{}^k)
		(\psi_\hc{}_{ j} \S_{\hd\he} \bm \l^j) \bm W^2
	\eol &\qquad
	+ \frac{1}{24} \eps^{\ha\hb\hc\hd\he} (\psi_\ha{}_j \S_{\hb\hc} \psi_\hd{}_k)
		(\psi_\he{}^{j} \bm \l^k) \bm W^2
	+ \frac{1}{96} \eps^{\ha\hb\hc\hd\he}
	(\psi_\ha{}_k \G_\hb \psi_\hc{}^{k})
		(\psi_\hd{}_j \psi_\he{}^j) \bm W^3 \Big\} \ ,
\end{align}
}
where we have defined
\be
\frac{\ri}{2} [{\bm \nabla}_{\hat{\g} \hat{\d}} \bm \l^{\hat{\g} k} , \bm \l^{\hat{\d}}_k] 
= \frac{\ri}{2} \bm \l^k \overleftrightarrow{\not \! \! {\bm\nabla}} \bm \l_k 
:= \frac{\ri}{2} \bm \l^k \! \not \! \! {\bm \nabla} \bm \l_k 
- \frac{\ri}{2} ( \! \not \! \! {\bm \nabla} \bm \l^k )\bm \l_k 
\ee
and introduced the bar-projected field strength and one-form:
\be 
\bm f_{\ha\hb} = 2 e_\ha{}^{\hat{m}} e_{\hb}{}^{\hat{n}} \big(
\partial_{[\hat{m}} \bm v_{\hat{n}]}
-\ri \bm v_{[\hm}\bm v_{\hn]}\big)
\ , \quad
\bm v_\ha := e_\ha{}^{\hat{m}}\bm v_{\hat{m}} 
\ , 
\quad \bm v_{\hat{m}} := \bm V_{\hat{m}}| \ .
\ee

The vector covariant derivatives of the component fields may be expressed in terms of the $\hat{\cD}_\ha$ derivatives and hatted component fields 
introduced in section \ref{WeylMultiplet}. For completeness we include the following component results:
\bsubeq
\begin{align}
\bm \nabla_\ha \bm W &= \hat{\bm \cD}_\ha \bm W - \frac{\ri}{2} \psi_\ha{}^\hga_k \bm \l_\hga^k
\ , \\
\bm \nabla^\ha \bm \nabla_\ha \bm W &= 
\hat{\bm \cD}^\ha \bm \nabla_\ha \bm W
- \frac{\ri}{2} \psi^\ha{}^\hbe_j \bm \nabla_\ha \bm \l_\hbe^j + \frac{\ri}{2} \psi_\hbe{}^\hga{}^\hbe_j W_{\hga\hde} \bm \l^{\hde j}
- \frac{8 \ri}{3} \psi_\hbe{}^\hga{}^\hbe_j \c_\hga^j \bm W
\non\\ &\qquad
+ \frac{\ri}{2} \psi_\hbe{}^\hga{}^{\hbe}{}_j [\bm\l_\hga^j, \bm W]
- 2 \hat{\frak{f}}_\ha{}^\ha \bm W 
+ \hf \hat\phi_\hbe{}^\hga{}^{\hbe j} \bm \l_{\hga j}
\non\\ &\qquad
+ \frac{20 \ri}{3} \c^{\hal i} \bm \l_{\hal i} + \frac{20}{3} D \bm W + \frac{1}{32} W^{\ha\hb} W_{\ha\hb} \bm W
\ , \\
\bm \nabla_\ha \bm \l_\hal^i &= \hat{\bm \cD}_\ha \bm \l_\hal^i 
- \psi_\ha{}^{\hbe i} (\bm F_{\hbe\hal} + W_{\hbe\hal} \bm W)
- \hf \psi_\ha{}_{\hal j} \bm X^{ij}
- \hf \psi_\ha{}^{\hbe i} \bm\nabla_{\hbe\hal} \bm W \non\\
&\qquad- \ri \hat{\phi}_\ha{}^i_\hal \bm W
+ \frac{1}{4} \tilde{W}_{\ha\hb\hc} (\S^{\hb\hc})_\hal{}^\hbe \bm \l_\hbe^i
- \frac{8}{3} (\G_\ha)_\hal{}^\hbe \c_\hbe^i \bm W
\ ,
\end{align}
\esubeq
where
\be \hat{\bm \cD}_\ha = \hat{\cD}_\ha -\ri \bm V_\ha \ .
\ee

Note that the covariant field strength may be expressed in terms of the bar-projected field strength. Performing the component projection of 
the identity
\be \bm F_{\hat{m}\hat{n}} = E_{\hat{m}}{}^{\hA} E_{\hat{n}}{}^{\hB} \bm F_{\hA\hB}(-1)^{\eps_{\hA} \eps_{\hB}} \ ,
\ee
we find
\be
\bm F_{\ha\hb}| = \bm f_{\ha\hb} 
+ \ri (\G_{[\ha})_\hal{}^\hbe \psi_{\hb]}{}^\hal_k \bm \l_\hbe^k
+ \frac{\ri}{2} \psi_{[\ha}{}^\hga_k \psi_{\hb]}{}^k_{\hga} \bm W \ .
\ee

It should be mentioned that the abelian Chern-Simons action can straightforwardly be read off of the action presented in this section.


\section{Supermultiplets with gauged central charge} \label{SGCC}

In the presence of a gauged central charge 
different off-shell multiplets in conformal supergravity 
become possible. For example, in 4D $\cN = 2$ conformal supergravity there exist so-called vector-tensor 
multiplets, which may be viewed as dual versions of the abelian vector multiplet and possess gauge two-forms.\footnote{See 
\cite{Novak1, Novak2} for a superspace description of all known off-shell vector-tensor multiplets in 4D $\cN = 2$ conformal supergravity.} 
The situation in 5D conformal supergravity is similar. There also exists a dual version of the abelian 
vector multiplet, which we refer to as the two-form multiplet. The off-shell 
multiplet was first constructed in \cite{Ohashi4} within the component approach 
and was shown that it may be generalized to a so-called large tensor multiplet that may be given a mass. 
Recently, two of us have shown how to describe both the two-form and large tensor multiplets in Minkowski superspace 
by making convenient use of superform formulations \cite{KN-CS5D}.

In this section we generalize the results of \cite{KN-CS5D} to conformal superspace. 
Firstly, we discuss how to gauge the central charge in conformal superspace. We then give the superform 
formulation for the linear multiplet with central charge and immediately derive its corresponding action principle. 
The action provides an important ingredient in constructing actions for multiplets with gauged central charge. 
Finally, we give the superform formulations for the gauge two-form and large tensor multiplets.


\subsection{Gauging a central charge in conformal superspace} \label{gauging}

We can introduce a central charge $\D$ in conformal superspace and gauge it using an abelian vector multiplet 
associated with a gauge connection $\cV$. Doing so requires that we follow a similar procedure 
as the one used in section \ref{SYM}. We can obtain the resulting structure by simply replacing the gauge connection $V$ and field strength 
$F$ with those associated with the central charge $\D$ as follows:
\be \ri V \rightarrow \cV \D \ , \quad \ri F \rightarrow \cF \D \ .
\ee
The central charge is required to commute with the covariant derivatives
\be [\D, \bm \nabla_{\hat{A}}] = 0
\ee
and annihilate both $\cV$ and $\cF$
\be  \quad \D \cV = 0 \ , \quad \D \cF = 0 \ .
\ee
The central charge gauge transformations of the covariant derivatives are
\be \d \bm \nabla_{\hat{A}} = [\L \D , \bm \nabla_{\hat{A}}] \implies  \d \cV_{\hat{A}} = \nabla_{\hat{A}} \L \ , \label{CCGTrans}
\ee
where the gauge parameter $\L$ is inert under the central charge, $\D \L = 0$.

We constrain the field strength $\cF$ formally the same way as $F$ but with $W$ replaced by $\cW$. 
The components of $\cF$ are given by the following:
\bsubeq \label{VGCC}
\begin{align}
\cF_{\hat{\a}}^i{}_{\hat{\b}}^j &= 2 \ri \eps^{ij} \eps_{\hat{\a} \hat{\b}} \cW \ , \\
\cF_{\hat{a}}{}_{\hat{\b}}^j &= - (\G_{\hat{a}})_{\hat{\b}}{}^{\hat{\g}} \nabla_{\hat{\g}}^j \cW \ , \\
\quad \cF_{\hat{a}\hat{b}} &= - \frac{\ri}{4} (\S_{\hat{a}\hat{b}})^{\hat{\a} \hat{\b}} (\nabla^k_{\hat{\a}} \nabla_{\hat{\b} k} - 4 \ri W_{\hal \hbe}) \cW \ ,
\end{align}
\esubeq
with $\cW$ constrained by the Bianchi identity
\be \nabla_{\hat{\a}}^{(i} \nabla_{\hat{\b}}^{j)} \cW = \frac{1}{4} \eps_{\hat{\a} \hat{\b}} \nabla^{\hat{\g} (i} \nabla_{\hat{\g}}^{j)} \cW \ . \label{VGCCBI}
\ee
The above results will be used in the remainder of this section.


\subsection{The linear multiplet with central charge}
\label{Linear-MultipletCC}

In this subsection we construct a superform formulation for 
the 5D linear multiplet with gauged central charge in conformal superspace, generalizing the one given in \cite{KN-CS5D}. Our 
approach is similar to the one adopted for the 4D $\cN = 2$ linear multiplet in conformal supergravity \cite{Butter:2012ze}.
We will show that the superform formulation naturally 
leads to the action principle based on a linear multiplet.


\subsubsection{Superform formulation for the linear multiplet} \label{SFlinear}

In \cite{Butter:2012ze}  a superform formulation for the 4D $\cN = 2$ linear multiplet
was found by extending the vielbein to 
include the one-form gauging the central charge. This leads to a system of superforms describing the linear multiplet. As in \cite{KN-CS5D} 
we instead start with a system of superforms that generalizes the one that appeared in \cite{Butter:2012ze}.

We introduce two primary superforms: a five-form $\tilde{\S}$ and a four-form $\Phi$. We require that they satisfy the superform equations
\be
\bm\nabla \tilde{\S} = F \wedge \Phi \ , \quad \bm\nabla \Phi = - \D \tilde{\S} \label{SUPERFORMEQNS}
\ee
and transform as scalars under the gauge transformations \eqref{CCGTrans}
\be \d \tilde{\S} = \L \D \tilde{\S} \ , 
\quad \d \Phi = \L \D \Phi \ .
\ee
The superforms $\tilde{\S}$ and $\Phi$ can now be related to the linear multiplet with central charge by imposing certain 
constraints. However, it will prove useful to first introduce some notation to deal with the component form of \eqref{SUPERFORMEQNS}.

We introduce indices that range over not just $\hat{A}$ but an additional bosonic coordinate, $\hat{\cA} = (\hat{A} , 6)$. Then we may 
rewrite eq. \eqref{SUPERFORMEQNS} in components as
\be \bm\nabla_{[\hat{\cA}} \S_{\hat{\cB} \hat{\cC} \hat{\cD} \hat{\cE} \hat{\cF} \}} 
+ \frac{5}{2} \scT_{[\hat{\cA} \hat{\cB}}{}^{\hat{\cG}} \S_{|\hat{\cG}| \hat{\cC} \hat{\cD} \hat{\cE} \hat{\cF} \} } = 0 \ , \label{BISigma}
\ee
where we have made the identifications
\be \scT_{\hat{A} \hat{B}}{}^6 = F_{\hat{A} \hat{B}} \ , \quad 
\scT_{6 \hat{B}}{}^{\hat{\cA}} = \scT_{\hat{B} 6}{}^{\hat{\cA}} = 0 \ , \quad \bm\de_6 = \D
\ee
and
\begin{align} \tilde{\S} &= \frac{1}{5!} E^{\hat{E}} \wedge E^{\hat{D}} \wedge E^{\hat{C}} \wedge E^{\hat{B}} \wedge E^{\hat{A}} \S_{\hat{A} \hat{B} \hat{C} \hat{D} \hat{E}} \ , \non\\
\Phi &= \frac{1}{4!} E^{\hat{D}} \wedge E^{\hat{C}} \wedge E^{\hat{B}} \wedge E^{\hat{A}} \S_{6 \hat{A} \hat{B} \hat{C} \hat{D}} \ .
\end{align}

We constrain the lowest dimension components by
\begin{align}
\S_\hal^i{}_\hbe^j{}_\hga^k{}_\hde^l{}_{\hat{\r}}^p
&= \S_{\hat{a}}{}_\hal^i{}_\hbe^j{}_\hga^k{}_\hde^l
= \S_{\hat{a} \hat{b}}{}_\hal^i{}_\hbe^j{}_\hga^k
= 
\S_{6}{}_\hal^i{}_\hbe^j{}_\hga^k{}_\hde^l
= 
\S_{6 \hat{a}}{}{}_\hbe^j{}_\hga^k{}_\hde^l = 0 \ , \non\\
\S_{6 \hat{a} \hat{b}}{}_\hal^i{}_\hbe^j &= 8 \ri (\S_{\hat{a} \hat{b}})_{\hat{\a} \hat{\b}} \cL^{ij} \ ,
\end{align}
and analyze eq. \eqref{BISigma}. The remaining components are completely determined as follows:
\bsubeq
\begin{align}
\S_{\hat{a} \hat{b} \hat{c}}{}_\hal^i{}_\hbe^j &=  4 \ri \eps_{\hat{a} \hat{b} \hat{c} \hat{d} \hat{e}} (\S^{\hat{d} \hat{e}})_{\hat{\a} \hat{\b}}  \cW \cL^{ij} \ , \\
\S_{6 \hat{a} \hat{b} \hat{c}}{}_\hal^i &= - \frac{2}{3} \eps_{\hat{a} \hat{b} \hat{c} \hat{d} \hat{e}} (\S^{\hat{d} \hat{e}})_{\hat{\a}}{}^{\hat{\b}} \bm \nabla_{\hat{\b} j}\cL^{ji} 
= - 2 \eps_{\ha\hb\hc\hd\he} (\S^{\hd \he})_\hal{}^\hbe \varphi_\hbe^i \ , \\
\S_{\hat{a} \hat{b} \hat{c} \hat{d}}{}_\hal^i &= - 2 \eps_{\hat{a} \hat{b} \hat{c} \hat{d} \hat{e}} (\G^{\hat{e}})_{\hat{\a}}{}^{\hat{\b}} ( \cW \varphi_\hbe^i + \ri \l_{\hbe j} \cL^{ji})  \ , \\
\S_{6 \hat{a} \hat{b} \hat{c} \hat{d}} &= \frac{\ri}{12} \eps_{\hat{a} \hat{b} \hat{c} \hat{d} \hat{e}} (\G^{\hat{e}})^{\hat{\a} \hat{\b}} \bm \nabla_{\hat{\a}}^i \bm\nabla_{\hat{\b}}^j \cL_{ij} 
\equiv \eps_{\ha\hb\hc\hd\he} \Phi^\he \ , \\
\S_{\hat{a} \hat{b} \hat{c} \hat{d} \hat{e}} &= \eps_{\hat{a} \hat{b} \hat{c} \hat{d} \hat{e}} ( \cW F + X^{ij} \cL_{ij} + 2 \l^{\hga k} \varphi_{\hga k} ) \ ,
\end{align}
\esubeq
where $\cL^{ij}$ satisfies the constraint for the linear multiplet
\be \bm\nabla_{\hat{\a}}^{(i} \cL^{jk)} = 0 \label{constLinear}
\ee
and we have introduced the superfields
\bsubeq
\begin{align}
\varphi_\hal^i &:= \frac{1}{3} \bm \nabla_{\hal j} \cL^{ij} \ , \\
F &:= \frac{\ri}{12} \bm\nabla^{\hga i} \bm\nabla_\hga^j \cL_{ij} = - \frac{\ri}{4} \bm\nabla^{\hga k} \varphi_{\hga k} \ .
\end{align}
\esubeq

The above superfields together with
\be \Phi_{\ha} = \frac{\ri}{12} (\G_{\hat{a}})^{\hat{\a} \hat{\b}} \bm \nabla_{\hat{\a}}^i \bm\nabla_{\hat{\b}}^j \cL_{ij} 
= - \frac{\ri}{4} (\G_\ha)^{\hal\hbe} \bm \nabla_\hal^k \varphi_{\hbe k}
\ee
satisfy the following useful identities:
\bsubeq
\bea
\bm\nabla_\hal^i \cL^{jk} &=& 2 \eps^{i(j} \varphi_\hal^{k)} \ , \\
\bm\nabla_\hal^i \varphi_\hbe^j &=& - \frac{\ri}{2} \eps^{ij} \eps_{\hal\hbe} F + \frac{\ri}{2} \eps^{ij} \Phi_{\hal\hbe} 
+ \ri \bm \nabla_{\hal\hbe} \cL^{ij} + \ri \eps_{\hal\hbe}  W \D \cL^{ij} \ , \\
\bm\nabla_\hal^i F &=& - 2 \bm \nabla_\hal{}^\hbe \varphi_\hbe^i - 2 \ri \l_{\hal j} \D \cL^{ij} - 6 W_{\hal\hbe} \varphi^{\hbe i} - 9 X_{\hal j} \cL^{ij} \ , \\
\bm\nabla_\hal^i \Phi_{\ha} &=& 4 (\S_{\ha\hb})_\hal{}^\hbe \bm \nabla^\hb \varphi_\hbe^i 
- 2 \ri (\G_\ha)_\hal{}^\hbe \l_{\hbe j} \D \cL^{ij} 
- 2 (\G_\ha)_\hal{}^\hbe  W \D \varphi_\hbe^i \non\\
&& - 4 (\G_\ha)_\hal{}^\hbe W_{\hbe \hga} \varphi^{\hga i} 
- 6 (\G_\ha)_\hal{}^\hbe X_{\hbe j} \cL^{ij} \ .
\eea
\esubeq
Using the additional identities
\bsubeq
\begin{align} S_\hal^i \varphi_\hbe^j &= - 6 \eps_{\hal \hbe} \cL^{ij} \ , \\
S_\hal^i F &= 6 \ri \varphi_\hal^i \ , \\
S_\hal^i \Phi_\hb &= - 8 \ri (\G_\hb)_\hal{}^\hbe \varphi_\hbe^i \ ,
\end{align}
\esubeq
one can check that $\tilde{\S}$ and $\Phi$ are primary.

The superform equations imply the differential condition on $\Phi_\ha$
\be \bm \nabla^\ha \Phi_\ha = \D (\cW F + X^{ij} \cL_{ij} + 2 \L^{\hga k} \varphi_{\hga k}) - 5 \ri X^{\hga k} \varphi_{\hga k} \ .
\ee

It should be mentioned that in the above the central charge transformation of $\cL^{ij}$ is arbitrary. If we 
instead require $\cL^{ij}$ to be inert under the central charge, $\D \cL^{ij}=0$, we have
\be \rd \Phi = 0
\ee
and $\cL^{ij}$ becomes an $\cO(2)$ multiplet already described in previous sections.


\subsubsection{Action principle}

Having derived the components of $\S_{\hat{\cA} \hat{\cB} \hat{\cC} \hat{\cD} \hat{\cE}}$, it is straightforward to construct a closed five-form.
It is
\be
\frak{J} = \tilde{\S} + V \wedge \Phi \ .
\ee
One can check that it is closed,
\be \rd J = \rd \tilde{\S} + V \wedge \rd \Phi + \rd V \wedge \Phi = \bm \nabla \tilde{\S} + V \wedge \D \tilde{\S} + V \wedge \bm\nabla \Phi + F \wedge \Phi = 0 \ ,
\ee
and it transforms by an exact form under the central charge transformations,
\begin{align}
\d_{\L} J &= \d_{\L} \S - \d_{\L} V \wedge \Phi - V \wedge \d_{\L} \Phi \non\\
&= \L \D \S - \rd \L \wedge \Phi - V \wedge (\L \D \Phi) = - \rd (\L \Phi) \ .
\end{align}
The corresponding action is found using eq. \eqref{ectoS} to be
\begin{align} \label{LinearCompAction}
S
 &=
 - \int \rd^5x \,e\, \Big(
v^\ha \phi_\ha
+W F
+X^{ij} \cL_{ij} 
+ 2 \l^{k} \varphi_{k}
\eol & \quad
- \psi_{\ha i} \G^\ha \varphi^i W
- \ri \psi_{\ha i} \G^\ha \l_j \cL^{ij}
+ \ri \psi_{\ha i} \S^{\ha\hb} \psi_{\hb j} W \cL^{ij}
\Big) \ ,
\end{align}
where all superfields appearing in the action are understood as their component projections and we have defined
\be v_\ha := e_\ha{}^\hm V_\hm| \ , \quad \phi^\ha := - \frac{1}{4!} e_{\hat{r}}{}^\ha \eps^{\hm\hn\hat{p}\hat{q}\hat{r}} \Phi_{\hm \hn \hat{p} \hat{q}}| \ .
\ee
For completeness we 
also give the component field projection of $\Phi_\ha$:
\be \Phi^\ha| 
 = \phi^\ha
 - 2 \psi_{\hat{b}}{}_k \S^{\ha \hb}  \varphi^k
 + \frac{3 \ri}{8} \eps^{\ha\hb\hc\hd\he} \psi_\hb{}_k \S_{\hc\hd} \psi_\he{}_l \cL^{kl}| \ .
 \ee


\subsection{Gauge two-form multiplet} \label{2FormMult}

In superspace, the two-form multiplet is described by a constrained real superfield $L$ that is coupled to the vector multiplet 
gauging the central charge \cite{K06, KN-CS5D}, similar to the 4D $\cN = 2$ vector-tensor multiplets. Here we show 
how a geometric formulation of the multiplet naturally leads to the constraints on $L$ in conformal supergravity. Our presentation 
is similar to the one given in \cite{KN-CS5D} in Minkowski superspace.

In this subsection we wish to describe couplings of the two-form multiplet to additional Yang-Mills multiplets $\bm W$. Therefore 
in what follows we make use of covariant derivatives which contain both the gauge connection gauging the central charge and the Yang-Mills 
gauge connection:
\be \bm \nabla = \rd - \cV \D - \ri \bm V \ , \quad \bm \nabla_\hA = \nabla_\hA - \cV_\hA \D - \ri \bm V_\hA \ .
\ee

We introduce 
a gauge two-form, $\tfB = \hf E^B E^A \tfB_{AB}$ and define its three-form 
field strength $\tfH$ by
 \be \tfH := \bm \nabla \tfB - \tr \big( \bm V \wedge \bm F - \frac{\ri}{3} \bm V \wedge \bm V \wedge \bm V \big) \ ,
\ee
where $\bm V$ and $\bm F$ are the Yang-Mills connection and field strength corresponding to the superfield $\bm W$.\footnote{The special 
case of $n$ abelian vector multiplets may be obtained by taking $\tr(V \wedge F) \rightarrow \eta_{IJ} V^I F^J$, where 
$\eta$ is a symmetric, $\eta_{IJ} = \eta_{JI}$, coupling constant and $V^I$ and $F^I$ are the gauge connections and field strengths of the abelian vector multiplets.} 
Here $\tfB$ is a gauge singlet but is not assumed to be annihilated by the central charge. The (infinitesimal) transformation law for the system of superforms is
\begin{align}
\d \cV &= \rd \L \ , \quad \D \L = 0 \ , \non\\
\d \bm V &= \rd \bm \t - \ri [\bm V, \bm \t] \ , \quad \D \t = 0 \ , \non\\
\d \tfB &= \L \D \tfB - \tr(\t \wedge \rd V) + \rd \Xi \ , \quad \D \Xi = 0 \ ,
\end{align}
where $\L$, $\bm \t$ and $\Xi$ generate the gauge transformations of $\cV$, $\bm V$ and $\tfB$, respectively. The field strength $\tfH$ transforms 
covariantly under the central charge transformations
\be \d \tfH = \L \D \tfH
\ee
and satisfies the Bianchi identity
\be \bm \nabla \tfH = - \cF \wedge \D \tfB - \tr (\bm F \wedge \bm F) \ . \label{VTBI1}
\ee

Again we can make use of the notation that was introduced in section \ref{SFlinear}. We extend the Bianchi identity by introducing an additional 
bosonic index, $\hat{\cA} = (\hat{A} , 6)$. This can be done because we also have the additional superform equation
\be \D \tfH = \bm \nabla (\D \tfB) \ . \label{VTBI2}
\ee
Combining the above equation with the Bianchi identity \eqref{VTBI1} gives
\be \bm \nabla_{[\hat{\cA}} \tfH_{\hat{\cB} \hat{\cC} \hat{\cD} \}} + \frac{3}{2} \scT_{[\hat{\cA} \hat{\cB}}{}^{\hat{\cE}} \tfH_{|\hat{\cE}| \hat{\cC} \hat{\cD} \}} 
+ \frac{3}{2} \tr(\bm F_{[\hat{\cA} \hat{\cB}} \bm F_{\hat{\cC} \hat{\cD} \}}) = 0 \ , \label{GBI1F}
\ee
where we have defined
\bsubeq \label{VTID}
\begin{align} \tfH_{6 \hat{A} \hat{B}} &:= \D \tfB_{\hat{A} \hat{B}} \ , \quad \bm F_{6 \hat{\cA}} = \bm F_{\hat{\cA} 6} = 0 \ , \\
\scT_{\hat{A} \hat{B}}{}^6 &:= \cF_{\hat{A} \hat{B}} \ , \quad \scT_{\hat{A} 6}{}^{\hat{\cB}} = \scT_{6 \hat{A}}{}^{\hat{\cB}} = 0 \ , \quad \cD_6 := \D \ .
\end{align}
\esubeq

Constraining the lowest components of $\tfH_{\hat{\cA} \hat{\cB} \hat{\cC}}$ by
\be \tfH_\hal^i{}_\hbe^j{}_\hga^k = 0 \ , \quad \tfH_{6}{}_\hal^i{}_\hbe^j = - 2 \ri \eps^{ij} \eps_{\hat{\a} \hat{\b}} L \label{2FConst}
\ee
fixes the remaining components of $\tfH_{\hat{\cA} \hat{\cB} \hat{\cC}}$. Analyzing eq. \eqref{GBI1F} by increasing dimension and subject to the constraints \eqref{2FConst} 
(and the identifications \eqref{VTID}) 
leads to the remaining components:
\bsubeq \label{TWOFORMCOMPS}
\bea
 \tfH_{\hat{a}}{}_\hbe^j{}_\hga^k &=& - 2 \ri \eps^{jk} \big(\G_{\hat{a}})_{\hat{\b} \hat{\g}} (\cW L- \tr(\bm W^2)\big) \ , \\
\tfH_{6 \hat{a}}{}_{\hat{\b}}^j &=& (\G_{\hat{a}})_{\hat{\b}}{}^{\hat{\g}} \bm \nabla_{\hat{\g}}^j L \ , \\
\tfH_{\hat{a} \hat{b}}{}_{\hat{\g}}^k &=&  2 (\S_{\hat{a} \hat{b}})_{\hat{\g}}{}^{\hat{\d}} \bm \nabla_{\hat{\d}}^k (\cW L - \tr (\bm W^2 )) \ , \\
\tfH_{6 \hat{a} \hat{b}} &=& \frac{\ri}{4} (\S_{\hat{a}\hat{b}})^{\hat{\a} \hat{\b}} (\bm \nabla^k_{\hat{\a}} \bm \nabla_{\hat{\b} k} - 4 \ri W_{\hal\hbe}) L \ , \\
\tfH_{\hat{a} \hat{b} \hat{c}} &=& - \frac{\ri}{8} \eps_{\hat{a} \hat{b} \hat{c} \hat{d} \hat{e}} (\S^{\hat{d} \hat{e}})^{\hat{\a} \hat{\b}} 
\Big( (\bm \nabla_{\hat{\a}}^k \bm \nabla_{\hat{\b} k} + 4 \ri W_{\hal\hbe}) \big(\cW L - \tr(\bm W^2) \big) \non\\
&&+ 2 (\bm \nabla_{\hat{\a}}^k \cW )\bm \nabla_{\hat{\b} k} L 
- 2 \tr\big((\bm \nabla_{\hat{\a}}^k \bm W) \bm \nabla_{\hat{\b} k} \bm W\big)\Big) \ ,
\eea
\esubeq
where $L$ satisfies the constraints
\bsubeq \label{2FORMCONST}
\begin{align} \bm \nabla_{\hat{\a}}^{(i} \bm \nabla_{\hat{\b}}^{j)} L &= \frac{1}{4} \eps_{\hat{\a} \hat{\b}} \bm \nabla^{\hat{\g} (i} \bm \nabla_{\hat{\g}}^{j)} L \ , \label{3.31a} \\
\bm\nabla^{\hat{\g} (i} \bm\nabla_{\hat{\g}}^{j)} \big(\cW L - \tr(\bm W^2) \big) &= 
- 2 (\bm\nabla^{\hat{\g} (i} \cW )\bm\nabla_{\hat{\g}}^{j)} L 
+ 2 \tr\big(( \bm\nabla^{\hat{\g} (i} \bm W) \bm \nabla_{\hat{\g}}^{j)} \bm W\big) \label{3.31b} \ .
\end{align}
\esubeq

To describe the action for the two-form multiplet one can use the composite linear multiplet\footnote{This superfield Lagrangian first appeared in \cite{K06} 
in Minkowski superspace.}
\be 
\cL^{ij} = \frac{\ri}{2} \big( 2 (\bm \nabla^{\hat{\a} (i} L) \bm \nabla_{\hat{\a}}^{j)} L + L \bm \nabla^{\hat{\a} (i} \bm \nabla_{\hat{\a}}^{j)} L \big) 
= \frac{\ri}{6 L} \bm \nabla^{ij} (L^3) \ .
\label{3.32}
\ee
Note that it is also possible to construct another linear multiplet
\be \cL^{ij} = \frac{\ri}{4} \big( 4 (\bm \nabla^{\hat{\a} (i} W )\bm \nabla_{\hat{\a}}^{j)} L 
+ W \bm \nabla^{\hat{\a} (i} \bm \nabla_{\hat{\a}}^{j)} L 
+ L \bm \nabla^{\hat{\a} (i} \bm \nabla_{\hat{\a}}^{j)} W \big) 
\ ,
\ee
which couples the two-form multiplet to a vector multiplet $W$. 
The corresponding component actions can be found in \cite{Ohashi4, Bergshoeff3}.


\subsection{Large tensor multiplet}\label{largeTensorMultiplet}

In \cite{Ohashi4} it was discovered that there also exists the large tensor multiplet, which consists of $16+16$ degrees of freedom. 
In superspace the large tensor multiplet may be viewed as a generalization of the gauge two-form multiplet in which the constraints 
\eqref{2FORMCONST} are weakened. To show this let $\cL$ be a superfield constrained in the same way as eq. \eqref{3.31a},
\be \bm\nabla_{\hat{\a}}^{(i} \bm\nabla_{\hat{\b}}^{j)} \cL 
= \frac{1}{4} \eps_{\hat{\a} \hat{\b}} \bm\nabla^{\hat{\g} (i} \bm\nabla_{\hat{\g}}^{j)} \cL \label{3.33} \ .
\ee
Requiring only the above constraint, it is possible to show that consistency requires us to have \cite{K06}
\bea 
0&=& \D \Big\{ 
\bm\nabla^{\hat{\g} (i} \bm\nabla_{\hat{\g}}^{j)} (\cW  \cL )  + 2 (\bm\nabla^{\hat{\g} (i} \cW) \bm\nabla_{\hat{\g}}^{j)}  
\cL \Big\}     
 \non \\
&=&\bm\nabla^{\hat{\g} (i} \bm\nabla_{\hat{\g}}^{j)} (\cW \D \cL )  
+ 2( \bm\nabla^{\hat{\g} (i} \cW )\bm\nabla_{\hat{\g}}^{j)} \D \cL  \ , ~~~~
\label{3.34}
\eea
which is automatically satisfied for the gauge two-form multiplet. 
Here we will take eq. \eqref{3.34} as a second constraint on $\cL$. 
The constraints \eqref{3.33} and \eqref{3.34} allow us to construct a superform 
framework describing the large tensor multiplet.

We begin by introducing a two-form\footnote{In this subsection $\cB$ will be used for the
two-form. It is unrelated to the three-form $\cB$
used for the $\cO(2)$ multiplet.}  $\cB$, transforming as
\be \d \cB = \L \D \cB + \rd \Xi \ , \quad \D \Xi = 0 \ ,
\ee
and an associated three form $\cH$
\be \cH = \bm\nabla \cB \ . \label{hatHSE}
\ee
Imposing the constraints
\be \cH_\hal^i{}_\hbe^j{}_\hga^k = 0 \ , \quad \cH_{6}{}_\hal^i{}_\hbe^j= - 2 \ri \eps^{ij} \eps_{\hat{\a}\hat{\b}} \D \cL
\ee
and solving the Bianchi identities yields the components of $\cH$:
\bsubeq
\begin{align} \cH_{\hat{a}}{}_\hbe^j{}_\hga^k &= - 2 \ri \eps^{jk} \big(\G_{\hat{a}})_{\hat{\b} \hat{\g}} \cW \D \cL \ , \\
\cH_{6 \hat{a}}{}_{\hat{\b}}^j &= (\G_{\hat{a}})_{\hat{\b}}{}^{\hat{\g}} \bm\nabla_{\hat{\g}}^j \D \cL \ , \\
\cH_{\hat{a} \hat{b}}{}_{\hat{\g}}^k &
=  2 (\S_{\hat{a} \hat{b}})_{\hat{\g}}{}^{\hat{\d}} \bm\nabla_{\hat{\d}}^k (\cW \D \cL) \ , \\
\cH_{6 \hat{a} \hat{b}} &= \frac{\ri}{4} (\S_{\hat{a}\hat{b}})^{\hat{\a} \hat{\b}} (\bm\nabla^k_{\hat{\a}} \bm\nabla_{\hat{\b} k} - 4 \ri W_{\hal\hbe}) \D \cL \ , \\
\cH_{\hat{a} \hat{b} \hat{c}} &= - \frac{\ri}{8} \eps_{\hat{a} \hat{b} \hat{c} \hat{d} \hat{e}} 
(\S^{\hat{d} \hat{e}})^{\hat{\a} \hat{\b}} \Big( (\bm\nabla_{\hat{\a}}^k \bm\nabla_{\hat{\b} k} + 4 \ri W_{\hal\hbe}) \big(\cW \D \cL\big) 
+ 2( \bm\nabla_{\hat{\a}}^k \cW) \bm\nabla_{\hat{\b} k}  \D \cL \Big) \ ,
\end{align}
\esubeq
where $\cL$ is constrained by eqs. \eqref{3.33} and \eqref{3.34} and $\cH_{6 \hat{A} \hat{B}} = \D \cB_{\hat{A} \hat{B}}$. There are still too many component fields and to eliminate them we 
impose the constraint
\be \cB_{\hat{\a}}^i{}_{\hat{\b}}^j = - 2 \ri \eps^{ij} \eps_{\hat{\a} \hat{\b}} \cL \ ,
\ee
which fixes the remaining components via eq. \eqref{hatHSE} as
\be \cB_{\hat{a}}{}_{\hat{\b}}^j = (\G_{\hat{a}})_{\hat{\b}}{}^{\hat{\g}} \bm\nabla_{\hat{\g}}^j \cL \ , 
\quad \cB_{\hat{a} \hat{b}} 
= \frac{\ri}{4} (\S_{\hat{a}\hat{b}})^{\hat{\a} \hat{\b}} (\bm\nabla^k_{\hat{\a}} \bm\nabla_{\hat{\b} k} - 4 \ri W_{\hal\hbe}) \cL \ .
\ee
At the highest dimension eq. \eqref{hatHSE} gives
\bea
3 (\bm\nabla'_{[\hat{a}} \cB_{\hat{b} \hat{c}]} - \scT_{[\ha\hb}{}^\hde_l \cB_{c]}{}_{\hde}^l) &=& - \frac{\ri}{8} \eps_{\hat{a} \hat{b} \hat{c} \hat{d} \hat{e}} 
(\S^{\hat{d} \hat{e}})^{\hat{\a} \hat{\b}} \D \Big( (\bm\nabla_{\hat{\a}}^k \bm\nabla_{\hat{\b} k} + 4 \ri W_{\hal\hbe}) \big(\cW \cL\big) 
\non\\
&&~~~~~~~~~~~~~~~~~~~~~~~~~
+ 2( \bm \nabla_{\hat{\a}}^k \cW )\bm\nabla_{\hat{\b} k} \cL \Big)  \ , \label{firstKO}
\eea
where 
\be \bm\nabla'_\ha = \bm\nabla_\ha - \frac{1}{8} \eps_{\ha\hb\hc\hd\he} W^{\hb\hc} M^{\hd\he} \ .
\ee
The conditions  \eqref{3.34} and \eqref{firstKO} are similar to the ones imposed 
in \cite{Ohashi4}
from requiring 
closure of the supersymmetry transformations. In contrast with the 
gauge two-form multiplet, which was based on the stronger constraints \eqref{2FORMCONST}, the component fields of the 
large tensor multiplet
\be \D \bm\nabla_\a^i \cL | \ , \quad \D^2 \cL |
\ee
are no longer composite.

We should remark that the above constraints can be naturally generalized 
to include couplings to the Yang-Mills multiplet. Furthermore, since $\cB$ possesses the gauge transformation law
\be \d \cB = \rd \Xi \ , \quad \D \Xi = 0 \ , 
\ee
one can always shift $\cL$ by an abelian vector multiplet
\be \cL \rightarrow \cL + c W \ ,
\ee
where $c$ is an arbitrary real coefficient. One can check that the 
constraints \eqref{3.33} and \eqref{3.34} are invariant 
under such transformations.

We can construct an action for an even number of large tensor multiplets $\cL^I$. 
To do so we make use of the superfield Lagrangian
\be \cL^{ij} = \cL_{\rm kin}^{ij} + \cL_{\rm mass}^{ij} \ ,
\ee
where
\bsubeq
\begin{align} \cL_{\rm mass}^{ij} &= \frac{\ri}{2} m_{IJ} \Big( 2( \bm\nabla^{\hat{\a} (i} \cL^I) \bm\nabla_{\hat{\a}}^{j)} \cL^J + \cL^I \bm\nabla^{\hat{\a} (i} \bm\nabla_{\hat{\a}}^{j)} \cL^J \Big) \ , \quad m_{IJ} = m_{JI} \ , \\
\cL_{\rm kin}^{ij} &= \frac{\ri}{4} k_{IJ} \Big( 2( \bm\nabla^{\hat{\a} (i} \cL^I) \overleftrightarrow{\D} \bm\nabla_{\hat{\a}}^{j)} \cL^J 
+ \cL^I \overleftrightarrow{\D} \bm\nabla^{\hat{\a} (i} \bm\nabla_{\hat{\a}}^{j)} \cL^J \Big) \ , \quad k_{IJ} = - k_{JI} \ .
\end{align}
\esubeq
The constant matrices $m_{IJ}$ and $k_{IJ}$ are assumed to be nonsingular. 
The Lagrangian $\cL^{ij}$ may be seen to be a linear multiplet.
The component action in supergravity is given in \cite{Ohashi4}.


\section{Off-shell (gauged) supergravity} \label{Off-shellSUGRA}

We now turn to an off-shell formulation for 5D minimal supergravity
obtained by coupling the Weyl multiplet to the following compensators:
(i) the vector multiplet; and (ii) the $\cO(2)$ multiplet.  
This is the 5D analogue of the off-shell formulation for 4D $\cN=2$
supergravity proposed by de Wit, Philippe and Van Proeyen \cite{deWPV}.\footnote{The 
4D $\cN=2$ supergravity formulation of \cite{deWPV} makes use of the $\cN=2$ improved
tensor multiplet constructed in terms of $\cN=1$ superfields in Minkowski superspace \cite{LR83} and then in terms of component fields in the locally supersymmetric case \cite{deWPV}.}
We will first describe the construction within superspace and then
briefly give the bosonic part of the component action.

\subsection{Superspace formulation}
The superfield Lagrangian  for 5D (gauged) supergravity is analogous to the one
for 4D $\cN=2$ supergravity \cite{K-08} and reads
\bea
\cL^{(2)}_{\rm SG} =  
\frac{1}{4} {V}\,{ H }_{\rm VM}^{(2)} 
+{G}^{(2)}  \ln \frac{{ G}^{(2)}}{{\rm i} \U^{(1)} \breve{\U}{}^{(1)}}      
+\k V G^{(2)}
\equiv \cL^{(2)}_{\rm V} + \cL^{(2)}_{\rm L} + \cL^{(2)}_{\rm VL}
~.
\label{SUGRA8.1}
\eea

In the first term,
${ H }_{\rm VM}^{(2)} $ denotes the 
composite $\cO(2)$ multiplet
\eqref{HVM}.
The superspace action generated by $\cL_{\rm V}^{(2)}$ then leads
to the abelian Chern-Simons action, but normalized
with the wrong sign (as usual for a compensator action)
and with an additional factor of 3 for later convenience (compare with eq.~\eqref{CSaction9.1}).

Modulo a similar overall sign, the second term in \eqref{SUGRA8.1}
denoted by $ \cL^{(2)}_{\rm L}$ 
describes the dynamics of the $\cO(2)$  multiplet or, equivalently, 
linear multiplet without central charge. 
The superfield ${ \U}^{(1)}(v)$ is a covariant weight-one {\it arctic} multiplet, 
and $\breve{\U}^{(1)} (v)$ its smile-conjugated {\it antarctic} superfield.
The action proves to be independent of ${ \U}^{(1)}$ and 
$\breve{\U}^{(1)}$ \cite{K-08}.

The BF term in \eqref{SUGRA8.1} denoted by $ \cL^{(2)}_{\rm VL}$ describes 
a supersymmetric cosmological term. For $\k=0$ the Lagrangian \eqref{SUGRA8.1}
describes pure Poincar\'e supergravity, while the case $\k \neq 0$ corresponds 
to gauged or anti-de Sitter supergravity.

Making use of 
\eqref{SGV},
the action generated by $\cL^{(2)}_{\rm V} $ 
may be rewritten as an integral over the full superspace,
\bea
S[ \cL^{(2)}_{\rm V} ]&=&
\frac{1}{2\pi} \oint (v, \rd v)
\int\rd^{5|8}z\, E\,  C^{(-4)}\cL^{(2)} 
=
\frac{1}{4} \int\rd^{5|8}z\, E\,   V_{ij} { H }_{\rm VM}^{ij} ~,~~~
\label{rep8.3}
\eea
with $V_{ij}$ being Mezincescu's prepotential. 
Applying \eqref{SGV} once more 
gives 
another representation 
\bea
S[ \cL^{(2)}_{\rm V} ]= 
\frac{1}{4}
 \int \rd^{5|8}z\, E\, 
 {\bm \O}_{\rm VM} W~,
 \label{rep8.4}
\eea
where we have introduced the primary superfield 
\bea
 {\bm \O}_{\rm VM} = 
\frac{\ri}{4}\Big(
 W\de^{ij} V_{ij}
-2  (\de^{\hal i}V_{ij})\de_\hal^{j}W
-2 V_{ij}\de^{ij} W \Big)~,
\eea
which is a prepotential for ${ H }_{\rm VM}^{(2)}$
in the sense of \eqref{def-G-0-a}.
The representations \eqref{rep8.3} and \eqref{rep8.4} allow us to compute the variation of 
$S[ \cL^{(2)}_{\rm V} ]$ induced by an arbitrary variation 
of the vector multiplet prepotential, either Mezincescu's or the tropical one,
\begin{subequations}
\bea
\d S[ \cL^{(2)}_{\rm V} ]
&=&
\frac{3}{4}\int \rd^{5|8}z\, E\, \d V_{ij}
{ H }_{\rm VM}^{ij}   \\
&=&
\frac{3}{8\pi} \oint (v, \rd v)
\int \rd^{5|8}z\, E\, C^{(-4)} \d V  { H }_{\rm VM}^{(2)}~.
\label{var8.6b}
\eea
\end{subequations}
Making use of \eqref{var8.6b}, we readily find the equation of motion for the vector
multiplet in the supergravity theory \eqref{SUGRA8.1} to be
\bea
 { H }_{\rm VM}^{(2)} + \frac{4\k}{3} G^{(2)} =0~.
 \label{eq-mot-vec}
 \eea

We next consider the action generated by $\cL^{(2)}_{\rm L} $. 
It may be rewritten as an integral over the  full superspace 
\bea
S[ \cL^{(2)}_{\rm L} ] =  \int \rd^{5|8}z\, E\,  \O \mathbb W ~,
\eea
where 
\bea 
\mathbb W := - \frac{\ri}{16 \pi} \oint (v, \rd v) \de^\pmd
\log \Big( \frac{G^\pd}{\ri \U^\pu \breve{\U}^\pu} \Big) 
\label{CVM8}
\eea
is a composite vector multiplet field strength obeying the Bianchi identity \eqref{6.20}.
The direct evaluation of $\mathbb W$ will be given in section \ref{subsection13.2}. 
The result is
\bea
\mathbb W = 
 \frac{\ri}{16} G \nabla^{\hal i} \nabla_\hal^j \Big(\frac{G_{ij}}{G^2}\Big) ~ .
\eea
It may be seen that varying the prepotential $\O$ leads to the following variation 
of the action:
\bea
\d S[ \cL^{(2)}_{\rm L} ] 
= 
 \int \rd^{5|8}z\, E\, \d \O \mathbb W ~.
\label{rep8.11}
\eea

Finally, we note that  the action generated by $\cL^{(2)}_{\rm VL} $
may also be rewritten as an integral over the  full superspace 
\bea
S[ \cL^{(2)}_{\rm VL} ] =  
\k \int \rd^{5|8}z\, E\,   V_{ij} G^{ij} 
= \k \int \rd^{5|8}z\, E\,  \O W~. 
\eea
As a result, the complete (gauged) supergravity action becomes
\begin{subequations}
\bea
S_{\rm SG}&=&  
 \int \rd^{5|8}z\, E\, \Big\{ 
\frac{1}{4} V_{ij} { H }_{\rm VM}^{ij} 
+\O \mathbb W  
+\k V_{ij} G^{ij} \Big\} \\
&=&
 \int \rd^{5|8}z\, E\,  \Big\{ 
\frac{1}{4} V_{ij} { H }_{\rm VM}^{ij} 
+ \O \mathbb W  
+\k \O W \Big\}~. \label{rep8.13b}
\eea
\end{subequations}
Now, from the relations \eqref{rep8.11} and \eqref{rep8.13b} we deduce 
the supergravity equation of motion for the $\cO(2)$ compensator:
\bea
\mathbb W + \k W =0~.
\label{eq-mot-ten}
\eea

The equation of motion for the Weyl multiplet is
\bea
G - W^3 =0~.
\label{eq-mot-Weyl}
\eea
It may be shown that, modulo gauge freedom,  the Weyl multiplet is described by a 
single unconstrained real prepotential $\frak U$.\footnote{This 
can be done in complete analogy with the case of 4D $\cN=2$ supergravity \cite{KT}.}
The equation \eqref{eq-mot-Weyl} is obtained by varying the supergravity action with respect to $\frak U$. 
The meaning of \eqref{eq-mot-Weyl} is that the supercurrent of pure supergravity 
is equal  to zero. 

In general, given a dynamical system involving (matter) superfields $\vf^i$ coupled
to the Weyl multiplet, the supercurrent of this theory is a dimension-3 primary real scalar superfield 
defined by 
\bea
\cT = \frac{\D }{\D \frak U}   S[\vf ] ~,
\eea
where $\D / {\D \frak U}$ denotes a covariantized variational derivative with respect to
$\frak U$. The variation $\D\frak U$ is a primary superfield with dimension $-2$. 
The supercurrent turns out to satisfy the conservation 
equation
\bea
\nabla^{ij} \cT=0
\eea
provided the dynamical superfields obey their equations of motion, 
$\d S[\vf ] / \d \vf^i = 0$. This follows from the fact that $\D \frak U$ is defined modulo gauge
transformations 
\bea
\D \frak U ~\to ~
\D \frak U +
\de^{ij}\O_{ij}
~,
\eea
where gauge parameter $\O_{ij}$  is a primary real isovector 
superfield with dimension $-3$.

It is an instructive exercise to prove that the left-hand side of  \eqref{eq-mot-Weyl}
obeys the constraint 
\bea
\nabla^{ij}\Big(G - W^3\Big) =0
\eea
provided the equations  \eqref{eq-mot-vec} and  \eqref{eq-mot-ten} hold. 

The supergravity equations of motion  \eqref{eq-mot-vec}, 
\eqref{eq-mot-ten} and  \eqref{eq-mot-Weyl}  appeared in \cite{KNT-M14}. 
 They are analogous to the superfield equations for 4D $\cN=2$ 
 (gauged) supergravity \cite{BK11,BK111}.

\subsection{Component formulation}
To complement the superspace discussion, we now present briefly the bosonic
part of the component action for gauged supergravity. The three superspace
actions given in \eqref{SUGRA8.1}
can be analyzed in components easily using results
given elsewhere in this paper. The first term $\cL_{\rm V}^{(2)}$ leads to
the wrong sign abelian Chern-Simons Lagrangian
\begin{align}
\cL_{\rm V} &= 
	\frac{1}{8} \eps^{\ha\hb\hc\hd\he} v_\ha f_{\hb\hc} f_{\hd\he}
	+ \frac{3}{4} W  f_{\hat{a} \hat{b}}  f^{\hat{a} \hat{b}}
	- \frac{3}{4} W  X^{ij}  X_{ij} 
	+ \frac{9}{4} w^{\ha\hb}  f_{\ha\hb}  W^2
	\eol & \quad
	+ \frac{3}{2} W (\hat\cD^\ha W) \hat \cD_\ha W
	- \frac{1}{8} \hat \cR W^3
	+ 4 D W^3
	+ \frac{39}{32} w^{\ha\hb} w_{\ha\hb}  W^3~,
\end{align}
where $\hat{\cD}_\ha$ is defined by eq. \eqref{hatDder}. 
The second term $\cL_{\rm L}^{(2)}$ leads to the $\cO(2)$ multiplet Lagrangian
\begin{align}\label{eq:TensorComp}
\cL_{\rm L} 
	&= \frac{1}{4} G^{-1}( \hat\cD_\ha G^{ij} )\hat\cD^\ha G_{ij}
	- \frac{1}{2} G^{-1} \phi^\ha \phi_\ha
	\eol & \quad
	+ \frac{1}{12} \veps^{\ha\hb\hc\hd\he} b_{\hc\hd\he} \Big(
	\frac{1}{2} G^{-3} (\hat\cD_\ha G_{ik}) (\hat\cD_\hb G_j{}^k) G^{ij}
		+ G^{-1} \hat R(J)_{\ha\hb}{}^{ij} G_{ij}
	\Big)
	\eol & \quad
	- \frac{1}{8G} F^2
	- \frac{3}{8} \hat\cR G
	- 4 D G
	- \frac{3}{32} w^{\ha\hb} w_{\ha\hb} G~.
\end{align}
This Lagrangian is analogous to the 4D improved tensor multiplet
Lagrangian \cite{deWPV} and shares similar features. In particular,
the second line of \eqref{eq:TensorComp}
involves a BF coupling between the three-form
$b_{\ha\hb\hc}$ and a composite two-form constructed from the tensor
multiplet scalars and the SU(2) gauge fields.
As discussed in \cite{deWPV}, this two-form is closed
but not exact: it has no SU(2)-invariant one-form potential.
The third superspace Lagrangian $\cL_{\rm VL}^{(2)}$ leads to the simple
expression
\begin{align}
\cL_{\rm VL}
	&= - \kappa W F
	- \kappa X^{ij} G_{ij}
	- 2 \kappa v_\ha \phi^\ha~.
\end{align}

We now combine all three Lagrangians and eliminate the auxiliary fields
using their equations of motion. The equation of motion for $D$ is
\begin{align}
W^3 - G = 0
\end{align}
and corresponds to the lowest component of the superfield equation of
motion \eqref{eq-mot-Weyl}. Similarly, the equations of motion for the vector multiplet
auxiliary $X^{ij}$ and the $\cO(2)$ multiplet auxiliary $F$ lead, respectively, to
\begin{align}
\frac{3}{2} W X_{ij} + \kappa G_{ij} &= 0~, \\
\frac{1}{4G} F + \kappa W &= 0~,
\end{align}
which correspond to the bosonic parts of the lowest components of \eqref{eq-mot-vec}
and \eqref{eq-mot-ten}, respectively. Finally, we must impose the equation of
motion for $w_{\ha\hb}$, which leads to
\begin{align}
w_{\ha\hb} W + f_{\ha\hb} = 0~.
\end{align}
This is actually the bosonic part of a higher component of
the Weyl superfield equation of motion; it can be extracted by applying
$\nabla_{(\hal}^k \nabla_{\hbe) k}$ to \eqref{eq-mot-Weyl} and taking
the lowest component.

After imposing each of these equations, we finally choose the Weyl gauge
$W = 1$. This leads to the component Lagrangian
\begin{align}
\cL_{\rm SG}
	&= -\frac{1}{2} \hat \cR
	+ \frac{1}{8} \veps^{\ha\hb\hc\hd\he} v_\ha f_{\hb\hc} f_{\hd\he}
	- \frac{3}{8} f_{\hat{a} \hat{b}}  f^{\hat{a} \hat{b}}
	+ \frac{8}{3} \kappa^2
	\eol & \quad
	+ \frac{1}{4} (\hat \cD_\ha G^{ij} )\hat \cD^\ha G_{ij}
	- \frac{1}{2} \phi^\ha \phi_\ha
	- 2 \kappa v_\ha \phi^\ha
	\eol & \quad
	+ \frac{1}{12} \veps^{\ha\hb\hc\hd\he} b_{\hc\hd\he} \Big(
	\frac{1}{2} (\hat\cD_\ha G_{ik} )(\hat\cD_\hb G_j{}^k) G^{ij}
		+ \hat R(J)_{\ha\hb}{}^{ij} G_{ij} \Big) \ .
\end{align}
The terms in the second and third lines turn out to lead to auxiliary fields.
The easiest way to see this is to adopt the SU(2) gauge
\begin{align}
G_{\1\2} = \ri~, \qquad G_{\1\1} = G_{\2\2} = 0~,
\end{align}
which breaks the R-symmetry group to U(1). Using
\begin{align}
\hat \cD_\ha G^{\1\1} = 2\ri\, \cV_\ha{}^{\1\1}~, \qquad
\hat \cD_\ha G^{\2\2} = -2\ri\, \cV_\ha{}^{\2\2}~, \qquad
\hat \cD_\ha G^{\1\2} = 0~,
\end{align}
the supergravity Lagrangian can be rewritten as
\begin{align}
\cL_{\rm SG}
	&= -\frac{1}{2} \hat \cR
	+ \frac{1}{8} \veps^{\ha\hb\hc\hd\he} v_\ha f_{\hb\hc} f_{\hd\he}
	- \frac{3}{8} f_{\hat{a} \hat{b}}  f^{\hat{a} \hat{b}}
	+ \frac{8}{3} \kappa^2
	\eol & \quad
	+ 2 \cV_\ha{}^{\1\1} \cV^\ha{}^{\2\2}
	- \frac{1}{2} \phi^\ha \phi_\ha
	- 2 (\kappa v_\ha + \ri \cV_\ha{}^{\1\2}) \phi^\ha~.
\end{align}
Now one introduces a Lagrange multiplier term $\phi^{\ha} \hat \cD_\ha \l $
to enforce the constraint on $\phi^{\ha}$; the field $\l$ is eaten by
$\cV_{\ha}{}^{\1\2}$, which fixes the remaining R-symmetry up to
a compensating $\kappa$-dependent transformation to counter the
graviphoton's gauge transformation. Integrating out $\phi_\ha$ then gives
\begin{align}
\cL_{\rm SG}
	&= -\frac{1}{2} \hat \cR
	+ \frac{1}{8} \veps^{\ha\hb\hc\hd\he} v_\ha f_{\hb\hc} f_{\hd\he}
	- \frac{3}{8} f_{\hat{a} \hat{b}}  f^{\hat{a} \hat{b}}
	+ \frac{8}{3} \kappa^2
	+ (\cV_\ha{}^{ij} + \kappa G^{ij} v_\ha)^2
\end{align}
where we have written the auxiliary one-forms in a way which holds for
any choice of constant $G^{ij}$. The equation of motion for this
auxiliary then fixes $\cV_\ha{}^{ij} = -\kappa G^{ij} v_\ha$, which is
ultimately responsible for the $\kappa$-dependent minimal coupling between the
gravitino and the graviphoton.
Note that the cosmological constant is given in these conventions by
\begin{align}
\L = -\frac{8}{3} \kappa^2 < 0~.
\end{align}


\section{Dilaton Weyl multiplets and superforms} \label{section11}

It is possible to construct variant formulations for conformal supergravity
by coupling the standard Weyl multiplet, which is described in sections 
\ref{setup} and \ref{WeylMultiplet}, 
to an on-shell abelian vector multiplet
with nowhere vanishing field strength, 
$ W \neq 0$. The field strength $W $ of such a vector multiplet
 satisfies the Bianchi identity \eqref{vector-Bianchi}
as well as the equation of motion 
\bea
{\mathbb H}^{ij} =0
\label{12.2}
\eea
derived from a gauge invariant action $S [W]$, 
see section \ref{subsection6.6}. 

In this section, we consider a special case of the equation of motion \eqref{12.2} 
that originates in 5D minimal supergravity with cosmological term
realized as conformal supergravity coupled to  
two compensators: (i) the vector multiplet; and (ii) the $\cO(2)$ multiplet. In this case 
 $\mathbb H^{ij} \equiv {H}_{\rm VM}^{ij}$, 
 where ${H}_{\rm VM}^{ij}$ denotes
the composite Yang-Mills $\cO(2)$ multiplet 
\eqref{YML}. In the superspace setting, the  supergravity equations of motion 
\cite{KNT-M14} are given by eqs. \eqref{eq-mot-vec},  \eqref{eq-mot-ten} and \eqref{eq-mot-Weyl}. 
In what follows, we will only use eq. \eqref{eq-mot-vec}.


\subsection{The dilaton Weyl multiplet} \label{dilWeylMultiplet}

The dilaton Weyl multiplet\footnote{The dilaton Weyl  multiplet 
 corresponds to the Nishino-Rajpoot version 
\cite{NR} of 5D $\cN=1$ Poincar\'e supergravity.} 
\cite{Ohashi3, Bergshoeff1} 
is equivalently described as the standard Weyl multiplet 
coupled to a vector multiplet compensator 
obeying the equation of motion 
\bea
{H}_{\rm VM}^{ij}=0~.
\label{12.5}
\eea
The formulation of this multiplet in SU(2) superspace was given in \cite{KT-M5D1}.
Eq. \eqref{12.5} in SU(2) superspace is equivalent to 
\be
S^{ij} = \frac{\rm i}{2W^2} \Big\{  
(\cD^{\hal (i}W)\cD_\hal^{j)} W+\hf W \cD^{ij} W\Big\}~.
\label{12.5SU(2)}
\ee

Equation \eqref{12.5} 
tells us that the matter fields of the super Weyl 
tensor $W_{\ha\hb}$ satisfy certain constraints that allow one to 
solve $W_{\ha\hb}$ in terms of a gauge two-form. To see this we 
make use of the equivalence between  vector  and  two-form multiplets 
on the mass shell.
We recall that the two-form multiplet was described 
in section \ref{2FormMult} and here we will use its superform realization.

Ignoring the Chern-Simons couplings to Yang-Mills multiplets, a two-form multiplet possesses 
a gauge two-form $\tfB$ with corresponding field strength
\be \tfH = \rd \tfB - \cV \wedge \D \tfB \ .
\ee
Imposing the on-shell condition for a two-form multiplet, $\D L = 0$, allows one 
to identify $L$ with a vector multiplet. In fact, we identify both the vector multiplet gauging the 
central charge and the two-form multiplet with the same vector multiplet. 
To do this we make the replacements
\be L \rightarrow - W \ , \quad \cW \rightarrow W \ , 
\ee
which requires
\be
\quad \D \tfB = F \ , \quad \cF = F \ , \quad \cV = V \ .
\ee
Note that the gauge transformations become
\be \d V = \rd \L \ , \quad \d \tfB = \L F + \rd \Xi \ .
\ee

The field strength $\tfH = \rd \tfB - V \wedge F$  
satisfies the Bianchi identity
\be \rd \tfH = - F \wedge F \ ,
\ee
or, equivalently,
\be \nabla_{[\hA} \tfH_{\hB\hC\hD \}} + \frac{3}{2} \scT_{[\hA \hB}{}^{\hE} \tfH_{|\hE|\hC\hD \} } 
+ \frac{3}{2} F_{[\hA \hB} F_{\hC\hD \} } = 0 \ .
\ee
The solution to the above Bianchi identity may be read off of eq. \eqref{TWOFORMCOMPS}:
\bsubeq \label{dilWsuperform}
\begin{align} 
\tfH_\hal^i{}_\hbe^j{}_\hga^k &= 0 \ , \\
\tfH_{\hat{a}}{}_\hbe^j{}_\hga^k &= 2 \ri \eps^{jk} \big(\G_{\hat{a}})_{\hat{\b} \hat{\g}} W^2 \ , \\
\tfH_{\hat{a} \hat{b}}{}_{\hat{\g}}^k &=  - 2 (\S_{\hat{a} \hat{b}})_{\hat{\g}}{}^{\hat{\d}} \nabla_{\hat{\d}}^k W^2 \ , \\
\tfH_{\hat{a} \hat{b} \hat{c}} &=  \frac{\ri}{8} \eps_{\hat{a} \hat{b} \hat{c} \hat{d} \hat{e}} (\S^{\hat{d} \hat{e}})^{\hat{\a} \hat{\b}} 
\Big( \nabla_{\hat{\a}}^k \nabla_{\hat{\b} k}  W^2 
+ 4 \ri W_{\hal\hbe} W^2 + 2 (\nabla_{\hat{\a}}^k W) \nabla_{\hat{\b} k} W \Big) \ .
\end{align}
\esubeq
The Bianchi identities are satisfied since we have the on-shell condition \eqref{12.5}, which is equivalent to
\bsubeq
\begin{align}
\nabla^{\hat{\g} (i} \nabla_{\hat{\g}}^{j)} W^2 &= - 2 (\nabla^{\hat{\g} (i} W) \nabla_{\hat{\g}}^{j)} W \ .
\end{align}
\esubeq

From the component $\tfH_{\ha\hb\hc}$ one finds the expression
for the super Weyl tensor
\be \label{dilWeylReplace}
W_{\ha\hb} = \frac{1}{3 W^2}
\Big( 
\frac{1}{6} \eps_{\ha\hb\hc\hd\he} \tfH^{\hc\hd\he} - 2 W F_{\ha\hb} - \ri (\S_{\ha\hb})^{\hal\hbe} \l_\hal^k \l_{\hbe k} 
\Big) \ .
\ee
Due to the above relation we see that we may instead choose 
the gauge two-form $\tfB_{\ha\hb}$ as a fundamental component field. This means that 
the matter fields in the standard Weyl multiplet become composite. They may be derived directly from the above superspace 
expression for $W_{\ha\hb}$.

Using the above relations we can replace the matter fields in the standard Weyl multiplet:
\be (W_{\ha\hb} , X_\hal^i , Y) \rightarrow ( W , \l_\hal^i, V_\hm , \tfB_{\hm\hn}) \ .
\ee
This leads to the dilaton Weyl multiplet, which only differs from the standard Weyl multiplet in the matter field 
content. 
One can check that both Weyl multiplets contain 32+32  degrees of freedom.

The construction of actions involving the dilaton Weyl multiplet may be readily obtained from those involving the 
standard Weyl multiplet upon making the replacements in this subsection. 
One can further construct actions by 
replacing any vector multiplet $\hat{W}$ with the components of the dilaton Weyl multiplet as follows:
\be (\hat{W} , \hat{\l}_\hal^i , \hat{V}_\hm , \hat{X}^{ij}) \rightarrow (W , \l_\hal^i , V_\hm , \frac{\ri}{2 W} \l^i \l^j) \ .
\ee

\subsection{The deformed dilaton Weyl multiplet}

The deformed Weyl multiplet \cite{CO} 
is equivalently described as the standard Weyl multiplet 
coupled to a vector multiplet compensator 
obeying the equation of motion 
\bea 
{H}_{\rm VM}^{ij} 
= 
- \frac{4 \k}{3} G^{ij}~,
\eea
which implies
\be X^{ij} = \frac{\ri}{2 W} \l^i \l^j - \frac{2 \k}{3 W} G^{ij} \ .
\ee
Here the $\cO(2)$ compensator $G^{ij}$ is considered as a background field.  

Just like in the previous case we can give the constrained system a geometric description. We now modify the 
superform equation to
\be \rd \tfH = - F \wedge F - \frac{4 \k}{3} \Phi \ ,
\ee
where
\be \tfH = \rd \tfB - V \wedge F - \frac{4 \k}{3} \cB \ .
\ee
Here $\cB$ is the gauge three-form for the $\cO(2)$ multiplet. From the above we see that we must 
modify the gauge transformation of $\tfB$ to be
\be \d \tfB = \L F + \rd \Xi + \frac{4 \k}{3} \r
\ee
since
\be \d \cB = \rd \r \ .
\ee

The solution is
\bsubeq 
\begin{align} 
\tfH_\hal^i{}_\hbe^j{}_\hga^k &= 0 \ , \\
\tfH_{\hat{a}}{}_\hbe^j{}_\hga^k &= 2 \ri \eps^{jk} \big(\G_{\hat{a}})_{\hat{\b} \hat{\g}} W^2 \ , \\
\tfH_{\hat{a} \hat{b}}{}_{\hat{\g}}^k &=  - 2 (\S_{\hat{a} \hat{b}})_{\hat{\g}}{}^{\hat{\d}} \nabla_{\hat{\d}}^k W^2
 \ , \\
\tfH_{\hat{a} \hat{b} \hat{c}} &=  \frac{\ri}{8} \eps_{\hat{a} \hat{b} \hat{c} \hat{d} \hat{e}} (\S^{\hat{d} \hat{e}})^{\hat{\a} \hat{\b}} 
\Big( \nabla_{\hat{\a}}^k \nabla_{\hat{\b} k}  W^2 
+ 4 \ri W_{\hal\hbe} W^2 + 2 (\nabla_{\hat{\a}}^k W) \nabla_{\hat{\b} k} W \Big) \ .
\end{align}
\esubeq

From the component $\tfH_{\ha\hb\hc}$ one finds the expression
\be 
W_{\ha\hb} = \frac{1}{3 W^2}
\Big( 
\frac{1}{6} \eps_{\ha\hb\hc\hd\he} \tfH^{\hc\hd\he} - 2 W F_{\ha\hb} - \ri (\S_{\ha\hb})^{\hal\hbe} \l_\hal^k \l_{\hbe k} 
\Big) \ .
\ee
Again the matter components of the Weyl multiplet may be replaced using the above expression. The above expression for $W_{\ha\hb}$ looks formally 
the same as eq. \eqref{dilWeylReplace}. However, 
it should be kept in mind that $W$ now satisfies the different on-shell constraint
\begin{align}
\nabla^{\hat{\g} (i} \nabla_{\hat{\g}}^{j)} W^2 &= - 2 (\nabla^{\hat{\g} (i} W) \nabla_{\hat{\g}}^{j)} W + \frac{16 \ri}{3} \k G^{ij}\ .
\end{align}


\subsection{The deformed dilaton Weyl multiplet with Chern-Simons couplings}

It was mentioned in \cite{KT-M08} that one can generalize the construction of the dilaton Weyl multiplet 
to include a system of abelian vector multiplets. Using a similar idea we generalize the deformed dilaton Weyl 
multiplet of the previous subsection in the presence of Yang-Mills couplings.

We now modify the superform equation to
\be \rd \tfH = - F \wedge F - \tr (\bm F \wedge \bm F) - \frac{4 \k}{3} \Phi \ , \label{CSdeformeddilWeyl}
\ee
where
\be \tfH = \rd \tfB - V \wedge F - \tr(\bm V \wedge \bm F 
- \frac{\ri}{3} \bm V \wedge \bm V \wedge \bm V) -\frac{4 \k}{3} \cB \ .
\ee
From the above we see that we must 
modify the gauge transformation of $\tfB$ to be
\be \d \tfB = \L F + \tr(\bm \t \rd \bm V) + \rd \Xi + \frac{4\k}{3} \r
\ee
since
\be \d \cB = \rd \r \ , \quad \d \bm V = \rd \bm \t - \ri [\bm V , \bm \t] \ .
\ee

The superform equation \eqref{CSdeformeddilWeyl} is solved by
\bsubeq \label{ModdilWsuperform}
\begin{align} 
\tfH_\hal^i{}_\hbe^j{}_\hga^k &= 0 \ , \\
\tfH_{\hat{a}}{}_\hbe^j{}_\hga^k &= 2 \ri \eps^{jk} (\G_{\hat{a}})_{\hat{\b} \hat{\g}} \big(W^2 + \tr (\bm W^2)\big)\ , \\
\tfH_{\hat{a} \hat{b}}{}_{\hat{\g}}^k &=  - 2 (\S_{\hat{a} \hat{b}})_{\hat{\g}}{}^{\hat{\d}} \bm \nabla_{\hat{\d}}^k \big( W^2 + \tr (\bm W^2) \big)
 \ , \\
\tfH_{\hat{a} \hat{b} \hat{c}} &=  \frac{\ri}{8} \eps_{\hat{a} \hat{b} \hat{c} \hat{d} \hat{e}} (\S^{\hat{d} \hat{e}})^{\hat{\a} \hat{\b}} 
\Big( \bm \nabla_{\hat{\a}}^k \bm \nabla_{\hat{\b} k}  \big( W^2 + \tr (\bm W^2) \big)
+ 4 \ri W_{\hal\hbe} W^2 \non\\
&\qquad\qquad + 2 \tr \big((\bm \nabla_{\hat{\a}}^k \bm W) \bm \nabla_{\hat{\b} k} \bm W\big) + 2 (\nabla_{\hat{\a}}^k W) \nabla_{\hat{\b} k} W \Big) \ ,
\end{align}
\esubeq
where $W$ satisfies the Bianchi identity
\begin{align}
\bm \nabla^{\hat{\g} (i} \bm \nabla_{\hat{\g}}^{j)} \big(W^2 + \tr (\bm W^2)\big) &= - 2 (\nabla^{\hat{\g} (i} W) \nabla_{\hat{\g}}^{j)} W
- 2 \tr\big((\bm \nabla^{\hat{\g} (i} \bm W) \bm \nabla_{\hat{\g}}^{j)} \bm W\big) + \frac{16 \ri}{3} \k G^{ij} \ ,
\end{align}
which implies
\be X^{ij} = \frac{\ri}{2 W} \big(\l^i \l^j + \tr (\bm \l^i \bm \l^j) \big) - \frac{1}{W} \tr (\bm W \bm X^{ij}) - \frac{2 \k}{3 W} G^{ij} \ .
\ee

From the component $\tfH_{\ha\hb\hc}$ one finds the expression
\begin{align} 
\label{ModdilWeylReplace}
3 \big(W^2 + \tr (\bm W^2) \big)W_{\ha\hb} &= 
\Big( 
\frac{1}{6} \eps_{\ha\hb\hc\hd\he} \tfH^{\hc\hd\he} - 2 W F_{\ha\hb} - 2 \tr (\bm W \bm F_{\ha\hb}) \non\\
&\qquad \qquad- \ri (\S_{\ha\hb})^{\hal\hbe} \l_\hal^k \l_{\hbe k} - \ri (\S_{\ha\hb})^{\hal\hbe} \tr(\bm\l_\hal^k \bm\l_{\hbe k})
\Big) \ .
\end{align}
If $\big(W^2 + \tr (\bm W^2) \big)$ does not vanish then we can again replace the matter fields of the Weyl multiplet with those of $W$.

The supersymmetry transformations of the gauge fields may be obtained from eq. \eqref{eq:WeylSUSY2}
upon using \eqref{ModdilWeylReplace}. We list the supersymmetry transformations of the 
matter fields below:
\begin{subequations}
\begin{align}
\d W &= \ri \xi_k \l^k \ , \\
\d \L_\hal^i &= 
2 \xi^{\hbe i} (F_{\hbe\hal} + W_{\hbe\hal} W) + \frac{\ri}{2 W} \xi_{\hal j} \big( \l^{(i} \l^{j)} + \tr (\bm \l^{(i} \bm \l^{j)}) \big)
- \frac{1}{W} \xi_{\hal j} \tr (\bm W \bm X^{ij}) \non\\
&\qquad- \frac{2\k}{3 W} \xi_{\hal j} G^{ij} - \xi^{\hbe i} \nabla_{\hbe\hal} W + 2 \ri \eta_\hal^i W \ , \\
\d V_\hm &=\ri \xi_k \G_\hm \l^k + \ri \xi_j \psi_\hm{}^j W \ , \\
\d \tfB_{\hm\hn} &= 2 \ri \xi_k \G_{[\hm} \psi_{\hn]}{}^k \big(W^2 + \tr(\bm W^2)\big) - 4 \ri \xi_k \S_{\hm\hn} \l^k W \non\\
&\qquad- 4 \ri \xi_k \S_{\hm\hn} \tr(\bm \l^k \bm W)
- 2 V_{[\hm} \d V_{\hn]} - 2 \tr \big( \bm V_{[\hm} \d \bm V_{\hn]} \big)
\ .
\end{align}
\end{subequations}

The superconformal field strengths are given by
\begin{subequations}
\begin{align}
F_{\ha\hb} &= 2 e_\ha{}^\hm e_\hb{}^\hn \partial_{[\hm} V_{\hn]}
- \ri \psi_{[\ha}{}_k \G_{\hb]} \l^k
+ \frac{\ri}{2} \psi_{[\ha}{}^\hga_k \psi_{\hb]}{}^k_{\hga} W \ , \\
\bm F_{\ha\hb} &= 2 e_\ha{}^\hm e_\hb{}^\hn (\partial_{[\hm} \bm V_{\hn]} - \ri \bm V_{[\hm} \bm V_{\hn]}) 
- \ri \psi_{[\ha}{}_k \G_{\hb]} \bm \l^k
+ \frac{\ri}{2} \psi_{[\ha}{}^\hga_k \psi_{\hb]}{}^k_{\hga} \bm W \ , \\
\tfH_{\ha\hb\hc} &= e_\ha{}^\hm e_\hb{}^\hn e_\hc{}^\hp \big(3 \partial_{[\hm} \tfB_{\hn\hp]} - 3 V_{[\hm} F_{\hn \hp]} - \tr (3 \bm V_{[\hm} \bm F_{\hn \hp]} + 2 \ri \bm V_{[\hm} \bm V_\hn \bm V_{\hp]}) 
- \frac{4 \k}{3} \cB_{\hm\hn\hp}\big) \non \\
&\qquad + \frac{3 \ri}{2} \psi_{[\ha}{}^k \G_\hb \psi_{\hc] k} \big(W^2 + \tr(\bm W^2)\big) \non\\
&\qquad - 6 \ri W \psi_{[\ha}{}^k \S_{\hb\hc]} \l_k - 6 \ri \ \tr \big( \bm W \psi_{[\ha}{}^k \S_{\hb\hc]} \bm \l_k \big) \ .
\end{align}
\end{subequations}

The supersymmetry transformations for the previous two cases (the dilaton Weyl and deformed dilaton Weyl multiplets) may be straightforwardly 
obtained from the above general results.


\section{Higher derivative couplings} \label{section12}

The superspace formalism developed in this paper offers 
more general tools
to construct composite primary multiplets
(that may be used, e.g., to generate higher derivative invariants) 
 than those which have so far been employed within 
 the component approaches
\cite{Zucker1,Zucker2,Zucker3,Ohashi1,Ohashi2,Ohashi3,Ohashi4,Bergshoeff1,Bergshoeff2,Bergshoeff3}. 
This will be demonstrated below.

\subsection{Composite primary multiplets and invariants}

In section \ref{N1SYM_and_PS} 
we derived two gauge prepotentials for the abelian vector multiplet:
(i) the tropical prepotential $V(v)$; and (ii) Mezincescu's prepotential $V_{ij}$. 
These constructions lead to two different procedures to generate 
composite vector multiplet field strengths. 

Associated with a composite weight-0 tropical multiplet $\mathbb V (v) $ is 
the following primary real scalar 
\bea 
\mathbb W_{\rm tropical} \equiv W[\mathbb V ]
= - \frac{\ri}{16 \pi} \oint (v, \rd v) \nabla^{(-2)} \mathbb V (v)~.
\label{13.1}
\eea
It obeys the Bianchi identity \eqref{6.20}. Thus we may think of 
$\mathbb W_{\rm tropical}$ as the field strength of a composite vector multiplet. 
An example is provided by  
\bea
\mathbb V = \frac{ H^{(2n)} }{ [ G^{(2)}]^n}~, \qquad \qquad n=1,2, \dots~, 
\label{13.2}
\eea
for an arbitrary real $\cO(2n)$ multiplet $H^{(2n)} (v)$ 
and an $\cO(2)$ multiplet $G^{(2)} (v)$ such that the scalar $G$ defined by 
\eqref{6.32} is nowhere vanishing. 
The existence of the latter is assumed in this section. 

Associated with a composite real isovector superfield 
$\mathbb V_{ij} $ 
with dimension $-2$ is 
the following primary real scalar 
\bea
\mathbb W_{\rm Mezincescu} \equiv W[\mathbb V_{ij}]
=
-\frac{3\ri}{40}\de_{ij}\D^{ijkl}\mathbb V_{kl}~.
\label{13.3}
\eea
It obeys the Bianchi identity \eqref{6.20}. 
As an example, we consider
\bea
\mathbb V_{ij} = \frac{G_{ij} } { G^{5/3 } }~.
\eea

In section \ref{section7} we derived 
the unconstrained prepotential $\O$ for the $\cO(2)$ multiplet. This construction leads to 
a procedure to generate 
composite  $\cO(2)$ multiplets. 
Associated with a composite primary dimensionless scalar $\mathbb N$ is 
the $\cO(2)$ multiplet 
\bea
{\mathbb G}^{(2)}
=v_iv_j {\mathbb G}^{ij}
\equiv G^{(2)} [ \mathbb N ]
&=&
-\frac{\ri}{8}\D^{(4)}\de^{(-2)}\mathbb N \quad \Longleftrightarrow \quad
\mathbb G^{ij}
=
-\frac{3\ri}{40}\D^{ijkl}\de_{kl}\mathbb N ~.~~~~~~
\label{13.5}
\eea
By construction, ${\mathbb G}^{ij}$ obeys the constraint \eqref{5.58}.
An example is provided by 
\bea
\mathbb N = \Big( \frac{ W^{\hal \hbe} W_{\hal \hbe}}{G^{2/3} } \Big)^n~,
\eea
for a positive integer $n$. Here $W_{\hal \hbe}$ is the super Weyl tensor. 

It is also possible to generate composite $\cO(4+n) $ multiplets by making use of 
the prepotential construction \eqref{7.34}, for any non-negative integer $n$. 
As an example, consider the case of an even integer $n =2m$.
Given  a composite 
$\cO(4+2m)$ multiplet, we can introduce a 
composite tropical multiplet of the form \eqref{13.2} and then make use of the latter 
to generate the composite vector multiplet field strength \eqref{13.1}.

As concerns  the component approaches
\cite{Zucker1,Zucker2,Zucker3,Ohashi1,Ohashi2,Ohashi3,Ohashi4,Bergshoeff1,Bergshoeff2,Bergshoeff3}, there is essentially only one regular procedure 
(the vector-tensor embedding) to generate
composite primary multiplets.
It is defined as follows:
Given a composite vector multiplet field strength $\mathbb W$
constrained by \eqref{6.20},  
the following superfield 
\bea
\mathbb  H^{ij}_{\rm VM} \equiv 
H^{ij}_{\rm VM} [\mathbb W ]
= \ri ( \nabla^{\hat \a (i} \mathbb W ) \, \nabla^{j)}_{\hat \a} \mathbb W 
+ \frac{\ri}{2} \mathbb W  \nabla^{\hal (i } \nabla^{j)}_\hal \mathbb W 
\label{composite-YM-G}
\eea
is a composite $\cO(2)$ multiplet. 

In addition, there exists the composite $\cO(2)$ multiplet 
 constructed by Hanaki, Ohashi and Tachikawa \cite{HOT} 
and associated with the Weyl multiplet.\footnote{This $\cO(2)$ multiplet 
was denoted $L^{ij}[\mathbf W^2]$ in \cite{HOT}.} 
In superspace, it is given in terms of  the super Weyl tensor
as in eq.~\eqref{2.43}.

We are in a position to generate supersymmetric invariants
given primary composite multiplets.
If the theory under consideration involves a dynamical vector 
multiplet, which is described by a tropical prepotential $V(v)$, 
and also possesses a composite $\cO(2)$ multiplet 
${\mathbb G}^{(2)}$, a supersymmetric BF invariant is generated by the Lagrangian
\bea
\cL^{(2)}_{\mathbb G} = V {\mathbb G}^{(2)}~.
\label{13.9}
\eea
If the theory involves a dynamical $\cO(2)$ 
multiplet, which is described by a prepotential $\O$, 
and possesses a composite vector multiplet field strength $\mathbb W$, 
then we are able to construct a supersymmetric invariant of the type
\eqref{action7.29} with the Lagrangian 
\bea
\cL_{\mathbb W} = \O \mathbb W~.
\eea
More generally, the action principles 
\eqref{InvarAc} and \eqref{action7.29} provide universal procedures
to generate supersymmetric invariants. 
For instance, supersymmetric $R^{4+2n}$ terms may be realized as full 
superspace invariants \eqref{action7.29} with 
\bea
\cL = \frac{ (W^{\hal \hbe} W_{\hal \hbe} )^2} {G} 
\Big( \frac{W[ \mathbb V ] }{G^{1/3}} \Big)^n~, \qquad
\mathbb V :=   \frac{ H^{(2)}_{\rm Weyl} }{G^{(2)} } ~, \qquad n=0,1, \dots~,
\eea
where $W[ \mathbb V ]$ is defined by \eqref{13.1}.


\subsection{Composite vector multiplets}\label{subsection13.2}

In this subsection we consider several examples of applying the rule
\eqref{13.1} to generate composite vector multiplets. Our results are 
inspired by the four-dimensional analysis in  \cite{BK11}.
Below we denote $\mathbb W_{\rm tropical} $ simply as $\mathbb W$.

Our first example is
\be 
\mathbb V = \log \Big( \frac{G^\pd}{\ri \U^\pu \breve{\U}^\pu} \Big) \ ,
\ee
where $\U^\pu$ is a weight-one arctic multiplet. 
The corresponding composite vector multiplet \eqref{13.1}
has already appeared in 
\eqref{CVM8}. It constitutes the equation of motion for the theory 
of a single $\cO(2) $ multiplet coupled to conformal supergravity.  
Evaluating the covariant derivatives gives
\be \mathbb W 
= 
- \frac{\ri}{16 \pi} \oint (v, \rd v) \Big( \frac{\de^\pmd G^\pd}{G^\pd} 
- \frac{(\nabla^{\pmu\hal} G^\pd)\nabla_\hal^{\pmu} G^\pd}{(G^\pd)^2} \Big) \ .
\ee
The contour integral can be explicitly evaluated.
To do so we make use of the identities
\bsubeq \label{14.14}
\begin{align} 
\nabla_\hal^\pmu G^\pd &= 2 \varphi_\hal^\pu = 2 \varphi_\hal^i v_i \ , \\
\de^\pmd G^\pd &= - 4 \ri F \ ,
\end{align}
\esubeq
where we have introduced the descendant superfields \eqref{O2superfieldComps}.
Then applying
the integration 
identities of \cite{BK11}, 
we obtain
\bea
\mathbb W 
&=& 
- \frac{1}{4\pi} \oint_C (v, \rd v) \Big( \frac{F}{G^\pd} - \ri \frac{\varphi^{\pu\hal}\varphi_\hal^{\pu}}{G^\pd} \Big)
= \frac{1}{4G} F - \frac{\ri}{8 G^3} \varphi^{\hal i} \varphi_\hal^j G_{ij} 
\non\\
&=&
 \frac{\ri}{48 G} \nabla^{\hal i} \nabla_{\hal}^j G_{ij}  
- \frac{\ri}{72 G^3}  G_{ij} (  \nabla^\hal_k G^{ik} ) \nabla_{\hal l} G^{jl} 
~.
\label{13.13}
\eea
${}$From the $S$-supersymmetry transformations of $\varphi_\hal^i$ and $F$,
\begin{align} 
S_\hal^i \varphi_\hbe^j = - 6 \eps_{\hal \hbe} G^{ij} 
~,~~~~~~
S_\hal^i F = 6 \ri \varphi_\hal^i \ ,
\end{align}
it is straightforward to explicitly check that $\mathbb W$ is primary.

It is an instructive exercise to show that the composite vector multiplet \eqref{13.13}
can also be rewritten in the following
 compact form
\bea
\mathbb W 
&=&
 \frac{\ri}{16} G \nabla^{\hal i} \nabla_\hal^j \Big(\frac{G_{ij}}{G^2}\Big)
 ~ .
\label{13.15}
\eea
This expression resembles the one in four dimensions  \cite{BK11}.
The vector multiplet \eqref{13.15} is actually well known. 
At the component level it was first derived by Zucker \cite{Zucker4}, 
using a brute force approach, as an extension of the construction for 
the improved $\cN=2$ tensor multiplet in four dimensions \cite{deWPV}.

As another example, we consider a composite tropical prepotential of the form
\bea
\mathbb V_n = \frac{ H^{(2n)} }{ [ G^{(2)}]^n}~, \qquad \qquad n=1,2, \dots~, 
\eea
where $H^{(2n)}$ is an arbitrary $\cO(2n)$ multiplet.
The
corresponding  composite vector multiplet 
\be 
\mathbb W_n = - \frac{\ri}{16 \pi} \oint_C (v, \rd v) \de^\pmd
\Big( \frac{H^{(2n)}}{(G^{(2)})^n} \Big) 
\ee
can be computed in complete analogy with the 4D $\cN = 2$ analysis in \cite{BK11}.
Evaluating the covariant derivatives gives
\begin{align} \mathbb W_n 
&= 
- \frac{\ri}{16 \pi} \oint_C (v, \rd v) \Big( 
\frac{2 n - 1}{2 n + 1} \frac{h^{(2 n - 2)}}{(G^{(2)})^n}
- \frac{8 n^2}{2 n + 1} \frac{\Psi^{(2 n - 1)} \varphi^{(1)}}{(G^{(2)})^{n+1}}
+ 4 n \ri H^{(2 n)} \frac{F}{(G^{(2)})^{n+1}} \non\\
&\qquad \quad + 4 n (n+1) H^{(2 n)} \frac{\varphi^{(1)} \varphi^{(1)}}{(G^{(2)})^{n+2}}
 \Big) \ ,
\end{align}
where we made use of the identities \eqref{14.14} and
\bsubeq
\begin{align} 
\nabla_\hal^- H^{(2n)} &= \frac{2 n}{2 n + 1} \Psi_\hal^{(2 n - 1)} \ , \\
\nabla^{(-2)} H^{(2 n)} &= \frac{2 n - 1}{2 n + 1} h^{2 n - 2} \ ,
\end{align}
\esubeq
with
\begin{align}
\Psi_\hal^{(2 n - 1)} &= \nabla_{\hal k} H^{k i_1 \cdots i_{2 n -1}} v_{i_1} \cdots v_{i_{2n - 1}} \ , \\
h^{(2 n -2)} &= \nabla_{kl} H^{kl i_1 \cdots i_{2 n - 2}} v_{i_1} \cdots v_{i_{2n - 2}}~.
\end{align}
Making use of the integration results of \cite{BK11} gives
\bea
\mathbb W_n 
&=& \frac{\ri (2 n)!}{2^{2 n + 2} (n!)^2} \Big(
\frac{n}{2 (2 n + 1)} \frac{h^{i_1 \cdots i_{2 n - 2}} G_{(i_1 i_2} \cdots G_{i_{2 n -3} i_{2 n -2})}}{G^{2 n - 1}} \non\\
&& - \frac{2 n^2}{2 n +1} \frac{\Psi^{i_1 \cdots i_{2 n - 1}} \varphi^{i_{2n}} G_{(i_1 i_2} \cdots G_{i_{2 n -1} i_{2n})}}{G^{2 n + 1}} \non\\
&& + \ri n \frac{F h^{i_1 \cdots i_{2 n}} G_{(i_1 i_2} \cdots G_{i_{2n - 1} i_{2 n})}}{G^{2 n + 1}} \non\\
&& + n (2 n + 1) \frac{h^{i_1 \cdots i_{2n}} \varphi^{i_{2n + 1}} \varphi^{i_{2 n + 2}} G_{(i_1 i_2} \cdots G_{i_{2n + 1} i_{2 n + 2})}}{G^{n+3}}
\Big)
~.
\eea
It turns out the above complex expression may be cast in the following simpler form
\begin{subequations} \label{13.23}
\bea
\mathbb W_n = 
\frac{\ri(2 n)!}{2^{2 n + 3} (n+1)! (n - 1)!} G \nabla_{ij} \cR_n{}^{ij} \ ,
\eea
where
\bea 
\cR_n{}^{ij} = \Big( \d^i_k \d_l^j - \frac{1}{2 G^2} G^{ij} G_{kl} \Big) \frac{H^{kl i_1 \cdots i_{2 n - 2}} G_{(i_1 i_2} \cdots G_{i_{2n - 3} i_{2 n - 2})}}{G^{(2 n)}} \ .
\eea
\end{subequations}

The composite vector multiplets  \eqref{13.23} 
are new for $n>1$. The choice $n=1$ is a special case in the family of 
composite tropical prepotentials of the form 
\bea
\mathbb V = \cF \big(H^{(2)}_A \big)~, \qquad A =1, \dots, m ~,
\label{13.24}
\eea
where $\cF (z_A)$ is 
a homogeneous function of degree zero,   $\cF (\l z_A) =\cF (z_A)$,
and $H^{(2)}_A $ are $\cO(2)$ multiplets, $A =1, \dots, m $.
The composite vector multiplet associated with \eqref{13.24}
can be computed in complete analogy with 
the 4D $\cN=2$ analysis in 
\cite{BK11} (the latter analysis was inspired by \cite{deWS}).

\subsection{Ricci squared $\cO(2)$ multiplet} \label{subsection14.3}

As discussed above, associated with the super Weyl tensor is the $\cO(2)$ 
multiplet \eqref{2.43}. In this subsection we discover one more $\cO(2)$
multiplet associated with the supergravity dynamical variables. 
Our analysis is inspired by the construction of chiral invariants in 4D $\cN=2$  
supergravity presented in \cite{BdeWKL}. 

In section \ref{prep-O2} we constructed the prepotential formulation for the $\cO(2)$ multiplet such that the prepotential is a
primary dimensionless real scalar  $\O$.
It turns out that this construction can be
generalized by replacing $\O$ with  $\log \F$
defined in terms of a primary nowhere vanishing real scalar $\F$ of dimension $q$:
\bea
S_\hal^i\log \F=0
~,~~~
\bbD\log \F
=
q
~.
\eea
Let us consider the superfield
\bea
{G}^{(2)} [\log \F ]
=
-\frac{\ri}{8}\D^{(4)}\de^{(-2)}\log \F
=
-\frac{3\ri}{40}v_iv_j\D^{ijkl}\de_{kl}\log \F
~.
\label{def-G-0tilde}
\eea
It follows 
that 
${G}^{(2)} [\log \F ]$ is analytic,
$\de_\hal^{(1)}{G}^{(2)} [\log \F ]=0$,
and of dimension $3$.
As demonstrated in section \ref{prep-O2}, 
the superfield  $G^{(2)} := G^{(2)} [\O] $
defined by \eqref{def-G-0}
 is primary $S_\hal^i G^{(2)}=0$. We observe that 
 exactly the same derivation holds for ${G}^{(2)} [\log \F ]$.
Indeed, in the case of $G^{(2)}$ we used  the fact that  $\bbD\O=0$.
In computing $S_\hal^i {G}^{(2)}[\log \F ] $, 
there may be extra terms due to the fact that
$\bbD\log \F=q \neq 0$. But it can be checked that all these terms are 
actually annihilated
by some operator acting on the constant $q$.  
Since $S_\hal^i {G}^{(2)}[\log \F ]=0$, we conclude that 
${G}^{(2)}[\log \F] $ is also an $\cO^{(2)}$ multiplet.

The reason why ${G}^{(2)}[\log \F]$ is of interest can be made clear
 once we consider the degauged versions of \eqref{def-G-0} and \eqref{def-G-0tilde}.
It is a straightforward, although tedious, exercise to apply the degauging procedure of section \ref{degauging}
in order to  express \eqref{def-G-0} and \eqref{def-G-0tilde} in SU(2) superspace.
Let us denote by
$
\cG^{(2)}
=
\frak O^{(2)}_6 \O
=v_iv_j\frak O^{ij}_6 \O
$
the degauged version of \eqref{def-G-0}.
Here the sixth-order differential operator 
$\frak O^{ij}_6=\frak O^{ji}_6$
 is constructed only in terms of $\cD_\hA$, $M_{\ha\hb}$, $J^{ij}$ and the torsion tensors of SU(2)
 superspace.
It can be obtained by iteratively
degauging the six $\de$-derivatives while moving to the right the $S_\hal^i$, $K_\ha$ and $\bbD$ operators
to use $S_\hal^i\O=K_\ha\O=\bbD\O=0$.
For the scope of this paper we do not need the explicit expression 
for $\frak O^{ij}_6$.
Since $G^{(2)}$ is an $\cO(2)$ multiplet, it holds 
by construction
that $\cD_\hal^{(1)}
\frak O^{(2)}_6
\O=0$ .
 
The result of degauging ${G}^{(2)}[\log \F] $, which we denote  $\cG^{(2)}[\log \F]$, 
is more interesting.
A straightforward but somewhat lengthy calculation leads to the following 
relation
\bea
\cG^{(2)}[\log \F] 
&=&
\frak O^{(2)}_6 
\log \F
-q\,H_{{\rm Ric}}^{(2)}
~.
\label{def-G-0tildedegauged}
\eea
The superfield $H_{{\rm Ric}}^{(2)}$ encodes all the contributions that arise
from using $\bbD\log \F=q$
 and is given by 
\bea
H_{{\rm Ric}}^{(2)}
&=&
-\frac{\ri}{128}\Big\{
\cD^{(2)}\cD^{(2)}\frak{F}^{(-2)}{}^{\hal}{}_\hal
-12\cD^{(2)}\big(\frak{F}^{(0)}{}^{\hal}{}_\hal\frak{F}^{(0)}{}^{\hbe}{}_\hbe\big)
+12\cD^{(2)}\big(\frak{F}^{(0)}{}^{\hal\hbe}\frak{F}^{(0)}_{\hal\hbe}\big)
\non\\
&&~
+12\Big(
3(\cD^{(1)\hal}\frak{F}^{(0)}_\hal{}^{[\hbe})\ve^{\hga\hde]}
+\ri(\G_\ha)^{\hbe\hga}\frak{F}^{(1)\hde}{}^{\ha}
\Big)
\big(
\cD_{\hbe}^{(1)}\frak{F}^{(0)}_{\hga\hde}
-\ri(\G^\hb)_{\hga\hde}\frak{F}^{(1)}_{\hbe\hb}
\big)
\non\\
&&~
-12\Big(
3\frak{F}^{(0)}{}^{\hal[\hbe}\ve^{\hga\hde]}
+\frak{F}^{(0)}{}^{\hde\hal}\ve^{\hbe\hga}
\Big)\cD_\hal^{(1)}\big(
\cD_{\hbe}^{(1)}\frak{F}^{(0)}_{\hga\hde}
-\ri(\G^\hb)_{\hga\hde}\frak{F}^{(1)}_{\hbe\hb}
\big)
\non\\
&&~
+8\cD^{(2)}(\frak{F}^{(2)}{}^{\hal}{}_\hal\frak{F}^{(-2)}{}^{\hbe}{}_\hbe)
+4\cD^{(1)\hal}\Big(
\frak{F}^{(2)}{}_\hal{}^{\hbe}\cD_\hbe^{(1)}\frak{F}^{(-2)}{}^{\hga}{}_\hga
-6\ri(\G^\ha)^{\hbe\hga}\frak{F}^{(1)}_{\hal\ha}\frak{F}^{(0)}_{\hbe\hga}
\Big)
\non\\
&&~
+16\frak{F}^{(2)}{}^{\hal\hbe}\Big(
-2\frak{F}^{(2)}_{\hal\hbe}\frak{F}^{(-2)}{}^{\hga}{}_\hga
+6\frak{F}^{(0)}_{\hal\hbe}\frak{F}^{(0)}{}^{\hga}{}_\hga
-3\frak{F}^{(0)}_{\hal}{}^{\hga}\frak{F}^{(0)}_{\hbe\hga}
\non\\
&&~~~~~~~~~~~~~~~~~
+3\frak{F}^{(0)}{}^\hga{}_{\hal}\frak{F}^{(0)}_{\hga\hbe}
+12\frak{F}^{(0)}{}_\hal{}^{\hga}\frak{F}^{(0)}_{\hga\hbe}
\Big)
\non\\
&&~
+16\frak{F}^{(2)}{}^{\hal}{}_{\hal}\Big(
2\frak{F}^{(2)}{}^{\hbe}{}_\hbe\frak{F}^{(-2)}{}^{\hga}{}_\hga
+3\frak{F}^{(0)}{}^{\hbe\hga}\frak{F}^{(0)}_{\hbe\hga}
-3\frak{F}^{(0)}{}^{\hbe }{}_{\hbe}\frak{F}^{(0)}{}^{\hga}{}_\hga
\Big)
\Big\}
~.~~~~~~~~~~~~
\label{calG2}
\eea
Here we have introduced the following superfields:
\bsubeq
\bea
\frak{F}^{(2)}_{\hal\hbe}
&:=&
v_iv_j\frak{F}_{\hal}^i{}_\hbe^j
~,
~~~
\frak{F}^{(0)}_{\hal\hbe}
:=
\frac{v_iu_j}{(v,u)}\frak{F}_{\hal}^i{}_\hbe^j
~,
~~~
\frak{F}^{(-2)}_{\hal\hbe}
:=
\frac{u_iu_j}{(v,u)^2}\frak{F}_{\hal}^i{}_\hbe^j
~,
\\
\cF_{\hal\hbe\hga}^{(1)}
&:=&
\cD_{\hal}^{(1)}\frak{F}^{(0)}_{\hbe\hga}
-\ri(\G^\hb)_{\hbe\hga}\frak{F}^{(1)}_{\hal\hb}
~.
\eea
\esubeq
What is remarkable about \eqref{def-G-0tildedegauged}
is that by construction $H_{{\rm Ric}}^{(2)}$ is a composite $\cO(2)$ 
multiplet\footnote{It should be pointed out that $H_{{\rm Ric}}^{(2)}$
is a non-primary $\cO(2)$ multiplet, since its super Weyl transformation law 
is inhomogeneous. }
constructed only in terms of the curvature tensors of SU(2) 
superspace;  it  is completely independent of $\log \F$.
As will be discussed in the next two subsections, 
$\cG^{(2)}[\log \F] $
gives rise to a supersymmetric extension of the Ricci squared action.

By construction, $H_{{\rm Ric}}^{(2)}$ is independent of $u_i$ and 
can be represented in the form
$H_{{\rm Ric}}^{(2)}=v_iv_j H_{{\rm Ric}}^{ij}$. From \eqref{calG2} we deduce
\bea
H_{{\rm Ric}}^{ij}
&=&
-\frac{\ri}{128}\Big\{
\frac{3}{5}\cD^{(ij}\cD^{kl)}\frak{F}^{\hal}_{ k}{}_{\hal l}
-\frac{36}{5}\cD^{(ij}\big(\frak{F}^{\hal k}{}_{\hal k}\frak{F}^{\hbe l)}{}_{\hbe l}\big)
+\frac{24}{5}\cD^{(ij}(\frak{F}^{\hal k}{}_\hal^{l)}\frak{F}^{\hbe}_{ k}{}_{\hbe l})
\non\\
&&~
+\frac{36}{5}\cD^{(ij}\big(\frak{F}^{\hal k}{}^{\hbe}_{ k}\frak{F}_{\hal}^{l)}{}_{\hbe l} \big)
+6\cD^{\hal (i}\Big(
\frac{2}{5}\frak{F}_\hal^j{}^{\hbe k}\cD_\hbe^{l)}\frak{F}^{\hga}_{k}{}_{\hga l}
-3\ri(\G^\ha)^{\hbe\hga}\frak{F}_{\hal}^j{}_{\ha}\frak{F}_{\hbe}^{k)}{}_{\hga k}
\Big)
\non\\
&&~
+27(\cD^{\hal (i}\frak{F}_\hal^j{}^{[\hbe}_k)\ve^{\hga\hde]}
\Big(
\frac{4}{5}\cD_{\hbe}^{k}\frak{F}_{\hga}^{l)}{}_{\hde l}
-\ri\frak{F}_{\hbe}^{k)}{}_{\hb}(\G^\hb)_{\hga\hde}
\Big)
\non\\
&&~
+9\Big(
3\frak{F}^{\hal (i}{}^{[\hbe}_k\ve^{\hga\hde]}
+\frak{F}^{\hde (i}{}^{\hal}_k\ve^{\hbe\hga}
\Big)
\Big(
\ri\cD_\hal^{j}\frak{F}_{\hbe}^{k)}{}_{\hb}(\G^\hb)_{\hga\hde}
-\frac{4}{5}\cD_\hal^j\cD_{\hbe}^{k}\frak{F}_{\hga}^{l)}{}_{\hde l}
\Big)
\non\\
&&~
+9\ri(\G_\ha)^{\hbe\hga}\frak{F}^{\hde (i}{}^{\ha}\cD_{\hbe}^{j}\frak{F}_{\hga}^{k)}{}_{\hde k}
-12(\G_\ha)_\hbe{}^{\hga}(\G_\hb)_{\hga\hde}\frak{F}^{\hde (i}{}^{\ha}\frak{F}^{\hbe j)}{}^{\hb}
\non\\
&&~
+\frac{48}{5}\frak{F}^{\hal (i}{}^{\hbe j}\Big(
6\frak{F}_{\hal}^k{}_{\hbe k}\frak{F}^{\hga l)}{}_{\hga l}
-2\frak{F}_{\hal}^k{}_{\hbe}^{l)}\frak{F}^{\hga }_{k}{}_{\hga l}
-3\frak{F}_{\hal}^k{}^{\hga}_{k}\frak{F}_{\hbe}^{l)}{}_{\hga l}
+3\frak{F}^{\hga k}{}_{\hal k}\frak{F}_{\hga}^{l)}{}_{\hbe l}
+12\frak{F}_\hal^k{}^{\hga}_{k}\frak{F}_{\hga}^{l)}{}_{\hbe l}
\Big)
\non\\
&&~
+\frac{48}{5}\frak{F}^{\hal (i}{}_{\hal}^j\Big(
2\frak{F}^{\hbe k}{}_\hbe^{l)}\frak{F}^{\hga}_{ k}{}_{\hga l}
+3\frak{F}^{\hbe k}{}^{\hga}_{k}\frak{F}_{\hbe}^{l)}{}_{\hga l}
-3\frak{F}^{\hbe k}{}_{\hbe k}\frak{F}^{\hga l)}{}_{\hga l}
\Big)
\Big\}
~.~~~~~~~~~
\label{calG2_2}
\eea
On the other hand, 
the condition that the expression \eqref{calG2} is independent of $u_i$
gives the  constraints
\bsubeq\label{14.33}
\bea
0
&=&
\cD^{(ij}\cD^{kl}\frak{F}^{\hal p}{}_\hal^{q)}
-4\cD^{(ij}\big(\frak{F}^{\hal k}{}_\hal^l\frak{F}^{\hbe p}{}_\hbe^{q)}\big)
+12\cD^{(ij}\big(\frak{F}^{\hal k}{}^{\hbe l}\frak{F}_{\hal}^p{}_{\hbe}^{q)}\big)
\non\\
&&
+4\cD^{\hal (i}\big(\frak{F}_\hal^j{}^{\hbe k}\cD_\hbe^{l}\frak{F}^{\hga p}{}_\hga^{q)}\big)
+36(\cD^{\hal (i}\frak{F}_\hal^j{}^{[\hbe k})\ve^{\hga\hde]}\cD_{\hbe}^{l}\frak{F}_{\hga}^p{}_{\hde}^{q)}
\non\\
&&
-12\Big(
3\frak{F}^{\hal (i}{}^{[\hbe j}\ve^{\hga\hde]}
+\frak{F}^{\hde (i}{}^{\hal j}\ve^{\hbe\hga}
\Big)\cD_\hal^k\cD_{\hbe}^{l}\frak{F}_{\hga}^p{}_\hde^{q)}
\non\\
&&
+16\Big(
7\frak{F}^{\hal (i}{}^{\hbe j}\frak{F}_{\hal}^k{}_{\hbe}^l\frak{F}^{\hga p}{}_\hga^{q)}
-\frak{F}^{\hal (i}{}_{\hal}^j\frak{F}^{\hbe k}{}_{\hbe}^l\frak{F}^{\hga p}{}_\hga^{q)}
+12\frak{F}^{\hal (i}{}^{\hbe j}\frak{F}_\hal^k{}^{\hga l}\frak{F}_{\hga}^p{}_{\hbe}^{q)}
\Big)
\eea
and
\bea
0&=&
\frac{5}{3}\cD^{(ij}\cD^{kl}\frak{F}^{\hal p)}{}_{\hal p}
-\frac{20}{3}\cD^{(ij}(\frak{F}^{\hal k}{}_\hal^l\frak{F}^{\hbe p)}{}_{\hbe p})
+20\cD^{(ij}\big(\frak{F}^{\hal k}{}^{\hbe l}\frak{F}_{\hal}^{p)}{}_{\hbe p}\big)
\non\\
&&
+4\cD^{\hal (i}\Big(
\frac{5}{3}\frak{F}_\hal^j{}^{\hbe k}\cD_\hbe^{l}\frak{F}^{\hga p)}{}_{\hga p}
-6\ri(\G^\ha)^{\hbe\hga}\frak{F}_{\hal}^j{}_{\ha}\frak{F}_{\hbe}^k{}_{\hga}^{l)}
\Big)
\non\\
&&
+6(\cD^{\hal (i}\frak{F}_\hal^j{}^{[\hbe k})\ve^{\hga\hde]}
\Big(
5\cD_{\hbe}^{l}\frak{F}_{\hga}^{p)}{}_{\hde p}
-6\ri\frak{F}_{\hbe}^{l)}{}_{\hb}(\G^\hb)_{\hga\hde}
\Big)
+30(\cD^{\hal (i}\frak{F}_\hal^j{}^{[\hbe}_{p})\ve^{\hga\hde]}\cD_{\hbe}^{k}\frak{F}_{\hga}^l{}_{\hde}^{p)}
\non\\
&&
+2\Big(
3\frak{F}^{\hal (i}{}^{[\hbe j}\ve^{\hga\hde]}
+\frak{F}^{\hde (i}{}^{\hal j}\ve^{\hbe\hga}
\Big)
\Big(
6\ri\cD_\hal^{k}\frak{F}_{\hbe}^{l)}{}_{\hb}(\G^\hb)_{\hga\hde}
-5\cD_\hal^k\cD_{\hbe}^{l}\frak{F}_{\hga}^{p)}{}_{\hde p}
\Big)
\non\\
&&
-10\Big(
3\frak{F}^{\hal (i}{}^{[\hbe}_{p}\ve^{\hga\hde]}
+\frak{F}^{\hde (i}{}^{\hal}_{p}\ve^{\hbe\hga}
\Big)\cD_\hal^j\cD_{\hbe}^{k}\frak{F}_{\hga}^l{}_\hde^{p)}
+12\ri(\G_\ha)^{\hbe\hga}\frak{F}^{\hde (i}{}^{\ha}\cD_{\hbe}^{j}\frak{F}_{\hga}^k{}_{\hde}^{l)}
\non\\
&&
+\frac{80}{3}\frak{F}^{\hal (i}{}^{\hbe j}\Big(
\frak{F}_{\hal}^k{}_{\hbe}^l\frak{F}^{\hga p)}{}_{\hga p}
+3\frak{F}_{\hal}^k{}_{\hbe p}\frak{F}^{\hga l}{}_\hga^{p)}
-9\frak{F}_{\hal}^k{}^{\hga l}\frak{F}_{\hbe}^{p)}{}_{\hga p}
+3\frak{F}_\hal^{k}{}^{\hga l}\frak{F}_{\hga}^{p)}{}_{\hbe p}
\Big)
\non\\
&&
+\frac{80}{3}\frak{F}^{\hal (i}{}_{\hal}^j\big(
3\frak{F}^{\hbe k}{}^{\hga l}\frak{F}_{\hbe}^{p)}{}_{\hga p}
-\frak{F}^{\hbe k}{}_{\hbe}^l\frak{F}^{\hga p)}{}_{\hga p}
\big)
~,
\eea
\esubeq
which have to be satisfied identically. 


\subsection{Supersymmetric $R^2$ invariants}

Supersymmetric extensions of the $R^2$ terms may be realized
 using the BF action principle \eqref{BFactionsection9}, in which the tropical prepotential 
corresponds to the vector multiplet compensator. 
There are three invariants associated with the Lagrangians 
\begin{subequations} \label{13.34} 
\bea
\cL^{(2)}_{\rm Weyl} &=& V H^{(2)}_{\rm Weyl}~,  \label{13.34a}  \\
\cL^{(2)}_{\rm Ric} &=& -V G^{(2)} [\log W]
~,  \label{13.34b}  \\
\cL^{(2)}_{\rm scal} &=& V H^{(2)}_{\rm VM} [\mathbb W ]~, 
\qquad 
\mathbb W =
 \frac{\ri}{16} G \nabla^{\hal i} \nabla_\hal^j \Big(\frac{G_{ij}}{G^2}\Big)~.
 \label{13.34c} 
\eea
\end{subequations}
The supersymmetric invariants associated with \eqref{13.34a} and
\eqref{13.34c} are known in the literature  \cite{HOT,BRS,OP1,OP2}.
At the component level, they generate the Weyl tensor squared and scalar curvature squared terms, respectively. 
The invariant associated with \eqref{13.34b} is new. 
At the component level, it turns out to generate
 the Ricci tensor squared term.
In order to achieve a better understanding of this invariant, 
it is useful to consider a special case when the vector multiplet compensator
$W$ obeys the equation \eqref{12.5}. As discussed in section 
\ref{dilWeylMultiplet}, this case corresponds to the dilaton Weyl 
multiplet. 


\subsection{The supersymmetric Ricci squared term  and the dilaton Weyl multiplet}

When dealing with the vector multiplet compensator, it is often convenient
to impose the gauge condition
\eqref{GCond}
which fixes  the local special conformal symmetry and 
eliminates the dilatation connection entirely, thus 
leading us to SU(2) superspace. 
In addition, the local dilatation symmetry can also be fixed by 
making the gauge choice
\bea
W=1~.
\label{W=1}
\eea

We recall that the Bianchi identity  for the vector multiplet \eqref{vector-Bianchi}
takes the following form in SU(2) superspace \cite{KT-M08}
\bea
\cD_{\hal}^{(i}\cD_{\hbe}^{j)}W
-{1\over 4}\ve_{\hal\hbe}\cD^{\hga(i}\cD_{\hga}^{j)} W 
= {\ri\over 2} C_{\hal\hbe}{}^{ij} W ~.~~~~~~
\label{W-BI}
\eea
Then choosing the gauge condition \eqref{W=1} gives 
\begin{subequations}\label{13.37}
\bea
C_\ha{}^{ij}=0~.
\eea
We also recall that the equation of motion for the vector multiplet \eqref{12.5}
turns into \eqref{12.5SU(2)} in SU(2) superspace.
Then imposing the gauge condition \eqref{W=1} gives 
\bea
S^{ij}=0 ~.
\eea
\end{subequations}

Under the conditions \eqref{13.37}, the algebra of covariant derivatives 
in SU(2) superspace simplifies drastically.
In particular, the anti-commutator of two spinor covariant derivatives becomes
\bea
\{ \cD_\hal^i , \cD_\hbe^j \} 
&=& 
-  2 \ri \eps^{ij} \cD_{\hal \hbe} 
- \ri \eps_{\hal \hbe} \eps^{ij} (W^{\hc\hd} + Y^{\hc\hd}) M_{\hc\hd} 
+ \frac{\ri}{2} \eps^{ij} \eps^{\ha \hb \hc \hd \he} (\G_\ha)_{\hal\hbe} Y_{\hb\hc} M_{\hd\he} 
\non\\ 
&&
- 12 \ri Y_{\hal\hbe} J^{ij}
~,
\eea
where $W_{\ha\hb}$ and $Y_{\ha\hb}$ 
satisfy the Bianchi identities
\bsubeq
\bea
\cD_\hga^k W_{\ha\hb}
&=&
W_{\ha\hb\hga}{}^k
-\frac{5}{2}(\S_{\ha\hb})_\hga{}^\hde\cY_\hde^{k}
~, 
\\
\cD_\hga^k Y_{\ha\hb}
&=&
2(\G_{{[}\ha})_{\hga}{}^{\hde}\cY_{\hb{]}\hde}{}^k
+(\S_{\ha\hb})_\hga{}^\hde \cY_{\hde}^k~.
\eea
\esubeq
Using the Bianchi identities, at dimension 2 we find the relations
\bsubeq
\bea
\cD_\hal^{(i}\cD_\hbe^{j)} W_{\ha\hb}
&=&
\frac{5}{2}\ve_{\hal\hbe}(\S_{\ha\hb})^{\hga\hde}\cD_{\hga}^{(i} \cY_{\hde}^{j)} 
-\frac{5}{8}\ve_{\ha\hb\hc\hd\he}(\G^\he)_{\hal\hbe}(\S^{\hc\hd})^{\hde\hrh}\cD_{\hde}^{(i} \cY_{\hrh}^{j)} 
~,
\\
\cD_\hal^{(i}\cD_\hbe^{j)}Y_{\ha\hb}
&=&
\frac{5}{4}\ve_{\hal\hbe}(\S_{\ha\hb})^{\hga\hde}\cD_{\hga}^{(i} \cY_{\hde}^{j)} 
+\frac{5}{8}
\ve_{\ha\hb\hc\hd\he}(\G^\he)_{\hal\hbe}(\S^{\hc\hd})^{\hga\hde}\cD_{\hga}^{(i} \cY_{\hde}^{j)} 
~.
\eea
\esubeq
Furthermore,  at dimension 5/2 we derive 
\bea
\cD_\hal^{(i}\cD^{jk)}Y_{\ha\hb}
=0
~,~~~~~~
\cD_\hal^{(i}\cD^{jk)}W_{\ha\hb}
=0
~.
\label{DDDX_W}
\eea
It can also be seen that the bivector $X_{\ha\hb}:=Y_{\ha\hb}+W_{\ha\hb}$ 
satisfies 
\bea
\cD_\hal^{(i}\cD_\hbe^{j)}X_{\ha\hb}
=\frac{1}{4}\ve_{\hal\hbe}\cD^{ij}X_{\ha\hb}
~.
\label{GR-YM}
\eea
This relation is reminiscent of the Bianchi identity  for the vector multiplet,
eq. \eqref{vector-Bianchi}. In the remainder of this section, we will refer
to the superspace geometry described as dilaton SU(2) superspace.

In the dilaton SU(2) superspace, the expressions \eqref{def-G-0tildedegauged}
and \eqref{calG2}
for the $\cO(2)$ multiplet on the right of  \eqref{13.34b}
proves to 
simplify drastically and takes the form:
\bea
- \cG^{(2)}[\log W] =
H_{\rm Ric}^{(2)}
&=&
\frac{15\ri}{8}  \Big\{
Y^{\hal\hbe}\cD_\hal^{(1)}\cY_{\hbe}^{(1)}
-\frac{2}{5} \cY^{(1)\ha\hal}\cY_{\ha\hal}^{(1)}
-\frac{1}{2}\cY^{(1)\hal}\cY_{\hal}^{(1)}
\Big\} ~.~~~~~~~
\label{13.43}
\eea
It is now easy to check that the constraints \eqref{14.33} are identically satisfied.
Now we are going to show that $H_{\rm Ric}^{(2)}$ can be represented 
as a linear combination of two different $\cO(2)$ multiplets. 

First of all, let us consider the Weyl squared $\cO(2)$ multiplet \eqref{2.43}. 
In the dilaton SU(2) superspace it may be rewritten as 
\bea
H^{(2)}_{\rm Weyl}
 = 
-\frac{15\ri}{4}\Big\{ 
W^{\hal\hbe} \cD_{\hal}^{(1)}\cY_{\hbe}^{(1)} 
+\frac{1}{15} W^{(1)\ha\hb\hal} W^{(1)}_{\ha\hb\hal}
+\frac{5}{4} \cY^{(1)\hal} \cY_\hal^{(1)}
\Big\}
~.
\label{13.44}
\eea
For the dilaton Weyl multiplet,  the BF Lagrangian 
\eqref{13.34a} 
generates a supersymmetric extension of the
$(C_{\ha\hb\hc\hd})^2+\frac{1}{6}R^2$ Lagrangian 
of \cite{HOT,OP1,OP2}.

A remarkable feature of the dilaton SU(2) superspace is that the 
relations  \eqref{DDDX_W} and \eqref{GR-YM} imply the existence of 
one more  $\cO(2)$ multiplet. It is 
\bsubeq \label{13.45}
\bea
H^{(2)}_{{\rm Riem}}
&:=&\frac{\ri}{4}
\Big\{
X^{\ha\hb}\cD^{(2)}X_{\ha\hb}
+2(\cD^{(1)\hal}X^{\ha\hb})\cD^{(1)}_\hal X_{\ha\hb}
\Big\}
~,
\label{GR3}
\\
&=&
\frac{15\ri}{2}
\Big\{
X^{\hga\hde}\cD_{\hga}^{(i} \cY_{\hde}{}^{j)} 
+\frac{1}{15}W^{(1)\ha\hb\hga}W^{(1)}_{\ha\hb\hga}
-\frac{2}{5}\cY^{(1)\ha\hal}\cY^{(1)}_{\ha\hal}
+\frac{3}{4}\cY^{(1)\hal} \cY^{(1)}_{\hal}
\Big\}
~.~~~~~~~~~~~~
\eea
\esubeq
One may check that
$\cD_\hal^{(1)} H^{(2)}_{{\rm Riem}}=0$.
The structure of $H^{(2)}_{{\rm Riem}}$
resembles 
the composite $\cO(2)$ multiplet built from a vector multiplet,
eq. \eqref{composite-YM-G}.
It turns out that 
the $\cO(2)$ multiplet \eqref{13.45}
generates the supersymmetric extension of the Riemann squared term, 
$(R_{\ha  \hb \hc\hd})^2$, 
constructed in \cite{BRS}. The construction of \cite{BRS}
was based on a map between the dilaton Weyl multiplet 
and the vector multiplet applied to the non-abelian Chern-Simons action.

${}$From the relations \eqref{13.43} -- \eqref{13.45}
we deduce
\bea
- \cG^{(2)}[\log W] 
&=&
\frac{1}{2}H^{(2)}_{\rm Weyl}
+\frac{1}{4} H^{(2)}_{\rm Riem}
~.
\label{13.46}
\eea
The important point is that the construction of \cite{BRS} and 
related works \cite{OP1,OP2} is defined only for the dilaton Weyl multiplet. 
Our Ricci squared $\cO(2) $ multiplet $ -G^{(2)}[\log W] $,
eq. \eqref{def-G-0tilde}, 
and the corresponding supersymmetric invariant generated by \eqref{13.34b}
makes use of the {\it standard} Weyl multiplet coupled to the {\it off-shell} vector multiplet
compensator.  
Eq. \eqref{13.46}
allows us to define
$H^{(2)}_{\rm Riem}$ for the standard Weyl multiplet
coupled to the off-shell vector multiplet
compensator:
\bea \label{compO2Riem}
H^{(2)}_{\rm Riem} = -4 \Big(G^{(2)} [\log W] + \hf H^{(2)}_{\rm Weyl}\Big).
\eea


\section{Concluding remarks}
\label{conclusion} 

The conformal superspace formalism in five dimensions
presented in this work combines the powerful features of the SU(2) superspace approach \cite{KT-M08} and the superconformal tensor calculus 
\cite{Ohashi3, Ohashi4,Bergshoeff1, Bergshoeff2}. Using this formalism 
we have reproduced practically all off-shell constructions derived so far.
 Most importantly, 
since the superspace setting offers more general off-shell multiplets than those employed in \cite{Ohashi3, Ohashi4,Bergshoeff1, Bergshoeff2}, 
we have developed novel tools to construct composite primary multiplets
and, as a consequence, to generate new higher-order off-shell invariants in supergravity. 
In addition to full superspace integrals,
we have introduced general techniques to build composite $\cO(2)$ and
vector multiplets, which in turn can be used in the universal
BF action. One particular example is the 
Ricci squared $\cO(2)$ multiplet
constructed in section \ref{subsection14.3}.\footnote{The construction 
of the Ricci squared $\cO(2)$ multiplet is analogous to that of
the nonlinear kinetic multiplet presented in \cite{BdeWKL}.} 

Prior to this paper, the superconformal tensor calculus was used to construct 
supersymmetric completions of $R^2$ terms. 
Hanaki, Ohashi and Tachikawa \cite{HOT} constructed the supersymmetric 
Weyl tensor squared term, while Ozkan and Pang \cite{OP2} 
constructed the supersymmetric scalar curvature squared term.
These invariants are generated by 
the Lagrangians \eqref{13.34a} and
\eqref{13.34c} respectively. 
An important feature of these invariants is that they
make use of the standard Weyl multiplet coupled to one or two conformal compensators, 
one of which is always the vector multiplet. Choosing the vector multiplet to be on-shell
leads one to a formulation in the dilaton Weyl multiplet. As concerns a supersymmetric completion of the Riemann squared term, it was constructed by 
Bergshoeff, Rosseel and Sezgin \cite{BRS} only  in the dilaton Weyl multiplet 
realization. However, a description 
of the supersymmetric Riemann squared action in the standard Weyl multiplet was completely unknown. 
Our paper has solved this problem with the use of the $\cO(2)$ multiplet
$G^{(2)} [\log W]$, eq. 
\eqref{def-G-0tilde},
which describes a supersymmetric 
Ricci squared invariant 
using the Lagrangian \eqref{13.34b}.
This invariant completes the description of the supersymmetric $R^2$ invariants
within the standard Weyl multiplet. 
In particular, the analogue 
of the supersymmetric Riemann squared action 
constructed in \cite{BRS} is generated by \eqref{compO2Riem}. 
We hope to 
elaborate further the component structure of the 
action generated by the Lagrangian \eqref{13.34b}
in another publication.

The main virtue of the SU(2) superspace approach \cite{KT-M08} and its 
extension given in our paper is 
that it offers
off-shell descriptions for the most general supergravity-matter systems.
Here we briefly comment on such off-shell descriptions.  
In section \ref{Off-shellSUGRA},
we discussed 
the two-derivative
supergravity action, corresponding to an $\cO(2)$ multiplet and an
abelian vector multiplet compensator. It is easy to generalize
this to include off-shell hypermultiplets.
One takes the same approach as in four dimensions \cite{K-08} and adds 
to the pure supergravity
Lagrangian \eqref{SUGRA8.1} a sigma model 
term\footnote{The normalization of \eqref{15.1}
is chosen so that in the super Weyl gauge $G = 1$, it reproduces a
canonically normalized sigma model.} resulting in
\bea
\cL^{(2)}_{\text{linear}} =  
\frac{1}{4} {V}{H}_{\rm VM}^{(2)} 
+{G}^{(2)}  \ln \frac{{ G}^{(2)}}{{\rm i} \U^{(1)} \breve{\U}{}^{(1)}}      
+\k V G^{(2)} + \frac{1}{2} G^{(2)} K(\U, \breve \U)~,
\label{15.1}
\eea
where $K(\U, \breve \U)$ depends on 
$n$ weight-zero arctic multiplets $\U^\cI$
and their smile-conjugate antarctic multiplets $\breve \U^{\bar \cI}$.
Here $K (\vf^\cI , \bar \vf^{\bar \cI})$ is chosen to be a real analytic function 
of  $n$ ordinary complex variables $\vf^\cI$ and their conjugates.
The action generated by the Lagrangian \eqref{15.1} proves
to be invariant under the K\"ahler transformations
\begin{align}\label{KahlerT}
K \rightarrow K + \L(\U) + \bar \L(\breve \U)
\end{align}
in accordance with eq. \eqref{6.14}.
This permits the identification of $K$ as the K\"ahler potential of 
a $2n$-dimensional K\"ahler manifold $\cM^{2n}$.

The Lagrangian \eqref{15.1} is reminiscent of the general 4D $\cN=1$
new minimal supergravity-matter Lagrangian, which similarly involves a
linear multiplet compensator coupled to a matter sector described by
a K\"ahler potential, see \cite{BK} for a review. 
As in that situation, it is possible here to perform
a duality transformation exchanging $G^{(2)}$ for
a weight-one arctic multiplet $\U^{(1)}$ 
and its smile-conjugate antarctic $\breve \U^{(1)}$. 
The analogous consideration in the case of 4D $\cN=2$ supergravity
was given in \cite{K-08}.
Following \cite{K-08}, the  Lagrangian dual to  \eqref{15.1}  is
\bea
\cL^{(2)}_{\text{hyper}} =  
\frac{1}{4} {V}{H}_{\rm VM}^{(2)} 
-2\ri \breve{\U}^{(1)} \re^{-\k V - \frac{1}{2}K(\U, \breve \U)} \U^{(1)}~.
\label{15.3}
\eea
Here the compensator $\U^{(1)}$ is charged under the U(1) gauge group
and transforms under the K\"ahler transformations 
\eqref{KahlerT} as $\U^{(1)} \rightarrow \re^{\L/2} \U^{(1)}$.

This supergravity-matter system may equivalently be described in terms of 
$(n+1)$ weight-one arctic multiplets $\U^{(1) I}$ and their conjugates
$\breve{\U}^{(1) \bar I}$ defined by
$\U^{(1) I} = \U^{(1)} \times (1, \U^\cI)$ for $I=0, \cdots, n$.
The corresponding Lagrangian is
\bsubeq \label{15.4.5}
\bea
\cL^{(2)}_{\text{hyper}} =  
\frac{1}{4} {V}{H}_{\rm VM}^{(2)} 
- 2\ri  \,\re^{-\k V}  \cK(\U^{(1)}, \breve{\U}^{(1)} )~.
\label{15.4}
\eea
Here $\cK(\U^{(1)}, \breve{\U}^{(1)} )$ obeys the homogeneity conditions
\bea
\U^{(1) I} \frac{\pa}{\pa \U^{(1) I} } \cK = \cK~, \qquad 
\breve{\U}^{(1) \bar I} \frac{\pa}{\pa \breve{\U}^{(1) \bar I}}\cK = \cK~.
\label{15.5}
\eea
\esubeq
In addition, $\cK (\vf^I , \bar \vf^{\bar I})$ is required to be real as a function 
of  $(n+1)$ ordinary complex variables $\vf^I$ and their conjugates.
Moreover, the action generated by the Lagrangian \eqref{15.4} is invariant under 
the gauge transformations 
\bea
\d V  = \l + \breve{\l} ~, \qquad \d \U^{(1) I} = \k \l \U^{(1) I}~,
\eea
with the gauge parameter $\l$ being an arbitrary weight-zero arctic 
multiplet. 

The Lagrangian \eqref{15.1} and each of its dual versions, \eqref{15.3}
and \eqref{15.4.5},
actually describes a large class of $4n$-dimensional
quaternion-K\"ahler sigma models that admit
a maximal $2n$-dimensional K\"ahler submanifold with K\"ahler
potential $K$ \cite{KLvU}. These sigma models also automatically
possess a quaternionic U(1) isometry.
To see this latter feature, one observes that the Lagrangian \eqref{15.1}
describes
a superconformal sigma model coupling the linear multiplet
$G^{(2)}$ to the $n$ weight-zero polar multiplets.
When the three-form in the linear multiplet is dualized,
the resulting scalar manifold is a hyperk\"ahler cone with a triholomorphic
U(1) isometry. When $G^{(2)}$ is gauge-fixed, the $(4n+4)$-dimensional hyperk\"ahler cone
becomes a $4n$-dimensional quaternion-K\"ahler space, and the triholomorphic
isometry descends to a quaternionic one.\footnote{The link between
triholomorphic isometries on the hyperk\"ahler cone (or Swann bundle)
and quaternionic
isometries on the quaternion-K\"ahler space is known from the mathematics
literature \cite{Swann}. It was discussed in a physics context
in \cite{dWRV}.}

The most general $4n$-dimensional
quaternion-K\"ahler sigma model is described by a very similar
supergravity-matter Lagrangian 
(for simplicity we switch off the cosmological constant)
\bea
\cL^{(2)}_{\text{hyper}} =  
\frac{1}{4} {V} {H}_{\rm VM}^{(2)} 
- 2    \cF(\U^{(1)}, \breve{\U}^{(1)} )~,
\label{15.7}
\eea
where $\cF(\U^{(1)}, \breve{\U}^{(1)} )$ obeys the homogeneity condition
\bea
\Big( \U^{(1) I} \frac{\pa}{\pa \U^{(1) I} } 
+ \breve{\U}^{(1) \bar I} \frac{\pa}{\pa \breve{\U}^{(1) \bar I}} \Big)\cF= 2 \cF~.
\label{15.8}
\eea
The dynamical system defined by eqs. \eqref{15.4} and \eqref{15.5} with $\k=0$ is 
a special case of the system under consideration. 
In the flat superspace limit, the Lagrangian 
$\cL^{(2)} =\cF(\U^{(1)}, \breve{\U}^{(1)} )$
describes the most general superconformal sigma model, with its target space being
an arbitrary hyperk\"ahler cone. If the stronger homogeneity conditions \eqref{15.5} hold, 
then the corresponding hyperk\"ahler cone possesses a triholomorphic isometry, 
which is associated with the rigid U(1) symmetry of the superfield Lagrangian
$ \U^{(1) I}  \to \re^{\ri \vf}  \U^{(1) I} $, with $\vf \in \mathbb R$. Similar issues
have been discussed in the case of the (3,0) supersymmetric sigma models in AdS$_3$ 
\cite{BKT-M}.

The Lagrangian \eqref{15.7} can be generalized to include additional abelian vector
multiplets in a straightforward way,
\begin{align}\label{GenVmHm}
\cL^{(2)} &= \frac{1}{4} C_{\rm abc} V^{\rm a} H^{(2)\rm bc} -2 \cF(\U^{(1)}, \breve \U^{(1)})~, \eol
H^{(2)\rm ab } &:=  \ri \,(\nabla^{(1)\hat \a } W^{\rm a} ) \, \nabla^{(1)}_{\hat \a} 
W^{\rm b}
+ \frac{\ri}{2} W^{(\rm a} \nabla^{(1)\hal  } \nabla^{(1)}_\hal W^{\rm b)}~,
\end{align}
for real constants $C_{\rm abc} = C_{\rm (abc)}$, as is well-known from the component literature.
The numerical factors chosen in front of the two terms in \eqref{GenVmHm} ensure
that the Weyl multiplet equation of motion and the canonical Weyl gauge
are respectively given by
\begin{align}
C(W) := C_{\rm abc} W^{\rm a} W^{\rm b} W^{\rm c} = \mathbb K~, \qquad C(W) = 1~,
\end{align}
where $\mathbb K$ is the hyperk\"ahler potential constructed from
$\cF$.\footnote{Our conventions for relating the hyperk\"ahler potential
to the Lagrangian $\cF$ are the same as in \cite{Butter:2014xua}.
There the potential was denoted $K$.}
The component reduction of the vector multiplet Lagrangian in
\eqref{GenVmHm} can be derived from the general result for the
non-abelian vector multiplet action given in section \ref{section9}.
The component reduction of the hypermultiplet sigma model
can be carried out
similarly
to the 4D $\cN=2$ case worked out in \cite{Butter:2014xua}. 

The SU(2) superspace approach to 5D conformal supergravity 
coupled to general matter systems \cite{KT-M08}
has been extended to locally supersymmetric theories in diverse dimensions:
4D $\cN=2$ supergravity \cite{KLRT-M_4D-2}, 
2D $\cN= (4,4)$ supergravity \cite{GTM_2D_SUGRA}, 
3D $\cN=3$ and $\cN=4$ supergravity theories \cite{KLT-M}, 
and 6D $\cN=(1,0)$ supergravity \cite{LT-M-2012}. 
In four dimensions, $\cN=2$ conformal superspace was
formulated in \cite{Butter4DN=2}, see also \cite{Butter:2012xg}.
In three dimensions, $\cN$-extended conformal superspace  
was described in \cite{BKNT-M1}. 
Interesting open problems are to develop conformal superspace 
settings in other cases such as 
the 2D $\cN= (4,4)$ and  6D $\cN=(1,0)$  ones.
\\


\noindent
{\bf Acknowledgements:}\\
The work of DB was supported by
ERC Advanced Grant No. 246974, ``{\it Supersymmetry: a window to
non-perturbative physics}''
and by the European Commission
Marie Curie International Incoming Fellowship grant no.
PIIF-GA-2012-627976.	
The work of SMK Êand JN was supported by the Australian Research Council, project No. DP0103925.
The work of GT-M and JN was supported by the Australian Research Council's Discovery Early Career Award (DECRA), project No. DE120101498.
The work of SMK, JN and GT-M was also supported by the Australian Research Council project DP140103925.


\appendix


\section{Notation and conventions} \label{NC}

Throughout the paper we follow the 5D notation and conventions in \cite{KL}. We summarize them here and include a number of 
useful identities.

The 5D gamma-matrices $\G_{\hat{a}} = (\G_a , \G_5)$, with $a = 0 , 1, 2, 3$, are defined by
\be \{ \G_{\hat{a}} , \G_{\hat{b}} \} = - 2 \eta_{\hat{a} \hat{b}} \mathds1 \ , \quad (\G_{\hat{a}})^\dag = \G_0 \G_{\hat{a}} \G_0 \ ,
\ee
where the Minkowski metric is
\be \eta_{\hat{a} \hat{b}} = {\rm diag}( -1 , 1, 1 , 1 , 1 ) \ .
\ee
We may choose a representation in which the gamma-matrices take the form 
\cite{WB,BK}
\be
(\G_a)_{\hat{\a}}{}^{\hat{\b}} =\left(
\begin{array}{cc}
0  & (\s_a)_{\a \bd} \\
(\tilde{\s}_a)^{\ad \b} &    0
\end{array}
\right) ~, \quad 
(\G_5)_{\hat{\a}}{}^{\hat{\b}} =\left(
\begin{array}{cc}
- \ri \d_\a^\b  & 0 \\
0 &  \ri \d^\ad_\bd
\end{array}
\right) 
\ee
and $\G_0 \G_1 \G_2 \G_3 \G_5 = \mathds1$.
The charge conjugation matrix, $C = (\eps^{\hat{\a} \hat{\b}})$, and its inverse, 
$C^{-1} = C^\dag = (\eps_{\hat{\a} \hat{\b}})$ are defined by
\be C \G_{\hat{a}} C^{-1} = (\G_{\hat{a}})^{\rm T} \ , \quad
\eps^{\hat{\a} \hat{\b}} = \left(
\begin{array}{cc}
\eps^{\a\b}  & 0 \\
0 &   - \eps_{\ad\bd}
\end{array} 
\right)
\ , \quad
\eps_{\hat{\a} \hat{\b}} =\left(
\begin{array}{cc}
\eps_{\a\b}  & 0 \\
0 &   - \eps^{\ad\bd}
\end{array}
\right) \ ,
\ee
where $\eps^{\hat{\a} \hat{\b}}$ and $\eps_{\hat{\a} \hat{\b}}$ are antisymmetric tensors which are used to raise and lower the four-component spinor indices.

A Dirac spinor, $\Psi = (\Psi_{\hat{\a}})$, and its Dirac conjugate, $\bar{\Psi} = ({\bar{\Psi}}^{\hat{\a}}) = \Psi^{\dag} \G_0$, decompose into two-component spinors 
as follows
\be \Psi_{\hat{\a}} = 
\left(
\begin{array}{c}
\psi_\a \\
{\bar{\phi}}^{\ad}
\end{array}
\right) 
\ , \quad
{\bar{\Psi}}_{\hat{\a}} = 
\left(
\begin{array}{cc}
\phi^\a , & {\bar{\psi}}_{\ad}
\end{array}
\right) \ .
\ee
One can combine ${\bar{\Psi}}^{\hat{\a}} = (\phi^\a , {\bar{\psi}}_{\ad})$ and $\Psi^{\hat{\a}} = \eps^{\hat{\a} \hat{\b}} \Psi_{\hat{\b}} = (\psi^\a , - {\bar{\phi}}_{\ad})$ into a 
$\rm SU(2)$ doublet,
\be 
\Psi^{\hat{\a}}_{i} = (\Psi^\a_i , - {\bar{\Psi}}_{\ad i} ) \ , \quad \overline{(\Psi^\a_i)} = {\bar{\Psi}}^{\ad i} \ , \quad i = \underline{1} , \underline{2} \ ,
\ee
with $\Psi^\a_{\underline{1}} = \phi^\a$ and $\Psi^\a_{\underline{2}} = \psi^\a$. It is understood that the $\rm SU(2)$ indices are raised and lowered by 
$\eps^{ij}$ and $\eps_{ij}$, $\eps^{\underline{1} \underline{2}} = \eps_{\2\1} = 1$, in the standard fashion: $\Psi^{\hat{\a} i} = \eps^{ij} \Psi_j^{\hat{\a}}$. The 
Dirac spinor $\Psi^i = (\Psi_{\hat{\a}}^i)$ satisfies the pseudo-Majorana reality condition $\bar{\Psi}_i{}^{\rm T} = C \Psi_i$. This can be concisely written as
\be (\Psi_{\hat{\a}}^i)^* = \Psi^{\hat{\a}}_i \ .
\ee
In defining products of spinors, we occasionally suppress spinor indices.
In such cases, the spinor indices should be understood as contracted from top
left to bottom right; that is, given $\chi_\hal$ and $\Psi_\hal$, we define
\begin{align}
\chi \Psi := \chi^\hal \Psi_\hal~, \quad
\chi \G^\ha \Psi := \chi^\hal (\G^\ha)_\hal{}^\hbe \Psi_\hbe~, \quad
\chi \G^\ha \G^{\hb} \Psi := \chi^\hal (\G^\ha)_\hal{}^\hbe (\G^{\hb})_\hbe{}^\hga \Psi_\hga~,
\end{align}
and so forth.

With the definition $\S_{\hat{a} \hat{b}} = - \S_{\hat{b} \hat{a}} = - \frac{1}{4} [\G_{\hat{a}} , \G_{\hat{b}}]$, the matrices $\{ \mathds1 , \G_{\hat{a}} , \S_{\hat{a} \hat{b}} \}$ form 
a basis in the space of $4 \times 4$ matrices. The matrices $\eps_{\hat{\a} \hat{\b}}$ and 
${(\G_{\hat{a}}})_{\hat{\a} \hat{\b}}$
are antisymmetric (with $\eps^{\hat{\a} \hat{\b}} (\G_{\hat{a}})_{\hat{\a} \hat{\b}} = 0$),
while the matrices 
$(\S_{\hat{a} \hat{b}})_{\hat{\a}\hbe}$
are symmetric.

It is useful to write explicitly the 4D reduction of these matrices
\begin{align}
(\G_a)_{\hat{\a} \hat{\b}} &=\left(
\begin{array}{cc}
0  & - (\s_a)_\a{}^\bd \\
(\s_a)_\b{}^\ad &    0
\end{array}
\right) ~, \quad 
(\G_5)_{\hat{\a} \hat{\b}} =\left(
\begin{array}{cc}
\ri \eps_{\a\b}  & 0 \\
0 &  \ri \eps^{\ad\bd}
\end{array}
\right) ~, \\
(\S_{ab})_{\hat{\a}}{}^{\hat{\b}} &=\left(
\begin{array}{cc}
(\s_{ab})_\a{}^\b  & 0 \\
0 &    (\tilde{\s}_{ab})^\ad{}_\bd
\end{array}
\right) ~, \quad 
(\S_{a5})_{\hat{\a}}{}^{\hat{\b}} =\left(
\begin{array}{cc}
0  & - \frac{\ri}{2} (\s_a)_{\a\bd} \\
\frac{\ri}{2} (\tilde{\s}_a)^{\ad\b} &  0
\end{array}
\right) ~, \\
(\S_{ab})_{\hat{\a} \hat{\b}} &=\left(
\begin{array}{cc}
(\s_{ab})_{\a \b}  & 0 \\
0 &    - (\tilde{\s}_{ab})^{\ad \bd}
\end{array}
\right) ~, \quad 
(\S_{a5})_{\hat{\a} \hat{\b}} =\left(
\begin{array}{cc}
0  & \frac{\ri}{2} (\s_a)_\a{}^\bd \\
\frac{\ri}{2} (\s_a)_\b{}^\ad &  0
\end{array}
\right) ~,
\end{align}
where $(\s_{ab})_\a{}^\b = - \frac{1}{4} (\s_a \tilde{\s}_b - \s_b \tilde{\s}_a)_\a{}^\b$ and $(\tilde{\s}_{ab})^\ad{}_\bd = - \frac{1}{4} (\tilde{\s}_a \s_b - \tilde{\s}_b \s_a)^\ad{}_\bd$~.

A 5-vector $V^{\hat{a}}$ and an antisymmetric tensor $F^{\hat{a}\hat{b}} = - F^{\hat{b} \hat{a}}$ can be equivalently represented as the bi-spinors $V = V^{\hat{a}} \G_{\hat{a}}$ 
and $F = \hf F^{\hat{a} \hat{b}} \S_{\hat{a} \hat{b}}$ respectively with the following symmetry properties
\be V_{\hat{\a} \hat{\b}} = - V_{\hat{\b} \hat{\a}} \ , \quad \eps^{\hat{\a} \hat{\b}} V_{\hat{\a} \hat{\b}} = 0 \ , \quad F_{\hat{\a} \hat{\b}} = F_{\hat{\b} \hat{\a}} \ .
\ee
The equivalent descriptions of $V_{\hat{a}}$ and $F_{\hat{a} \hat{b}}$ by $V_{\hat{\a} \hat{\b}}$ and $F_{\hat{\a} \hat{\b}}$ are explicitly related as follows:
\bsubeq
\begin{align} V_{\hat{\a} \hat{\b}} &= V^{\hat{a}} (\G_{\hat{a}})_{\hat{\a} \hat{\b}} \ , \quad V_{\hat{a}} = - \frac{1}{4} (\G_{\hat{a}})^{\hat{\a} \hat{\b}} V_{\hat{\a} \hat{\b}} \ , \\
F_{\hat{\a}\hat{\b}} &= \hf F^{\hat{a} \hat{b}} (\S_{\hat{a} \hat{b}})_{\hat{\a} \hat{\b}} \ , \quad F_{\hat{a} \hat{b}} = (\S_{\hat{a} \hat{b}})^{\hat{\a} \hat{\b}} F_{\hat{\a} \hat{\b}} \ .
\end{align}
\esubeq
This means that we may decompose an arbitrary tensor with two spinor indices, $T_{\hat{\a} \hat{\b}}$, as follows
\be T_{\hat{\a} \hat{\b}} = \hf (\S^{\hat{a} \hat{b}})_{\hat{\a} \hat{\b}} (\S_{\hat{a} \hat{b}})^{\hat{\g} \hat{\d}} T_{\hat{\g}\hat{\d}}
- \frac{1}{4} \big( (\G^{\hat{a}})_{\hat{\a} \hat{\b}} (\G_{\hat{a}})^{\hat{\g} \hat{\d}} + \eps_{\hat{\a} \hat{\b}} \eps^{\hat{\g} \hat{\d}} \big) T_{\hat{\g} \hat{\d}} \ .
\ee
These results may be checked using the identities
\begin{align}
\eps_{\hat{\a} \hat{\b} \hat{\g} \hat{\d}} &= \eps_{\hat{\a}\hat{\b}} \eps_{\hat{\g} \hat{\d}} + \eps_{\hat{\a} \hat{\g}} \eps_{\hat{\d} \hat{\b}} 
+ \eps_{\hat{\a} \hat{\d}} \eps_{\hat{\b}\hat{\g}} \non\\
&= \hf (\G^{\hat{a}})_{\hat{\a} \hat{\b}} (\G_{\hat{a}})_{\hat{\g} \hat{\d}} + \hf \eps_{\hat{\a} \hat{\b}} \eps_{\hat{\g} \hat{\d}} \ ,
\end{align}
where $\eps_{\hat{\a} \hat{\b} \hat{\g} \hat{\d}}$ is the completely antisymmetric fourth-rank tensor.

The conjugation rules give
\be (\eps_{\hat{\a} \hat{\b}})^{*} = - \eps^{\hat{\a} \hat{\b}} \ , \quad (V_{\hat{\a} \hat{\b}})^{*} = V^{\hat{\a} \hat{\b}} \ , \quad (F_{\hat{\a} \hat{\b}})^* = F^{\hat{\a} \hat{\b}} \ ,
\ee
provided $V^{\hat{a}}$ and $F^{\hat{a} \hat{b}}$ are real.

One can derive a number of identities involving the contraction of vector indices. These are listed below:
\bsubeq
\begin{align}
(\G^{\hat{a}})_{\hat{\a} \hat{\b}} (\G_{\hat{a}})_{\hat{\g} \hat{\d}} &= \eps_{\hat{\a} \hat{\b}} \eps_{\hat{\g} \hat{\d}} - 2 \eps_{\hat{\a} \hat{\g}} \eps_{\hat{\b} \hat{\d}} + 2 \eps_{\hat{\a} \hat{\d}} \eps_{\hat{\b} \hat{\g}} \ , \\
(\S_{\hat{a} \hat{b}})_{\hat{\a} \hat{\b}} (\G^{\hat{b}})_{\hat{\g} \hat{\d}} &= \hf \big( (\G_{\hat{a}})_{\hat{\a} \hat{\d}} \eps_{\hat{\b} \hat{\g}} - (\G_{\hat{a}})_{\hat{\a} \hat{\g}} \eps_{\hat{\b} \hat{\d}} 
+ (\G_{\hat{a}})_{\hat{\b} \hat{\d}} \eps_{\hat{\a} \hat{\g}} - (\G_{\hat{a}})_{\hat{\b} \hat{\g}} \eps_{\hat{\a} \hat{\d}}\big) \ , \\
(\S^{\hat{a} \hat{b}})_{\hat{\a} \hat{\b}} (\S_{\hat{a} \hat{b}})_{\hat{\g} \hat{\d}} &= \eps_{\hat{\a} \hat{\g}} \eps_{\hat{\b} \hat{\d}} + \eps_{\hat{\a} \hat{\d}} \eps_{\hat{\b} \hat{\g}} \ , \\
\eps_{\hat{a} \hat{b} \hat{c} \hat{d} \hat{e}} (\G^{\hat{c}})_{\hat{\a} \hat{\b}} (\S^{\hat{d}\hat{e}})_{\hat{\g} \hat{\d}} &= 2 \eps_{\hat{\a} \hat{\b}} (\S_{\hat{a} \hat{b}})_{\hat{\g} \hat{\d}} 
+ 2 \eps_{\hat{\g} \hat{\a}} (\S_{\hat{a} \hat{b}})_{\hat{\b} \hat{\d}} 
+ 2 \eps_{\hat{\d} \hat{\a}} (\S_{\hat{a} \hat{b}})_{\hat{\b} \hat{\g}} \non\\
&\quad - 2 \eps_{\hat{\g} \hat{\b}} (\S_{\hat{a} \hat{b}})_{\hat{\a} \hat{\d}} 
- 2 \eps_{\hat{\d} \hat{\b}} (\S_{\hat{a} \hat{b}})_{\hat{\a} \hat{\g}}  \ , \\
\eps_{\hat{a} \hat{b} \hat{c} \hat{d} \hat{e}} (\S^{\hat{b} \hat{c}})_{\hat{\a} \hat{\b}} (\S^{\hat{d} \hat{e}})_{\hat{\g} \hat{\d}} &= (\G_{\hat{a}})_{\hat{\a} \hat{\g}} \eps_{\hat{\b} \hat{\d}}
+ (\G_{\hat{a}})_{\hat{\a} \hat{\d}} \eps_{\hat{\b} \hat{\g}}
+ (\G_{\hat{a}})_{\hat{\b} \hat{\g}} \eps_{\hat{\a} \hat{\d}}
+ (\G_{\hat{a}})_{\hat{\b} \hat{\d}} \eps_{\hat{\a} \hat{\g}} \ ,
\end{align}
\esubeq
where the Levi-Civita tensor $\eps_{\hat{a} \hat{b} \hat{c} \hat{d} \hat{e}}$ is defined to be completely antisymmetric with normalization
\be \eps_{01235} = - \eps^{01235} = 1 \ .
\ee
The Levi-Civita tensor also satisfies the useful identity
\be \eps^{\hat{a}_1 \cdots \hat{a}_r \hat{a}_{r+1} \cdots a_5} \eps_{\hat{b}_1 \cdots \hat{b}_{r} \hat{a}_{r+1} \cdots a_5} = - r! (5 - r)! \d^{[\hat{a}_1}_{\hat{b}_1} \cdots \d^{\hat{a}_r]}_{\hat{b}_r} \ .
\ee

Some other useful relations are given by
\bsubeq
\begin{align}
(\G^{[\hat{a}})_{\hat{\a} \hat{\b}} (\G^{\hat{b}]})_{\hat{\g} \hat{\d}} &= \eps_{\hat{\a} \hat{\g}} (\S^{\hat{a} \hat{b}})_{\hat{\b} \hat{\d}} + \eps_{\hat{\b} \hat{\d}} (\S^{\hat{a} \hat{b}})_{\hat{\a} \hat{\g}} 
- \eps_{\hat{\a} \hat{\d}} (\S^{\hat{a} \hat{b}})_{\hat{\b} \hat{\g}} - \eps_{\hat{\b} \hat{\g}} (\S^{\hat{a} \hat{b}})_{\hat{\a} \hat{\d}} \ , \\
(\S^{\hat{c} [\hat{a} })_{\hat{\a} \hat{\b}} (\S_{\hat{c}}{}^{\hat{b}]})_{\hat{\g} \hat{\d}} &= -\frac{1}{4} \big( \eps_{\hat{\a} \hat{\g}} (\S^{\hat{a}\hat{b}})_{\hat{\b} \hat{\d}}
+ \eps_{\hat{\a} \hat{\d}} (\S^{\hat{a}\hat{b}})_{\hat{\b} \hat{\g}} + \eps_{\hat{\b} \hat{\g}} (\S^{\hat{a}\hat{b}})_{\hat{\a} \hat{\d}}
+ \eps_{\hat{\b} \hat{\d}} (\S^{\hat{a}\hat{b}})_{\hat{\a} \hat{\g}} \big)
~,
\end{align}
\esubeq
and
\bsubeq
\bea
(\G^{\hat{a}}\G^{\hat{b}})_{\hat{\a}}{}^{\hat{\b}} 
&=& 
(\G^{\hat{a}})_{\hat{\a}}{}^{\hat{\g}} (\G^{\hat{b}})_{\hat{\g}}{}^{\hat{\b}} 
=
- \eta^{\hat{a} \hat{b}} \d^{\hat{\b}}_{\hat{\a}} - 2 (\S^{\hat{a} \hat{b}})_{\hat{\a}}{}^{\hat{\b}} \ , \\
(\G^{\hat{a}}\G^\hb\G^\hc)_{\hat{\a}}{}^{\hat{\b}} 
&=&
 \big( - \eta^{\hat{a} \hat{b}} \eta^{\hat{c} \hat{d}} 
+ \eta^{\hat{c} \hat{a}} \eta^{\hat{b} \hat{d}} 
- \eta^{\hat{b} \hat{c}} \eta^{\hat{a} \hat{d}}\big) (\G_{\hat{d}})_{\hat{\a}}{}^{\hat{\b}} 
+ \eps^{\hat{a} \hat{b} \hat{c} \hat{d} \hat{e}} (\S_{\hat{d} \hat{e}})_{\hat{\a}}{}^{\hat{\b}} 
\ , ~~~~~~
\label{threeG}
\\
(\G^\ha\G^\hb\G^\hc\G^\hd)_{\hat{\a}}{}^{\hat{\b}} 
&=& (\eta^{\hat{a} \hat{b}} \eta^{\hat{c} \hat{d}} - \eta^{\hat{a} \hat{c}} \eta^{\hat{b} \hat{d}} + \eta^{\hat{a} \hat{d}} \eta^{\hat{b} \hat{c}}) \d_{\hat{\a}}^{\hat{\b}}
- \eps^{\hat{a} \hat{b} \hat{c} \hat{d} \hat{e}} (\G_{\hat{e}})_{\hat{\a}}{}^{\hat{\b}} 
\non\\
&& + 2 \eta^{\hat{a} \hat{b}} (\S^{\hat{c} \hat{d}})_{\hat{\a}}{}^{\hat{\b}} 
-2 \eta^{\hat{a} \hat{c}} (\S^{\hat{b} \hat{d}})_{\hat{\a}}{}^{\hat{\b}} 
+2 \eta^{\hat{b} \hat{c}} (\S^{\hat{a} \hat{d}})_{\hat{\a}}{}^{\hat{\b}} 
\non\\
&&
 +2 \eta^{\hat{d} \hat{c}} (\S^{\hat{a} \hat{b}})_{\hat{\a}}{}^{\hat{\b}} 
 -2 \eta^{\hat{d} \hat{b}} (\S^{\hat{a} \hat{c}})_{\hat{\a}}{}^{\hat{\b}} 
 +2 \eta^{\hat{d} \hat{a}} (\S^{\hat{b} \hat{c}})_{\hat{\a}}{}^{\hat{\b}}
~, 
\label{fourG}\\
(\G^{\ha}\G^\hb\G^\hc\G^\hd\G^\he)_\hal{}^\hbe
&=&
\ve^{\ha\hb\hc\hd\he}\d_\hal^\hbe+
(\G^\ha)_\hal{}^\hbe
(\eta^{\hb\hc}\eta^{\hd\he}-\eta^{\hb\hd}\eta^{\hc\he}+\eta^{\hc\hd}\eta^{\hb\he})
\non\\
&&
+(\G^\hb)_\hal{}^\hbe
(-\eta^{\hc\hd}\eta^{\he\ha}+\eta^{\hc\he}\eta^{\hd\ha}-\eta^{\hd\he}\eta^{\hc\ha})
\non\\
&&
+(\G^\hc)_\hal{}^\hbe
(\eta^{\hd\he}\eta^{\ha\hb}-\eta^{\hd\ha}\eta^{\he\hb}+\eta^{\he\ha}\eta^{\hd\hb})
\non\\
&&
+(\G^\hd)_\hal{}^\hbe
(-\eta^{\he\ha}\eta^{\hb\hc}+\eta^{\he\hb}\eta^{\ha\hc}-\eta^{\ha\hb}\eta^{\he\hc})
\non\\
&&
+(\G^\he)_\hal{}^\hbe
(\eta^{\ha\hb}\eta^{\hc\hd}-\eta^{\hc\ha}\eta^{\hb\hd}+\eta^{\hb\hc}\eta^{\ha\hd})
+2\ve^{\ha\hb\hc\hd\hm}(\S_\hm{}^\he)_\hal{}^\hbe
\non\\
&&
+(\S_{\hm\hn})_{\hal}{}^{\hbe}\Big(-\eta^{\ha\hb}\ve^{\hc\hd\he\hm\hn}
+\eta^{\hc\ha}\ve^{\hb\hd\he\hm\hn}
-\eta^{\hb\hc}\ve^{\ha\hd\he\hm\hn}
\non\\
&&~~~~~~~~~~~~~~~
-\eta^{\hd\ha}\ve^{\hb\hc\he\hm\hn}
+\eta^{\hd\hb}\ve^{\ha\hc\he\hm\hn}
-\eta^{\hd\hc}\ve^{\ha\hb\he\hm\hn}
\Big)
~.
\label{someGamma19}
\eea
\esubeq


\section{The conformal Killing supervector fields of ${\mathbb R}^{5|8}$}
\label{KVF}

The 5D superconformal algebra $\rm F^2 (4)$ \cite{Nahm} 
can be identified with 
the algebra of conformal Killing supervector fields
of 5D $\cN=1$ Minkowski superspace \cite{K06}. 
In this appendix we spell out this construction. 

Simple
Minkowski superspace in five dimensions, $\mathds R^{5|8}$,
 is parametrized by coordinates $z^{\hA} = (x^\ha , \q^\hal_i)$. The flat covariant 
derivatives
$D_{\hA} = (\partial_\ha , D_\hal^i)$
\bea
\pa_\ha:=\frac{\pa}{\pa x^\ha}~,~~~~~~
D_\hal^i
:=
\frac{\pa}{\pa\q^\hal_i}
-\ri(\G^\hb)_{\hal\hbe}\q^{\hbe i}\pa_\hb
~,
\eea
 satisfy the algebra:
\bea
\{ D_\hal^i , D_\hbe^j \} = - 2 \ri (\G^\ha)_{\hal\hbe} \partial_\ha 
~, 
~~~~~~
{[}\partial_\ha , D_{\hbe}^j{]} = 0 
~,~~~~~~
{[}\pa_\ha,\pa_\hb{]}=0
~.
\eea
The spinor covariant 
derivatives satisfy the reality condition $(D_\hal^i F)^*=-(-1)^{\ve(F)}D^\hal_i\overline{F}$
with $F$ an arbitrary superfield of Grassmann parity $\ve(F)$.

According to \cite{K06}, 
the conformal Killing supervector fields
\be \xi = \bar{\xi} = \xi^\ha(z) \partial_\ha + \xi^\hal_i(z) D_\hal^i
\ee
are defined to satisfy
\be [\xi , D_\hal^i ] = - (D_\hal^i \xi^\hbe_j) D_\hbe^j \ , \label{2.22}
\ee
which implies the fundamental equation
\be D_\hal^i \xi_\ha = 2 \ri (\G_\ha)_\hal{}^\hbe \xi_\hbe^i \ . \label{2.23}
\ee
From eq. \eqref{2.23} one finds
\be \eps^{ij} (\G_\ha)_{\hal\hbe} \partial^\ha \xi^\hb = (\G^\hb)_{\hal \hga} D_\hbe^j \xi^{\hga i} + (\G^\hb)_{\hbe\hga} D_\hal^i \xi^{\hga j} \ ,
\ee
which gives us the usual equation for a conformal Killing vector field 
\be \partial_{(\ha} \xi_{\hb)}= \frac{1}{5} \eta_{\ha\hb} \partial^\hc \xi_\hc \ . \label{BosonicKilling}
\ee

The conformal Killing vector acts on the spinor covariant derivatives as
\be [\xi , D_\hal^i] = - \omega_\hal{}^\hbe D_\hbe^i + \L^{ij} D_{\hal j} - \hf \s D_\hal^i  \ ,
\ee
where the parameters $\omega_{\hal\hbe}$, $\s$ and $\L^{ij}$ are given by the following expressions:
\bsubeq
\begin{align}
\omega_{\hal\hbe} &:= \hf D_{(\hal}^k \xi_{\hbe) k} = \hf (\S^{\ha\hb})_{\hal\hbe} \partial_{\ha} \xi_\hb \ ,\\
\s &:= \frac{1}{4} D^{\hal}_k \xi_{\hal}^k = \frac{1}{5} \partial^\ha \xi_\ha \ , \\
\L^{ij} &:= \frac{1}{4} D_\hga^{(i} \xi^{\hga j)} \ .
\end{align}
\esubeq

As a consequence of eq. \eqref{BosonicKilling} we find the parameters satisfy the identities
\bsubeq
\begin{align}
\partial_{\ha} \omega_{\hb \hc} &= - 2 \eta_{\ha [\hb} \partial_{\hc]} \s \ , \\
\partial_{\ha} \partial_{\hb} \xi_{\hc} &= - \eta_{\ha\hb} \partial_\hc \s + 2 \eta_{\hc (\ha} \partial_{\hb )} \s \ .
\end{align}
\esubeq
Furthermore, as a consequence of eq. \eqref{2.23} we also find
\bsubeq
\begin{align}
D_\hga^k \omega_{\hal\hbe} &= -2 \eps_{\hga (\hal} D_{\hbe)}^k \s 
\ , \\
D_\hal^i \L^{jk} &= 3 \eps^{i(j} D_\hal^{k)} \s \ ,
\end{align}
\esubeq
where $\s$ obeys
\be D_\hal^i D_\hbe^j \s = -  \ri \eps^{ij} (\G^\ha)_{\hal\hbe} \partial_\ha \s \ ,
\ee
and
\be \partial_\ha D_\hbe^j \s = 0 \ .
\ee

The above results tell us that we can parametrize superconformal Killing vectors as follows
\be \xi \equiv \xi(\L(P)^\ha , \L(Q)^\hal_i \ , \L(M)_{\ha\hb} , \L(\gD) , \L(K)^\ha , \L(S)^{\hal i}) \ ,
\ee
where we have defined
\bsubeq
\begin{align} \L(P)^\ha &:= - \xi^\ha |_{x = \theta = 0} \ , \quad \L(Q)^\hal_i = - \xi^\hal_i|_{x = \theta = 0} \ , \\
\L(M)_{\ha\hb} &:= \omega_{\ha\hb}|_{x = \theta = 0} \ , \quad \L(\gD) := \s|_{x = \theta = 0} \ , \\
\L(K)_\ha &:= - \hf \partial_\ha \s |_{x = \theta = 0} \ , \quad \L(S)^{\hal i} := - \hf D^{\hal i} \s |_{x = \theta = 0} \ .
\end{align}
\esubeq
The commutator of two superconformal Killing vectors,
\be \xi = \xi(\L(P)^\ha , \L(Q)^\hal_i \ , \L(M)_{\ha\hb} , \L(\gD) , \L(K)^\ha , \L(S)^{\hal i})
\ee
and
\be \tilde{\xi} = \xi(\tilde{\L}(P)^\ha , \tilde{\L}(Q)^\hal_i , \tilde{\L}(M)_{\ha\hb} , \tilde{\L}(\gD) , \tilde{\L}(K)_\ha , \tilde{\L}(S)^{\hal i}) \ ,
\ee
is another superconformal Killing vector given by
\begin{align} 
[\xi , \tilde{\xi}] =&\, (\xi^\ha \partial_\ha \tilde{\xi}^\hb - \tilde{\xi}^\ha \partial_\ha \xi^\hb 
+ \xi^\hal_i D_\hal^i \tilde{\xi}^\hb - \tilde{\xi}^\hal_i D_\hal^i \xi^\hb + 2 \ri \xi^\hal_k \tilde{\xi}^{\hbe k} (\G^\hb)_{\hal \hbe}) \partial_\hb \non\\
& + (\xi^\ha \partial_\ha \tilde{\xi}^\hbe_j - \tilde{\xi}^\ha \partial_\ha \xi^\hbe_j + \xi^\hal_i D_\hal^i \tilde{\xi}^\hbe_j - \tilde{\xi}^\hal_i D_\hal^i \xi^\hbe_j) D_\hbe^j \non\\
\equiv&\, \xi( \hat{\L}^\ha(P), \hat{\L}^\hal_i(Q) , \hat{\L}(M)_{\ha\hb} , \hat{\L}(\gD) , \hat{\L}(K)^\ha , \hat{\L}(S)^{\hal i}) \ ,
\end{align}
where
\bsubeq
\begin{align}
\hat{\L}^{\ha}(P) :=&\,  \L(P)^\hb \tilde{\L}_\hb{}^\ha + \L(P)^\ha \tilde{\L}(\gD) - 2 \ri \L(Q)^{\hal}_k \tilde{\L}(Q)^{\hbe k} (\G^\ha)_{\hal \hbe} \non\\
&- \tilde{\L}(P)^\hb \L_\hb{}^\ha - \tilde{\L}(P)^\ha \L(\gD)  \ , \\
\hat{\L}^\hal_i(Q) :=&\, - \ri (\G_\ha)^{\hal \hbe} \L(P)^\ha \tilde{\L}(S)_{\hbe i} + \L(Q)^\hbe_i \tilde{\L}(M)_\hbe{}^\hal + \hf \L(Q)^\hal_i \tilde{\L}(\gD) + \L(Q)^\hal_j \tilde{\L}(J)^j{}_i \non\\
& + \ri (\G_\ha)^{\hal \hbe} \tilde{\L}(P)^\ha \L(S)_{\hbe i} - \tilde{\L}(Q)^\hbe_i \L(M)_\hbe{}^\hal - \hf \tilde{\L}(Q)^\hal_i \L(\gD) - \tilde{\L}(Q)^\hal_j \L(J)^j{}_i \ , \\
\hat{\L}(M)_{\ha\hb} :=&\, 2 \L(M)^\hc{}_{[\ha} \tilde{\L}(M)_{\hb] \hc} - 4 \L(P)_{[\ha} \tilde{\L}(K)_{\hb]} + 4 \tilde{\L}(P)_{[\ha} \L(K)_{\hb]} \ , \\
\hat{\L}(\gD) :=&\, 2 \L(P)^\ha \tilde{\L}(K)_\ha - 2 \tilde{\L}(P)^\ha \L(K)_\ha + 2 \L(S)^{\hal i} \tilde{\L}(Q)_{\hal i} - 2 \tilde{\L}(S)^{\hal i} \L(Q)_{\hal i} \ , \\
\hat{\L}(K)^\ha :=&\, \L(M)^{\ha \hb} \tilde{\L}(K)_\hb + \L(\gD) \tilde{\L}(K)^\ha - 2 \ri \L(S)^\hal_k \tilde{\L}(S)^{\hbe k} (\G^\ha)_{\hal \hbe} \non\\
&- \tilde{\L}(M)^{\ha \hb} \L(K)_\hb - \tilde{\L}(\gD) \L(K)^\ha  \ , \\
\hat{\L}(S)^{\hal i} :=&\, \ri (\G_\ha)^{\hal \hbe} \L(K)^\ha \tilde{\L}(Q)_{\hbe}^i + \L(S)^{\hbe i} \tilde{\L}(M)_\hbe{}^\hal - \hf \L(S)^{\hal i} \tilde{\L}(\gD) + \L(S)^\hal_j \tilde{\L}(J)^{j i} \non\\
& - \ri (\G_\ha)^{\hal \hbe} \tilde{\L}(K)^\ha \L(Q)_{\hbe}^i - \tilde{\L}(S)^{\hbe i} \L(M)_\hbe{}^\hal + \hf \tilde{\L}(S)^{\hal i} \L(\gD) - \tilde{\L}(S)^\hal_j \L(J)^{j i} \ .
\end{align}
\esubeq

Associating with the superconformal Killing vector $\xi$ the transformation
\be \d_\xi = \L(P)^\ha P_\ha + \L(Q)^\hal_i Q_\hal^i + \hf \L(M)^{\ha\hb} M_{\ha\hb} + \L(\gD) \gD + \L(K)^\ha K_\ha + \L(S)^{\hal i} S_{\hal i}
\ee
and comparing to the above gives us the superconformal algebra \eqref{SCA}.


\section{Modified superspace algebra} \label{app:ModifiedSuperspace}
In section \ref{WeylMultiplet}, we introduced a modified
definition of the composite vector connections. It is actually
possible to introduce this redefinition directly within the context
of superspace. The modified superspace vector derivative is
\begin{align}
\hat \nabla_\ha &= \nabla_\ha
	- \frac{1}{4} \tilde W_{\ha\hb\hc} M^{\hb\hc}
	+ \frac{1}{8} X^{\hbe i} (\Gamma_\ha)_\hbe{}^\hal S_{\hal i}
	+\frac{1}{64}\big(  Y+3 W^{\hb\hc}W_{\hb\hc}\big) K_\ha
	\eol & \quad
	- \frac{1}{4} (\nabla^\hc \tilde W_{\hc\ha}{}^\hb) K_\hb
	- \frac{1}{4} W_{\ha\hd} W^{\hb\hd} K_\hb
	~.
\end{align}
The new vector derivative possesses a deformed $S$-supersymmetry
transformation, but it retains the original $K$-transformation,
\begin{align}
[S_{\hbe i}, \hat\nabla_\ha] &= 
	\ri (\Gamma_\ha)_\hbe{}^\hal \nabla_{\hal i}
	- \frac{1}{2} W_\ha{}^\hb (\Gamma_\hb)_\hbe{}^\hal S_{\hal i}
	+ \frac{\ri}{8} (\G_\ha \G^\hb)_\hbe{}^\hga X_{\hga i} K_\hb
	- \frac{\ri}{4} W_{\ha\hb\hbe i} K^\hb~,
	\\
[K_\hb, \hat\nabla_\ha] &= 2 \eta_{\ha\hb} \mathbb D + 2 M_{\ha\hb}~.
\end{align}
The spinor derivative remains unchanged,
$\hat\nabla_\hal^i = \nabla_\hal^i$.

The new curvature tensors, given in their general form as
\begin{align}
[\hat\nabla_\hA, \hat\nabla_\hB]
	&= - \hat{\scT}_{\hA\hB}{}^\hC \hat \nabla_\hC
	- \frac{1}{2} \hat{\mathscr{R}}(M)_{\hA\hB}{}^{\hc\hd} M_{\hc\hd}
	- \hat{\mathscr{R}}(D)_{\hA\hB} \mathbb D
	\eol & \quad
	- \hat{\mathscr{R}}(J)_{\hA\hB}{}^{ij} J_{ij}
	- \hat{\mathscr{R}}(S)_{\hA\hB}{}^{\hal i} S_{\hal i}
	- \hat{\mathscr{R}}(K)_{\hA\hB}{}^\hc K_\hc~,
\end{align}
can be found by direct computation.
For the algebra of two spinor derivatives, we find 
\begin{subequations}
\begin{align}
\hat \scT_{\hal}^i{}_\hbe^j{}^\hc &= 2 \ri \veps^{ij} (\Gamma^\hc)_{\hal\hbe}~, \\
\hat \scT_{\hal}^i{}_\hbe^j{}^\hga_k &= 0~, \\
\hat {\mathscr{R}}(M)_{\hal}^i{}_\hbe^j{}^{\hc\hd}
	&= 2 \ri \veps^{ij} \veps_{\hal\hbe} W^{\hc\hd}
		+ \ri \veps^{ij} (\Gamma_\hb)_{\hal\hbe} \tilde W^{\hb\hc\hd} ~, \\
\hat {\mathscr{R}}(\mathbb D)_{\hal}^i{}_\hbe^j &= 0~, \\
\hat {\mathscr{R}}(J)_{\hal}^i{}_\hbe^j{}^{kl} &= 0~, \\
\hat {\mathscr{R}}(S)_{\hal}^i{}_\hbe^j{}^{\hga k}
	&= \frac{3\ri}{4} \veps^{ij} \veps_{\hal\hbe} X^{\hga k}
		+ \ri \veps^{ij} \delta^\hga_{[\hal} X_{\hbe]}^k ~, \\
\hat {\mathscr{R}}(K)_{\hal}^i{}_\hbe^j{}^{\hc}
	&= - \frac{\ri}{2} \eps^{ij} \eps_{\hal\hbe} \hat \nabla^\hb W_\hb{}^\hc
	+ \frac{\ri}{2} \eps^{ij} (\Gamma^\ha)_{\hal\hbe} \hat\nabla^\hd \tilde W_{\hd\ha}{}^\hc
	- \frac{\ri}{32}  \eps^{ij} (\Gamma^\hc)_{\hal\hbe} Y
	\eol & \quad
	+ \frac{\ri}{4} \eps^{ij} \eps_{\hal\hbe} \tilde W_\hc{}^{\hd\he} W_{\hd\he}
	+ \frac{\ri}{2} \eps^{ij} (\Gamma^\ha)_{\hal\hbe}
	\Big(W_{\ha\hd} W^{\hc\hd} 
	- \frac{3}{16} W^{\hb\hd}W_{\hb\hd} \delta_\ha{}^\hc	
	\Big)~.
\end{align}
\end{subequations}
The spinor-vector commutators lead to
\begin{subequations}
\begin{align}
\hat \scT_{\hb}{}_{\hal}^i{}^\hc &= 0~, \\
\hat \scT_{\hb}{}_{\hal}^i{}^\hga_k
	&= \frac{1}{4} \delta^i_k \Big(
		3(\Gamma_\hb)_\hal{}^\hbe W_\hbe{}^\hga
		- W_\hal{}^\hbe (\Gamma_\hb)_\hbe{}^\hga
	\Big)~, \\
\hat {\mathscr{R}}(D)_{\hb}{}_{\hal}^i &= -\frac{1}{4} (\Gamma_\hb)_\hal{}^\hga X_\hga^i~, \\
\hat {\mathscr{R}}(J)_{\hb}{}_\hal^i{}^{jk} &= -\frac{3}{4}(\Gamma_\hb)_\hal{}^\hga \eps^{i(j} X_\hga^{k)}~, \\
\hat {\mathscr{R}}(M)_\hb{}_\hal^i{}^{\hc\hd} 
	&= -(\Gamma_\hb)_\hal{}^\hga W^{\hc\hd}{}_\hga^i 
	- \frac{1}{4} \veps_\hb{}^{\hc\hd\he\hat{f}} W_{\he\hat{f}}{}_\hal^i 
	+ \hf \delta_\hb^{[\hc} (\Gamma^{\hd]})_\hal{}^\hga X_\hga^i ~,\\
\hat {\mathscr{R}}(S)_{\hb}{}_{ \hal}^i{}^{\hga j}
	&=
	\frac{1}{16} X_{\hc\hd}{}^{ij} (\S^{\hc\hd} \G_\hb - 2 \G_\hb \S^{\hc\hd})_\hal{}^\hga
	\eol & \quad
	- \frac{3\ri}{8} \veps^{ij}\, \hat\nabla_{[\hb} W_{\hc\hd]} (\S^{\hc\hd})_\hal{}^\hga 
	- \frac{\ri}{8} \veps^{ij}\,\hat \nabla_\hd W^{\hd\hc} (\S_{\hc\hb})_\hal{}^\hga 
	\eol & \quad
	+ \frac{3\ri}{16} \veps^{ij} \,\hat\nabla^\hd W_{\hd\hb} \,\delta_\hal^\hga
	- \frac{\ri}{8} \veps^{ij}\,\hat\nabla^\hc \tilde W_{\hc\hb}{}^\hd (\G_\hd)_\hal{}^\hga
	\eol & \quad
	+ \frac{\ri}{16} \veps^{ij} \tilde W^{\hc\hd\he} W_{\hd\he} (\S_{\hc\hb})_\hal{}^\hga
	- \frac{3\ri}{32} \veps^{ij} \tilde W_{\hb\hd\he} W^{\hd\he} \delta_\hal{}^\hga
	\eol & \quad
	+ \frac{\ri}{4} \veps^{ij}\, W_{\hb\hd} W^{\hc\hd} (\G_\hc)_\hal{}^\hga 
	- \frac{3\ri}{64} \veps^{ij}\,W^{\hc\hd}W_{\hc\hd} (\G_\hb)_\hal{}^\hga ~, \\
\hat {\mathscr{R}}(K)_{\hb}{}_{ \hal}^i{}^\hc &=
	\frac{1}{6} (\G^\hc)_\hal{}^\hbe \hat\nabla^\hd W_{\hd\hb}{}_\hbe^i
	+ \frac{1}{12} (\G_\hb)_\hal{}^\hbe \hat\nabla^\hd W{}_\hd{}^{\hc}{}_\hbe^i
	+ \frac{1}{6} \hat{\nabla}_\hal{}^\hbe W_\hb{}^\hc{}_\hbe^i
	- \frac{1}{24} \veps_\hb{}^{\hc\hd\he\hat{f}} \hat\nabla_\hd W_{\he\hat{f}}{}_\hal^i
	\eol & \quad
	+ \frac{1}{8} (\G^\hc)_\hal{}^\hbe \hat\nabla_\hb X^i_\hbe
	+ \frac{1}{64} W^{\hd\he} (3 \G_\hb \S_{\hd\he} \G^\hc - \S_{\hd\he} \G_\hb \G^\hc)_\hal{}^\hbe X_\hbe^i
	\eol & \quad
	- \frac{1}{48} \tilde W_{\hb\hd\he} (\G^\hc)_\hal{}^\hbe W^{\hd\he}{}_\hbe^i
	+ \frac{1}{8} \delta_\hb{}^\hc W^{\hd\he} W_{\hd\he}{}_\hal^i
	\eol & \quad
	+ \frac{1}{12} (\S_\hb{}^\hc)_\hal{}^\hbe W_{\hd\he}{}_\hbe^i W^{\hd\he}
	- \frac{1}{12} W^{\hd\he} (\S_{\hd\he})_\hal{}^\hbe W_{\hb}{}^{\hc}{}_\hbe{}^i
	\eol & \quad
	+ \frac{13}{48} W_{\hb\hd} W^{\hd\hc}{}_\hal^i
	+ \frac{11}{48} W_{\hb\hd}{}_\hal^i W^{\hd\hc}
	- \frac{13}{96} (\G_\hb)_\hal{}^\hbe W_{\hd\he}{}_\hbe^i \, \tilde W^{\hd\he\hc}
	~.
\end{align}
\end{subequations}
The vector-vector commutator is given by
\begin{subequations}
\begin{align}
\hat \scT_{\ha\hb}{}^\hc &= 0~, \\
\hat \scT_{\ha\hb}{}^\hal_i &= -\frac{\ri}{2} W_{\ha\hb}{}^\hal_i~, \\
\hat {\mathscr{R}}(D)_{\ha\hb} &= 0~, \\
\hat {\mathscr{R}}(J)_{\ha\hb}{}^{ij} &= -\frac{3\ri}{4} X_{\ha\hb}{}^{ij}~, \\
\hat {\mathscr{R}}(M)_{\ha\hb}{}^{\hc\hd} &=
	-\frac{1}{4} (\Sigma_{\ha\hb})^{\hal\hbe}
	(\Sigma^{\hc\hd})^{\hga \hde}
	\Big( \ri W_{\hal\hbe\hga\hde}
		+ 3 W_{(\hal \hbe} W_{\hga \hde)}
	\Big)~, \label{eq:newRMab} \\
\hat {\mathscr{R}}(S)_{\ha\hb}{}_\hal^i
	&=
	- \frac{1}{2} \hat{\nabla}_\hal{}^\hbe W_{\ha\hb\hbe}{}^i
	- \frac{1}{2} (\G_{[\ha })_\hal{}^\hbe \hat\nabla^\hc W_{\hb]\hc \hbe}{}^i
	\eol & \quad
	- \frac{1}{8} W_\hal{}^\hbe W_{\ha\hb\hbe}{}^i
	+ \frac{1}{16} (\S_{\ha\hb})_\hal{}^\hbe W^{\hc\hd} W_{\hc\hd\hbe}{}^i
	+ \frac{3}{8} W^\hc{}_{[\ha } W_{\hb]\hc\hal}{}^i~, \\
\hat {\mathscr{R}}(K)_{\ha\hb}{}^\hc &=
	\frac{1}{4} \hat \nabla_d \hat {\mathscr{R}}(M)_{\ha\hb}{}^{\hc\hd}
	- \frac{\ri}{16} W_{\ha b}{}^\hal_j (\G^\hc)_\hal{}^\hbe X_\hbe^j
	- \frac{\ri}{8} W_{\hd [\ha}{}^\hal_j (\G_{\hb]})_\hal{}^\hbe W^{\hc\hd}_\hbe{}^j
	\eol & \quad
	+ \frac{\ri}{8} W_{\ha \hd }{}^\hal_i (\G^\hc)_\hal{}^\hbe
		W_\hb{}^{\hd}{}_\hbe{}^i~.
\end{align}
\end{subequations}


\section{Conventions for 5D conformal supergravity} \label{app:Conventions}

For the convenience of the reader, we provide in Table \ref{tab:ConvComp}
a brief translation scheme between our conventions and the other groups'.
A similar table may be found in \cite{OP2}.

\begin{table*}[!htb]
\centering
\renewcommand{\arraystretch}{1.4}
\resizebox{16cm}{!}{
\begin{tabular}{@{}cccc@{}} \toprule
Our conventions & de Wit and Katmadas & Bergshoeff et al. & Fujita et al. \\ \midrule
$\eta^{\ha\hb}$ & $\eta^{ab}$ & $\eta^{ab}$ & $-\eta^{ab}$ \\
$\Gamma^\ha$ & $-\ri \gamma^a$ & $\ri \gamma^a$ & $\gamma^a$ \\
$\S^{\ha\hb}$ & $\frac{1}{2} \gamma^{ab}$ & $\frac{1}{2} \gamma^{ab}$ & $-\frac{1}{2} \gamma^{ab}$ \\
$\veps^{\ha\hb\hc\hd\he}$ & $-\ri\veps^{abcde}$ & $-\veps^{abcde}$ & $\veps^{abcde}$ \\
\midrule
$\psi_\hm{}^i$ & $\psi_\mu{}^i$ & $\psi_\mu{}^i$ & $2 \,\psi_\mu{}^i$ \\
$\cV_\hm{}^i{}_j$ & $-\frac{1}{2} V_\mu{}_j{}^i$ & $-V_\mu{}^i{}_j$ & $V_\mu{}^i{}_j$ \\
$\hat\omega_\hm{}^{\ha\hb}$ & $\omega_\mu{}^{ab}$ & $-\omega_\mu{}^{ab}$ & $-\omega_\mu{}^{ab}$ \\
$\ri \,\hat\phi_\hm{}^i$ & $\phi_\mu{}^i$ &
	$-\phi_\mu{}^i + \frac{1}{3} T_{ab} \gamma^{ab} \psi_\mu{}^i$ &
	$2 \,\phi_\mu{}^i - \frac{2}{3} v_{ab} \gamma^{ab} \psi_\mu{}^i$ \\
$\hat f_\hm{}^\ha$ & $-f_\mu{}^a + \frac{1}{3} \psi_{\mu i} \gamma^a \chi^i$ &
	$-f_\mu{}^a + \frac{1}{3} \psi_{\mu}{}^i \gamma^a \chi_i$ &
	$-f_\mu{}^a + \frac{\ri}{24} \psi_\mu{}^i \gamma^a \chi_i$ \\ 
\midrule
$w^{\ha\hb}$ & $-4 T^{ab}$ & $\frac{16}{3} T^{ab}$ & $\frac{4}{3} v^{ab}$ \\
$\chi^i$ & $\chi^i$ & $\chi^i$ & $\frac{1}{32} \chi^i$ \\
$D$ & $D$ & $D$ & $\frac{1}{16} (D - \frac{8}{3} v^{ab} v_{ab})$ \\
\midrule
$\hat R(Q)_{\ha\hb}{}^i$ & $\frac{1}{2} R(Q)_{ab}{}^i$ & 
	$\frac{1}{2} \hat R(Q)_{ab}{}^i$ &
	$\hat R(Q)_{ab}{}^i$ \\
$\hat R(M)_{\ha\hb}{}^{\hc\hd}$ &
	$R(M)_{ab}{}^{cd}$ &
	$-\hat R(M)_{ab}{}^{cd}$ &
	$-\hat R(M)_{ab}{}^{cd}$ \\
$\hat R(J)_{\ha\hb}{}^i{}_j$ & 
	$-\frac{1}{2} R(\cV)_{ab}{}_j{}^i$ & 
	$-\hat R(V)_{ab}{}^i{}_j$ & 
	$\hat R(U)_{ab}{}^i{}_j$ \\
$\ri\, \hat R(S)_{\ha\hb}{}^i$ & $\frac{1}{2} R(S)_{ab}{}^i$ &
	$-\frac{1}{2} \hat R(S)_{ab}{}^i + \frac{1}{6} T_{cd} \gamma^{cd} \hat R(Q)_{ab}{}^i$ &
	$\hat R(S)_{ab}{}^i - \frac{1}{3} v_{cd} \gamma^{cd} \hat R(Q)_{ab}{}^i$\\
$\hat R(K)_{\ha\hb}{}^\hc$ & $-R(K)_{ab}{}^c + \frac{1}{3} R(Q)_{ab i} \gamma^c \chi^i$
	& $-\hat R(K)_{ab}{}^c + \frac{1}{3} \hat R(Q)_{ab}{}^i \gamma^c \chi_i$
	& $-\hat R(K)_{ab}{}^c + \frac{\ri}{24} \hat R(Q)_{ab}{}^i \gamma^c \chi_i$ \\
\bottomrule
\end{tabular}}
\caption{Conventions for Weyl multiplet}\label{tab:ConvComp}
\end{table*}

We must be careful to note that the definitions of supersymmetry are
different between the various groups, with the differences amounting
not only to normalizations but also to additional field-dependent $S$
and $K$ transformations in the definition of $\delta_Q$. In other words,
given a transformation $\delta_Q + \delta_S + \delta_K$ in our conventions
with respective parameters $\xi_\hal^i$, $\eta_\hal^i$ and $\L_K^\ha$,
we will find a transformation $\delta'_Q + \delta'_S + \delta'_K$ with
new parameters $\eps^i$, $\eta'^i$ and $\L_K'^a$ given in 
Table \ref{tab:N2SusyConventions}.

\begin{table*}[!hbt]
\centering
\renewcommand{\arraystretch}{1.4}
\begin{tabular}{@{}cccc@{}} \toprule
de Wit and Katmadas & Bergshoeff et al. & Fujita et al. \\ \midrule
$\eps^i = 2\,\xi^i$ & $\eps^i = 2\,\xi^i$ & $\eps^i =\xi^i$ \\
$\eta'^i = 2\ri \eta^i$
	& $\eta'^i = -2\ri \eta^i + \frac{2}{3} T_{ab} \gamma^{ab} \xi^i$ 
	& $\eta'^i = \ri\eta^i + \frac{1}{3} v_{ab} \gamma^{ab} \xi^i$ \\
$\L_K'^a = -\L_K^a + \frac{2}{3} \xi_i \gamma^a \chi^i$
	& $\L_K'^a = -\L_K^a + \frac{2}{3} \xi^i \gamma^a \chi_i$
	& $\L_K'^a = -\L_K^a + \frac{\ri}{24} \xi^i \gamma^a \chi_i$ \\
\bottomrule
\end{tabular}
\caption{Conventions for $\delta_Q+\delta_S+\delta_K$}\label{tab:N2SusyConventions}
\end{table*}

It should be emphasized that each group uses the same
vector derivative $D_a$, corresponding to our $\hat\nabla_\ha$, modulo
differing overall normalizations of the superconformal generators.
The additional gravitino-dependent terms in the $S$-supersymmetry
and special conformal connections in Table \ref{tab:ConvComp}
cancel against additional terms found within $\delta_Q$, so that the
vector derivative is unchanged.

For completeness, we also give in Table \ref{tab:ConvVect}
the relation between our conventions for the vector multiplet and the other
groups.
\begin{table*}[!hbt]
\centering
\renewcommand{\arraystretch}{1.4}
\begin{tabular}{@{}cccc@{}} \toprule
Our conventions & de Wit and Katmadas & Bergshoeff et al. & Fujita et al. \\ \midrule
$W$ & $\sigma$ & $-\sigma$ & $M$ \\
$\l^i$ & $\Omega^i$ & $\psi^i$ & $-2\Omega^i$ \\
$X^{ij}$ & $2 Y^{ij}$ & $-2 Y^{ij}$ & $2Y^{ij}$ \\
\bottomrule
\end{tabular}
\caption{Conventions for vector multiplet}\label{tab:ConvVect}
\end{table*}



\section{The $\cO(2)$ multiplet prepotential from harmonic superspace}
\label{HarmonicG}

In this appendix we use the harmonic superspace techniques
\cite{GIOS} extended to the 5D $\cN=1$ super-Poincar\'e case
(see \cite{KL,Zupnik99} for the technical details regarding the $\cN=1$ harmonic 
superspace in five dimensions) to derive a prepotential formulation 
for the $\cO(2)$ multiplet. In Appendix  \ref{AppendixH},
the same techniques will be used to derive 
unconstrained prepotentials for the $\cO(4+n)$ multiplets, $n=0,1,\dots$, 
in 5D  $\cN=1$ Minkowski superspace.\footnote{The 
harmonic and projective superspace descriptions of the $\cO(n)$ multiplets 
are completely equivalent \cite{K98}.}
 
We consider an $\cO(2)$ multiplet $G^{ij}(z) $ in 5D  $\cN=1$ Minkowski superspace
and associate with it the
analytic superfield $G^{++} (z, u^+)= G^{ij}(z) u^+_i u^+_j$. The latter is constrained
by 
\bea
D^+_\hal G^{++} = 0 ~, \qquad D^{++} G^{++} =0
~,
\eea
where $D^{++}:=u^{+i}{\pa}/{\pa u^{-i}}$.
As in the 4D $\cN=2$ super-Poincar\'e case \cite{GIOS85},
the analytic projector on the space of $\cO(2) $ multiplets\footnote{This projector plays an important role 
in computing the one-loop effective action for $\cN=4$ SYM in four dimensions
\cite{KuzMcA-1}.}
 is 
\bea
\P^{(2,2)}_{\rm L}(\z_1 ,\z_2) = - (\hat{D}^+_1)^4 (\hat{D}^+_2)^4 \,\frac{1}{\Box}
\frac{\d^{5|8} (z_1 -z_2)  }{(u^+_1 u^+_2)^2} ~,
\label{1.2}
\eea
where 
\bea
(\hat{D}^+)^4 &=& -\frac{1}{ 32}  (\hat{D}^+)^2 
\,  (\hat{D}^+)^2~, 
\qquad (\hat{D}^+)^2 = D^{+ \hat \a} D^+_{\hat \a}~,
\eea
and $\z$ denotes the coordinates of the analytic subspace. 
The  properties of
$\P^{(2,2)}_{\rm T}(\z_1 , \z_2)$ are:
\begin{subequations}
\bea
 D^{+ \hat{\a}}_1 \P^{(2,2)}_{\rm T}(\z_1 , \z_2)
&=&  D^{+ \hat{\a}}_2 \P^{(2,2)}_{\rm T}(\z_1 , \z_2) =0~, \label{prop1} \\
  D^{++}_1  \P^{(2,2)}_{\rm T}(\z_1 , \z_2)
&=& D^{++}_2  \P^{(2,2)}_{\rm T}(\z_1 , \z_2)=0~, \label{prop2}\\
 \int {\rm d} \z^{(-4)}_3\,
\P^{(2,2)}_{\rm T}(\z_1 , \z_3) \, \P^{(2,2)}_{\rm T}(\z_3 , \z_2)
&=& \P^{(2,2)}_{\rm T}(\z_1 , \z_2)~, \label{prop3}\\
\Big(\P^{(2,2)}_{\rm T}(\z_1 , \z_2)\Big)^{\rm T} &=&
\P^{(2,2)}_{\rm T}(\z_2 , \z_1)~.\label{prop4}
\eea
\end{subequations}
For any $\cO (2)$ multiplet $G^{++}$ we have
\bea
G^{++} (z_1, u_1^+) \equiv
G^{++} (\z_1) = \int  {\rm d} \zeta_2^{(-4)}\,\P^{(2,2)}_{\rm L}(\z_1 ,\z_2) G^{++}(\z_2)~.
\label{1.4_0}
\eea
Introduce a superfield $\X^{--}(z,u)$ such that $(\hat{D}^+)^4 \X^{--}=G^{++}$. 
Then we can rewrite \eqref{1.4_0} as follows
\bea
G^{++} (\z_1) = \int  \rd^{5|8}z_2 \,\rd u_2 \,\P^{(2,2)}_{\rm L}(\z_1 ,\z_2) 
\X^{--}(z_2, u_2)~.
\label{1.4}
\eea

In the expression \eqref{1.2} we represent 
\bea
(\hat{D}^+_2)^4  \d^{5|8} (z_1 -z_2)  
&=& -\frac{1}{ 32}  (\hat{D}_2^+)^2 
  (\hat{D}_2^+)^2 \d^{5|8}(z_1 - z_2) \non \\
&=&  -\frac{1}{ 32}  (\hat{D}_2^+)^2 
 u_2^{+i} u_2^{+j} \hat{D}_{2\,ij} \d^{5|8}(z_1 - z_2) \non \\
&=&  -\frac{1}{ 32}  (\hat{D}_2^+)^2 
 u_2^{+i} u_2^{+j} \hat{D}_{1\,ij} \d^{5|8}(z_1 - z_2) \non \\
 &=&  -\frac{1}{ 32} u_2^{+i} u_2^{+j} \hat{D}_{1\,ij} (\hat{D}_2^+)^2 
  \d^{5|8}(z_1 - z_2)~.
\eea
We plug this expression in \eqref{1.2} and make use of the identity
\bea
\J^+_2 = (u^+_1 u^+_2) \, \J^-_1   - (u^-_1 u^+_2) \, \J^+_1~, \qquad \quad
\J^\pm = \J^i \,u^\pm_i
\eea
in conjunction with $D_1^{\hal+}(\hat{D}_1^+)^4=(\hat{D}_1^+)^4D_1^{\hal+}=0$. This gives 
\bea
\P^{(2,2)}_{\rm L}(\z_1 ,\z_2) = \frac{1}{32} (\hat{D}^+_1)^4 (\hat{D}^-_1)^2
  (\hat{D}_2^+)^2 \frac{1}{\Box}
\d^{5|8} (z_1 -z_2)  ~.
\eea
As a result, relation \eqref{1.4} becomes equivalent to
\bea
G^{++} (z, u^+) = (\hat{D}^+)^4 (\hat{D}^-)^2 \O(z)~.
\eea


\section{Gauge freedom for the  $\cO(2)$ multiplet}
\label{gauge-invariance-G}

Let us show that the gauge transformation of the $\cO(2)$ multiplet prepotential $\O$,
eq.~\eqref{gauge-O2-0}, leaves invariant the superfield $G^{(2)}$ defined by 
\eqref{def-G-0}.
We need to prove that the  superfield
\bea
\O(B)
=
-\frac{\ri}{2}\de_\hal^k \de_\hbe^lB^{\hal\hbe}{}_{kl}
~,~~~~~~
B_{\hal\hbe}{}^{ij}=(\G^\ha)_{\hal\hbe}B_\ha{}^{ij}
~,
\label{E.1}
\eea
is annihilated by the operator $\ri\D^{ijkl}\de_{kl}$.
It is useful to employ the equivalent expression for $\O(B)$ given by
\bsubeq
\bea
\O(B)
&=&
-\frac{\ri}{2}\de_\hal^\pu\Big( \de_\hbe^\pu B^{\hal\hbe}{}^{(-2)}
-2\de_\hbe^\pmu B^{\hal\hbe}{}^{(0)}\Big)
\non\\
&&
-\frac{\ri}{2}\de_\hal^\pmu \de_\hbe^\pmu B^{\hal\hbe}{}^{(2)}
+\de_{\hal\hbe} B^{\hal\hbe}{}^{(0)}
~,
\\
B_\ha^{(2)}&:=&v_iv_jB_\ha{}^{ij}
~,~~~
B_\ha^{(0)}:=\frac{v_iu_j}{(v,u)}B_\ha{}^{ij}
~,~~~
B_\ha^{(-2)}:=\frac{u_iu_j}{(v,u)^2}B_\ha{}^{ij}
~.
\eea
\esubeq
By using 
\bea
\de^{(-2)}
\de_\hal^\pmu \de_\hbe^\pmu
=-\frac{1}{4}\ve_{\hal\hbe}\de^{(-2)}\de^{(-2)}
~,
\eea
and $\D^{(4)}\de_\hal^\pu=0$, we obtain
\bea
\ri\D^{(4)}\de^{(-2)}\Omega(B)
&=&
-\hf\D^{(4)}{[}\de_\hal^\pu,\de^{(-2)}{]}\big( \de_\hbe^\pu B^{\hal\hbe}{}^{(-2)}
-2 \de_\hbe^\pmu B^{\hal\hbe}{}^{(0)}\big)
\non\\
&&
+\ri\D^{(4)}\de^{(-2)}\de_{\hal\hbe} B^{\hal\hbe}{}^{(0)}
~.
\label{various-eqs-1}
\eea
By making use of
\bea
{[}\nabla_{\hal\hbe} , \nabla_\hga^j {]} = (\G^\ha)_{\hal\hbe}(\G_\ha)_{\hga\hde} {[}\mathscr{W},\de^{\hde j} {] }
~,~~~~~~
{[}\nabla_{\hal\hbe} , \nabla^{\hbe j} {] }
=
-5 {[}\mathscr{W},\nabla_\hal^j {]}
~,
\eea
it can be seen that
\bea
{[}\de_\hal^\pu,\de^{(-2)}{]}
&=&
-4 \ri \Big(
 \de^{\hga\pmu} \nabla_{\hal\hga} 
+  \de_\hal^\pmu  \mathscr{W}
-2   {[}\mathscr{W},\de_\hal^\pmu{]}
\Big)
~.
\label{various-eqs-2}
\eea
Note that in performing this calculation we will keep implicit as long as possible the expression \eqref{def-W}
for the operator $\mathscr{W}$ in the covariant derivative algebra
\eqref{algebra-W}.
Plugging eq. \eqref{various-eqs-2} into \eqref{various-eqs-1}, after some algebra one can obtain
\bea
&&\ri\D^{(4)}\de^{(-2)}\Omega(B)
=
\ri\D^{(4)}\Big\{
\tr{[}\G^\ha\G^\hb\G^\hc{]}\Big(4\ri\de_\ha \nabla_\hb B_\hc^{(-2)}
-(\G_\hb)^{\hga'\hbe'} \de_{\hga'}^\pmu \de_{\hbe'}^\pmu \nabla_\ha B_\hc^{(0)}\Big)
\non\\
&&~~~~~~~~~~~~
-\hf\tr{[}\G^\ha\G^\hb\G_\ha{]}\de^{\hrh\pmu}\Big(
{[}\mathscr{W}, \de_\hrh^\pu{]}B_\hb^{(-2)}
-2 {[}\mathscr{W}, \de_\hrh^\pmu{]}B_\hb^{(0)}\Big)
\non\\
&&~~~~~~~~~~~~
-\hf\tr{[}\G^\ha\G^\hb\G_\ha\G^\hc{]}(\G_\hb)_{\hrh\hta}\de^{\hrh\pmu}\Big(
{[}\mathscr{W}, \de^{\hta \pu}{]}B_\hc^{(-2)}
-2 {[}\mathscr{W}, \de^{\hta \pmu}{]}B_\hc^{(0)}\Big)
\non\\
&&~~~~~~~~~~~~
-4\ri{[}\mathscr{W} ,\nabla^{\hal\hbe}{]}B_{\hal\hbe}^{(-2)}
-8\ri\nabla^{\hal\hbe}\mathscr{W} B_{\hal\hbe}^{(-2)}
+2\de_\hal^\pmu {[} \mathscr{W},\de_\hbe^\pu{]}B^{\hal\hbe}{}^{(-2)}
\non\\
&&~~~~~~~~~~~~
-  4 \de_\hal^\pmu \de_\hbe^\pmu \mathscr{W} B^{\hal\hbe}{}^{(0)}
-4\{\de_\hal^\pu,   {[}\de_\hbe^\pmu,\mathscr{W}{]}\}B^{\hal\hbe}{}^{(-2)}
\non\\
&&~~~~~~~~~~~~
+4 \de_\hal^\pmu {[} \mathscr{W},\de_\hbe^\pmu{]}B^{\hal\hbe}{}^{(0)}
-8\{  {[}\mathscr{W},\de_\hal^\pmu{]},\de_\hbe^\pmu\}B^{\hal\hbe}{}^{(0)}
\Big\}
~.~~~~~~~
\label{various-eqs-3}
\eea
Some terms in the previous expression are identically zero.
First of all note that due to \eqref{threeG} we have
\bea
\tr{[\G_\ha\G_\hb\G_\hc]}=0
~.
\label{various-eqs-4}
\eea
Then the first two lines in \eqref{various-eqs-3} are zero.
Moreover, the Bianchi identity \eqref{WalbeBI} implies
\bea
(\G_\ha)^{\hal\hbe}\{  {[}\mathscr{W},\de_\hal^\pmu{]},\de_\hbe^\pmu\}
=0
~,
\eea
which removes the last term in 
\eqref{various-eqs-3}.
Once we use
\bea
\tr{[}\G^{\hat{a}}\G^{\hat{b}} \G^{\hat{c}} \G^{\hat{d}} {]}
= 
4(\eta^{\hat{a} \hat{b}} \eta^{\hat{c} \hat{d}} - \eta^{\hat{a} \hat{c}} \eta^{\hat{b} \hat{d}} 
+ \eta^{\hat{a} \hat{d}} \eta^{\hat{b} \hat{c}}) 
~,~~~~~~
\tr{[}\G^{\hat{a}}\G^{\hat{b}} \G_{\hat{a}} \G^{\hat{c}} {]}
= 
-12 \eta^{\hb\hc} 
~,~~~~~~
\eea
which follow from \eqref{fourG},
\eqref{various-eqs-3} can be brought to the following form:
\bea
\ri\D^{(4)}\de^{(-2)}\Omega(B)
&=&
\ri\D^{(4)}\Big\{
-4\ri{[}\mathscr{W} ,\nabla^{\hal\hbe}{]}B_{\hal\hbe}^{(-2)}
-8\ri\de^{\hal\hbe}  \mathscr{W} B_{\hal\hbe}^{(-2)}
\non\\
&&~~~
+8\de_\hal^\pmu {[} \mathscr{W},\de_\hbe^\pu{]}B^{\hal\hbe}{}^{(-2)}
-4\{\de_\hal^\pu,   {[}\de_\hbe^\pmu,\mathscr{W}{]}\}B^{\hal\hbe}{}^{(-2)}
\non\\
&&~~~
-8 \de_\hal^\pmu {[}\mathscr{W},\de_\hbe^\pmu{]}B^{\hal\hbe}{}^{(0)}
-  4 \de_\hal^\pmu \de_\hbe^\pmu \mathscr{W} B^{\hal\hbe}{}^{(0)}
\Big\}
~.~~~~~~
\label{various-eqs-5}
\eea
As a next step, we can simplify the second term in the second line.
In fact, the Bianchi identity \eqref{WalbeBI} implies
\bea
(\G_\ha)^{\hal\hbe}\{\de_\hal^\pu,   {[}\de_\hbe^\pmu,\mathscr{W}{]}\}
=
-(\G_\ha)^{\hal\hbe}\{\de_\hal^\pmu,   {[}\de_\hbe^\pu,\mathscr{W}{]}\}
~,
\eea
which together with the super-Jacobi identity, can be used to derive the following result
\bea
(\G_\ha)^{\hal\hbe}\{\de_\hal^\pu,   {[}\de_\hbe^\pmu,\mathscr{W}{]}\}
=
-\hf(\G_\ha)^{\hal\hbe}{[}\mathscr{W},\{\de_\hal^\pu,\de_\hbe^\pmu\}{]}
=
-\ri(\G_\ha)^{\hal\hbe}{[}\mathscr{W},\de_{\hal\hbe}{]}
~.~~~~~~
\eea
If we use this expression in \eqref{various-eqs-5}, we arrive at the simple result
\bea
\ri\D^{(4)}\de^{(-2)}\Omega(B)
&=&
\ri\D^{(4)}\Big\{
8\de_\hal^\pmu {[} \mathscr{W},\de_\hbe^\pu{]}B^{\hal\hbe}{}^{\pmd}
-8 \de_\hal^\pmu {[}\mathscr{W},\de_\hbe^\pmu{]}B^{\hal\hbe}{}^{\pz}
\non\\
&&~~~~~~~
+4 \de_\hal^\pmu \de_\hbe^\pu  \mathscr{W} B^{\hal\hbe}{}^{\pmd}
-  4 \de_\hal^\pmu \de_\hbe^\pmu \mathscr{W} B^{\hal\hbe}{}^{\pz}
\Big\}
\non\\
\ri\D^{(4)}\de^{(-2)}\Omega(B)
&=&
-\frac{4\ri u_iu_j}{(v,u)^2}\D^{(4)}\Big\{
2\de_\hal^{(i} {[} \mathscr{W},\de_{\hbe k}{]}B^{\hal\hbe}{}^{j)k}
+\de_\hal^{(i} \de_{\hbe k}  \mathscr{W} B^{\hal\hbe}{}^{j)k}
\Big\}
~.~~~~~~~~
\label{various-eqs-6}
\eea
Now we use the explicit expression of $\mathscr{W}$ and obtain
\bsubeq
\bea
\mathscr{W} B^{\hal\hbe}{}^{ij}
&=&
2W^{[\hal}{}_{\hga}B^{\hbe]\hga}{}^{ij}
~,
\\
{[}\mathscr{W},\de_{\hbe k}{]} B^{\hal\hbe}{}^{jk}
&=& 
- W_{\hbe\hde} \nabla^{\hde}_k B^{\hal\hbe}{}^{jk}
- 5X_{\hbe k} B^{\hal\hbe}{}^{jk}
=
-\nabla^{\hga}_k W_{\hga\hbe}  B^{\hal\hbe}{}^{jk}
~.
\eea
\esubeq
Equation \eqref{various-eqs-6} then becomes
\bea
\ri\D^{(4)}\de^{(-2)}\Omega(B)
&=&
-\frac{8\ri u_iu_j}{(v,u)^2}\D^{(4)}\de_\hal^{i} \de_{\hbe k} W^{(\hal}{}_{\hga}B^{\hbe)\hga}{}^{jk}
\non\\
\ri\D^{(4)}\de^{(-2)}\Omega(B)
&=&
8\ri\D^{(4)}\de_{(\hal}^\pu\de_{\hbe)}^\pmu W^{(\hal}{}_{\hga}B^{\hbe)\hga}{}^{(-2)}
\equiv0
~.
\eea
This completes the proof that
the operator $\D^{(4)}\de^{(-2)}$
annihilates 
the superfield \eqref{E.1}.


\section{Prepotentials for  $\cO(4+n)$ multiplets, $n=0,1,\dots$, from harmonic superspace}
\label{AppendixH}

Here we consider an $\cO(4)$ multiplet 
$G^{(4)}(z,u) =G^{ijkl} (z)u^+_i u^+_j u^+_k u^+_l$
realized in 5D $\cN=1$ harmonic superspace, 
\bea
D^+_\hal G^{(4)} =0~, \qquad D^{++} G^{(4)} =0~.
\eea
It may be represented as 
\bea
G^{(4)} (u)= (\hat{D}^+)^4 V(u)~, 
\eea
where  
\bea
V(u) = V_0 +\sum_{n=1}^\infty V^{(i_1 \dots  i_{2n})} 
u^+_{i_1} \dots u^+_{i_n} u^-_{i_{n+1}} \dots u^-_{i_{2n}} 
\equiv V_0 + {\frak V} (u) 
\eea
obeys the equation
\bea
D^{++} V = D^{++} {\frak V}=D^{+\hal } \S^+_\hal ~,
\eea
for some spinor superfield $\S^+_\hal (u)$.
We note that $V(u)$ is defined modulo abelian gauge transformations
of the form:
\bea
 V \to \widetilde{V} = \widetilde{V}_0 + \widetilde{{\frak V}}:= V+D^{+\hal } \l^{-}_\hal~,
\eea
where $ \l^{-}_\hal (u) $ is arbitrary. We now consider the following harmonic equation
\bea
D^{++} \l^{-}_\hal = -\S^+_\hal~, 
\eea
with $\S^+_\hal$ given. This equation proves to have a unique solution $\l^-_\hal (u)$.
Upon applying the above gauge transformation, we obtain
\bea
D^{++} \widetilde{V} = 0 \quad \Longrightarrow \quad \widetilde{V} = \widetilde{V}_0~.
\eea 
As a result, the $\cO(4) $ multiplet can always be represented in the form
\bea
G^{(4)} (u)= (\hat{D}^+)^4 V~, 
\eea
with the prepotential $V$ being harmonic independent.

Given a non-negative integer $n=1,2,\dots$,
consider an $\cO(4+n)$ multiplet 
\bea
G^{(4+n)}(z,u) =G^{i_1 \dots i_{4+n} } (z)
u^+_{i_1} \dots u^+_{i_{4+n}}~,
\eea
realized in 5D $\cN=1$ harmonic superspace, 
\bea
D^+_\hal G^{(4+n)} =0~, \qquad D^{++} G^{(4+n)} =0~.
\eea
The superfield $G^{(4+n)}$ may be represented as 
\bea
G^{(4+n)} (u)= (\hat{D}^+)^4 V^{(n)} (u)~, 
\eea
where  
\bea
V^{(n)}(u) &=& V_{0}{}^{i_1 \dots i_n} u^+_{i_1} \dots u^+_{i_n}
 +\sum_{m=1}^\infty V^{(i_1 \dots  i_{n+2m})} 
u^+_{i_1} \dots u^+_{i_{n+m}} u^-_{i_{n+m+1}} \dots u^-_{i_{n+2m}} \non \\
& \equiv & V_0^{(n)}(u)   + {\frak V}^{(n)} (u) 
\eea
obeys the equation
\bea
D^{++}  V^{(n)}= D^{++} {\frak V}^{(n)}=D^{+\hal } \S^{(n+1)}_\hal ~,
\eea
for some harmonic superfield $ \S^{(n+1)}_\hal (u)$. By construction, 
the prepotential $V^{(n)} $ is defined modulo gauge transformations
\bea
V^{(n)}  \to \widetilde{V}^{(n)}  = \widetilde{V}_0^{(n)} +\widetilde{\frak V}^{(n)} 
:= V^{(n)} + D^{+\hal} \l^{(n-1)}_\hal~,
\eea 
for an arbitrary harmonic superfield $\l^{(n-1)}_\hal (u)$.
It is  possible to choose the gauge parameter $\l^{(n-1)}_\hal (u)$
to be a solution of the harmonic equation
\bea
D^{++} \l^{(n-1)}_\hal = - \S^{(n+1)}_\hal ~.
\eea
Such a solution always exists and is not unique for $n>0$. Upon applying such a
finite gauge transformation, we observe that the transformed prepotential 
$\widetilde{V}^{(n)} (u)$ is characterized by
\bea
D^{++} \widetilde{\frak V}^{(n)}  =0~.
\eea
We conclude that the $\cO(4+n)$ multiplet can be represented in the form:
\bea
G^{(4+n)} (u)= (\hat{D}^+)^4 V^{(n)} (u)~, \qquad
V^{(n)}(u) = V^{i_1 \dots i_n} u^+_{i_1} \dots u^+_{i_n}~.
\eea


\begin{footnotesize}

\end{footnotesize}

\end{document}